# Liquid $^4$He: contributions to first principles theory of quantized vortices, thermohydrodynamic properties, and the $\lambda$ transition

H.W. Jackson
6780 Calmbank Avenue
La Verne, California 91750
Telephone: (909) 596-1938

**Abstract**

Liquid $^4$He has been studied extensively for almost a century, but there are still a number of outstanding weak or missing links in our comprehension of it. This paper reviews some of the principal paths taken in previous research and then proceeds to fill gaps and create an integrated picture with more complete understanding through first principles treatment of a realistic model that starts with a microscopic, atomistic description of the liquid. Newly derived results for vortex cores and thermohydrodynamic properties for a two-fluid model are used to show that interacting quantized vortices may produce a $\lambda$ anomaly in specific heat near the superfluid transition where flow properties change. The nature of the order in the superfluid state is explained. Experimental support for new calculations is exhibited, and a unique specific heat experiment is proposed to test predictions of the theory. Relevance of the theory to modern research in cosmology, astrophysics, and Bose-Einstein condensates is discussed.

Contents

## 1. Introduction

Physicists have studied liquid $^4$He for almost a century, but our understanding of it is still incomplete, with a number of conspicuous weak or missing links. This paper reviews some of the principal paths taken in previous research and contains new results that help fill major gaps so that a coherent theory emerges. Relevance of this theory to topics in cosmology, astrophysics, and Bose-Einstein condensates is discussed in a context of research in the modern era.

The main objective of the paper is to explain how microscopic theory that accounts for the two-fluid model also accounts for properties of quantized vortices and the $\lambda$ transition. Experimental support for the theory is discussed and a unique specific heat experiment is proposed to test new predictions. An explanation of the physical mechanism for the $\lambda$ transition based on new calculations is presented. An *Overview* and *Historical Perspective* contained in the next two Sections explain our approach and motivation in more detail.

### *1.1 Overview*

A realistic model of liquid helium is treated in this paper by a first principles theory that is grounded in well-established quantum mechanical, correlated basis function methods and standard statistical mechanics. The goal is to better understand properties of quantized vortices and the $\lambda$ transition, and thermohydrodynamic properties that link them, through calculations based on a microscopic, atomistic description of the liquid.

New calculations are reported here for quantized vortex lines and circular vortex rings of arbitrary radius in liquid $^4$He at $T$ = 0 K and also at finite temperatures. The results indicate that no *large* vortex rings and no vortex lines should exist in He II, the superfluid phase, if the liquid is in *complete* thermodynamic equilibrium and the container walls are stationary. A first principles theory of the $\lambda$ transition for the liquid in complete thermodynamic equilibrium is developed and presented here. It is used to make quantitative predictions for some properties in a crossover region from smooth background to strongly critical. Lessons learned from treatment of the crossover region are applied in extending the theory to the strongly critical region, where most of the analysis is more qualitative.

The theory advances a detailed explanation of the physical mechanism of the $\lambda$ transition. There are two particularly notable features of the theory. One is that the theory shows how a singularity in the specific heat at $T_\lambda$ is linked to the disappearance of superfluidity. The other is the identification of the nature of the ordering that is present in He II and is absent in He I, the normal fluid phase.

As a possible means of validating the predictions, an experiment is proposed to produce a superfluid sample that is free of all large vortex rings and all vortex lines at some low temperature and then to measure the specific heat as the temperature increases and the liquid passes through the superfluid transition. Implications of possible outcomes of that experiment for understanding the λ transition are discussed in the context of our new theory.

Analysis in this paper argues for the idea that almost all existing measurements of specific heat in liquid $^4$He, some of which are cited in Sec. 7.4, are for metastable states where there are pinned line vortices, and it is explained how these vortices may significantly affect the results of those measurements near $T_\lambda$. At first sight this idea seems to conflict with other theories, such as those based on scaling or on a Landau-Ginzburg-Wilson Hamiltonian, that appear to be currently widely accepted. However, it is suggested in this paper that the new theory may be compatible with those other theories.

Properties of both stationary and flowing liquid $^4$He are calculated by applying statistical mechanics methods, elaborated in earlier work, to a model that accounts more accurately for interactions among rotons. Close agreement between theoretical and experimental results for entropy, specific heat and superfluid density up to temperatures somewhat below the presently observed λ temperature supports the credibility of our approach in treating quantized vortices at finite temperatures and the λ transition.

The calculated Helmholtz potential that includes flow velocities up to the Landau critical velocity is at the heart of our treatment of quantized vortices at finite temperatures. Consideration of thermodynamic stability, constraint on superfluid velocity by the postulated requirement of quantization of circulation, and a principle of constrained instability is a unique feature of our theory of the λ transition.

Results of calculations in this paper for large vortex rings are in close agreement with experimental observations by Rayfield and Reif and by Steingart and Glaberson, further supporting the credibility of our approach. Both stationary and moving large vortex rings are treated. Properties of small vortex rings are calculated for the first time and the results indicate that small vortex rings are distinct from rotons. Both rotons and vortex rings play major roles in the new theory of the λ transition. The true momentum of vortex rings, distinct from their impulse, is calculated using new methods.

The theory shows how the vortex model for liquid $^4$He, where the λ line is a line of critical points, generates pictures similar to those for an ordinary fluid near its isolated critical point where there are drops or bubbles.

Some features of He I are studied using the new theory. One striking feature is that if the liquid starts out free of vortex lines and large vortex rings below $T_\lambda$, then a catastrophic event is predicted in which large vortex rings suddenly appear above $T_\lambda$ as the temperature of a finite size sample increases.

Other topics treated in the paper include explanations, based on theoretical and experimental results, of the essence of superfluidity and of a visualizable model of rotons as density fluctuation excitations. Also, the physical interpretation of negative superfluid density is explained.

Methods and results of this work present interesting possibilities for applications in other fields of research, as described in the paper, including Bose-Einstein condensates, neutron stars, and cosmology.

Historical background outlined next provides an aid for viewing the new results in terms of a coherent picture based on a vast body of research on $^4$He and related physics that has spanned more than a century. This historical account tracks developments on several topics that merge into the picture, including superfluidity and vortices in $^4$He, quantum mechanical many-body theory, and physics of λ-type phase transitions.

### *1.2 Historical perspective*

Discovery of helium through its characteristic spectral lines in radiation from the sun in 1869 by Janssen and Lockyer [1] initiated an era of research that has fueled the imagination of physicists and challenged their resourcefulness and analytical abilities since that singular event. In 1895, $^4$He was found on earth by Ramsay [1] in gases expelled by pitchblend, and a race to liquefy helium was subsequently joined by the leading cryogenic laboratories of Europe [2]. The competition was won in 1908 by Kamerlingh Onnes [3] at Leiden. For many years thereafter, progress in exploring properties of liquid $^4$He was almost entirely experimental. Particularly significant milestones in that exploration were measurements of a λ-type behavior in specific heat near 2.2 K by Keesom and Clusius and by Keesom and Keesom [4-6] in 1932, and discovery by Kapitza and by Allen and Misener [7-9] of superfluidity, revealed in flow with almost no viscosity at low velocities in 1938.

On the theoretical front, an early attempt by Fröhlich [10] in 1937 at explaining the λ-phenomenon in liquid helium as an order-disorder transition was based on a crystalline model in which large zero point motion was invoked to account for the liquid behavior. This approach was discredited in 1938 by London [11,12], who then argued that a degenerate Bose-Einstein gas might serve as a useful model that accounts, at least crudely, for both the specific heat anomaly and superfluidity. In a brilliant display of physical insight, Tisza [13] almost immediately abstracted the idea of a two-fluid model for He II from London's ideas and accounted for a number of unusual thermohydrodynamic properties of liquid $^4$He, including prediction of temperature oscillations, which are now called second sound.

Landau [14,15] independently developed a two-fluid model, first published in 1941 that could also explain a number of thermohydrodynamic properties of He II, including second sound. In a remarkable feat, Landau [16,15] proposed the now-famous phonon-roton elementary excitation energy spectrum that bears his name. Calculations based on that spectrum accounted semi-quantitatively for the experimentally observed specific heat from low temperatures up to about 1.8 K, several tenths of a Kelvin below $T_\lambda$. Over the same temperature range Landau's theory also accounted semi-quantitatively for the normal fluid density and superfluid density that occur in the two-fluid model. Landau did not predict a phase transition with λ behavior in specific heat for any temperature. However, by an interpolation procedure Landau [14] did predict an end to superfluidity around 2.3 K, in fairly good agreement with the temperature of the λ point, which at that time had been established experimentally to be around 2.19 K.

A complete set of non-linear thermohydrodynamic equations for a two-fluid model was derived in 1950 by Zilsel [17,18], who used a phenomenological Lagrangian that he had constructed, and a variational principle that had been formulated by Eckart [19]. The two-fluid model of liquid $^4$He was derived from microscopic theory by Jackson [20,21] in the 1970's. The results produced by that microscopic theory are in agreement with results found by Zilsel, and they are described in more detail in the literature [20-23] and later in this paper.

In addition to its superfluidity, helium is unique in another fundamental way that demanded explanation. Helium remains liquid at pressures below about 25 atmospheres even at the lowest temperatures where observations have been made. Analysis based on a combination of theory and experiment suggest it will not solidify, but will remain liquid, in the limit of 0 K [24]. It was recognized that this exceptional behavior must be grounded in quantum mechanics and that the liquidity was likely associated with the weakness of the attractive part of the interatomic potential and the low atomic mass, both of which favor large zero-point motion. It was also recognized that a fundamental understanding of the phenomenological two-fluid models, such as proposed by Tisza and Landau, very likely requires quantum mechanical calculation of the energy spectrum.

Early attempts at quantum mechanical calculation of the energy spectrum of liquid $^4$He were aimed at qualitative understanding and were based on models that assumed weakly interacting particles, or low densities, or both. Bogoliubov [25] was one of the first to give a systematic quantum mechanical treatment of a weakly interacting, low density Bose fluid. In his approach, the problem is greatly simplified by the assumption that most of the atoms occupy a condensate of zero momentum orbitals for the liquid in its ground state and also for the liquid in any excited state. Bogoliubov derived formulas for the energy spectrum that seemed to be qualitatively consistent with the phonon-roton structure postulated by Landau, but the formulas could not be applied directly to liquid $^4$He.

Many other authors [26-28] treated the weakly interacting or low density Bose fluid by a variety of techniques. The results were all similar to those obtained by Bogoliubov. Improved formulas involving a condensate were derived by others [29-31]. However, in striving for self-consistency in the calculations with respect to the condensate fraction and the excitation spectrum while using interaction potentials qualitatively similar to those for helium atoms, rough agreement with the Landau spectrum degraded substantially, making this approach less useful for understanding liquid $^4$He.

Fortunately there was another approach to a quantum mechanical, microscopic treatment of liquid $^4$He that could treat a model with realistic interatomic potentials and atomic densities in the range that is observed experimentally. Feynman [32] led the way with this approach. Variational methods were used by Feynman to treat low excited states, and the calculations yielded results in rough agreement with the phonon-roton spectrum postulated by Landau [16].

Feynman and Cohen [33] used an improved class of trial functions and obtained moderately good agreement with the Landau spectrum in the roton region. Both of these variational calculations relied on input from experimental x-ray data for the liquid structure function at low temperatures in characterizing ground state properties that appeared in formulas for excited state energy levels. In each of those variational calculations, only single excitation states were treated. However, arguments involving wave packet states were given to justify application of the results to a multiple elementary excitation picture. The wave functions used by those authors could be represented by a ground state wave function multiplied by another factor for the specific state being considered. The ground state wave function factor is the major consideration that enables the treatment of realistic models of strongly interacting high-density systems like liquid $^4$He. Bijl [34] had previously arrived at a similar structure for excited state wave functions for weakly interacting, low-density systems, but in Bijl's theory an approximation to the ground state wave function occurs as a factor. The dispersion curve found by Feynman in his first variational treatment is now commonly called the Bijl-Feynman spectrum.

A generalization of the two-factor wave function structure, containing a correlation factor that is either the ground state wave function or an approximation to it, is the defining feature of what is now called the correlated basis function (CBF) formalism [20,35-38].

A multiple excitation model was developed by Jackson and Feenberg [36-39]. Starting with the space coordinate representation of the Hamiltonian and momentum operators, they evaluated the dominant contributions for matrix elements for an infinite set of CBF wave functions and subsequently diagonalized the Hamiltonian. The results they found included the Bijl-Feynman excitation spectrum for a realistic model of liquid $^4$He [36-38] and the Bogoliubov results for weakly interacting, low-density systems [38-39]. The formulas they derived [36-39] are also applicable to a broad range of particle interaction strengths and particle densities. Their formulas and methods could be adapted and used in model studies for Bose-Einstein condensates in ultracold gases of sodium and other elements that are of current interest [40-45,45a]. A unifying element in their work [36-39] that makes possible the simultaneous treatment of a broad range of physical systems is the generalized normalization integral. This integral was originally introduced by Iwamoto and Yamada [46] in nuclear physics.

The multiple excitation theory was also treated by Campbell [46a] and Campbell and Feenberg [46b]. Their approach used a generalized normalization integral too, but their procedure for evaluating matrix elements was different from that used by Jackson and Feenberg. Formulas for energy and momentum eigenvalues are the same in the two approaches. An important result in their work was an optimization procedure for calculating a correlation factor that is an approximation to the ground state wave function. Use of the optimized correlation factor greatly simplified the theory. Their optimization procedure so far has been applied to helium liquids [46a,46b]. It should also be useful in treating Bose-Einstein condensates in ultracold gases as the experimental conditions reach the range of higher atomic densities and interaction strengths where Bogoliubov's approximate theory loses accuracy.

CBF wave functions were used in calculating corrections to the Bijl-Feynman spectrum. Brillouin-Wigner perturbation theory calculations were carried out through second order by Jackson and Feenberg [47] and through third order by Lee and Lee [48]. Campbell and Pinski [49] used an optimized version of a Bijl-Dingle-Jastrow (BDJ) type correlation factor [50] in recalculating the three CBF approximations to the phonon-roton spectrum, so that results could more meaningfully be compared with each other and with experimental results from inelastic neutron scattering. These dispersion curves are discussed in the body of this paper. They indicate the promise that CBF formalism holds for quantitatively understanding liquid $^4$He starting from microscopic theory.

One strength of the CBF approach is that in some calculations the ground state and excited states can be dealt with separately. Another strength is that the CBF approach allows experimental results to be used in a simple way to simplify the theory and improve accuracy of calculated properties. In fact, in this paper we will take advantage of these features in some instances. For a complete microscopic theory, ground state properties must be calculated. Variational treatments of the ground state properties for BDJ [37] functions that can be applied directly in CBF theory have been carried out by several investigators [46a,51-54] using a variety of procedures.

Next we shall outline additional paths of research that progressed mainly in parallel with each other and which may be merged to help form a more complete understanding of prominent properties of liquid $^4$He.

In 1949 Onsager [55] proposed that hydrodynamic circulation, $\oint \boldsymbol{v}_s \bullet d\boldsymbol{l}$, would be quantized in integral multiples of $h/m$, where $\boldsymbol{v}_s$ is the superfluid velocity, $h$ is Planck's constant and $m$ is the mass of a $^4$He atom. Onsager further pointed out that a consequence of this condition is that the distribution of vorticity is discrete rather than continuous, whenever it occurs in liquid $^4$He. In 1955 Feynman [56] independently came to similar conclusions, and further indicated how a realistic model of quantized vortices in liquid $^4$He may be treated quantum mechanically. Variational calculations for realistic models of rectilinear vortices at $T = 0$ K that are consistent with Feynman's ideas were reported by Chester, Metz, and Reatto [57] in 1968, by Jackson [20] in 1979, and by Vitiello, Reatto, Chester, and Kalos [58] in 1996. When the technique described by Jackson [20] is applied to a class of trial functions described later in this paper, the results for vortex energy found in the three treatments are in close agreement with each other despite the fact that the calculation procedures used were different.

Experimental evidence for quantized vortex lines was first found by Hall and Vinen [59], who used second sound as a probe. Rayfield and Reif [60] found convincing evidence for quantized vortex rings created by fast ions and studied their properties with the aid of a time of flight spectrometer. Images of discrete rectilinear vortex lines in rotating liquid $^4$He at temperatures near $T = 0$ K were made by Williams and Packard [61,62], thereby indisputably establishing the reality of quantized vortex lines.

At finite temperatures the core of a quantized vortex is surrounded by a mantle where the number density of rotons is very high. Vinen [63] recognized this early in his investigations. Glaberson, Strayer, and Donnelly [64] made calculations based on an elementary excitation model that supported this picture. Calculations by Jackson [20] based on free energy considerations supported this picture. The present paper extends that work and shows how the temperature dependence of that mantle plays a major role in accounting for the λ transition in liquid $^4$He.

Critical behavior in λ transitions that occur along the boundary line that separates He I and He II on a pressure-temperature phase diagram has been studied extensively in the past, and the λ transition is one of the main subjects treated later in this paper. In order to view this research in a larger framework, we recount next a brief history of λ transitions in a broad context, and along the way indicate why these transitions are of widespread interest to physicists.

Criticality in fluids was discovered first by Cagniard de la Tour [65] in 1822 during experiments on alcohol contained in the barrel of a canon. In this case, critical behavior in a λ transition was associated with liquid-gas equilibrium. Cagniard de la Tour's experiments laid the foundation for further study of λ transitions, where specific heat has a characteristic shape similar to the Greek letter λ. A chain of experiments involving many researchers [66] elucidated implications of de la Tour's work and resulted in creation of the field of cryogenics and eventually led to the experiment in which Onnes [67] liquefied $^4$He in 1908.

The λ-type specific heat anomaly has been observed in many substances besides liquid helium, e.g., in fluids near liquid-gas critical points, in ferromagnetic materials at their Curie points, in certain binary alloys ($\beta$-brass, for example), and in methane, ammonium chloride, and hydrochloric acid [68]. There is a widely held view that λ transitions are "order-disorder" phenomena, and that understanding them for one system will aid in understanding others, and this has led to widespread interest in them [68].

Among all the substances in which this type anomaly has been observed, liquid $^4$He has been the most extensively studied experimentally, partly because the greatest accuracy can be achieved there. Since the first measurements showing the λ specific heat anomaly in liquid $^4$He were carried out by Keesom and Clusius [4] and by Keesom and Keesom [5] in 1932, major milestone measurements were reported by Buckingham and Fairbank [69] in 1961 and by Lipa, Swanson, Nissen, Chui and Israelsson in 1996 [70,70a]. The latter authors conducted their experiment in space to reduce effects of gravity. And they advanced the frontiers of thermometry by using high-resolution thermometers that could measure temperature changes as small as a nanokelvin.

On the theoretical front in understanding λ transitions, in 1872 van der Waal's [71,72] gave the first widely accepted explanation of critical points connected with liquid-gas equilibrium with the aid of an approximate equation of state formula that now bears his name. A huge body of theoretical research devoted to critical behavior in λ transitions has been developed since van der Waal's original work. A towering achievement in this field is Onsager's [73] exact solution of the Ising model for a two-dimensional system of spins. The present understanding [72,74] of λ transitions is based largely on phenomenological theories, highly idealized models that can be solved exactly, and idealized but more realistic models that so far have been solved only approximately. The approach to understanding λ transitions introduced in the body of this paper, where first principles theory is applied to a realistic, microscopic model of liquid $^4$He, may initiate a new path for studying critical phenomena in a general context and complement existing methods.

The historical perspective that we have sketched so far is obviously of limited scope. It establishes a basis for showing how the new work in this paper meshes with an existing vast amount of research on liquid $^4$He, especially superfluidity, the two-fluid model, quantized vortices, and the λ transition. Of necessity, much important research has not been included in our review. However, because of the great interest and massive research effort that has been focused on Bose-Einstein condensation in liquid $^4$He, we will briefly discuss this subject. Many researchers believe that Bose-Einstein condensation is very likely a key to understanding superfluidity and the λ transition in liquid $^4$He. We wish to emphasize that our entire theory, including the vortex model of the λ transition, is applicable whether a Bose-Einstein condensate is absent or present in liquid $^4$He. On the other hand, explanations that rely on Bose-Einstein condensation cannot be ruled out as alternatives to our theory by existing results of research. We do not know of any existing work by others that starts with a realistic microscopic theory and attempts to explain both the two-fluid model for superfluidity and the λ transition with the aid of a Bose-Einstein condensate. However, it should be noted that Hohenberg and Martin [74a] have developed a microscopic theory of superfluid $^4$He that involved a condensate, but they did not treat the λ transition. That theory differs significantly from ours in its assumptions and method. Also, it should be noted that some existing theories of the λ transition assume the existence of a condensate [74b,74c]. With these qualifications before us, we will now summarize some of the work on Bose-Einstein condensation in liquid $^4$He.

Reasoning from an analogy with an ideal Bose-Einstein gas, London [11,74d] suggested that order due to condensation in momentum space, associated with a Bose-Einstein condensate, may be responsible for superfluidity and the λ transition in liquid $^4$He. Penrose [74e], and Penrose and Onsager [74f], formulated this idea of ordering in terms of a Fourier transform of the one-particle density matrix $\rho_1$ in the space coordinate representation for models with realistic atomic densities and interactions in liquid $^4$He. Yang [74g] has called this ordering concept *off-diagonal long-range order* (ODLRO). Penrose and Onsager [74f] estimated that the fraction of $^4$He atoms in the condensate for the liquid ground state is 8%. McMillan [50] performed Monte Carlo calculations based on a BDJ approximation to the ground state wave function and found a condensate fraction of 11%.

In 1966, Hohenberg and Platzman [74h] showed that scattering of fast neutrons from liquid $^4$He may yield direct experimental evidence for a Bose-Einstein condensate. They argued that in such scattering the atom struck by the neutron would obey essentially billiard ball physics, and the dynamic structure function $S(Q,\omega)$ would exhibit a narrow peak, due to the condensate, superimposed on a broader peak, due to the non-condensed atoms. In the limit of high momentum transfer, where $Q$ approaches infinity, the impulse approximation (IA) for $S(Q,\omega)$ is applicable and the narrow peak is a delta function centered at the free particle recoil energy $\hbar^2 Q^2/2m$, corresponding to $n(p)$ at $p$ = 0 in the momentum distribution for $^4$He atoms of mass $m$. The coefficient of the delta function is proportional to the fraction of atoms in the condensate. At finite $Q$, the narrow peak will be somewhat broadened by final state effects (FSE) associated with scattering of the struck atom by other atoms in the liquid. The paper by Hohenberg and Platzman [74h] launched an era of extensive experimental [74i-74q] and theoretical [53,75,75a-75m] studies in pursuit of validation of the concept of Bose-Einstein condensate in liquid $^4$He.

The two-part structure in $S(Q,\omega)$ predicted to occur if a condensate is present has never shown up distinctly in any of the experiments carried out so far. Final state effects, complications associated with instrumental resolution in time of flight spectrometers, and statistical uncertainty associated with weak incident neutron flux at high momentum in the case of reactor-based experiments are among the main mechanisms that are believed to contribute to masking evidence for a narrow peak signaling the presence of a condensate.

A number of measures have been taken to deal with these difficulties. Detailed theories of FSE have been developed by Gersch and his collaborators [75e-75g] and by Silver [75j]. Spallation sources [74n,74o] have been used to obtain higher incident neutron fluxes at high momentum. Sophisticated time-of-flight spectrometer techniques have been applied [74o] to help reduce statistical uncertainty. The theories of FSE [75e-75g,75j] as well as sum rules have been applied in analyzing experimental observations. Theoretical results for momentum distributions of $^4$He atoms, including the condensate fraction, have been calculated using various Monte Carlo techniques [50,75,75a,75b,75l] and other methods [75c]. These results have been combined with the IA and FSE theoretical results to construct forms for $S(Q,\omega)$ that have been compared with experiment. Sokol, Silver, and Clark [75k] have given a critical discussion of approximations and techniques used in the Monte Carlo calculations.

Application of these and other theoretical techniques in analyzing experimental data have yielded estimates for the condensate fraction that fall in two ranges. Estimates of the condensate fraction based on analysis of experiments using reactor neutrons and a triple axis crystal spectrometer have usually been less than 3% near $T = 1.2$ K and $SVP$ [74j,75e,75f,75h,75i]. Estimates of the condensate fraction based on neutrons from spallation sources and time-of-flight spectrometers have usually fallen in the range 8% - 11% in the vicinity of 1 K and below, and at *SVP* [74n,75o,75l,75m]. The actual scattering data from both types of experiment are consistent with the absence of a condensate [74o, 75h,75m].

Finally, to end this review, it should be noted that the idea that quantized vortices may account for the λ transition in liquid $^4$He has a long history that dates back to early work of Onsager [55] and Feynman [56]. Others, including Byckling [75n], Popov [75o], and Williams [75p] have also studied this idea.

### *1.3 Resources for supplemental information*

A number of resources have been of special help to the author in the research reported here, and may also be useful in supplying background information to the reader. Feenberg's [37] book contains detailed applications of correlated basis function methods to realistic models of $^4$He and $^3$He liquids and to idealized models. London's [6] book contains descriptions of some early experiments and also his own work and profound insights on liquid helium. Donnelly's [76] book and Fetter's [76a] article provide clear expositions of elements of hydrodynamics, including classical and semiclassical treatments of vortices, and overviews of experiments and treatments of vortices based on models for weakly interacting particles. Callen's [77] book contains treatments of thermodynamic stability and higher order phase transitions that include λ transitions. Books by Stanley [72] and by Binney, Dowrick, Fisher, and Newman [74] contain useful expositions of theory of phase transitions and critical phenomena, including treatment of a broad range of model systems and methods of calculation.

## 2. Elements of the new theory

The theory described here applies the quantum mechanical correlated basis function formalism to a system of $N$ strongly interacting $^4$He atoms at experimentally observed densities. In the space coordinate representation for atoms, Hamiltonian and momentum operators can be written as

$$\hat{H} = -\frac{\hbar^2}{2m}\sum_i \Delta_i + \frac{1}{2}\sum_{i \neq j} V\left(r_{ij}\right) \tag{2.1}$$

$$\hat{\boldsymbol{P}} = \frac{\hbar}{i}\sum_i \nabla_i \ . \tag{2.2}$$

The indices $i$ and $j$ range over all atoms in the liquid. $V(r)$ may be a Lennard-Jones pair potential with strongly repulsive behavior at small separation distance $r$, for example. Also, $m$ is the mass of a $^4$He atom. Furthermore, it is assumed that all wave functions satisfy boundary conditions that are required for $\hat{H}$ and $\hat{\boldsymbol{P}}$ to be Hermitian, e.g., cyclic or vanishing wave functions at wall boundaries.

The essential property of all CBF wave functions is that they can be expressed in the form

$$\psi = F\psi_0 \ . \tag{2.3}$$

The correlation function $\psi_0$ is either the ground state wave function for the liquid or an approximation to it. A particularly useful approximate form for $\psi_0$ is a Bijl-Dingle-Jastrow (BDJ) [78] function consisting of a product over all distinct pairs of pair factors $\exp \frac{1}{2}U\left(r_{ij}\right)$for the liquid at rest. The model function $F$ is specific to the quantum state being described. Some explicit forms of $F$ are in the literature [20,36-38] and other forms are given later in this paper.

For low excited states, approximate forms of energy and momentum eigenstate wave functions and eigenvalues have been calculated and the results fit into a picture of non-interacting elementary excitations [36-38]. Calculations of the Landau spectrum were refined over a long period of time by many workers [32,33,36-38,48,49] and the best theoretical results are in close agreement with neutron scattering measurements [79]. Theoretical and experimental curves in Fig. 1 reproduced from a paper by Campbell and Pinski [49] display some of the milestone results achieved in progress toward the close agreement alluded to above.

High-excited states require inclusion of interactions among excitations, and also imposition of an upper bound on the number of elementary excitations [20,32]. Methods for incorporating these effects have been developed in earlier work [20] and are applied later in this paper. It is assumed that Eq. (2.3) still holds when excitations interact, and the energy $E$ and momentum $\mathbf{P}$ eigenvalues for the liquid containing elementary excitations (rotons and phonons) are as follows when the superfluid is at rest:

$$E = E_0 + \sum_i n_i \varepsilon_i + \frac{1}{2N} \sum_{i,j} f_{ij} n_i n_j \tag{2.4}$$

$$\boldsymbol{P} = \sum_i n_i \boldsymbol{p}_i \ . \tag{2.5}$$

The indices $i$ and $j$ range over all elementary excitation states.

For definiteness, we now assume $\psi_0$ to be the exact ground state wave function for the liquid at rest. Then

$$\begin{aligned} \hat{H}\psi_0 &= E_0\psi_0 \quad , \quad \hat{H}\psi = E\psi \\ \hat{\boldsymbol{P}}\psi_0 &= 0 \quad , \quad \hat{\boldsymbol{P}}\psi = \boldsymbol{P}\,\psi \ . \end{aligned} \tag{2.6}$$

The basis set in the CBF formalism has been extended in earlier work to include uniform superfluid flow where the correlation factor may be viewed as moving and the wave functions are in the form

$$\psi' = F\psi_0' = F\exp\left[i\frac{\boldsymbol{P}_0\bullet R}{\hbar}\right]\psi_0 \ . \tag{2.7}$$

$\boldsymbol{P}_0$ is called the correlated momentum and $\boldsymbol{R}$ is the center of mass coordinate for the atoms in the liquid: $\boldsymbol{R} = (1/N)\sum_i \boldsymbol{r}_i$ . The theory [20,22,23] identifies the superfluid velocity $\boldsymbol{v}_s$ as

$$\boldsymbol{v}_s = \frac{\boldsymbol{P}_0}{M} \ , \tag{2.8}$$

where $M = Nm$ is the mass of the liquid. Using these equations one can readily derive formulas for energy and momentum eigenvalues applicable to the model of uniform superfluid flow and interacting elementary excitations. These formulas are given in Eqs. (4.1) and (4.2) of this paper.

Statistical mechanical properties for this model can be deduced with the aid of the microcanonical entropy formula

$$S = k\sum_i\left[\left(1+n_i\right)\ln\left(1+n_i\right) - n_i\ln n_i\right] \ , \tag{2.9}$$

where $i$ is an index for elementary excitation states and $k$ is the Boltzmann constant. This program has been carried out in earlier work [20], where formulas for internal energy and other thermodynamic potentials were derived as well as formulas for their various partial derivatives.

A certain Legendre transform of the internal energy was identified [20,21] as the Lagrangian thermodynamic potential. That potential, assumed to hold also for local thermodynamic equilibrium, provided a means for applying Eckart's [19] variational principle and deriving the entire set of non-linear equations for the two-fluid model, including equations of motion for the superfluid and normal fluid, and conservation laws for energy and momentum [17,20,21,80]. Furthermore, that approach produced formulas, in terms of elementary excitation properties, for all quantities that appear in those relations, such as pressure, chemical potentials, superfluid and normal fluid densities and entropy density [20,21]. That approach provided a microscopic foundation for phenomenological models developed earlier by Zilsel [17,80] and independently by Lhuillier, Francois, and Karatchentzeff (LFK) [81]. In addition, that approach resolved objections to Zilsel's treatment that had been raised by Temperley [82] and independently by Dingle [83], and seemingly inhibited its general acceptance. Also, results based on the microscopic approach demonstrated [21] that the phenomenological models of Zilsel and LFK were equivalent, which was previously believed not to be the case [81]. Since local thermodynamic equilibrium is assumed in all of these approaches, the term macroscopic equations is sometimes associated with the two-fluid model, although the model is frequently applied on very small spatial scales, in some cases near atomic dimensions. We shall sometimes apply it on small scales in this paper also.

## 2.1 *Understanding superfluidity in liquid ${}^4$He*

Vorticity $\boldsymbol{\omega}$ in the superfluid velocity field $\boldsymbol{v}_s$, given by

$$\boldsymbol{\omega} = \nabla \times \boldsymbol{v}_s \ , \tag{2.10}$$

plays an important role in understanding superfluidity in ${}^4$He, including why the superfluid component can sometimes move without drag or impediment through tubes or in certain other situations. Much has already been written about this subject, but its foundations have usually been considered by careful authors to be speculative to some extent. The following discussion aims to make the foundations more secure.

In a phenomenological treatment of liquid ${}^4$He, Zilsel [17,80] derived the result

$$\nabla \times \boldsymbol{v}_s = 0 \tag{2.11}$$

along with equations of motion and conservation equations in a two fluid model. It is Eq. (2.11) that we will focus on now. In earlier work [20,21], Jackson showed that a microscopic theory of liquid ${}^4$He, where only Landau excitations (phonons and rotons) were considered, produced a two-fluid model in agreement with Zilsel's [17,80] results, including Eq. (2.11). The microscopic theory resolved objections [82,83] to some intermediate steps in Zilsel's treatment, a matter alluded to earlier. Thus the microscopic theory established a more secure foundation for Zilsel's results, and especially for Eq. (2.11).

London [84] had recognized that Eq. (2.11) could explain the absence of a shearing viscosity term $\eta_s \nabla \times \nabla \times \boldsymbol{v}_s$. London's observation implied that a continuous drag on the liquid due to viscosity near a solid wall, e.g., in a boundary layer, would be absent, and this absence is a necessary, but not sufficient, condition for superfluidity. We note here that in the derivation of the two-fluid model from microscopic theory as well as in Zilsel's treatment, in the normal fluid, $\nabla \times \boldsymbol{v}_n \neq 0$ in general even when the viscosity $\eta_n$ = 0. The property that normal fluid transports entropy but superfluid does not is the feature that accounts for this difference in behavior with respect to vorticity. Returning to London, we note that he was able to deduce from Eq. (2.11) much more about superfluidity than just the absence of a boundary layer. Paraphrasing the words of London [84], if $\nabla \times \boldsymbol{v}_s = 0$ holds, the absence of the shearing viscosity term $\eta_s \nabla \times \nabla \times \boldsymbol{v}_s$ does not depend on the absence of interactions as a result of $\eta_s$ = 0, an incredible result, but rather the term is absent because $\nabla \times \boldsymbol{v}_s = 0$.

In London's words: "The mere absence of states into which the superflow could be tempted to dissipate would then be the guarantee for the stability of the superflow. This would be very similar to the stability of an isolated quantum state. In superfluid helium a kind of *macroscopic quantum state* describing the state of the whole superfluid proper would appear as completely determined by its macroscopic boundary conditions."

The following discussion can usefully supplement London's ideas on superfluidity. For simplicity we consider only the case where normal fluid velocity $v_n$ = 0. In any microscopic theory, including the CBF theory that starts with the Schrödinger equation, boundary conditions must be imposed on all allowed wave functions to guarantee that the Hamiltonian and momentum operators are Hermitian. For example, at a solid wall one commonly requires all wave functions to vanish at the boundary. This frequently has led to calculation of a healing length where the atomic density decreases to zero in the liquid as the wall is approached. However, experimental evidence indicates that the first one or two layers of $^4$He are at such high density that they would be solid, because they are compacted there due to van der Waal's attraction with the wall material. So if the system of helium atoms is macroscopically at rest, the wave function must be zero at the interface of *solid* $^4$*He* with the wall, but the atomic density must then rise to some high level characteristic of the solid $^4$He phase. As distance from the wall increases, the atomic density must then *decrease* to that of the bulk liquid, and *not increase* as in usual healing length models. It seems extremely unlikely that atomic density falls to zero at the interface between liquid and solid $^4$He.

A microscopic theory of superfluidity [20-23] consistent with London's macroscopic quantum state can be constructed if the $^4$He atoms in the solid layers are immobilized as far as macroscopic motion is concerned, and if the atomic density is non-vanishing and essentially equal to that of bulk liquid next to the solid layers. Cyclic boundary conditions may then be used with box normalization to treat the ground state $\psi_0$ *for the liquid* at rest.

So far we have described the $^4$He system that is macroscopically at rest. Now suppose the superfluid (think of the flowing ground state wave function as a correlation factor in CBF theory) is caused to flow parallel to the wall with velocity $\boldsymbol{v}_s$. If the helium atoms in the liquid ground state next to the solid helium layers are partly restrained due to atomic interactions or if there is momentum transferred between the liquid layers moving at different velocities, like what is believed to happen in the boundary layer of ordinary liquids, then this will induce motion that violates the condition $\nabla \times \boldsymbol{v}_s = 0$ in Eq. (2.11). The condition $\nabla \times \boldsymbol{v}_s = 0$ can be explained if one considers that the superfluid acts like a large molecule and that energy cannot be transferred to it by the solid helium layers and the wall continuously, atom by atom, but only in steps that correspond to excitation of the superfluid as a whole. At the Landau critical velocity $v_c$ the flowing ground state becomes degenerate in the sense that the ground state flowing at velocity $v_s$ possesses the same energy as the state of flow at velocity $v_s$ where a roton is present. At any velocity where $v_s \geq v_c$, motion of the ground state, and in turn the correlation factor in CBF theory, no longer represents "adiabatic transformations of a single non-degenerate quantum state," which is London's [85] criterion for superfluid flow.

It is well known that the Landau [14,16] critical velocity $v_c$ is much higher than the upper limit for pure superfluid flow in most situations in He II. However, London's [85] criterion for superfluidity is more general than Landau's, as explained below. However, they coincide when applied to Landau excitations alone. This is where Onsager's [55] postulate of quantization of circulation and the related hypothesis of Onsager and Feynman [56] of quantized vortices are important for understanding superfluidity, as illustrated in the following example.

Experiments show that for He II flow through tubes, some threshold velocity $V_c$ exists such that only at lower velocities is superfluidity present in its pure sense, i.e., with no dissipation. Critical velocity in orifices has been an extensively studied subject, and the work of Vijfeijken, Wahlraven, and Staas [86] provides one of the most careful treatments. I will draw from their work in the following discussion. They considered a quantized vortex ring in an infinitely long tube having a circular cross section and calculated the energy of the ring taking into account its interaction with the tube wall, which is equivalent to taking into account the image vortex ring. (Viscosity plays no role here.) Then they proposed and studied a condition for $V_c$ that was determined by mechanical stability. Here I depart from their analysis and propose that the criterion for breakdown of pure superfluid flow at $V_c$ occurs when the lowest energy vortex ring inside the tube, taking into account its image, in a *background* uniform flow field $v_s$ becomes degenerate with the state having uniform velocity $v_s$ throughout the tube. I will call this the London critical velocity $V_c$ for flow through a tube because it is determined by London's [85] criterion for superfluidity mentioned earlier. In the CBF description of the liquid, the macroscopic quantum state is represented by the correlation factor with flow.

The explanation of superfluidity presented here, where it is specified that $v_n = 0$, can be somewhat generalized and summarized as follows. The atoms in solid layers of $^4$He next to a boundary always remain macroscopically at rest with respect to the walls. In any macroscopic motion parallel to the solid layers, when the parallel velocity component is less than a critical velocity $V_c$, the liquid layers in the ground state move without being impeded by interactions with the solid layers or with the wall itself. Below $V_c$ there is pure superfluid flow, where energy is not even partly transferred from macroscopic translational motion into internal motion of the liquid. $V_c$ is determined by a condition for degeneracy of the flowing ground state wave function as a correlation factor in a CBF description of the liquid. This correlation factor represents London's macroscopic non-degenerate wave function. Degeneracy is associated with quantized vortices and their images in the walls. The critical velocity considered here is not temperature dependent since it is a property of the correlation factor. However, $V_c$ is pressure dependent because the correlation factor is pressure dependent. It should be mentioned that very near the $\lambda$ point, a temperature dependent critical velocity is expected to occur that has a smaller value than the $V_c$ we considered above. That temperature dependent critical velocity $v_0$ is associated with *stability properties* of the liquid in which there is local thermodynamic equilibrium. That velocity $v_0$ will be discussed in later parts of this paper.

Putterman [87] has discussed the seeming conflict between the boundary condition that the macroscopic wave function vanishes at boundaries and the Landau two-fluid equations. The explanation we have proposed for superfluidity in $^4$He, where the correlation factor is identified as the macroscopic wave function and the *density of the liquid* at the interface with solid $^4$He is essentially the same as for the bulk, may resolve that apparent conflict.

### 2.2 *Comments on stability and Helmholtz potential*

Now we turn to discussion of other results that follow from the CBF based theory. One significant result is recognition that $v_s$ is not an intrinsically intensive variable, like pressure, but rather an intensive variable, like density, formed by the ratio of two extensive variables (see Eq. (2.8)). This character of $v_s$ is important in determining stability properties of the liquid and in understanding small oscillations such as second sound near vortex cores and near the superfluid transition temperature.

The Helmholtz potential for the flowing liquid is another Legendre transform of the internal energy. It has been evaluated and studied for certain models in earlier work [20], and it is treated by an improved model later in this paper. That Helmholtz potential plays an important part in determining properties of quantized vortices at finite temperatures, as will become clear in Secs. 5 - 7.

## 3. New developments in the theory

In this Section, energy, core structure, momentum, and velocity of quantized vortices at $T$ = 0 K are calculated using simple, realistic models for vortex cores and first principles theory.

### 3.1 *Quantized vortices*

Properties of a vortex core are determined in this theory by a variational calculation in which the expectation value of $\hat{H} - E_0$ is minimized with respect to a class of parameterized CBF trial wave functions. All wave functions in the class are *assumed* to be orthogonal to $\psi_0$. The superfluid velocity $\boldsymbol{v}_s$ satisfies the condition $\oint \boldsymbol{v}_s \bullet d\boldsymbol{l} = n\,h/m$ where $n$ is a non-negative integer in accordance with Onsager's [5] quantization of circulation hypothesis. The atomic density for each function in the class is zero at the ideally thin line where the velocity is singular, and joins smoothly at the core boundary with the assumed constant density outside the core. The assumption of constant density outside the core has been studied and it is expected to introduce errors that are small but not quite negligible for the calculations in this paper. Estimates for $T$ = 0 K based on isothermal compressibility and the equation of motion for superfluid $^4$He under steady state conditions show that the mass density is about 2 or 3% less at the core boundary than at points far from the core. Each of the trial functions must decrease to zero fast enough as the singularity is approached to insure that the expectation value of the Hamiltonian converges. The variational calculation just gives an upper limit on energy, but if the class of trial functions is judiciously chosen, core energy and other properties determined by the minimization procedure could be quite close to actual physical values.

In our variational calculations a vortex core boundary is fixed by the condition that the superfluid velocity $v_s$ attains the Landau critical velocity $v_c$ there. Then outside the boundary the flowing correlation function is non-degenerate and the superfluid flows without dissipation. It is postulated in this theory that inside the core the correlation function varies so rapidly, in phase or magnitude or sterically, that degeneracy is avoided there also.

The difficult problem of dealing with a transition region where, say, a roton is half inside and half outside the core is circumvented by assumptions made in this theory. If details of that transition region are indeed important, then the theory may need improvement. The model of the core is postulated to hold for both $T$ = 0 K and finite $T$; consequently we are assuming there are no rotons or phonons inside the core at any temperature. Subject to the validity of our postulate, the model of the vortices and the entire treatment of liquid $^4$He described here is consistent with London's idea that superfluid flow is associated with an adiabatic transformation of a single non-degenerate quantum state. In addition to being compatible with that idea of London's, our postulate for the absence of rotons and phonons inside the vortex core having a boundary fixed by $v_s = v_c$ finds support in agreement between predictions of the theory with experimental measurements on large vortex rings by Rayfield and Reif [60] and by Steingart and Glaberson [88,89]. That agreement is demonstrated in Section 3.3.8 of this paper.

### *3.2 Rectilinear vortex*

The basic relations used in our variational calculation for a rectilinear vortex are contained in earlier work [20], and so only a few details will be given here. Consider the class of CBF trial functions, Eq. (2.3), where the model function is

$$F = \prod_j e^{i\theta_j} P(\rho_j) \ , \tag{3.1}$$

$j$ ranges over all the atoms, and $(\rho,\theta,z)$ are cylindrical coordinates. The atomic density $n$ and a function $P_0$ inside the core are specified by

$$\frac{n(\rho)}{n_0} = \left[\sin\frac{\pi\rho}{2r_0}\right]^{2N} \tag{3.2}$$

and

$$P_0(\rho) = \left(\frac{n(\rho)}{n_0}\right)^{\frac{1}{2}} \ , \tag{3.3}$$

where $r_0$ is the radius of the core. Also, $n_0$ and $P_0$ are constant outside the core. Quantum mechanics determines the connection between $P(\rho)$ and $n(\rho)$ through an integral equation given below. $P_0(\rho)$ is a zeroth order approximation to $P(\rho)$. The parameter $N$ is varied to determine the minimum value of energy $\varepsilon$ described below. For most of this paper we use the zero pressure values for average density $n_0$ = 0.0218 atom Å$^{-3}$, Landau critical velocity $v_c \simeq 58.0$ ms$^{-1}$, and core radius $r_0$ = 2.80Å. The atomic density in and near the core is shown in Fig. 2 for several values of *N*. There it can be seen that the core density profile can be varied from an almost filled core to an almost empty core as *N* increases.

The excitation energy $\varepsilon$ is given by

$$\varepsilon = \frac{\langle \psi | \hat{H} - E_0 | \psi \rangle}{\langle \psi | \psi \rangle} = \frac{\hbar^2}{2m} \int n(\rho_1) \left[ \left( \frac{P'}{P} \right)^2 + \frac{1}{\rho_1^2} \right] d^3 r_1 \tag{3.4}$$

and the functions $P(\rho)$ and $n(\rho)$ are related by

$$\frac{d}{d\rho_1} \ln P^2(\rho_1) = \frac{d}{d\rho_1} \ln n(\rho_1) - \int d^3 r_2 \left[ n(\rho_2) - n_0 \right] g(r_{12}) \frac{\partial U(r_{12})}{\partial \rho_1} \tag{3.5}$$

as shown in Ref. [20]. The approximate forms for the pair distribution function $g(r)$ and the BDJ trial function exponent $U(r)$ used here are the same as in Ref. [20].

Numerical calculations based on these equations yielded the values for vortex core energy per unit length for several values of *N*, shown in Fig. 3. The lines are just a guide for the eye. The method we used to arrive at the core energy will be explained shortly.

The dressed vortex line core energy $\varepsilon_C$ in Fig. 3 was calculated with $P(\rho)$, and the bare vortex line core energy $\varepsilon_0$ was calculated with $P_0(\rho)$. The difference between $\varepsilon_C$ and $\varepsilon_0$ is due to effects of atomic correlations associated with $g(r)$. The minimum dressed vortex line energy occurs at $N$ = 0.50, where $\varepsilon_C$ = 1.68 K Å$^{-1}$. Clearly, these may be only approximate values for the minimum because we used a discrete set of *N* values.

Radial functions are shown in Fig. 4 for $N$ = 0.50. The figure shows that $P(\rho)$ extends beyond $r_0$. Effects of this "halo" are included in the values of the core energy $\varepsilon_C$ displayed in Fig. 3. One advantage of treating the "halo" in this way is that for any normalization radius $r_N$, the energy per unit length of vortex line can be evaluated by adding $\varepsilon_C$ to the kinetic energy of liquid having uniform mass density in the region between $r_0$ and $r_N$. For $r_N = 6.0$ Å, the vortex line energy is 2.31 K Å$^{-1}$. The corresponding energy found by Chester, Metz, and Reatto [57] using a rather different variational approach was about 2.375 K Å$^{-1}$ for $r_N = 6.0$ Å. Vitiello, Reatto, Chester, and Kalos [58] performed a Monte Carlo variational calculation and found 2.25 K Å$^{-1}$ for $r_N = 6.0$ Å. The close agreement of results obtained by these three rather different approaches promotes confidence in the calculated value for energy.

It is important to note that using our model, one can show that an empty core corresponds to the limit $N \to \infty$ and $\varepsilon_C \to \infty$, a result alluded to later in this paper.

## *3.3 Circular vortex rings*

The vortex core boundary is toroidal for large rings, and we will call these *open rings*. However, for sufficiently small rings, in the plane of the singularity, the fluid velocity on the axis of the ring is never as small as the Landau critical velocity. We will call these *closed rings* because there is no "hole" in the middle of the external core boundary, and the external boundary of the very small rings are spheroidal, as shown explicitly a little later.

For a single ring, which may be open or closed, the wave function $\psi$ and the model function $F$ for a CBF trial function are given by

$$\psi = F\psi_0 \tag{3.6}$$

where

$$F = \prod_{1 \le j \le N} \varphi\left(\mathbf{r}_j\right) \tag{3.7}$$

and

$$\varphi\left(\boldsymbol{r}_j\right) = P\left(\boldsymbol{r}_j\right)\exp\left[\frac{-im\Phi\left(\boldsymbol{r}_j\right)}{\hbar}\right] . \tag{3.8}$$

$P(\boldsymbol{r}_j)$ and $\Phi(\boldsymbol{r}_j)$ are real-valued functions. The superfluid velocity at $\boldsymbol{r}_j$ is given by

$$\boldsymbol{v}_s\left(\boldsymbol{r}_j\right) = -\nabla\Phi\left(\boldsymbol{r}_j\right) . \tag{3.9}$$

The normal fluid velocity $v_n = 0$ in these variational calculations.

In specifying functions that characterize a ring vortex we shall use coordinate systems aligned with the ring as shown in Fig. 5. There is cylindrical symmetry about the $z$-axis. Figures 5a and 5b show the $x$–$z$ plane projection of a right-handed system of Cartesian coordinates $(x, y, z)$ and two overlaid coordinate systems utilized in the calculations. One is a cylindrical system $(r, \theta, z)$ having its origin at the center of the ring and $z$ aligned with the ring vortex axis. The other is a local plane coordinate system $\left(\rho, \chi\right)$ centered on the vortex singularity in any plane where $\theta$ is constant. In Figs. 5a and 5b, the $\left(\rho, \chi\right)$ coordinates are displayed in the $x$–$z$ plane where $\theta = 0$. The reference line where $\chi = 0$ should be clear from Figs. 5a and 5b. We will represent $P\left(\boldsymbol{r}\right)$ in terms of the coordinates as

$$P\left(\boldsymbol{r}\right) = P\left(\rho, \chi\right) . \tag{3.10}$$

The velocity potential $\Phi(\boldsymbol{r})$ is given by [76]

$$\Phi(\boldsymbol{r}) = \frac{K}{4\pi} \int d^2 S' \, \hat{n} \bullet \frac{(\boldsymbol{r} - \boldsymbol{r}')}{|\boldsymbol{r} - \boldsymbol{r}'|^3} \ . \tag{3.11}$$

The surface integral is over the plane circular area of radius *R*, where *R* is the vortex ring radius. Throughout this paper, we shall assume *K* = *h*/*m*, corresponding to a single quantum of circulation for a ring.

At this point it is useful to introduce auxiliary functions $P_0(\rho, \chi)$ and $n_a(\rho, \chi)$ that hold for both open and closed rings,

$$\frac{n(\rho, \chi)}{n_0} = P_0^2(\rho, \chi) \tag{3.12}$$

$$\frac{n_a(\rho, \chi)}{n_0} = \left[ \sin \frac{\pi\rho}{2 r_0(\chi)} \right]^{2N} \ . \tag{3.13}$$

$P_0$ is a zeroth approximation to *P*.

For an open ring, the core boundary radius $r_0(\chi)$ is determined by the condition $v_s = v_c$ there; and the atomic density is given as follows for our variational calculation

$$\frac{n(\rho, \chi)}{n_0} = \begin{cases} \dfrac{n_a(\rho, \chi)}{n_0} & \text{for } 0 \le \rho \le r_0(\chi) \\ 1 & \text{for } \rho > r_0(\chi) \end{cases} \ . \tag{3.14}$$

For a closed ring it is useful to consider three regions. The atomic density is given as follows for our variational calculation:

$$\frac{n(\rho, \chi)}{n_0} = \begin{cases} \dfrac{n_a(\rho, \chi)}{n_0} & \text{for region 1} \\ [S(\chi)] \dfrac{n_a(\rho, \chi)}{n_0} & \text{for region 2} \\ 1 & \text{for region 3} \end{cases} \ . \tag{3.15}$$

The function *S* is specified as follows and it produces smooth behavior of $n(\rho, \chi)$ inside the core and at its boundary,

$$S(\chi) = \left[ 1 - \frac{A}{2} \left( 1 + \cos \frac{\pi Z_0(\chi)}{z_a} \right) \right]^2 \ . \tag{3.16}$$

The parameters $A$, $Z_0(\chi)$, and $z_a$ are indicated graphically in Fig. 5. The amplitude $A$ determines the amount of depression or elevation of atomic density on the $z$-axis inside the core. Also, $z_a$ is the position where the external core boundary meets the $z$-axis, and $Z_0(\chi)$ is the coordinate where a ray at angle $\chi$ meets the axis.

Region 1 refers to that range of angle $\chi$ where $v_s = v_c$ determines the external boundary $r_0(\chi)$ of the core, the core itself being at $0 \le \rho \le r_0(\chi)$. Region 2 refers to that range of angle $\chi$ where $v_s > v_c$ on the $z$ – axis. For ring radius $R$ we shall define $r_0(\chi)$ by $r_0(\chi) = R/\cos\chi$ in region 2, and in this region the core is located at $0 \le \rho \le r_0(\chi)$. Region 3 refers to the region external to the core and where $v_s < v_c$.

The variational calculation is treated in detail in Appendix A, and it consists of seeking the minimum value of the excitation energy $\varepsilon$, defined by

$$\varepsilon = \frac{\langle \psi | \hat{H} - E_0 | \psi \rangle}{\langle \psi | \psi \rangle} \ , \tag{3.17}$$

when one varies the parameter $N$ (for open rings) or $N$ and $A$ (for closed rings). The formula for $\varepsilon$ in Eq. (3.17) takes into account that the wave function $\psi$ given by Eqs. (3.6) - (3.8) are not normalized.

### *3.3.1 Vortex core boundaries*

The first step in the variational calculation is to find the external core boundary $r_0(\chi)$, determined by the condition $v_s = v_c$. Some results are shown in Figs. 6 - 8. In Fig. 6 one can see that the vortex rings are closed for $R$ = 2.0 Å and 5.0 Å, but open for $R$ = 15.0 Å. Fig. 7 shows a closed ring at $R$ = 8.0 Å, and Fig. 8 shows an open ring at $R$ = 9.0 Å.

### *3.3.2 Energy*

The variational calculation yielded the optimum value $N$ = 0.50 for every ring radius $R$ that was examined, for both open and closed rings. This result coincides with the optimum value of $N$ found in the variational calculation for a rectilinear vortex. It is noted here that a discrete set of values of $N$ were used in the variational calculation, and accordingly $N$ = 0.50 should be interpreted within this context.

The ring vortex energy vs $R$ for $T$ = 0 K is shown in Fig. 9 for ring sizes up to $R$ = 20 Å, but results have also been calculated for $R$ = 50 Å and 100 Å and are shown in the inset to Fig. 9. $\varepsilon_C$ is the vortex ring core energy associated with the region where $n(r)$ departs from $n_0$. The core energy $\varepsilon_C$ includes effects associated with the atomic pair distribution function $g(r)$ that reaches outside the core region. $\varepsilon_E$ is the kinetic energy where the atomic density is $n_0$, external to the core. $\varepsilon_V$ is the total energy of the vortex ring, equal to the sum of $\varepsilon_C$ and $\varepsilon_E$.

Numerical integration was used in obtaining these variational results. The fluid velocity was evaluated using the analogue of the Biot-Savart formula that occurs in electromagnetic theory (see Eq. (A34)). The depletion of atoms in the core region was taken into account, as described in Appendix A. Also, evaluation of all functions near the singular line was treated accurately, as described in Appendix A.

The peak in vortex energy $\varepsilon_C$ near $R$ = 8 Å, shown in Fig. 9, is associated with the transition from closed to open ring structure indicated in Figs. 7 and 8. Calculations for $R$ as small as 0.1 Å were made, although it is questionable whether rings having $R$ less than about 2.0 Å are physically realizable. The numerical results indicate that the core energy curve $\varepsilon_C$ crosses the external energy curve $\varepsilon_E$ when $R \simeq 23$ Å. The core energy is much larger that the external energy for $R$ below about 12 Å.

### *3.3.3 Momentum*

It is almost universally accepted that the momentum of a liquid containing a vortex ring or a solid body cannot be uniquely determined by integrating the momentum density over the liquid volume. Some authors [90] say that the momentum is indeterminate, others [86] say it is zero, and others [76,91] say that its value depends on the size or shape of the container. Nevertheless the following discussion explains how the momentum of the liquid containing a vortex ring can be *uniquely* determined by calculation for the models treated in this paper.

Here we consider a liquid in a container having rigid walls that are not rotating or otherwise moving. We specify that for every time instant $t$ the velocity can be calculated from a velocity potential that is consistent with the rigid wall boundary condition, which requires that the fluid velocity component normal to a wall is zero. (One can, for example, use images to impose the boundary condition; then terms due to images must be included in the velocity potential. In some cases this would require an infinite set of images.) In the case of a vortex ring, the velocity potential is taken to be single valued, a condition that is imposed by quantization of circulation and the mathematical device of a barrier that spans the circular line where the velocity is singular. First we appeal to classical hydrodynamics, where one can show that kinematically a velocity field derived from a velocity potential can accommodate an incompressible liquid, say with density $\rho_0$ *that completely fills the container*. *In this circumstance*, it is intuitively obvious, but can also be shown with the aid of the continuity equation that the center of mass of the liquid does not change in time. This stationary condition implies that the total momentum of the liquid is zero in this case. Using that result together with another application of the continuity equation and an integration by parts, it can be shown that for every time instant $t$

$$\int_V \boldsymbol{v}(\boldsymbol{r})\, d^3r = 0 \ , \tag{3.18}$$

where $V$ is the entire region inside the container.

Next suppose that for some instant of time the mass density $\rho(\boldsymbol{r})$ deviates from $\rho_0$ only in some localized region $R$, e.g., a vortex core. Further suppose that the velocity field $\boldsymbol{v}(\boldsymbol{r})$ is the same potential flow considered before. Then the momentum of the mass defect is the integral of $(\rho_0 - \rho(\boldsymbol{r}))\ \boldsymbol{v}(\boldsymbol{r})$ over the region and this can be easily evaluated. This integral must be the negative of the momentum of the remaining liquid since the sum of the two must be zero according to the earlier analysis. This shows how one can calculate the momentum of a liquid containing a vortex ring with a core where there is reduced or excess mass and where the density is zero on the line where velocity is singular.

Turning to the quantum mechanical treatment of a vortex ring, one can evaluate the expectation values of the mass density and current density operators

$$\hat{\rho}(\boldsymbol{r}) = m\sum_j \delta\left(\boldsymbol{r} - \boldsymbol{r}_j\right) \tag{3.19}$$

$$\hat{\boldsymbol{j}}(\boldsymbol{r}) = \frac{\hbar}{2i}\sum_j \left(\nabla_j \delta\left(\boldsymbol{r} - \boldsymbol{r}_j\right) + \delta\left(\boldsymbol{r} - \boldsymbol{r}_j\right)\nabla_j\right) \tag{3.20}$$

with respect to the trial functions for vortex rings (see Eq. (3.6)), and show that

$$\boldsymbol{j}(\boldsymbol{r}) = \langle \psi_{vor} | \hat{\boldsymbol{j}}(\boldsymbol{r}) | \psi_{vor} \rangle = \rho(\boldsymbol{r})\boldsymbol{v}(\boldsymbol{r}) \tag{3.21}$$

where $|\psi_{vor}\rangle$ is normalized to 1 and $\boldsymbol{v}(\boldsymbol{r})$ is given by Eq. (3.9). Integrating Eq. (3.21) over the entire volume $V$ of the liquid and using Eq. (3.20), one obtains

$$\boldsymbol{P} \equiv \langle \psi_{vor} | \hat{\boldsymbol{P}} | \psi_{vor} \rangle = \int_V \rho(\boldsymbol{r})\boldsymbol{v}(\boldsymbol{r})\, d^3r \ . \tag{3.22}$$

One can also evaluate

$$\boldsymbol{j}_0(\boldsymbol{r}) \ = {}_0\langle \psi_{vor} | \hat{\boldsymbol{j}}(\boldsymbol{r}) | \psi_{vor} \rangle_0 = \rho_0 \boldsymbol{v}(\boldsymbol{r}) \ , \tag{3.23}$$

where $|\psi_{vor}\rangle_0$ is the normalized trial function for a vortex ring where the zero subscript means $P(\boldsymbol{r}) \equiv 1$ in Eq. (3.8). One finds in this case that the momentum $\boldsymbol{P}_0$ satisfies

$$\boldsymbol{P}_0 \equiv {}_0\langle \psi_{vor} | \hat{\boldsymbol{P}} | \psi_{vor} \rangle_0 = \rho_0 \int_V \boldsymbol{v}(\boldsymbol{r})\, d^3r = 0 \tag{3.24}$$

where the last equality was established in the analysis of a classical fluid, (Eq. (3.18)). The quantum mechanical relations in Eqs. (3.21) – (3.24) show that expectation values coincide with classical hydrodynamics results in the case of potential flow, and so the mass defect method can be also used to evaluate momentum of a vortex ring in our quantum mechanical treatment.

Our approach avoids difficulties that others have pointed out in assigning a unique momentum to a vortex ring [76,86] and avoids ambiguities that occur in attempts at identifying impulse of a vortex ring with its momentum [76]. Validity of this method is supported by a demonstration that the semiclassical formula for the velocity of a large vortex ring having a hollow core can be computed as the ratio of the momentum of the mass defect in the rest frame of the container walls to the mass defect itself. Known formulas [92] for the fluid velocity near the vortex ring singularity can be used in computing the momentum of the mass defect needed in that demonstration.

This mass defect method was used to calculate the momentum as a function of vortex ring radius $R$ and some of the results are displayed in Fig. 10. Calculations were also made for $R$ = 50 A and 100 A, and the results are shown in the inset in Fig. 10.

### *3.3.4 Small vortex ring dispersion curve*

Elimination of $R$ between calculated data sets plotted in Figs. 9 and 10 produce the data points for energy versus momentum for vortex rings displayed in Fig. 11. The maximum value of $R$ for the points displayed in Fig. 11 is 6 Å, but this value of $R$ has no special significance and was chosen for clarity of the graphical presentation. For radius $R$ = 6 Å, the vortex ring wave number is $k$ = 2.17 Å$^{-1}$. The Landau elementary excitation spectrum observed in neutron scattering measurements [79] is also shown in Fig. 11. The results in Fig. 11 show that rotons are not simply small vortex rings. The energy spectrum for vortex rings is clearly distinct from the Landau dispersion curve for phonons and rotons. This result of our new theory resolves the issue of whether rotons are small vortex rings. This issue has been the subject of much speculation, analysis and debate for more than 50 years, dating back to some of the earliest discussions of quantized vortices and rotons by Onsager [55,93,94] and Feynman [56].

The models of quantized vortex rings, including small closed rings that we have considered are easy to visualize. Even though Feynman [56] has elaborated on some features of a physical model of rotons, the discussion of these excitations in Appendix B is a useful supplement to his ideas and an aid in understanding and visualizing them.

The variational procedure will home in on energy eigenstates of the system, and the vortex rings we have treated so far are not expected to coincide exactly with vortex rings that have a constant velocity of translation like those that are treated in classical hydrodynamics. However, we expect that there is a close connection between those two kinds of vortex rings. This matter will be further discussed after the treatment of large vortex rings.

### *3.3.5 Large vortex rings*

Here we consider circular vortices having a ring radius much greater than the core radius. Known formulas [92] provide accurate approximations for the fluid velocity near the singularity for such rings. Using those formulas one can derive an equation for the locus of points where the magnitude of $\boldsymbol{v} = (\boldsymbol{v}_s - \boldsymbol{V})$, the local fluid velocity as viewed in a reference frame that moves with a constant translational velocity $\boldsymbol{V}$ such that the vortex ring appears to be stationary, is constant. There are two solutions of that equation, which will be given in Eqs. (3.25) and (3.26). Viewed as a cross section that includes the vortex axis, each solution is a circular locus where the center of the circle is displaced from the singularity.

One solution, for $V = 0$, was found to account accurately (less than 1% deviation) for the boundary of the approximate energy eigenstate ring at $R = 100$ Å treated numerically in the calculations described in Sec. 3.3.1. Studies carried out with the aid of the velocity formulas reveal the momentum *flux density* is not identically zero across the core boundary for an approximate energy eigenstate vortex ring. However, one can show that at least for the optimum trial function where $N = 0.50$ (refer to Eq. (3.14)) the *net* flux through the boundary is zero. One interesting consequence of this result is that one can then use the continuity equation and show that the total mass inside the vortex core is stationary in a reference frame that is at rest with respect to the container walls. This is a property that is shared with an *exact* energy eigenstate. It is conjectured that this stationary property also holds for the approximate energy eigenstates that are the small vortex rings found using the variational procedure. Further study of this matter would be useful.

The second solution of the locus problem for large vortex rings gives an equation that contains the translational velocity magnitude $V$. In Fetter's [95] treatment of the locus problem, a similar equation was found for arbitrary constant velocity magnitude at the core boundary, but his equation also contained an additive constant. Fetter used the Bernoulli equation and a condition of constant pressure at the core boundary. The additive constant mentioned above was treated as a contribution to pressure in Fetter's approach, and was therefore inconsequential. Fetter treated a hollow core model, and this involved deriving a differential equation for the streamline at the core boundary. For that model the momentum flux density through the core boundary is identically zero as calculated for the reference frame where the vortex ring appears stationary. We assume that this latter condition also holds for large moving vortex rings in our theory where the core is not hollow. In both the hollow core model and our model, a simultaneous solution of the differential equation for the streamline and the algebraic equation for constant magnitude of fluid velocity at the core boundary produced two important formulas. One is for the translational velocity, a well-known result of classical hydrodynamics, given later in Eq. (3.35). The other shows that the core boundary is a circle with its center displaced slightly from the singularity, as given in what follows (see Eq. (3.26)).

In terms of the local plane coordinate system $(\rho,\chi)$ described earlier, the circular locus of radius $a$ corresponding to constant velocity magnitude, $v$, is as follows for each of the two cases considered (here we write $\rho = r_0(\chi)$ for the core boundary).

$$\text{For} \quad V = 0: \quad r_0(\chi) = a - \frac{a^2}{2R} \ln \frac{8R}{a} \cos\chi \;. \tag{3.25}$$

$$\text{For} \quad V \neq 0: \quad r_0(\chi) = a - \frac{a^2}{4R}\cos\chi \ . \tag{3.26}$$

In each of these equations, the coefficient of cos $\chi$ is the distance by which the center of the circular locus of radius $a$ is displaced from the vortex ring singularity.

The result Eq. (3.26) calls for some comment. In arriving at the formula for $r_0(\chi)$ in Eq. (3.26), where $V \neq 0$, from the approach based on CBF formalism, we have treated $\boldsymbol{V}$ as being equal to the velocity $\boldsymbol{v}_n$ of the normal fluid and used the condition

$$\varepsilon(p) + \boldsymbol{p}\bullet(\boldsymbol{v}_s - \boldsymbol{v}_n) = 0 \tag{3.27}$$

as the criterion that fixes the core boundary. This replaces the criterion

$$\varepsilon(p) + \boldsymbol{p}\bullet\boldsymbol{v}_s = 0 \tag{3.28}$$

that was used in the case where $\boldsymbol{v}_n = 0$. The condition in Eq. (3.27) locates the threshold for degeneracy of the lowest eigenstate with superfluid flow for the free energy function

$$K = E' - \boldsymbol{v}_n \bullet \boldsymbol{P}' \tag{3.29}$$

for fixed $\boldsymbol{v}_n$. $K$ is defined for all microscopic eigenstates of the system containing rotons and phonons with $E'$ and $\boldsymbol{P}'$ given by Eqs. (4.1) and (4.2).

At the present level of my understanding, this threshold condition on $K$ that leads to the criterion in Eq. (3.27) seems to introduce a new postulate of the theory that generalizes Landau's and London's criteria for superfluidity when both $\boldsymbol{v}_n$ and $\boldsymbol{v}_s$ are possibly different from zero. Since the ground state at rest defines a special reference frame that all states and velocities are measured with respect to, Eq. (3.27) is not simply a consequence of Eq. (3.28) and Galilean invariance. Even when the liquid contains a moving vortex ring, that special reference frame remains at rest with respect to the stationary walls of the container.

A study of a large vortex ring at $R$ = 100 Å radius has been carried out using both numerical calculations and analytic formulas and several important conclusions applicable at $T$ = 0 K to all large vortex rings can be drawn from it, including the following:

i) The core energy of large stationary vortex rings (approximate energy eigenstate, $V = 0$) calculated variationally is essentially the same as that found using the energy per unit length 1.68 K/Å determined by variational calculations for a rectilinear vortex;

ii) If the atomic density profile in Eqs. (3.13) and (3.14) with $N = 0.50$ is used with $r_0(\chi)$ determined by the classical formula for the core boundary for a hollow core vortex ring, Eq. (3.26), then the core energy is essentially the same as for a stationary vortex ring where $r_0(\chi)$ is given by Eq. (3.25);

iii) For vortex rings having $R \geq 100$ Å, the energy of a stationary vortex ring at $T$ = 0 K is given by

$$\varepsilon_V = \frac{1}{2}\rho K^2 R\left[\left(\ln\frac{8R}{a} - 2\right) + \alpha_c\right], \quad (3.30)$$

where $K = h/m$ and $a$ = 2.80 Å corresponding to Landau critical velocity $v_c \simeq 58$ m/s. The core energy is accounted for by $\alpha_c = 2.02$, which corresponds to $\varepsilon_c/2\pi R$ = 1.68 KÅ$^{-1}$. Eq. (3.30) gives good agreement with computer results for $R$ as small as 50 Å;

iv) The external energy (i.e., outside the core) for a *stationary vortex ring* is given accurately by the semiclassical formula for a *hollow core moving ring*, the term in parentheses in Eq. (3.30);

v) A moving vortex ring whose core boundary is given by Eq. (3.26) has its singularity offset from the center of the circular locus by a smaller distance than does a stationary ring. Therefore the core energy for a moving ring will be even better approximated than for a stationary ring by the core energy of a rectilinear vortex. Therefore Eq. (3.30) should also be accurate for a moving vortex ring, where again one finds theoretically $a$ = 2.80 Å. These results will be useful in comparing this theory with experimental results found by Rayfield and Reif, a matter treated in Sec. 3.3.8.

It is noted here that Roberts and Grant [95a] calculated properties of large circular vortex rings in a Bose condensate, and those rings are in motion. Comparing our results in Eq. (3.30) with their energy formula, one can deduce that the core energy that they found corresponds to a value of $\alpha_c = 0.385$ in Eq. (3.30). The core radius $a$ in their theory is a healing length which is computed from a criterion that is different from that used to compute $a$ in our theory. If one used their value of $\alpha_c$ as an approximation instead of the value that we found, the statistical mechanical properties, including the specific heat, calculated for liquid $^4$He near $T_\lambda$ would be substantially different from those that we calculate in this paper. The influence of the supposed lower core energy on the individual vortex ring total free energy and other statistical mechanical properties can be understood with the aid of Figs. 19a – 19c and the discussion of these figures in Sec. 5.3.1. The lower core energy would result in a much higher number density of vortex rings and a much wider distribution of statistically important ring sizes than we find.

The Bose condensate model used by Roberts and Grant is an extreme idealization that implicitly assumes that the liquid consists of weakly interacting atoms or has low atomic density, or both, and that $T$ = 0 K.

### *3.3.6 Momentum of a large stationary vortex ring*

The density in Eq. (3.14) with $N$ = 0.50 in Eq. (3.13) together with known formulas [95] for the fluid velocity near the singularity have been used in deriving a formula for the momentum density of the mass defect in the core of a stationary vortex ring. Integration over the interior of the core can be carried out analytically. The mass defect method yields the result that the momentum $\boldsymbol{P}_0$ of the liquid containing the vortex ring is equal in magnitude but opposite in direction to the result found by integration over the core region having outer boundary $r_0(\chi)$ given by Eq. (3.26). The magnitude $P_0$ is given by

$$P_0 = \rho K \pi a^2 \left[ \alpha \ln \frac{8R}{a} - \gamma \right] \tag{3.31}$$

where

$$\alpha = 1 - \frac{1}{\pi} - \frac{4}{\pi^2} \simeq 0.276405 \tag{3.32}$$

$$\beta = Si\left(\tfrac{\pi}{2}\right) - 1 \simeq 0.37076 \tag{3.33}$$

$$\gamma = \frac{1}{2} + \frac{2}{\pi^2}\left(2\beta - 3\right) \simeq 0.042337 \ . \tag{3.34}$$

*S*i(x) is the sine integral [96]. As in Eq. (3.30), we shall use $a = 2.80$ Å in Eq. (3.31).

The direction of the momentum, $\hat{\boldsymbol{P}}_0$, is opposite to that of the dipole moment of the ring as determined by usual conventions in hydrodynamics; as a consequence the unit vector $\hat{\boldsymbol{P}}_0$ is opposite to the fluid flow direction on the axis of the ring.

### *3.3.7 Velocity of translation of a moving vortex ring*

The semi-classical formula for the velocity magnitude *V* of a large vortex ring having a hollow core is [76]

$$V = \frac{K}{4\pi R}\left[ \ln \frac{8R}{a} - \frac{1}{2} \right] . \tag{3.35}$$

We will show that the average velocity magnitude $\bar{V}$ of fluid inside the core region for a large vortex ring is given by Eq. (3.35) even when there is non-uniform density there. $\bar{V}$ is given by

$$\bar{V} = \frac{\left| \int d^3r\, \boldsymbol{v}_s(\boldsymbol{r}) \right|}{\int d^3r} \ , \tag{3.36}$$

where the integrals are over the core region and the core boundary is given by Eq. (3.26). Here it is specified that $\boldsymbol{v}_s(\boldsymbol{r})$ is given by

$$\boldsymbol{v}_s(\boldsymbol{r}) = \frac{\boldsymbol{j}(\boldsymbol{r})}{\rho(\boldsymbol{r})} = \frac{\rho(\boldsymbol{r})\boldsymbol{v}_s(\boldsymbol{r})}{\rho(\boldsymbol{r})} \tag{3.37}$$

for both classical and quantum mechanical treatments discussed earlier. The equality $\bar{V} = V$ is a simple consequence of a result mentioned in Sec. 3.3.3 for a calculation involving the mass defect method for the core momentum and mass in the case of a large vortex ring having a hollow core. The combination of these results for motion external and internal to the core indicates that Eq. (3.35) is applicable to a large vortex ring where the core has non-uniform density. We note that $\boldsymbol{j}(\boldsymbol{r})$ and $\rho(\boldsymbol{r})$ in Eq. (3.37) are expectation values of current density and mass density operators in the quantum mechanical case and there is no assumption that the operators themselves commute.

### *3.3.8 Large vortex rings: theory and experiment*

The results discussed so far indicate that at $T = 0$ K the energy and velocity of a large moving vortex ring is given to good approximation by Eq. (3.30) and (3.35) respectively. Energy versus radius plots in Fig. 12 are based on Eq. (3.30). These plots show that the core energy is substantial compared to the external energy even for vortex rings having $R$ as great as 10,000 Å, about the maximum size observed in experiments by Rayfield and Reif [60]. The velocity versus radius plot, shown in Fig. 13, is based on Eq. (3.35). In all of these calculations the core radius $a = 2.80$ Å is determined by the condition that the fluid velocity magnitude $|\boldsymbol{v}_s - \boldsymbol{V}|$ is equal to the Landau critical velocity magnitude $v_c \simeq 58$ m/s at the core boundary.

Numerical evaluation of vortex ring energy $\varepsilon_V$ and velocity magnitude $V$ for a set of $R$ values was used to construct the velocity versus energy curve displayed in Fig. 14. Also shown in Fig. 14 are experimental points observed by Rayfield and Reif [60] who used a charged carrier technique and a time of flight spectrometer to study large vortex rings in the liquid at 0.28 K and SVP. The good agreement provides considerable support for the theory developed here. No adjustable parameters were used in obtaining this agreement. Figures 12 and 13 display data that were used in constructing Fig. 14. The velocity versus radius plot in Fig. 13 enables one to refine the estimates of the sizes of vortex rings that Rayfield and Reif observed.

Rayfield and Reif [60] used semiclassical hydrodynamics formulas for a vortex ring having rotating *solid* core in comparing their results to theory, and they obtained a good fit to their experimental points. That model led them to a useful form for displaying their data and identifying important features of their observations. However a rotating solid core is not kinematically possible in a *vortex ring*. Donnelly [97] has also argued against the solid core model in a vortex ring.

In later work, Rayfield [98] used a hollow core model. However, the hollow core model is not consistent with quantum mechanics. The core can be completely hollow only if there is an infinite potential barrier at the core boundary, and there is no barrier of that sort that is known to exist in liquid $^4$He. On the other hand, if the hollow core model is obtained as the *limit* for an infinitely steep atomic density profile just inside the core boundary, then quantum mechanical calculation gives infinite core energy for all values of $R$, a result that we alluded to in Sec. 3.2. These considerations indicate that the schematic models used by Rayfield and Reif (and others) are not adequate for accounting for the properties of vortices in liquid $^4$He, and they further show how our new theory overcomes the deficiencies of those models.

Additional support for the theory developed here can be found in experimental results obtained by Steingart and Glaberson [88,89] who measured the pressure dependence of the relative core radius $a/a_0$ at $T = 0.368$ K. Here $a$ is the radius at pressure $p$ and $a_0$ the radius at $p = 0$. At $p = 10.5$ atm they found $a/a_0 = 1.15$ and at $p = 24.3$ atm, $a/a_0 = 1.29$. On the other hand, the roton spectrum for several pressures has been measured using neutron scattering [99]. The Landau critical velocity can be extracted from those measurements and the vortex core radius can be computed using the formula $a = h/2\pi m v_c$. The lowest pressure for the neutron scattering data was at 1 atm so we use $a_1$, at $p = 1$ atm for the reference level. The results are $p = 10$ atm, $a/a_1 = 1.10$; $p = 24$ atm, $a/a_1 = 1.29$.

The good agreement of these results with the pressure dependence reported by Steingart and Glaberson gives further support to the theory developed here. It is a testimony to the physical insight of Steingart and Glaberson [100] that they recognized that the pressure dependence of the core radius is determined by the Landau critical velocity despite the fact that the models they used in analyzing their data gave values of the core radius about one-third of the radius that corresponds to the Landau critical velocity.

The vortex core model that we have used is a central feature in the theoretically calculated properties of vortex rings, properties that are in close agreement with results obtained from experiments on large vortex rings, as we have just discussed. That agreement helps build confidence in the underlying postulates on which the theory of smaller rings as well as rectilinear vortices also rely. On the other hand, regions outside the core can be treated with the two-fluid model. In earlier work [20-23] it was shown how the two-fluid model could be derived from microscopic theory that starts with correlated basis function formalism. Certain features of the two-fluid model will be developed further in the next Section in preparation for treating vortices at finite temperatures, including near $T_\lambda$.

## 4. Statistical mechanics of flowing $^4$He

Statistical mechanical formulas for properties of liquid $^4$He treated in this Section provide a basis for calculating properties of quantized vortices at finite temperatures and for studying both the superfluid transition associated with flow properties, and the λ transition associated with an anomaly in specific heat. Most of the formulas and techniques needed here are available from earlier work [20], but the way they are applied in calculations are different in important ways. For efficiency of presentation we will draw freely from that earlier work while explaining new aspects of the calculations and results in detail.

We first consider a model for interacting elementary excitations where the superfluid is at rest and the wave functions, and energy and momentum eigenvalues are given respectively by Eqs. (2.3) - (2.5). We specify that the interaction term involving $f_{ij}$ is such that $f_{ij} = -\gamma\varepsilon_0$ for $i$ and $j$ restricted to rotons or a subset of rotons, and $f_{ij} = 0$ otherwise. There is an upper limit $\mathcal{n}$ on the number of rotons that can occur in the liquid. Values will be assigned to $\gamma$ and $\mathcal{n}$ shortly.

For uniform superfluid flow, the wave functions are given by Eq. (2.7) where $\boldsymbol{P}_0$ and $\boldsymbol{v}_s$ satisfy Eq. (2.8), and the energy and momentum eigenvalues found by operating on $\psi'$ with $\hat{H}$ and $\hat{P}$ in Eqs. (2.1) and (2.2) are

$$E' = E_0 + \frac{1}{2} \mathcal{N} \; m v_s^2 + \boldsymbol{v}_s \bullet \sum_i n_i p_i + \sum_i n_i \varepsilon_i - \frac{\gamma \varepsilon_0}{2\mathcal{N}} \sum_{i,j}{}' n_i n_j \tag{4.1}$$

$$\boldsymbol{P}' = \mathcal{N} \; m \boldsymbol{v}_s + \sum_i n_i \boldsymbol{p}_i \; . \tag{4.2}$$

The prime on the summation in Eq. (4.1) indicates restriction to the same set of rotons described above for the superfluid at rest. Also, $\gamma$ is the roton-roton coupling strength and $\varepsilon_0$ is the roton gap energy. The Helmholtz potential $F = E' - TS$ for the condition of thermodynamic equilibrium with superfluid velocity $v_s \geq 0$ and normal fluid velocity $v_n = 0$ is given by

$$F(v_s) = E_0 + \frac{1}{2} \mathcal{N} \; m v_s^2 - kT \sum_i \ln(1 + n_i) + \frac{1}{2} \mathcal{N} \; \gamma \varepsilon_0 \left( \frac{1}{\mathcal{N}} \sum_i{}' n_i \right)^2 \tag{4.3}$$

where for phonons

$$n_i = \left\{ \exp\left[ \beta \left( \varepsilon_i + \boldsymbol{p}_i \bullet \boldsymbol{v}_s \right) \right] - 1 \right\}^{-1} \tag{4.4}$$

and for rotons

$$n_i = \left\{ \exp\left[ \beta \left( \varepsilon_i + \boldsymbol{p}_i \bullet \boldsymbol{v}_s - \frac{\varepsilon_0 \gamma}{\mathcal{N}} \sum_i{}' n_i - \alpha \right) \right] - 1 \right\}^{-1} \; . \tag{4.5}$$

The parameter $\alpha$ is a roton chemical potential that is determined by the constraint

$$\sum_i{}' n_i \leq \mathcal{n} \; . \tag{4.6}$$

The summation $\Sigma'$ is just over the rotons whose occupation numbers $n_i$ are given by Eq. (4.5). An iteration procedure was used to solve Eqs. (4.4) and (4.6) simultaneously for $\alpha$ and $n_i$. Then the solution values of $n_i$ were substituted into Eq. (4.1) to evaluate isotherms of $F(v_s)$.

Now we must digress to explain some notation and approximations. $T_c$ is the temperature where the superfluid transition occurs when only Landau excitations consisting of phonons and rotons contribute to the normal fluid density. A formula for this normal fluid density is given in Eqs. (4.7) and (4.8). $T_\lambda$ is the temperature where the specific heat has its peak value. It is assumed that $T_c = T_\lambda = 2.172\text{K}$[1] in this paper up to and including Sec. 7.1.1 and 7.1.2. These latter two Sections deal with an approximation that we call the crossover model, where $T_c$ is shown by calculation to coincide with $T_\lambda$. However, in more accurate approximations, which we call complete models, treated in Sec. 7.1.3 and 7.3, the correct condition $T_\lambda < T_c$ is taken into account. $T_\lambda$ is very close to $T_c$. The complete models treat effects of thermally excited vortices more accurately than the crossover model does. Despite the numerical equivalence of $T_c$ and $T_\lambda$ in the crossover model, we have endeavored to use the symbol $T_c$ whenever the property of the liquid being considered relies primarily on the "unrenormalized" superfluid transition at $T_c$. This practice will make it easier to understand the discussion of the complete models in Sec. 7.

Figure 15 shows a set of isotherms of $F(v_s)$ computed taking into account phonons and interacting rotons only, and not including vortex rings. For comparison one can refer to Fig. 5 of the Ref. [20] where isotherms of $F(v_s)$ are displayed for non-interacting excitations without a constraint on the total number of rotons. Focusing again on the interacting roton case, we observe that all of the isotherms terminate at the Landau critical velocity $v_c$ because of the degeneracy there of the correlation factor when only phonons and rotons are taken into account. That degeneracy is discussed in Sec. 2.1. For $T < T_c$ it can be shown that $F(v_s) - F(v_s = 0) = \frac{1}{2} V \rho_s(0) v_s^2$ per atom on each isotherm for a range of values of $v_s$ that starts at $v_s = 0$, and the isotherm bends upward there. (Note: $V$ = volume. Also $\rho_s(0) \equiv \rho_s(v_s = 0)$; see Eqs. (4.7) – (4.9).) As $v_s$ increases further, each isotherm also reaches an inflection point at a velocity we will call $v_0$. Beyond $v_0$ the isotherm bends downward. Each of these isotherms also reaches a maximum value at some velocity $v_M$ that is greater than $v_0$. The velocities $v_0$ and $v_M$ decrease to $v_s = 0$ as temperature approaches $T_c$ = 2.172 K from below. At temperatures above $T_c$ there is no inflection point and each isotherm bends downward for all values of $v_s$. The isotherms of $F(v_s)$ for non-interacting excitations shown in Fig. 5 of the Ref. [20] indicate similar behavior, but there the inflection point disappears at approximately $T = 2.42$ K. The velocity $v_0$ plays an important role in the theory of the λ-transition for liquid $^4$He in complete thermodynamic equilibrium. $v_0$ is the order parameter in this theory and it governs the fluctuations that are responsible for the singularity in the specific heat.

[1] The value $T_\lambda = 2.172$ K at SVP used in this paper is referred to the 1958 $^4$He vapor pressure scale of temperatures. The experimental data that we refer to in comparison between theory and experiment used this scale. For conversion to more modern temperature scales, the reader should consult a current reference such as: F. Pobell, *Matter and Methods at Low Temperatures* (Springer-Verlag, Berlin, 1992).

The shape of isotherms of $F/N\,k$ versus $v_s$ in Fig. 15 may seem surprising or suspect, but it can be readily understood by examining parts of $F = E' - TS$. To begin with, notice that at high values of $v_s$ occupation numbers $n_i$ in Eq. (4.5) are large for a set of roton states having momentum opposite to $v_s$. This is particularly easy to see when rotons are non-interacting and their total number is not constrained. However, it is also true even in the absence of these conditions. That set of rotons causes entropy to increase as $v_s$ increases. The energy $\left(E' - E_0\right)$ (see Eq. (4.1)) remains positive for all values of $v_s$ and grows in magnitude as $v_s$ increases. However, as $v_s$ increases, the negative term ($-TS$) also grows in magnitude and eventually outweighs the positive term $\left(E' - E_0\right)$ with the result that at sufficiently high velocities the value of $F$ is less than the value of $F$ at $v_s = 0$. The occupation numbers also increase with temperature at any value of $v_s$. Also Fig. 15 shows that for $T > T_c$, $F\left(v_s\right) < F\left(v_s = 0\right)$ for every value of superfluid velocity where $v_s > 0$.

The behavior of isotherms of $F$ versus $v_s$ under conditions of uniform flow in a homogeneous liquid suggest that the liquid should perhaps have a preference for localized flow in vortex structures when the liquid is in local thermodynamic equilibrium, and this preference should increase with temperature. The density fluctuation excitations (rotons and phonons) that at finite temperature dress the topological excitations represented by vortices may then play an enhanced role in thermal properties of the liquid even when there is no macroscopic background velocity $V_s$. Furthermore, these observations suggest a possible link between the superfluid transition and anomalous behavior of heat capacity where the isotherms change in character. These possibilities provided the main motivation for much of the research reported in this paper.

For the model treated in Fig. 15 and throughout the remainder of this paper, unless otherwise stated, the value of the interaction strength $\gamma$ is selected to be such that the superfluid transition occurs at 2.172 K, the experimentally observed value at *SVP*. Stated more precisely, we select the value of $\gamma$ to be that which enforces the condition $\rho_s(0) = 0$ at $v_s = 0$ for SVP when $T = 2.172$ K and only phonons and rotons are included in the excitation spectrum and roton chemical potential $\alpha = 0$. Formulas for $\rho_s(0)$ are given in Eqs. (4.7) – (4.9). This prescription gave $\gamma = 0.840$ and $\sum_i' n_i \big/ N = 0.131$ at $T = T_c = 2.172$ K for the case where rotons are specified to be those Landau excitations whose energy is observed in neutron scattering [79] to be greater or equal to the gap energy 8.70 K. These excitations correspond to rotons having momentum $p \geq p_c$ where $p_c/\hbar = 0.474$ Å$^{-1}$. In our model calculations, the roton gap at $T$ = $T_c$ = 2.172 K *was lowered* in this case by the amount $\Delta/k = (\varepsilon_0/k)\gamma\left(\sum_i{}' n_i \middle/ N\right) = 0.957$ K.

We also studied another case for fixing the excitations that interact, again called rotons here, along with the coupling constant $\gamma$ so that together they produced $\rho_s(0) = 0$ at $T$ = $T_c$ = 2.172 K. In this case we took $p_c/\hbar = 1.58$ Å$^{-1}$, corresponding to the inflection point between relative maximum and minimum of the Landau spectrum [79]. The computed value of $\gamma$ for this case was $\gamma$ = 0.957 and $\sum_i{}' n_i \big/ N = 0.119$ at $T_\lambda$. In this case the roton gap at $T$ = $T_c$ was lowered by $\Delta/k = (\varepsilon_0/k)\gamma\left(\sum_i{}' n_i \middle/ N\right) = 0.991$ K. It is expected that the two cases would give about the same results for computed thermodynamic properties of the liquid, and we chose to use the case where $p_c/\hbar = 0.474$ Å$^{-1}$, $\gamma$ = 0.840 in our further calculations.

Formulas for normal fluid density and superfluid density are given [20], respectively, by

$$\rho_n\left(\boldsymbol{v}_n - \boldsymbol{v}_s\right) = \frac{1}{V}\sum_i n_i \boldsymbol{p}_i \tag{4.7}$$

$$\rho = \rho_s + \rho_n \tag{4.8}$$

where $\rho$ is the mass density, and the occupation numbers $n_i$ are given by Eqs. (4.4) and (4.5) when $\boldsymbol{v}_n$ = 0. Using Eq. (4.7) and Taylor series expansions of the $n_i$ in Eqs. (4.4) and (4.5) about $v_s$ = 0, one finds the following formula for $\rho_n(0)$, the normal fluid density in the limit where $v_s \to 0$:

$$\rho_n(0) = \frac{4\pi\beta}{3h^3}\int_0^\infty dp\, p^4 n_0(p)\left[n_0(p)+1\right]\,. \tag{4.9}$$

The density of states and conversion of a sum to an integral have been introduced in the usual way for closely spaced momentum eigenvalues for elementary excitations. The subscript zero on $n_0$ refers to evaluation at $v_s$ = 0. The formula for $\rho_n(0)$ in Eq. (4.9) also holds for non-interacting, unconstrained excitations provided that one sets $\gamma$ = 0 and $\alpha$ = 0 in $n_0(p)$. In that case, Eq. (4.9) is essentially the same as Landau's formula for normal fluid density at $v_s$ = $v_n$ = 0. Using Eqs. (4.8) and (4.9) one finds that when neutron scattering data are used to evaluate $n_0(p)$ in the non-interacting excitation case, $\rho_s(0)$ = 0 at approximately 2.42 K [20]. For the interacting, constrained excitation case, $\rho_s(0)$ = 0 at 2.172 K. It is not accidental that these superfluid transition temperatures coincide with the qualitative change of behavior of $F(v_s)$ discussed earlier, and this matter is further elucidated in the reference [20].

It is noteworthy that evaluation of Eq. (4.9) with the bare particle spectrum $\varepsilon(p) = p^2/2m$ and $\rho$ = 0.145 g cm$^{-3}$, the observed density of liquid $^4$He at essentially zero pressure, yields $\rho_n(0) = \rho$ and $\rho_s(0) = 0$ at $T_c$ = 3.13 K. This is the same transition temperature that London found for a perfect Bose-Einstein gas. It is easy to generate London's formula for $T_c$ from Eq. (4.9) using the free particle spectrum. It appears that one of the main reasons for believing that a Bose-Einstein condensate is involved in superfluidity of liquid $^4$He was the closeness of the calculated $T_c$ for a perfect gas to the transition temperature $T_\lambda$ for liquid $^4$He. Here we see a relationship between London's $T_c$ and the measured value of $T_\lambda$ at *SVP* from a point of view that does not explicitly involve a condensate. At the same time we see a relationship between Landau's and London's theories of superfluidity that was not obvious previously.

Next we shall explain how the value of $\alpha$ was fixed in the formulas for $n_i$ used in calculating the isotherms of $F(v_s)$ that appear in Fig. 15. Before giving the detailed explanation, we observe that if $\gamma > 0$ there must be an upper limit on the number of interacting rotons that can occur because otherwise the exponent in the denominator could sometimes be negative and that would lead to unphysical behavior. We believe our method of fixing $n$ and in turn $\alpha$ is the least arbitrary way of doing this.

When $\gamma = 0.840$, the value found earlier, computer calculations revealed that for $v_s$ = 0 there is a solution of Eqs. (4.5) and (4.6) for $T \leq 2.31$ K if one takes $\alpha = 0$ at those temperatures. However, for any temperature higher than 2.31 K, there is a solution only if $\alpha < 0$. Uncertainty in the threshold temperature $T_{t\alpha}$ = 2.31 K due to finite mesh sizes used in the computer calculations is estimated to be about ± 0.01 K. Based on these results, the value of the constraint $n/N$ on the ratio of the number of rotons $n$ to the number of atoms $N$ was fixed at 0.2838, the value computed with $\alpha = 0$ when $T$ = 2.31 K. This in turn fixed the constraint $n/N$ for all velocities and all temperatures, and enabled solution of Eqs. (4.5) and (4.6) and evaluation of all thermodynamic properties such as $F(v_s)$ for the entire range of interest for parameters $v_s$ and $T$. This method of fixing $n/N$ and calculating $\alpha$ self-consistently preserves the Legendre transform structure of the statistical mechanical potentials, as discussed in Ref. [20].

Thermodynamic properties for a model of liquid $^4$He that includes only phonons and constrained, interacting rotons has been studied in considerable detail in Ref. [20]. It was found that the specific heat is divergent at a threshold temperature $T_{t\alpha}$ (which was called $T_t$ in Ref. [20]) where the roton chemical potential $\alpha$ acquires non-zero values. The specific heat divergence at $T_{t\alpha}$ is rendered unobservable when vortex rings are taken into account in our model because of a catastrophic increase in vortex rings that occurs at a temperature $T_t$ that is less than $T_{t\alpha}$. The abundance of vortex rings is expected to be the dominant factor that determines specific heat and other thermodynamic properties for $T > T_t$. The vortex ring catastrophe is treated in Sec. 7.5.

Other criteria for fixing the value of $n/N$, where $n/N < 0.2838$, were studied. A particularly significant case set $n/N = 0.131$, the computed value of $\sum_i' n_i \Big/ N$ at $T = T_c$ = 2.172 K when $\gamma$ = 0.840 and $p_c/\hbar = 0.474$ Å$^{-1}$ and $\alpha$ = 0. For this case, for the liquid at rest ($v_s$ = 0) with specified number of atoms $N$, the number of interacting rotons could not increase further with temperature after the ratio $\sum_i' n_i \Big/ N$ reached its value at $T = T_c$ = 2.172 K. Calculations for this case showed that $\rho_s(0)$ would be positive at temperatures above $T_c$ and so the second sound velocity would be a real number there, indicating that second sound oscillations would be present above $T_c$. Propagating second sound oscillations above $T_\lambda$ (note that we are now supposing $T_c = T_\lambda$) are not observed to occur experimentally, thus eliminating this method for fixing $n/N$.

The relative superfluid density $\rho_s/\rho$ at $v_s$ = 0, evaluated using Eqs. (4.4) through (4.9), is displayed in Fig. 16 along with experimental values based on second and fourth sound measurements. Agreement is within 1% up to about 2.00 K. At 2.00 K the discrepancy is about 3%, and at 2.05 K it is about 7%. The discrepancy is much larger over much of the intervening temperature range up to $T_\lambda$. The calculated curve is shown in Fig. 16 for some temperatures above $T_c$, where $\rho_s/\rho < 0$. The occurrence of $\rho_s < 0$ for some values of $v_s$ and $v_n$ in our theory has been discussed in Ref. [20] and will also be discussed in Sec. 4.1.

Confidence in our theoretical approach is supported by the good agreement between calculated and experimental values of $\rho_s/\rho$ found here over most of the temperature range up to $T_\lambda$. However, the remaining discrepancies indicate that there is a need for further improvement of the theory near $T_\lambda$. We will return to this matter later in Secs. 5 - 7.

### *4.1 Physical interpretation of negative superfluid density*

The formula for $\rho_n$ in Eq. (4.7) was written down by Khalatnikov [101] in his extension of Landau's [14] work. We have also derived that formula in earlier work on microscopic theory of the two fluid model [20], where we considered conditions where the velocities $v_s$ and $v_n$ are not necessarily zero. In that earlier work we also showed that when $v_n$ = 0,

$$\frac{\partial F}{\partial v_s} = V\left(\rho - \rho_n\right) v_s = V \rho_s v_s \ . \tag{4.10}$$

Although in that earlier treatment [20] we established Eq. (4.10) just for the case where the excitations were unconstrained ($\alpha = 0$) and non-interacting ($\gamma = 0$), one can also show that Eq. (4.10) holds when those restrictions are not imposed.

With the aid of Eq. (4.10) one can readily show that $\rho_s = 0$ on each isotherm of $F(v_s)/Nk$ in Fig. 15 at the value of $v_s$ where the isotherm has its relative maximum value. We call that velocity $v_M$ in agreement with our earlier discussion in Sec. 4. For $v_s > v_M$, the superfluid density $\rho_s$ is negative. The necessity of allowing $\rho_s < 0$ for some velocities and temperatures was discussed in some of our earlier work [20], and in fact one finds that $\rho_s < 0$ at all values of $v_s$ when $T > T_c$. In Fig. 16 we displayed $\rho_s < 0$ at $v_s = 0$ for a small range of temperatures above $T_c$. For the present, we just want to explain the physical interpretation of $\rho_s < 0$ and comment on why it does not violate statistical mechanics of the two-fluid model.

Suppose $v_n = 0$ in Eq. (4.7) and let $\boldsymbol{G}$ represent the momentum of the rotons and phonons in thermodynamic equilibrium. Then

$$\begin{aligned} -\rho_n \boldsymbol{v}_s &= \frac{1}{V}\sum_i n_i \boldsymbol{p}_i \\ &= \frac{1}{V}\boldsymbol{G} \end{aligned} \tag{4.11}$$

where $n_i$ is given by Eqs. (4.4), (4.5).

This shows that the thermal average $\boldsymbol{G}$ is always oppositely directed to $\boldsymbol{v}_s$. From Eq. (2.8) we obtain

$$\frac{\boldsymbol{P}_0}{V} = \rho \boldsymbol{v}_s \ , \tag{4.12}$$

so the correlated momentum $\boldsymbol{P}_0$ is always in the same direction as $\boldsymbol{v}_s$. The total momentum $\boldsymbol{P}'$ of the fluid in thermodynamic equilibrium is given by Eq. (4.2), so

$$\begin{aligned} \boldsymbol{P}' &= \left[\boldsymbol{P}_0 + \boldsymbol{G}\right] \\ &= V\left[\rho \boldsymbol{v}_s - \rho_n \boldsymbol{v}_s\right] \\ &= V\left[\rho_s \boldsymbol{v}_s\right]. \end{aligned} \tag{4.13}$$

If $\rho_s > 0$, then the total momentum $\boldsymbol{P}'$ is in the same direction as $\boldsymbol{v}_s$ and the magnitude of $\boldsymbol{P}_0$ must be greater than the magnitude of $\boldsymbol{G}$. If $\rho_s < 0$, the total momentum $\boldsymbol{P}'$ is in the opposite direction to $\boldsymbol{v}_s$, and the magnitude of $\boldsymbol{P}_0$ must be less than the magnitude of $\boldsymbol{G}$. So $\rho_s > 0$ or $\rho_s < 0$ is just determined by whether correlated momentum or elementary excitation momentum gives the dominant contribution to the total momentum.

Even though $\rho_s < 0$ may offend our intuition, there is no problem in understanding its physical interpretation. It is also relevant to note that detailed analysis [20,21] shows that $\rho_s$ is not a natural variable of *any* thermodynamic potential that applies to the *whole* liquid, and there is no conflict with statistical mechanics when $\rho_s < 0$. However, our intuition does not completely misguide us in the question we examined here. There is a matter of thermodynamic stability that is important, but for $T < T_c$, it is related to the *inflection* point *not* the *maximum* in $F$ versus $v_s$. However, stability considerations are important for $T > T_c$ at all values of $v_s$, beginning with the maximum in $F$ versus $v_s$ at $v_s = 0$. We will deal with stability in some detail later in this paper.

*4.2 Specific entropy and specific heat due to phonons and rotons*

Specific entropy for rotons and phonons, evaluated at $v_s = 0$, $v_n = 0$, can be calculated using Eq. (2.9) and (4.4) - (4.6). Then the specific heat at constant volume, $C_V$, can be found using

$$C_V = T\frac{\partial S}{\partial T} \quad . \tag{4.14}$$

For non-interacting $(\gamma = 0)$, unconstrained $(\alpha = 0)$ excitations we have evaluated specific heat based on Landau's theory for non-interacting excitation, but instead of parameters found by Landau [16] we used the excitation spectrum from neutron scattering [79]. There is no critical behavior found in those results, which are shown in Fig. 17. The agreement between theory and experiment is only semi-quantitative.

For phonons and interacting, constrained rotons, the results for specific entropy and specific heat found using Eqs. (2.9) and (4.14) are displayed in Fig. 18, where experimental measurements are also shown. The constraint on rotons can be satisfied with $\alpha = 0$ in the temperature range shown in Fig. 18. Agreement between theory and experiment seems remarkably good when one takes into consideration that the only adjustable parameter used in our theory was the value of roton interaction strength $\gamma$ that was selected in a way that ensured $\rho_s(0) = 0$ at $T = 2.172$ K. Over much of the temperature range shown, agreement between theory and experiment is closer than 3%. However, the specific heat curves for theory and experiment start to deviate significantly around 2.157 K, which is about 15 mK below $T_\lambda$, where critical behavior is indicated experimentally by rapidly increasing specific heat. The plots in Fig. 18 for both theoretical and experimental data were terminated at 2.170 K, about 2 millikelvin below $T_\lambda$, where specific heat theoretical and experimental values in $\mathrm{Jg^{-1}K^{-1}}$ were 10.26 and 13.08, respectively. Data that were not plotted show that at $(T_\lambda - T) =$ 1.3 microkelvin, the specific heat theoretical and experimental values are 10.58 and 22.0 $\mathrm{Jg^{-1}K^{-1}}$, respectively. The theoretical values of $C_V$ change a small amount and to good approximation can be treated as constant in the 2 millikelvin region just below $T_\lambda$. Rotons and phonons in the model, on which the calculated values in Fig. 18 are based, are manifestly not accurately accounting for the rapid increase in specific heat in the critical region, but they just provide a background term in that region. It will be seen in a later part of this paper that vortices may produce a divergent specific heat to be added to that background as the superfluid transition temperature 2.172 K is approached from below.

## 5. Statistical mechanics of non-interacting vortex rings

Statistical mechanical properties of quantized vortices can be calculated by straightforward extension of results found in earlier parts of this paper. This extension will be used in developing a first principles theory of the $\lambda$ transition for liquid $^4$He in complete thermodynamic equilibrium. In the remainder of this paper we shall take $v_n = 0$ unless it is explicitly stated to be otherwise

Our treatment of a system of quantized vortices at finite temperatures is based on the following simple ideas. Quantization of circulation and assumption that non-zero vorticity occurs only on ideally thin straight lines or circular rings enable determination of fluid velocity at every position ***r*** inside the container. The container is assumed to have rigid walls, which fix the boundary conditions. In our model, we include in a basis set, that is assumed to be complete, the topological excitations consisting of individual vortices that are approximate (treated here as exact) eigenstates of energy and momentum as well as density fluctuation excitations consisting of rotons and phonons, like those treated by Landau. It is postulated that density fluctuation excitations do not occur inside vortex cores whose boundaries are fixed by the Landau critical velocity so that the flowing ground state is non-degenerate inside the cores and London's criterion for superfluidity is satisfied inside the cores as well as outside.

In our theory, the liquid *outside the cores* can be imagined as partitioned into cells that are so small that the velocity and average particle density are essentially constant throughout any cell. In each cell the internal energy density per $^4$He atom is the same as the internal energy for a large system with uniformly flowing superfluid, divided by the number of atoms. Entropy density and Helmholtz free energy density are also treated in this way by our model. This approximation is an expression of an assumption of local thermodynamic equilibrium in the liquid. In our model we take the atomic density per unit volume, $n_0$, to be constant outside the vortex cores.

Using this cell model in combination with the postulate for vortex cores, one can arrive at the following formula for the free energy $F_N$ of a system of $N$ quantized vortices that are in any specified configuration:

$$\begin{aligned} F_N &= F_{CN} + \int d^3r\, n_0 f_N(\boldsymbol{r}) \\ &= F_{CN} + F_{EN} \end{aligned} \tag{5.1}$$

where the integral is over the liquid region outside the cores. $F_{EN}$ is the free energy contribution by atoms external to the cores, measured above a background described next. $F_N$ is added to a background term $\mathcal{N}\ Ak$ to obtain the total free energy of the liquid. $\mathcal{N}$ is the total number of [4]He atoms in the liquid. For temperature $T$, $Ak$ is the free energy per atom for the liquid, evaluated at $v_s = 0$, taking into account phonons and rotons only. The value of $A$ is given by the ordinate where an isotherm intercepts the vertical axis in Fig. 15. The factor $f_N(\boldsymbol{r})$ is the free energy per atom at $\boldsymbol{r}$ measured above background $Ak$. The factor $f_N(\boldsymbol{r})$ relies on the assumption of local thermodynamic equilibrium when the velocity at $\boldsymbol{r}$, namely $\boldsymbol{v}(\boldsymbol{r})$, is determined by quantization of circulation and the specified configuration of quantized vortices. $F_{CN}$ is the energy of vortex cores measured above the background $\mathcal{N}\ Ak$.

There are a number of subtle issues that must be carefully treated when one applies Eq. (5.1). Those issues will be identified and treated in applications of Eq. (5.1) to several cases in what follows.

### *5.1 Free energy of one vortex ring*

Here we consider the case where $N = 1$ in Eq. (5.1). The energy $F_{C1}$ is the sum of just two terms in this case. One is the core energy at $T = 0$ K, e.g., shown as $\varepsilon_C$ in Fig. 9. $F_{C1}$ is a function of vortex radius $R$. We note further that the *magnitude* of vortex momentum is a function of $R$. See Fig. 10 for an illustration. Therefore $F_{C1}$ is a function of the magnitude but not the direction, of vortex momentum. The second term in $F_{C1}$ is obtained by multiplying the number $\mathcal{N}_c$ of atoms that remain in the core by $Ak$. The second term is based on an assumption that atoms displaced from the core are distributed uniformly over the region outside the core where velocity $v_s$ is very nearly zero almost everywhere, and the specification that the core energy $F_{C1}$ is referred to the background free energy density $Ak$ and not the ground state energy density $E_0/\mathcal{N}k$, at $v_s = 0$, the reference level in Fig. 15.

For one vortex ring of specified radius and orientation that is not near a boundary (so that images can be neglected), both the velocity $\boldsymbol{v}(\boldsymbol{r})$ and free energy density $f_1(\boldsymbol{r})$ can be readily evaluated using the definitions we have given, and taking $N = 1$ in Eq. (5.1), we have

$$F_1 = F_{C1} + F_{E1} \ . \tag{5.2}$$

Some numerical results for the quantities in Eq. (5.2) will be given in Sec. 5.3.1, after we have developed some further statistical mechanics results.

### *5.2 Statistical mechanics of a system of non-interacting vortex rings*

The partition function $Z_{N0}$ for $N$ non-interacting vortex rings is given by

$$\begin{aligned} Z_{N0} &= \frac{1}{N!} h^{-3N} \int d^3R_1\, d^3R_2 ... d^3R_N \int d^3P_1\, d^3P_2 ... d^3P_N\ e^{-\beta F_N} \\ &= \frac{1}{N!}\left[ V h^{-3} \int d^3P\, e^{-\beta F_1(P)} \right]^N . \end{aligned} \tag{5.3}$$

The grand partition function $\Omega_0$ is given by

$$\Omega_0 = \sum_{0 \le N \le \infty} e^{\beta\mu N} Z_{N0} \ , \tag{5.4}$$

where it is implied that the limit $\mu \to 0$ is taken finally to obtain observable thermodynamic properties in this instance where the number of vortices is not fixed by physical conditions.

Combining Eqs. (5.3) and (5.4) and then carrying out the sum over $N$, one finds

$$\Omega_0 = \exp\left\{ \frac{Ve^{\beta\mu}}{h^3} \int d^3P \ e^{-\beta F_1(P)} \right\} \ . \tag{5.5}$$

The grand canonical potential $W_0$ is given by

$$\begin{aligned} W_0 &= -kT \ln \Omega_0 \\ &= \frac{kTVe^{\beta\mu}}{h^3} \int d^3P \, e^{-\beta F_1(P)} \ . \end{aligned} \tag{5.6}$$

Using standard statistical mechanics methods we find the expectation value of $N$:

$$\begin{aligned} \langle N \rangle &= -\frac{\partial W_0}{\partial \mu} \\ &= \frac{Ve^{\beta\mu}}{h^3} \int d^3P \ e^{-\beta F_1(P)} \end{aligned} \tag{5.7}$$

and taking the limit $\mu \to 0$ one finds that the expected density of vortex rings is

$$\frac{\langle N \rangle}{V} = \frac{4\pi}{h^3} \int_{R_m}^{\infty} dR \ \left( P^2(R) \left| \frac{dP}{dR} \right| \right) e^{-\beta F_1(P(R))} \tag{5.8}$$

where *P*(*R*) is the magnitude of momentum of a vortex ring having a radius *R*. The mean square fluctuation in number of vortex rings is given by

$$\begin{aligned} \left\langle \left( N - \langle N \rangle^2 \right) \right\rangle &= -\frac{1}{\beta} \frac{\partial^2 W_0}{\partial \mu^2} \\ &= \langle N \rangle \, . \end{aligned} \tag{5.9}$$

The value of $R_m$ in Eq. (5.8) is taken to be 2 Å in computer calculations for statistical mechanical properties since physical intuition suggests that vortex rings with radius smaller than this are unlikely to occur in nature. Smaller rings could easily be included in the formulas. Density of states considerations suggest that they would not affect the results significantly.

Using the result in Eq. (5.8), we will call $L$ the average spacing between vortex rings, where

$$L = \left( \frac{V}{\langle N \rangle} \right)^{\frac{1}{3}} . \tag{5.10}$$

Recalling that $P$ is a single valued function of $R$, we will introduce a normalized distribution function $D(R)$ as

$$\begin{aligned} D(R) &= \frac{1}{C} P^2 \left| \frac{dP}{dR} \right| e^{-\beta F_1(P(R))} \\ &= \frac{N(R)}{N_{tot}} \end{aligned} \tag{5.11}$$

where

$$C = \int_{R_m}^{\infty} dR\, P^2 \left| \frac{dP}{dR} \right| e^{-\beta F_1(P(R))} . \tag{5.12}$$

The ensemble average value of $R$, viz, $<R>$ is given by

$$\langle R \rangle = \int_{R_m}^{\infty} dR\, R\, D(R) . \tag{5.13}$$

Formulas for entropy $S$ and heat capacity $C_V$ at constant volume of non-interacting vortex rings are given respectively by

$$S = -\frac{\partial W_0}{\partial T} \tag{5.14}$$

$$C_V = T \left( \frac{\partial S}{\partial T} \right)_V \tag{5.15}$$

where $W_0$ is given by Eq. (5.6). The variables held constant in computing $S$ from Eq. (5.14) should be clear from the earlier discussion.

Note that Eqs. (5.6) and (5.7) imply that for non-interacting rings, we have

$$W_0 = -kT \langle N \rangle , \tag{5.16}$$

a result that is useful in numerical evaluation of $S$ and $C_V$.

### *5.3 Numerical results*

Some numerical results based on these statistical mechanics formulas have been obtained, and they will be treated next.

#### *5.3.1 Free energy components for individual vortex rings*

Numerical results for free energy components for individual vortex rings, corresponding to terms in Eq. (5.2), are displayed in Figs. 19a - 19c, for $T$ = 2.100, 2.160 and 2.190 K. The core free energy $F_{C1}$ changes very little with temperature in this range. However, the external free energy $F_{E1}$ is a strong function of temperature. Referring to the Fig. 15 for free energy per particle at different velocities, one can see how the behavior of isotherms of free energy density at the higher velocities produces the rapid decrease in $F_{E1}$ in Figs. 19a-19c as temperature increases. However, $F_1$ remains large for all vortex rings in the ranges of temperature and radius shown in Figs. 19a - 19c.

$F_{E1}$ is negative for all values of $R$ in the temperature range near $T_c$ covered by Figs. 19a – 19c. It is the contribution from the core energy $F_{C1}$ that makes $F_1$ large, and the exponential factor exp [– $\beta F_1$] in the integrand of Eq. (5.8) small, so that the vortex ring density given by Eq. (5.8) remains small for temperatures near $T_c$. If it were not for this large core energy, the liquid would be essentially filled with vortex rings even below $T_c$, and the statistical mechanical properties of the liquid, including specific heat, would be strongly affected by this flood of vortices near $T_c$.

In Sec. 7.5, it will be shown that in a finite size container, a catastrophe threshold temperature is reached where for large vortex rings, having radius about as big as dimensions of the container, $F_1$ approaches zero over a narrow temperature range. As the size of the container increases without limit, the stated threshold temperature approaches $T_c$ from above.

The cusp in $F_1$ near $R$ = 8 Å is due to the cusp behavior of $\varepsilon_V$ near 8 Å in Fig. 9, which is in turn associated with transition from closed to open vortex rings for radius $R$ in that neighborhood.

#### *5.3.2 Distribution function and average radius*

Numerical results for vortex ring distribution $D(R)$ based on Eqs. (5.11) and (5.12) are displayed in Fig. 20 for a temperature somewhat below $T_c$, viz $T$ = 2.160 K. The plot of $\text{Log}_{10}\, D(R)$ implies that $D(R)$ itself has a narrow peak for rings having $R$ near $R$ = 2 Å, and only rings of this small size are present in sufficient numbers to contribute significantly to thermodynamic properties at this temperature. Additional results for a range of temperatures between about 2.1 and 2.2 K, which includes $T_c$, indicate that Fig. 20 is representative of the physical situation in a finite size system until the threshold temperature mentioned earlier is approached. Near that threshold, $D(R)$ develops a second narrow peak for large values of $R$ that broadens to include smaller vortex rings as temperature increases. Further explanation of this situation will be given in Sec. 7.5 in a discussion of a catastrophic event above $T_c$.

The cusp in $D(R)$ near $R$ = 8 Å in Fig. 20 is associated with the cusp in $F_1$ shown in Fig. 19b, and similar behavior occurs at higher temperatures. The average radius $\langle R \rangle$ is very close to 2 Å for all temperatures up to the neighborhood of the catastrophe threshold temperature.

### *5.3.3 Average spacing between vortex rings*

Numerical results for average spacing L between vortex rings based on Eqs. (5.8) and (5.10) are shown in Fig. 21 for a range of temperatures that includes $T_c$. Vortex rings having $2 \le R \le 100$ Å were taken into account in the calculations, but only those near $R$ = 2 Å were important statistically. The small relative change in the vortex ring length spacing $L$ as the temperature increases is due to the fact that the slowly varying function $F_{C1}$ gives the dominant contribution to $F_1$. Figure 21 shows that the average vortex ring spacing is about 145 Å near $T_c$ = 2.172 K.

### *5.3.4 Specific entropy and specific heat for non-interacting vortex rings*

Numerical results for specific entropy for non-interacting vortex rings based on Eq. (5.14) are displayed in Fig. 22. The local deviations from the trend line are unphysical structure that is associated with inaccuracies in numerically evaluating the exponent $F_1$ that occurs in $\langle N \rangle$ in Eq. (5.8) and consequently in the grand canonical potential $W_0$ in Eq. (5.16). Those inaccuracies are magnified through exponentiation in the formula for $W_0$, and further magnified by numerically evaluating the derivative of $W_0$ that appears in the formula for entropy $S$ in Eq. (5.14). The trend line can be used to roughly evaluate the specific heat $C_V$ due to non-interacting vortex rings using Eq. (5.15), and one finds that for the region shown, $C_V \simeq 5$ mJg$^{-1}$K$^{-1}$. This value of $C_V$ is so small that $C_V$ is unimportant physically, and it just adds a small contribution to the background value due to phonons and rotons alone. That background may be treated as a baseline when treating the λ transition associated with interactions among vortex rings in the next Section of this paper.

## 6. Vortex ring interactions and the λ transition for a simplified model

In this Section it will be shown that vortex ring interactions result in a λ transition in liquid $^4$He when the liquid is in complete thermodynamic equilibrium and treated with a simplified model. Some of the approximations made in that model will eventually be shown to be accurate only in a crossover region that is at least moderately far below $T_\lambda$ and that does not include the region immediately next to $T_\lambda$. In the simplified model the specific heat is singular at the temperature where superfluid density that takes into account only rotons and phonons vanishes. In order to simplify the notation, we will define $v \equiv v_s$.

Formulas in Eq. (5.1) are applicable to a system of interacting vortex rings provided that $f_N(\boldsymbol{r})$, $F_{EN}$, $F_{CN}$, and $\int d^3r$ are evaluated properly. We will make a number of approximations in our first study of these formulas. The effect of some of these approximations is to limit the temperature range of high accuracy of calculated results to the crossover region. What we learn using these approximations will be helpful in our subsequent development, in Sec. 7, of a more complete theory for the critical region that includes the temperature $T_\lambda$ where the specific heat peak occurs.

Using Eqs. (4.1) – (4.6), one can show that Taylor's series expansion about $v = 0$ can be used to define the free energy per atom as

$$\frac{1}{k}\bar{f}(v) \equiv \frac{F(v)}{kN} = A + Bv^2 + Cv^4 + \ldots \tag{6.1}$$

where only even powers of $v$ occur in the remainder terms. The terms shown explicitly in Eq. (6.1) are sufficient for explaining our method for treating vortex ring free energy in the region external to vortex cores. When superfluid velocity $\boldsymbol{v}$ is known at position $\boldsymbol{r}$, we shall define $f(\boldsymbol{r})$ as

$$f(\boldsymbol{r}) = \bar{f}(v(\boldsymbol{r})) - Ak \ . \tag{6.2}$$

Suppose that superfluid velocity at $\boldsymbol{r}$ is due to a fixed configuration of $N$ vortex rings whose positions $\boldsymbol{R}_1, \boldsymbol{R}_2, \ldots \boldsymbol{R}_N$, and momenta $\boldsymbol{P}_1, \boldsymbol{P}_2, \ldots \boldsymbol{P}_N$ are specified. Then

$$\boldsymbol{v}(\boldsymbol{r}) = \sum_i \boldsymbol{v}_i(\boldsymbol{r}) \tag{6.3}$$

and

$$v^2(\boldsymbol{r}) = \sum_i v_i^2(\boldsymbol{r}) + 2\sum_{i<j} \boldsymbol{v}_i(\boldsymbol{r}) \bullet \boldsymbol{v}_j(\boldsymbol{r}) \ . \tag{6.4}$$

The indices $i$ and $j$ range over the $N$ vortex rings.

In the expression for $v^4(\boldsymbol{r})$ there are terms having one through four indices. As an approximation we shall simply neglect the three and four index terms. The one and two index terms in $v^4(\boldsymbol{r})$ give the following result, subject to the stated approximation:

$$\begin{aligned} v^4(\boldsymbol{r}) &= \sum_i v_i^4(\boldsymbol{r}) \\ &+ \sum_{i<j} \left\{ 2v_i^2(\boldsymbol{r}) v_j^2(\boldsymbol{r}) + 4\left(\boldsymbol{v}_i(\boldsymbol{r}) \bullet \boldsymbol{v}_j(\boldsymbol{r})\right)^2 + 4\left(\boldsymbol{v}_i(\boldsymbol{r}) \bullet \boldsymbol{v}_j(\boldsymbol{r})\right) v_i^2(\boldsymbol{r}) + 4\left(\boldsymbol{v}_i(\boldsymbol{r}) \bullet \boldsymbol{v}_j(\boldsymbol{r})\right) v_j^2(\boldsymbol{r}) \right\} . \end{aligned} \tag{6.5}$$

When $v^2(\boldsymbol{r})$ and $v^4(\boldsymbol{r})$ in Eqs. (6.4) and (6.5) are substituted into Eqs. (6.1) and (6.2), the single index terms can be identified with contributions to the external free energy of individual vortices, the same as if the rings were non-interacting. When those terms are supplemented by single index terms that would appear in the remainder indicated by ellipsis in Eq. (6.1), the totality of single index terms is treated by using the exact expression for $F_{E1}$ (see Eqs. (5.1) – (5.2)) and summing over the $N$ vortex rings. These are the external self-energy terms $F_{ESN}$ for the $N$ vortex rings.

The two index terms in Eqs. (6.4) and (6.5) contribute to the external interaction free energy having coefficients *B* and *C*, respectively, in Eq. (6.1).

We will represent the external free energy for $N$ rings as follows:

$$F_{EN} = F_{ESN} + F_{EIN} \tag{6.6}$$

where

$$F_{ESN} = \sum_{1 \le i \le N} F_{E1}(i) \tag{6.7}$$

and $F_{E1}(i)$ is the external energy for the *i*th ring. Formulas for the external interaction terms $F_{EIN}$ are given later in Eqs. (6.11) – (6.15).

Next we shall consider the vortex core contributions to free energy, $F_{CN}$, (see Eq. (5.1) and explanation that follows it) when more than one vortex ring is present. We have already dealt with vortex core self-energy (see Sec. 5.1) for an individual vortex ring, and the core self-energy for $N$ rings is obtained simply by addition. However, there is also a core interaction term for two rings. This interaction term involves the velocity field due to ring *j*, acting as a source, when the field is evaluated at the core of another ring, *i*. If we assume that rings *i* and *j* are fairly far apart, then the velocity at the core of ring *i* due to the distant ring *j* can be treated as constant over the core region of *i*. The assumption of large separation will be justified a little later in this paper. The core interaction terms are evaluated in a reference frame where the correlation function $\psi_0$ corresponds to the ground state of the liquid at rest. Let $\boldsymbol{P}_{MI}(i)$ represent momentum of the mass inside the *i*th core, given by

$$\boldsymbol{P}_{MI}(i) = P_{MI}(i)\hat{\boldsymbol{M}}_i = m\int_{V_{Ci}} d^3r\, n_i(\boldsymbol{r})\,\boldsymbol{v}_i(\boldsymbol{r}) \tag{6.8}$$

where $V_{Ci}$ is the core volume of the *i*th vortex ring and $n_i(\boldsymbol{r})$ and $\boldsymbol{v}_i(\boldsymbol{r})$ are atomic density and superfluid velocity for the *i*th vortex ring treated as if the rings were non-interacting. The core free energy due to interaction of $\boldsymbol{P}_{MI}(i)$ with the velocity field having ring *j* as a source is $\boldsymbol{P}_{MI}(i)\bullet\boldsymbol{v}_j(\boldsymbol{R}_i)$. There is another contribution to core free energy where the roles of rings *i* and *j* are interchanged. Generalizing these results for two vortex rings to a system of $N$ rings, we find the following formula for core free energy $F_{CN}$ that includes both self-energy and core interaction contributions:

$$F_{CN} = \sum_{1 \le i \le N} F_{C1}(i) + \sum_{1 \le i < j \le N} F_{CI}(i,j) \tag{6.9}$$

where

$$F_{CI}(i,j) = \boldsymbol{P}_{MI}(i) \bullet \boldsymbol{v}_j(\boldsymbol{R}_i) + \boldsymbol{P}_{MI}(j) \bullet \boldsymbol{v}_i(\boldsymbol{R}_j) \ . \tag{6.10}$$

The single index term for the core free energy $F_{C1}(i)$ and external free energy $F_{E1}(i)$ for the ith ring combine to give the self-energy $F_1(i)$ as in Eq. (5.2). $F_1(i)$ is a function of the radius $R$ of the singularity in the ring, but does not depend on orientation or position of the ring. The ring radius $R_i$ completely determines the magnitude $P_i$ of the ring momentum and so instead of $F_1(i)$ we shall write $F_1(P_i)$.

The two-index terms combine to give the vortex ring interaction free energy $F_{IN}$, as follows

$$F_{IN} = F_{CIN} + F_{EIN} \tag{6.11}$$

where

$$F_{CIN} = \sum_{1 \le i < j \le N} F_{CI}(i,j) \ , \qquad F_{EIN} = \sum_{1 \le i < j \le N} F_{EI}(i,j) \tag{6.12}$$

$$F_{EI}(i,j) = F_{EB}(i,j) + F_{EC}(i,j) \tag{6.13}$$

$$F_{EB}(i,j) = 2kBn_0 \int d^3r \left\{ \boldsymbol{v}_i(\boldsymbol{r}) \bullet \boldsymbol{v}_j(\boldsymbol{r}) \right\}' \tag{6.14}$$

$$F_{EC}(i,j) = kCn_0 \int d^3r \left\{ 2v_i^2(\boldsymbol{r}) v_j^2(\boldsymbol{r}) + 4\left(\boldsymbol{v}_i(\boldsymbol{r}) \bullet \boldsymbol{v}_j(\boldsymbol{r})\right)^2 + 4\left(\boldsymbol{v}_i(\boldsymbol{r}) \bullet \boldsymbol{v}_j(\boldsymbol{r})\right)\left(v_i^2(\boldsymbol{r}) + v_j^2(\boldsymbol{r})\right) \right\}' \ . \tag{6.15}$$

The precise meaning of the prime on the brackets in these equations will be given in Sec. 6.4.1. We simply note here that the prime implies that non-zero contributions to $F_{EB}(i,j)$ and $F_{EC}(i,j)$ are due to positions $\boldsymbol{r}$ that are far from the centers $\boldsymbol{R}_i$ and $\boldsymbol{R}_j$ of the two vortex rings *i* and *j*. We also note here that the earlier analysis for non-interacting vortex rings indicated that only small rings contribute significantly to statistical mechanical properties and that the average separation of those rings is large compared to the ring radii. These observations provide justification for using a dipole approximation for each of the velocity terms, such as $v_i(\boldsymbol{r})$ and $v_j(\boldsymbol{R}_i)$, that occur in $F_{IN}$. Then, for example, for $\boldsymbol{R}$ measured from the center of a vortex ring, we have

$$\boldsymbol{v}(\boldsymbol{R}) = -\frac{1}{4\pi} M \left[ \frac{\hat{\boldsymbol{M}} - 3\hat{\boldsymbol{M}} \bullet \hat{\boldsymbol{R}}\, \hat{\boldsymbol{R}}}{\boldsymbol{R}^3} \right] \tag{6.16}$$

where $M$ is the strength of the dipole moment of a singly quantized vortex ring,

$$M = K\pi R^2 \tag{6.17}$$

where $K = h/m$ and $R$ is the radius of the ring. For the vortex ring model that we are considering, where the liquid mass density in the core is nearly everywhere less than the average density of the liquid, the direction $\hat{\boldsymbol{P}}$ of vortex ring momentum is related to direction $\hat{\boldsymbol{M}}$ of the vortex ring dipole moment by

$$\hat{\boldsymbol{P}} = -\hat{\boldsymbol{M}} \; . \tag{6.18}$$

Here we remind the reader that we are using the term "vortex ring momentum" to mean the momentum of the liquid when a vortex ring is present.

In what follows we shall sometimes use a relation implied by Eq. (6.18), viz. $\int d^2\hat{P} = \int d^2\hat{M}$ , when the integrals are over 4π steradians.

Using the foregoing relations, one finds that the free energy $F_N$ for a system of $N$ interacting vortex rings having specified positions $\boldsymbol{R}_1, \boldsymbol{R}_2, \ldots \boldsymbol{R}_N$ and momenta $\boldsymbol{P}_1, \boldsymbol{P}_2, \ldots \boldsymbol{P}_N$ is given by

$$\begin{aligned} &F_N\left(\boldsymbol{R}_1, \boldsymbol{R}_2, \ldots \boldsymbol{R}_N; \boldsymbol{P}_1, \boldsymbol{P}_2, \ldots \boldsymbol{P}_N\right) \\ &= \left\{ F_{SN}\left(P_1, P_2, \ldots P_N\right) + F_{IN}\left(\boldsymbol{R}_1, \boldsymbol{R}_2, \ldots \boldsymbol{R}_N; \boldsymbol{P}_1, \boldsymbol{P}_2, \ldots \boldsymbol{P}_N\right) \right\} . \end{aligned} \tag{6.19}$$

Here

$$F_{SN}\left(P_1, P_2, \ldots P_N\right) = \sum_{1 \le i \le N} F_1\left(P_i\right) \tag{6.20}$$

represents the self-energy terms and $F_{IN}$ is the interaction free energy given by Eqs. (6.11) – (6.13) and the interaction terms in Eqs. (6.9) and (6.10).

### *6.1 The partition function for interacting vortex rings*

The partition function $Z_N$ for $N$ *interacting* vortex rings is given by the right hand side of the first line of Eq. (5.3) provided that one evaluates $F_N$ as in Eqs. (6.19), (6.20). It will be shown later that the interaction terms are small near the transition temperature, and so it is useful to expand the interaction contribution to the exponential in Eq. (5.3). One obtains

$$Z_N = \frac{1}{N!} h^{-3N} \int d^3R_1\, d^3R_2 \ldots d^3R_N \int d^3P_1\, d^3P_2 \ldots d^3P_N$$
$$\times e^{-\beta F_{SN}} \left\{ 1 + \left(-\beta F_{IN}\right) + \frac{1}{2}\left(-\beta F_{IN}\right)^2 + \ldots \right\} . \tag{6.21}$$

For any function $X$ let $\langle X \rangle$ be an expectation value computed using the distribution function for non-interacting vortex rings, as follows:

$$\langle X \rangle = \frac{\int d^3R_1\, d^3R_2 \ldots d^3R_N \int d^3P_1\, d^3P_2 \ldots d^3P_N\; e^{-\beta F_{SN}} X}{\int d^3R_1\, d^3R_2 \ldots d^3R_N \int d^3P_1\, d^3P_2 \ldots d^3P_N\; e^{-\beta F_{SN}}} . \tag{6.22}$$

Then $Z_N$ can be written as

$$Z_N = e^{-\beta \bar{F}_{N0}} \left[ 1 + \left\langle -\beta F_{IN} \right\rangle + \left\langle \tfrac{1}{2}\left(-\beta F_{IN}\right)^2 \right\rangle + \ldots \right] . \tag{6.23}$$

Here $\bar{F}_{N0}$ is the free energy for a thermally distributed system of $N$ non-interacting vortex rings. It is related to $Z_{N0}$ in Eqs. (5.3) and (5.4) as follows:

$$\bar{F}_{N0} = -kT \ln Z_{N0} . \tag{6.24}$$

Before proceeding with evaluation of $Z_N$ we will deal with the region of integration over $d^3r$ in interaction terms, indicated by the prime notation in formulas for $F_{EB}(i,j)$ and $F_{EC}(i,j)$ in Eqs. (6.14) and (6.15), respectively.

### *6.1.1 Region of integration*

The region of integration over $d^3r$ in the interaction terms (see Eqs. (5.1), (6.6), and (6.11)) for quantized vortices is of central importance in exhibiting the critical properties of the liquid near the λ transition temperature. It will be seen that this region is related to thermodynamic stability, as indicated later in the discussion in Secs. 6.1.2 – 6.1.5.

### *6.1.2 Thermodynamic stability*

There is an inflection point at some velocity, called $v_0$ here, in each of the isotherms where $0 < T < T_c$, as one can see in Fig. 15. The isotherm is concave upward for $v_s < v_0$ and concave downward for $v_s > v_0$. According to ordinary thermodynamic criteria, the liquid is stable against small fluctuations in $v_s$ for $0 < v_s < v_0$ and unstable for $v_s > v_0$. These stability properties against small fluctuations can be derived by considering $d^2u$, the second order differential of internal energy per unit mass and requiring that it be non-negative, as shown by Callen [102] and derived earlier by Tisza [103]. Ordinary systems in complete thermodynamic equilibrium do not occur in nature under conditions where they are unstable against small fluctuations.

If a quantized vortex having vorticity concentrated in an ideally thin path occurs in the liquid, then the velocity exceeds $v_0$ for some spatial region as the singularity is approached, and that region of instability expands to large distances as $T$ approaches $T_c$. What is the correct way to treat the liquid in the region where it is ostensibly unstable and still outside the vortex core? The answer that we propose is based on the following principle.

### *6.1.3 Principle of constrained instability*

This principle is a postulate of our theory. It is concerned with different behavior of the liquid in regions that are unstable or stable as determined by ordinary thermodynamic criteria with respect to fluctuations in the superfluid velocity $v_s$. Here we consider the case where there are no externally induced flows in the liquid, although it must be possible to state a similar principle for more general conditions. We assume that the superfluid velocity is everywhere consistent with quantization of circulation. Then specification of a configuration of quantized vortex rings, including their sizes, orientations, and positions determines the superfluid velocity throughout the liquid. We may picture fluctuations in $\boldsymbol{v}_s$ at position $\boldsymbol{r}$ associated with small changes of orientation of a vortex ring relative to another ring as one example. Such fluctuations will be consistent with quantization of circulation.

The *principle of constrained instability* asserts that in an unstable region the free energy of the liquid is determined by local thermodynamic equilibrium at the local value of $\boldsymbol{v}_s$ that is determined by quantization of circulation and the vortex ring configuration. Furthermore, the principle asserts that in this unstable region there is no restoring force that counteracts small fluctuations in $\boldsymbol{v}_s$. The principle further states that in a stable region the free energy of the liquid is determined by local thermodynamic equilibrium at the local value of $\boldsymbol{v}_s$ and that a restoring force counteracts small fluctuations in $\boldsymbol{v}_s$ there, just as in ordinary thermodynamics. The restoring force is a consequence of an increase in free energy above the thermodynamic equilibrium value when $\boldsymbol{v}_s$ fluctuates away from its value at equilibrium.

Implications of this principle for evaluation of self-energy and interaction terms will be explained next.

### *6.1.4 Self-energy terms*

For an individual, single quantum vortex ring at position $\boldsymbol{R}$, having specified radius and orientation, which implies specified momentum, the velocity is completely determined throughout the liquid by the quantization of circulation. For such a ring the velocity cannot fluctuate even in a region where according to ordinary thermodynamic criteria the liquid would be unstable with respect to fluctuations in $v_s$. It is as if a ball was in equilibrium at the peak of a mountain and could not roll down to the valley below and attain a state of lower energy because the ball was constrained by a spike driven through it and into the mountain. In the case of liquid $^4$He, it is as if $\boldsymbol{v}_s$ were a mechanical variable in the sense treated in earlier work [20]. Even if the vortex ring were perturbed, the self-energy of that ring would not change because the velocity field would follow the ring instantaneously since the model we are considering neglects retardation effects. Only the parameters that specify the orientation and position of the ring would change and the self-energy of a thermal distribution of non-interacting rings would not be affected by perturbations. On this basis, we assume that the self-energy terms treated in Eq. (6.7) are still applicable even when perturbations are present.

*6.1.5 Vortex ring interaction terms*

Here we are interested in vortex ring pair interactions, and at first we will deal with external terms in Eqs. (6.14) and (6.15) which will be analyzed next. Any position $\boldsymbol{r}$ where the resultant velocity magnitude $v_s(\boldsymbol{r})$ is such that $v_s(\boldsymbol{r}) > v_0$ will be in the unstable region and will not contribute to a restoring force that counteracts a perturbation of the vortex pair. Only in the stable region will there be a contribution to a restoring force that counteracts perturbations that tend to change the configuration of the vortex pair. Except very near $T_c$ it is easy to see that the boundary of the spatial regions wherein $v_s > v_0$ is determined by $v_s$ in locations where velocity varies rapidly as a function of distance from the closest vortex ring for rings that are at least moderately far apart. Then a boundary point at a radius $R_0$ from the closest ring is approximately the same as that determined by the individual rings. This case is illustrated in Fig. 23a. As an approximation we shall use this same method for determining the boundary of the unstable regions for a vortex pair under more general conditions. These conditions include $T$ near $T_c$ and, alternatively, closely spaced rings at any $T < T_c$. The boundary for these cases is illustrated in Fig. 23b.

For a ring having dipole moment $\boldsymbol{M}$, we find from Eq. (6.16) that very near $T_c$, for both closed and open rings, the radius from the center of the ring determined by $v_0$ is given by

$$R(\theta) = \left[ \frac{1}{4\pi} M \frac{\left(1 + 3\left(\hat{\boldsymbol{M}} \bullet \hat{\boldsymbol{R}}\right)^2\right)^{\frac{1}{2}}}{v_0} \right]^{\frac{1}{3}} . \tag{6.25}$$

The boundary of the stable region has a small angle dependence, but as an approximation we shall take the boundary to be the sphere with the radius $R_0$ where $\hat{\boldsymbol{M}} \bullet \hat{\boldsymbol{R}} = 1$; then

$$R_0 = \left\{ \frac{M}{2\pi v_0} \right\}^{\frac{1}{3}} . \tag{6.26}$$

As an aside, it should be mentioned here that the mantle region whose outer boundary is determined by $R(\theta)$ and $v_0$ is important in understanding second sound absorption and certain ion propagation experiments in superfluid helium. This subject is treated in Appendix C.

Evaluation of $Z_N$ in Eq. (6.23) could now be carried out in principle for very general conditions since all relevant quantities have been defined. However, we shall henceforth proceed by assuming that to good approximation the dipole moment $M$ is the same for all vortex rings that are statistically important for temperatures up to and slightly beyond $T_c$ in a liquid in a finite container (but just up to $T_c$ in the limit of an infinite container). This is consistent with our earlier treatment of non-interacting vortex rings for a limited range $R \leq$ 100 Å of variational wave functions where $R \simeq 2$ Å was found to be statistically dominant. The component $P_{MI}$ of inside mass momentum in the direction $\hat{\boldsymbol{M}}$ in Eq. (6.8) is then also taken to be the same for all vortex rings that are statistically important.

### *6.2 Order parameter*

The velocity $v_0$ at which the inflection point occurs in the free energy for temperatures near, but somewhat below, $T_c$ can be evaluated with the aid of Eq. (6.1). The condition $\partial^2 \bar{f}\left(v_s\right) / \partial v_s \partial v_s = 0$ for leading order terms yields

$$v_0 = \left(-\frac{B}{6C}\right)^{\frac{1}{2}} . \tag{6.27}$$

In a straightforward calculation [104] one can start with the free energy $F(v_s)$ in Eq. (4.3), make a Taylor's series expansion about $v_s = 0$, and show that the $B$ coefficient in Eq. (6.1) is related to the superfluid density found from Eqs. (4.8) and (4.9) by

$$B = \frac{1}{2kn_0}\rho_s\left(0\right) . \tag{6.28}$$

It is found that near $T_c$,

$$\frac{\rho_s\left(0\right)}{n_0 m} = \alpha\left(T_c - T\right) . \tag{6.29}$$

Furthermore, near $T_c$ the coefficient $C$ (in Eq. (6.1) is a slowly varying function of $T$. Numerical calculations that use the coupling strength $\gamma$ specified earlier for low pressure, yield the following results for $\alpha$ and $C$ evaluated at $T_c$:

$$\alpha = 3.56 \text{ K}^{-1} \tag{6.30}$$

$$C = -0.1848 \times 10^{-6} \text{ K}^{-1}\,\text{m}^{-4}\,\text{s}^{4} . \tag{6.31}$$

The velocity $v_0$ governs the extent of the region ($v_s < v_0$) in which the liquid is stable against small fluctuations in $v_s$, and we shall call $v_0$ the *order parameter* in this theory. It is greater than zero below $T_c$ and has imaginary values above $T_c$, where there is no inflection point in $\bar{f}(v_s)$ and where $v_0$ is not a physically meaningful parameter. It is readily seen that near $T_c$,

$$v_0 \propto \left(\rho_s(0)\right)^{\frac{1}{2}} \propto \left(T_c - T\right)^{\frac{1}{2}} \tag{6.32}$$

The kind of order produced by the action of $v_0$ will be explained a little later.

### 6.3 Correlation length

For the vortex ring model that we are considering, where there are no vortex lines present for the liquid in complete thermodynamic equilibrium, there is a unique correlation length, $R_0$, that is the distance from the center of a vortex ring to its mantle boundary where $v_s = v_0$. This correlation length temperature dependence can be found using Eqs. (6.17) and (6.26) - (6.31). For $M$ corresponding to $R$ = 2 Å, one finds

$$R_0 = a_0\left(T_c - T\right)^{-\frac{1}{6}} \tag{6.33}$$

where

$$a_0 = 4.156 \ \text{Å}\,\text{K}^{\frac{1}{6}} \ . \tag{6.34}$$

The nature of the correlations will be explained in Sec. 7.1.1.

### 6.4 Critical behavior near $T_c$

Evaluation of $Z_N$ in Eq. (6.23) provides means for examining possible critical behavior near $T_c$. We will eventually show that a portion of the partition function term $\left\langle \frac{1}{2}\left(-\beta F_{IN}\right)^2 \right\rangle$ is responsible for a singularity in specific heat, and so we shall deal with that item next. Later we will examine other parts of $Z_N$. Although evaluation of $Z_N$ may at first seem like a formidable task given the complexity of $F_{IN}$, we will see that at least through second order terms all relevant integrals can be carried out analytically.

Referring to Eqs. (6.11) and (6.12) one finds a “cross” term when the expression $\left\langle \frac{1}{2}\left(-\beta F_{IN}\right)^2 \right\rangle$ is expanded, and we will call it $Y_{IN}$. Then we have

$$Y_{IN} = \beta^2 \left\langle \sum_{i<j}\sum_{k<l} F_{CI}(i,j) F_{EI}(k,l) \right\rangle \ . \tag{6.35}$$

With the aid of Eq. (6.18), one can write

$$\int d^3P = \int P^2\,dP \int d^2\hat{\boldsymbol{P}} = \int P^2\,dP \int d^2\hat{\boldsymbol{M}} \ , \tag{6.36}$$

the angular integrals ranging over 4π steradians. Only the terms where $i = k$ and $j = \ell$ are non-zero when one integrates over momentum directions in Eq. (6.35). This can be readily verified using Eqs. (6.8) and (6.11) – (6.18). Using the assumptions stated earlier that $M$ and $P_{MI}$ are essentially the same for all vortex rings that are statistically important and that each of the $N(N-1)/2$ distinct pairs $(i,j)$ contributes the same amount to $Y_{IN}$, one can write Eq. (6.35) as

$$Y_{IN} = \frac{\beta^2 N(N-1)\int d^3R_1\, d^3R_2 \int d^3P_1\, d^3P_2\; e^{-\beta[F_1(P_1)+F_1(P_2)]}F_{CI}(1,2)F_{EI}(1,2)}{2\int d^3R_1\, d^3R_2 \int d^3P_1\, d^3P_2\; e^{-\beta[F_1(P_1)+F_1(P_2)]}} \ . \tag{6.37}$$

In the denominator we have indicated explicitly that the self-energy is a function of the magnitude $P_1$ of vortex ring momentum and similarly for ring 2. Integrals over phase space for vortex rings 3 thru $N$ cancel in the numerator and denominator of Eq. (6.22) when that equation is applied to the term involving rings 1 and 2, for example. Using expressions specified earlier for quantities that occur in Eq. (6.37), one finds that Eq. (6.37) can be written as

$$Y_{IN} = Y_{IN,B} + Y_{IN,C} \tag{6.38}$$

where

$$\begin{aligned} Y_{IN,B} &= \beta^2 \frac{N(N-1)}{2} \frac{1}{\left[\int d^3R \int d^3P\, e^{-\beta F_1(P)}\right]^2} \\ &\times\left\{\int d^3R_1\, d^3R_2 \int d^3P_1\, d^3P_2\; e^{-\beta[F_1(P_1)+F_2(P_2)]}\, 2P_{MI}\left[\hat{\boldsymbol{M}}_1\bullet\boldsymbol{v}_2(\boldsymbol{R}_1)\right] 2kBn_0 \int d^3r \left[\boldsymbol{v}_1(\boldsymbol{r})\bullet\boldsymbol{v}_2(\boldsymbol{r})\right]'\right\} \end{aligned} \tag{6.39}$$

$$\begin{aligned} Y_{IN,C} &= \beta^2 \frac{N(N-1)}{2} \frac{1}{\left[\int d^3R \int d^3P\, e^{-\beta F_1(P)}\right]^2} \\ &\times\left\{\begin{array}{l} \int d^3R_1\, d^3R_2 \int d^3P_1\, d^3P_2\; e^{-\beta[F_1(P_1)+F_2(P_2)]}\, 2P_{MI}\left[\hat{\boldsymbol{M}}_1\bullet\boldsymbol{v}_2(\boldsymbol{R}_1)\right] k_B C n_0 \times \\ \int d^3r\left[2\boldsymbol{v}_1^2(\boldsymbol{r})\boldsymbol{v}_2^2(\boldsymbol{r}) + 4\left(\boldsymbol{v}_1(\boldsymbol{r})\bullet\boldsymbol{v}_2(\boldsymbol{r})\right)^2 + 4\left(\boldsymbol{v}_1(\boldsymbol{r})\bullet\boldsymbol{v}_2(\boldsymbol{r})\right)\left(\boldsymbol{v}_1^2(\boldsymbol{r})+\boldsymbol{v}_2^2(\boldsymbol{r})\right)\right]' \end{array}\right\} . \end{aligned} \tag{6.40}$$

Only the $Y_{IN,B}$ contribution to $Y_{IN}$ will be fully evaluated in this paper, and I assume that this will give the more important contribution to $Y_{IN}$. However, a method for completely evaluating $Y_{IN,C}$ will also be indicated. In the absence of a complete answer, dimensional analysis will be used to establish that $Y_{IN,C}$ has the same dependence on ($T_c$ – $T$) as $Y_{IN,B}$, subject to assumptions made in this theory.

*6.4.1 Evaluation of quadratic interaction integral*

The first step in treating $Y_{IN,B}$ in Eq. (6.39) is to evaluate the quadratic interaction integral $J$, where

$$J \equiv \int d^3r \left[\boldsymbol{v}_1(\boldsymbol{r}) \bullet \boldsymbol{v}_2(\boldsymbol{r})\right]' \quad . \tag{6.41}$$

The prime on the bracket means that the integrand is zero for $|\boldsymbol{r} - \boldsymbol{R}_1| < R_0$ or $|\boldsymbol{r} - \boldsymbol{R}_2| < R_0$ or both. The excluded volume is in the unstable region where there is no restoring force to counteract a perturbation of the vortex pair, as discussed earlier.

The component of fluid velocity normal to the walls of the container due to either vortex ring is zero. This condition is imposed by images that are implied, but not represented explicitly, here. A dipole approximation for vortex rings is applicable because the ring radius is much less than $R_0$ near $T_\lambda$, the temperature range considered here. The velocity potential for vortex ring 1 (image implied but not shown) is

$$\Phi_1(\boldsymbol{r}) = \frac{M\,\hat{\boldsymbol{M}}_1 \bullet (\boldsymbol{r} - \boldsymbol{R}_1)}{4\pi|\boldsymbol{r} - \boldsymbol{R}_1|^3} \quad . \tag{6.42}$$

The velocity at $\boldsymbol{r}$ due to ring 1 is

$$\begin{aligned} \boldsymbol{v}_1(\boldsymbol{r}) &= -\nabla_r \Phi_1(\boldsymbol{r}) \\ &= -\frac{M}{4\pi}\left[\frac{\hat{\mathrm{M}}_1}{|\boldsymbol{r} - \boldsymbol{R}_1|^3} - \frac{3\hat{\mathrm{M}}_1 \bullet (\boldsymbol{r} - \boldsymbol{R}_1)(\boldsymbol{r} - \boldsymbol{R}_1)}{|\boldsymbol{r} - \boldsymbol{R}_1|^5}\right]. \end{aligned} \tag{6.43}$$

Similar expressions hold for ring 2.

Refer to Fig. 23 for geometry and notation used in evaluating $J$. Surfaces $S_1$ and $S_2$ are spheres determined by $R_0$ for rings 1 and 2, respectively. $S_0$ is the liquid surface at the container walls and is not shown in Fig. 23. Gauss's divergence theorem is used to evaluate $J$; then

$$
\begin{aligned}
J &= \int d^3r \left[ \boldsymbol{v}_1(\boldsymbol{r}) \bullet \boldsymbol{v}_2(\boldsymbol{r}) \right]' \\
&= \int d^3r \left[ \left(-\nabla\Phi_1(\boldsymbol{r})\right) \bullet \left(-\nabla\Phi_2(\boldsymbol{r})\right) \right]' \\
&= \frac{1}{2} \int d^3r \, \nabla \bullet \left[ \Phi_1 \nabla\Phi_2 + \Phi_2 \nabla\Phi_1 \right]' \\
&= \frac{1}{2} \int\limits_{S_0+S_1+S_2} d^2a \; \hat{\boldsymbol{n}} \bullet \left[ \Phi_1 \nabla\Phi_2 + \Phi_2 \nabla\Phi_1 \right] .
\end{aligned}
\tag{6.44}
$$

The integral over $S_0$ vanishes due to effects of images. The following notation is introduced:

$$J_{1a} = \frac{1}{2} \int\limits_{S_1} d^2a \; \hat{\boldsymbol{n}} \bullet \left[ \Phi_1 \nabla\Phi_2 \right] \tag{6.45}$$

$$J_{1b} = \frac{1}{2} \int\limits_{S_1} d^2a \; \hat{\boldsymbol{n}} \bullet \left[ \Phi_2 \nabla\Phi_1 \right] \tag{6.46}$$

$$J_{2a} = \frac{1}{2} \int\limits_{S_2} d^2a \; \hat{\boldsymbol{n}} \bullet \left[ \Phi_1 \nabla\Phi_2 \right] \tag{6.47}$$

$$J_{2b} = \frac{1}{2} \int\limits_{S_2} d^2a \; \hat{\boldsymbol{n}} \bullet \left[ \Phi_2 \nabla\Phi_1 \right] \; . \tag{6.48}$$

Now Eq. (6.44) can be expressed as

$$J = J_1 + J_2 \tag{6.49}$$

where

$$J_1 = J_{1a} + J_{1b} \quad , \quad J_2 = J_{2a} + J_{2b} \; . \tag{6.50}$$

The following notation (see Fig. 23) will be used,

$$\boldsymbol{s} = \boldsymbol{r} - \boldsymbol{R}_1 \quad , \quad \boldsymbol{t} = \boldsymbol{r} - \boldsymbol{R}_2 \quad , \quad \boldsymbol{u} = \boldsymbol{R}_2 - \boldsymbol{R}_1 \; . \tag{6.51}$$

Note that on $S_1$, $\hat{\boldsymbol{n}} = -\hat{\boldsymbol{s}}$ and $s = R_0$, and on $S_2$, $\hat{\boldsymbol{n}} = -\hat{\boldsymbol{t}}$ and $t = R_0$. Using Eq. (6.51), one obtains the following relations:

$$\boldsymbol{t} = \boldsymbol{s} - \boldsymbol{u} \tag{6.52}$$

$$t^2 = s^2 + u^2 - 2\boldsymbol{s} \bullet \boldsymbol{u} = s^2 + u^2 - 2su\cos\theta \tag{6.53}$$

$$\boldsymbol{s} \bullet \boldsymbol{u} = \frac{1}{2}\left(s^2 + u^2 - t^2\right) \tag{6.54}$$

$$\hat{\boldsymbol{s}} \bullet \hat{\boldsymbol{u}} = \frac{s^2 + u^2 - t^2}{2su} = \cos\theta \ . \tag{6.55}$$

For $s$ and $u$ held constant (like on $S_1$),

$$\sin\theta \, d\theta = \frac{1}{us} t \, dt \ . \tag{6.56}$$

Also,

$$\boldsymbol{s} = \boldsymbol{t} + \boldsymbol{u} \tag{6.57}$$

$$\boldsymbol{t} \bullet \boldsymbol{u} = \frac{1}{2}\left(s^2 - t^2 - u^2\right) \tag{6.58}$$

$$\hat{\boldsymbol{t}} \bullet \hat{\boldsymbol{u}} = \frac{\left(s^2 - t^2 - u^2\right)}{2tu} \ . \tag{6.59}$$

Furthermore,

$$\boldsymbol{t} = \boldsymbol{s} - \boldsymbol{u} \tag{6.60}$$

$$\boldsymbol{s} \bullet \boldsymbol{t} = \frac{1}{2}\left(s^2 - u^2 + t^2\right) \tag{6.61}$$

$$\hat{\boldsymbol{s}} \bullet \hat{\boldsymbol{t}} = \frac{s^2 - u^2 + t^2}{2st} \ . \tag{6.62}$$

$J_{1a}$ can now be evaluated, and the first step is to write it as follows:

$$\begin{aligned} J_{1a} &= \frac{1}{2} s^2 \int_{\theta=0}^{\pi} d\theta \sin\theta \int_{\varphi=0}^{2\pi} d\varphi \, \Phi_1 \hat{\boldsymbol{s}} \bullet \left(-\nabla\Phi_2\right) \\ &= -\frac{M^2}{2(4\pi)^2} \frac{1}{us} \int_{t_1}^{t_2} dt \, \frac{1}{t^2} \int_{\varphi=0}^{2\pi} d\varphi \left(\hat{\boldsymbol{s}} \bullet \hat{\boldsymbol{M}}_1\right)\left[\hat{\boldsymbol{s}} \bullet \hat{\boldsymbol{M}}_2 - 3\left(\hat{\boldsymbol{s}} \bullet \hat{\boldsymbol{t}}\right)\left(\hat{\boldsymbol{t}} \bullet \hat{\boldsymbol{M}}_2\right)\right]. \end{aligned} \tag{6.63}$$

We now introduce a Cartesian coordinate system and also spherical coordinates consistent with Fig. 23, and where the polar axis is in the $\hat{\boldsymbol{u}}$-direction. The $x$-axis is chosen so that $\hat{\boldsymbol{M}}_1$ is in the $x$-$z$ plane. Then we can represent the relevant vectors as follows:

$$\hat{\boldsymbol{M}}_1 = \sin\delta \, \hat{\boldsymbol{x}} + \cos\delta \, \hat{\boldsymbol{z}} \tag{6.64}$$

$$\hat{\boldsymbol{M}}_2 = \sin\alpha\cos\beta \, \hat{\boldsymbol{x}} + \sin\alpha\sin\beta \, \hat{\boldsymbol{y}} + \cos\alpha \, \hat{\boldsymbol{z}} \tag{6.65}$$

$$\hat{\boldsymbol{s}} = \sin\theta\cos\varphi \, \hat{\boldsymbol{x}} + \sin\theta\sin\varphi \, \hat{\boldsymbol{y}} + \cos\theta \, \hat{\boldsymbol{z}} \tag{6.66}$$

$$\hat{\boldsymbol{t}} = \frac{\boldsymbol{t}}{t} = \frac{1}{t}\left(s\,\hat{\boldsymbol{s}} - u\,\hat{\boldsymbol{u}}\right) \tag{6.67}$$

$$\hat{\boldsymbol{u}} = \hat{\boldsymbol{z}} \ . \tag{6.68}$$

Using Eqs. (6.64) – (6.68), one can easily carry out the integration over $\varphi$ that occurs in Eq. (6.63). Then after some simple algebra and use of Eq. (6.55) the result can be expressed as

$$J_{1a} = -\frac{M^2}{2(4\pi)^2}\frac{2\pi}{u\,s^2} \times \left\{ \begin{array}{l} \frac{1}{2}\sin\delta\,\sin\alpha\,\cos\beta \int_{t_1}^{t_2} dt\,\frac{s}{t^2}\left[1 - \frac{\left(s^2+u^2-t^2\right)^2}{4s^2u^2}\right]\left[1 - \frac{3s}{t}\frac{\left(s^2-u^2+t^2\right)}{2st}\right] \\ +\cos\delta\,\cos\alpha \int_{t_1}^{t_2} dt\,\frac{s}{t^2}\left[\begin{array}{l}\frac{\left(s^2+u^2-t^2\right)^2}{4s^2u^2}\left(1-\frac{3s}{t}\frac{\left(s^2-u^2+t^2\right)}{2st}\right) \\ +\frac{\left(s^2+u^2-t^2\right)}{2su}\frac{3u}{t}\frac{\left(s^2-u^2+t^2\right)}{2st}\end{array}\right] \end{array} \right\} . \tag{6.69}$$

Turning next to evaluation of $J_{1b}$, we can write Eq. (6.46) as

$$\begin{aligned} J_{1b} &= \frac{M^2}{2}s^2\int_{\theta=0}^{\pi} d\theta\,\sin\theta \int_{\varphi=0}^{2\pi} d\varphi\;\Phi_2\;\hat{\boldsymbol{s}}\bullet\left(-\nabla\Phi_1\right) \\ &= \frac{M^2}{2(4\pi)^2}\frac{2}{u\,s^2}\int_{t_1}^{t_2} dt \int_{\varphi=0}^{2\pi} d\varphi\,\frac{1}{t}\left(\hat{\boldsymbol{t}}\bullet\hat{\boldsymbol{M}}_2\right)\left(\hat{\boldsymbol{s}}\bullet\hat{\boldsymbol{M}}_1\right). \end{aligned} \tag{6.70}$$

Using Eqs. (6.55), (6.56) and (6.64) – (6.68) and integrating over $\varphi$, one finds that $J_{1b}$ can be expressed as

$$J_{1b} = -\frac{M^2}{2(4\pi)^2}\frac{2\pi}{u\,s^2}\left\{ \begin{array}{l} -\sin\delta\,\sin\alpha\,\cos\beta\int_{t_1}^{t_2} dt\,\frac{s}{t^2}\left[1-\frac{\left(s^2+u^2-t^2\right)^2}{4s^2u^2}\right] \\ +\,2\cos\delta\,\cos\alpha\int_{t_1}^{t_2} dt\,\frac{s}{t^2}\left[\frac{u}{s}\frac{\left(s^2+u^2-t^2\right)}{2su} - \frac{\left(s^2+u^2-t^2\right)^2}{4s^2u^2}\right] \end{array}\right\} . \tag{6.71}$$

The following result for $J_1$ can then be obtained with the aid of Eqs. (6.64), (6.65), and (6.68):

$$J_1 = -\frac{M^2}{4\pi}\left[\hat{\boldsymbol{M}}_1 \bullet \hat{\boldsymbol{M}}_2 - 3\left(\hat{\boldsymbol{M}}_1 \bullet \hat{\boldsymbol{u}}\right)\left(\hat{\boldsymbol{M}}_2 \bullet \hat{\boldsymbol{u}}\right)\right] K_{1a} + \frac{M^2}{4\pi}\left(\hat{\boldsymbol{M}}_1 \bullet \hat{\boldsymbol{u}}\right)\left(\hat{\boldsymbol{M}}_2 \bullet \hat{\boldsymbol{u}}\right) K_{1b} \tag{6.72}$$

where

$$K_{1a} = \int_{t_1}^{t_2} dt \; f\left(s,t,u\right) \tag{6.73}$$

$$K_{1b} = \int_{t_1}^{t_2} dt \; g\left(s,t,u\right) \tag{6.74}$$

and

$$f\left(s,t,u\right) = -\frac{1}{8ust^2}\left[1 - \frac{\left(s^2 + u^2 - t^2\right)^2}{4s^2u^2}\right]\left[1 + \frac{3s}{t}\frac{\left(s^2 - u^2 + t^2\right)}{2st}\right] \tag{6.75}$$

$$g\left(s,t,u\right) = -\frac{1}{4ust^2}\left\{\frac{\left(s^2 + u^2 - t^2\right)^2}{2s^2}\left[2 + \frac{3s}{t}\frac{\left(s^2 - u^2 + t^2\right)}{2st}\right] - \left[1 + \frac{3s}{t}\frac{\left(s^2 - u^2 + t^2\right)}{2st}\right]\right\} \; . \tag{6.76}$$

Notice that $J_1$ in Eq. (6.72) is given in terms of quantities that are independent of the choice of the $x$, $y$, $z$ axes and the corresponding spherical coordinates. This observation is useful in evaluating $J_2$, where the coordinate systems can be chosen with orientations most useful for that problem. It is then easy to see that $J_2 = J_1$, and then

$$J = 2J_1 \; . \tag{6.77}$$

The integration limits, $t_1$, and $t_2$, in Eqs. (6.73), (6.74) can be easily found with the aid of Fig. 23, and the results are as follows.

$$\text{For} \quad s < \frac{1}{2}u: \quad t_1 = u - s \, , \quad t_2 = u + s \; . \tag{6.78}$$

$$\text{For} \quad s > \frac{1}{2}u: \quad t_1 = s \, , \quad t_2 = u + s \; . \tag{6.79}$$

The second step in evaluating $Y_{IN,B}$ is integration over the momentum directions $\hat{\boldsymbol{P}}_1$ and $\hat{\boldsymbol{P}}_2$ in the numerator of Eq. (6.39). Since $\hat{\boldsymbol{P}}_1 = -\hat{\boldsymbol{M}}_1$ and $\hat{\boldsymbol{P}}_2 = -\hat{\boldsymbol{M}}_2$, we can equivalently integrate over the directions $\hat{\boldsymbol{M}}_1$ and $\hat{\boldsymbol{M}}_2$. Toward that end, we define a function $I_B$ as follows,

$$I_B \equiv \int d^2\hat{\boldsymbol{M}}_1 \int d^2\hat{\boldsymbol{M}}_2 \left[\hat{\boldsymbol{M}}_1 \bullet \boldsymbol{v}_2\left(\boldsymbol{R}_1\right)\right]\int d^3r\left[\boldsymbol{v}_1\left(\boldsymbol{r}\right)\bullet \boldsymbol{v}_2\left(\boldsymbol{r}\right)\right]' \ . \tag{6.80}$$

Using Eqs. (6.43), (6.44), (6.72) and (6.77), we can express $I_B$ as

$$\begin{aligned} I_B = &\int d^2\hat{\boldsymbol{M}}_1 \int d^2\hat{\boldsymbol{M}}_2 \left(-\frac{M}{4\pi}\right)\left[\frac{\hat{\boldsymbol{M}}_1 \bullet \hat{\boldsymbol{M}}_2 - 3\left(\hat{\boldsymbol{M}}_1 \bullet \hat{\boldsymbol{u}}\right)\left(\hat{\boldsymbol{M}}_2 \bullet \hat{\boldsymbol{u}}\right)}{u^3}\right] \\ &\times 2\left\{-\frac{M^2}{4\pi}\left[\hat{\boldsymbol{M}}_1 \bullet \hat{\boldsymbol{M}}_2 - 3\left(\hat{\boldsymbol{M}}_1 \bullet \hat{\boldsymbol{u}}\right)\left(\hat{\boldsymbol{M}}_2 \bullet \hat{\boldsymbol{u}}\right)\right]K_{1a} + \frac{M^2}{4\pi}\left(\hat{\boldsymbol{M}}_1 \bullet \hat{\boldsymbol{u}}\right)\left(\hat{\boldsymbol{M}}_2 \bullet \hat{\boldsymbol{u}}\right)K_{1b}\right\}. \end{aligned} \tag{6.81}$$

It is now useful to introduce quantities $A$ and $B$ as follows,

$$A \equiv \int d^2\hat{\boldsymbol{M}}_1 \int d^2\hat{\boldsymbol{M}}_2 \left[\hat{\boldsymbol{M}}_1 \bullet \hat{\boldsymbol{M}}_2 - 3\left(\hat{\boldsymbol{M}}_1 \bullet \hat{\boldsymbol{u}}\right)\left(\hat{\boldsymbol{M}}_2 \bullet \hat{\boldsymbol{u}}\right)\right]^2 \tag{6.82}$$

$$B \equiv \int d^2\hat{\boldsymbol{M}}_1 \int d^2\hat{\boldsymbol{M}}_2 \left[\hat{\boldsymbol{M}}_1 \bullet \hat{\boldsymbol{M}}_2 - 3\left(\hat{\boldsymbol{M}}_1 \bullet \hat{\boldsymbol{u}}\right)\left(\hat{\boldsymbol{M}}_2 \bullet \hat{\boldsymbol{u}}\right)\right]\left[\left(\hat{\boldsymbol{M}}_1 \bullet \hat{\boldsymbol{u}}\right)\left(\hat{\boldsymbol{M}}_2 \bullet \hat{\boldsymbol{u}}\right)\right] \ . \tag{6.83}$$

The following results can be easily obtained with the aid of spherical coordinates (all integrals range over 4π steradians):

$$C_1 \equiv \int d^2\hat{\boldsymbol{M}}_1 \int d^2\hat{\boldsymbol{M}}_2 \left(\hat{\boldsymbol{M}}_1 \bullet \hat{\boldsymbol{M}}_2\right)^2 = \frac{1}{3}\left(4\pi\right)^2 \tag{6.84}$$

$$C_2 \equiv \int d^2\hat{\boldsymbol{M}}_1 \int d^2\hat{\boldsymbol{M}}_2 \left(\hat{\boldsymbol{M}}_1 \bullet \hat{\boldsymbol{M}}_2\right)\left(\hat{\boldsymbol{M}}_1 \bullet \hat{\boldsymbol{u}}\right)\left(\hat{\boldsymbol{M}}_2 \bullet \hat{\boldsymbol{u}}\right) = \frac{1}{9}\left(4\pi\right)^2 \tag{6.85}$$

$$C_3 \equiv \int d^2\hat{\boldsymbol{M}}_1 \int d^2\hat{\boldsymbol{M}}_2 \left(\hat{\boldsymbol{M}}_1 \bullet \hat{\boldsymbol{u}}\right)^2\left(\hat{\boldsymbol{M}}_2 \bullet \hat{\boldsymbol{u}}\right)^2 = \frac{1}{9}\left(4\pi\right)^2 \ . \tag{6.86}$$

Then

$$A = C_1 - 6C_2 + 9C_3 = \frac{2}{3}\left(4\pi\right)^2 \tag{6.87}$$

$$B = C_2 - 3C_3 = -\frac{2}{9}\left(4\pi\right)^2 \ . \tag{6.88}$$

Using Eqs. (6.81), (6.82) – (6.88) and noting that the variables in $I_B$ are $u$ and $s$, one can express Eq. (6.80) as

$$I_B\left(u,s\right) = \frac{M^3}{u^3}\left[\frac{4}{3}K_{1a} + \frac{4}{9}K_{1b}\right] \ . \tag{6.89}$$

The third step in treating $Y_{IN,B}$ is evaluation of $\int d^3R_2 = \int d^3u$ . Using Eq. (6.89) one can see that there is no dependence on the direction of $\boldsymbol{u}$ in the integrand of Eq. (6.39), and the following factor $Q_B$ occurs in $Y_{IN,B}$ :

$$\begin{aligned} Q_B &= 4\pi \int_{u_c}^{\infty} du \; u^2 I_B(u,s) \\ &= 4\pi M^3 \int_{u_c}^{\infty} du \; \frac{1}{u}\left[\frac{4}{3}K_{1a} + \frac{4}{9}K_{1b}\right]. \end{aligned} \tag{6.90}$$

Because we have used a dipole approximation for a vortex ring, we must impose a lower limit on the separation distance $u$ and we have called that cut-off distance $u_c$ in Eq. (6.90).

The function $Q_B$ was evaluated analytically for an arbitrary value of $u_c$. There are many terms in the general result that involve $u_c$ and those terms would ultimately result in complicated temperature dependence of specific heat near $T_c$ if $u_c$ were a fixed number. However, there is a particular choice of $u_c$ that causes all of that complication to disappear. That particular choice is $u_c = s = R_0$, and that is what we shall use in our further calculations. This choice of $u_c$ implies that there is no restoring force against small perturbations in the relative configuration of two vortex rings when the core of either one is in the zone of instability established by the other.

Using Eqs. (6.73), (6.74), (6.78), (6.79), and (6.90) and $u_c = s$, one can obtain the following expression for $Q_B$ as a function of $s$.

$$Q_B(s) = 4\pi M^3 \left( \begin{aligned} &\int_{s}^{2s} du \; \frac{1}{u} \int_{s}^{u+s} dt \left[\frac{4}{3} f(s,t,u) + \frac{4}{9} g(s,t,u)\right] \\ &+ \int_{2s}^{\infty} du \; \frac{1}{u} \int_{u-s}^{u+s} dt \left[\frac{4}{3} f(s,t,u) + \frac{4}{9} g(s,t,u)\right] \end{aligned} \right). \tag{6.91}$$

The function $Q_B(s)$ that was found using the symbolic integration capabilities of the computer software Mathematica is

$$Q_B(s) = 4\pi M^3 \frac{1}{s^3} a_B \tag{6.92}$$

where

$$a_B = \frac{-307 + 2560\ln 2 - 1280\ln 3}{720} + \frac{-351 + 320\ln 3}{180} \simeq 0.08812 \; . \tag{6.93}$$

The remaining steps in evaluating $Y_{IN,B}$ in Eq. (6.39) can now be readily carried out. The integrals over momentum magnitudes $P_1$ and $P_2$ in the numerator cancel similar integrals in the denominator. Integrals over $d^2\hat{P}_1$ and $d^2\hat{P}_2$ in the denominator contribute a factor $1/(4\pi)^2$. The integral $\int d^3R_1$ in the numerator and $\left[\int d^3R\right]^2$ in the denominator contribute a net factor of $1/V$. Finally, one obtains the following result, where we now use $s = R_0$ as specified earlier:

$$Y_{IN,B} = -N\beta P_{MI}\left\{-\beta\frac{M^3}{4\pi}\frac{(N-1)}{V}a_B\, 2n_0 k\, B\frac{1}{R_0^3}\right\} . \qquad (6.94)$$

Using Eqs. (6.28), (6.29), (6.33), (6.94), one can exhibit the dependence of $Y_{IN,B}$ on $(T_c - T)$, as follows:

$$Y_{IN,B} = -N\beta P_{MI}\left\{-\beta\frac{M^3}{4\pi}\frac{(N-1)}{V}a_B\, n_0 m\alpha\, a_0^{-3}\left(T_c - T\right)^{\frac{3}{2}}\right\} . \qquad (6.95)$$

### *6.4.2 Quartic interaction integral*

The integral $\int d^3r$ for the quartic interaction integral in the expression for $Y_{IN,C}$ in Eq. (6.40) can be carried out using as variables of integration $s$, $t$, $\varphi$, which are parameters that were defined in connection with evaluation of $\int d^3r$ in Eq. (6.39). Now the divergence theorem cannot be used. The integral over $\varphi$ can be easily carried out first if one uses the coordinates for vectors indicated in Eqs. (6.64) - (6.68), and then the remaining integrals over $s$ and $t$ can be carried out. It is expected that the results of that integration will be expressible in terms of quantities that are independent of the choice of coordinate system, as was the case for $J_1$ in Eq. (6.72). Now there will be a very large number of terms that occur in the result. However, it is expected that the symbolic integration capabilities of Mathematica can treat all of them satisfactorily. The remaining steps in evaluating $Y_{IN,C}$ can then be performed by methods similar to those in evaluating $Y_{IN,B}$. Dimensional analysis described in the next Section indicates that the dependence of $Y_{IN,C}$ on $(T_c - T)$ will be $(T_c - T)^{3/2}$, the same dependence found in Eq. (6.95) for $Y_{IN,B}$. It is anticipated that the contribution of $Y_{IN,C}$ to $Y_{IN}$ will be small compared to that of $Y_{IN,B}$, but confirmation must wait on explicit evaluation of $Y_{IN,C}$.

### *6.4.3 Dimensional analysis of interaction terms*

$R_0$ is the only variable having dimension of length that emerges from carrying out the integrals $\int d^3R_2\int d^2\hat{P}_1\int d^2\hat{P}_2\int d^3r$ (or in certain cases $\int d^3R_2\int d^2\hat{P}_1\int d^2\hat{P}_2\int d^3r\int d^3r'$) in terms that occur thru second order in the interaction free energy, such as in $Y_{IN,B}$ and $Y_{IN,C}$, and others that can be identified with the aid of Eq. (6.21). Now we will give an algorithm for determining the exponent of $R_0$ in Eq. (6.94) by dimensional analysis. That algorithm can then be used to analyze the other interaction contributions and their dependence on $(T_c - T)$ near $T_c$ for those terms that do not vanish as a consequence of angular integrals $\int d^2\hat{P}_1 \int d^2\hat{P}_2$ .

The algorithm is the following. Each factor of $\boldsymbol{v}$ in an expression such as $Y_{IN,B}$ in Eq. (6.39) contributes a factor of $R_0^{-3}$ to the expression, and the volume elements $d^3r$ and $d^3R$ $d^3R_2$ each contribute a factor of $R_0^3$. The product of all of these factors gives the resultant $R_0^n$, where $n$ is an integer that would result from actually carrying out the integrals. The dependence of that product on $(T_c - T)$ can then be found using Eq. (6.33) along with the temperature dependence of $(T_c - T)$ for any factor of $B$ (see Eqs. (6.28), (6.29)) that may be present. Terms that do not involve $\int d^3r$ as a factor have no dependence on $R_0$ and no dependence on $(T_c - T)$. (The cut-off distance $u_c$ is assumed to be a constant not dependent on $(T_c - T)$ for terms that do not involve $\int d^3r$ .)

In the case of $Y_{IN,B}$ in Eq. (6.39), the algorithm gives $Y_{IN,B}$ varying as $(B R_0^{-3})$, consistent with Eq. (6.94), and $Y_{IN,B}$ varying as $(T_c - T)^{3/2}$, consistent with Eq. (6.95). One can readily verify that application of the algorithm to $Y_{IN,C}$ in Eq. (6.40) gives the dependences $R_0^{-9}$ and $(T_c - T)^{3/2}$.

Now consider the first order term $< -\beta F_{IN} >$ in Eq. (6.23). The quantity $F_{IN}$ is given by Eqs. (6.10) – (6.18), and the meaning of the brackets < > is given by Eq. (6.22). One can readily see that angular integrals $\int d^2\hat{P}_1 d^2\hat{P}_2$ produce null results for terms $< F_{CI} >$ and $< F_{EB} >$. Applying the algorithm to $< F_{EC} >$, one finds that its dependence on $R_0$ is given by $R_0^{-6}$, and the dependence of $< F_{EC} >$ on $(T_c - T)$ is given by $(T_c - T)$. Anticipating a result that will be justified later, we note that the first derivative $d\langle F_{EC}\rangle / dT$ contributes to the entropy. Also, $d^2\langle F_{EC}\rangle / dT^2$ contributes formally to specific heat, and we see that this gives a null result for the contribution by $< F_{EC} >$ below $T_c$. Now we have dealt with all contributions that the first order term $< -\beta F_{IN} >$ makes to specific heat and found that they give a null contribution to specific heat.

Turning to the second order term $\left\langle \frac{1}{2}\left(-\beta F_{IN}\right)^2 \right\rangle$ in Eq. (6.23), we again use Eqs. (6.10) – (6.18), and Eq. (6.22). We have already dealt in detail with the cross term $2F_{CI}F_{EI}$ in the expanded form $F_{IN}^2 = \left(F_{CI}^2 + 2F_{CI}F_{EI} + F_{EI}^2\right)$. Using the algorithm we find that $< F_{CI}^2 >$ has no dependence on $R_0$. This term will contribute to the free energy both below and above $T_c$, varying smoothly through $T_\lambda$ and producing no singularity in specific heat at $T_c$.

Using the algorithm to evaluate $< F_{EI}^2 >$, we write it as $\left\langle F_{EB}^2 + 2F_{EB}F_{EC} + F_{EC}^2 \right\rangle$ and find that the dependence on $(T_c - T)$ is the same for each of the terms. Let us consider the term $\left\langle F_{EC}^2 \right\rangle$, where the following factors occur: $d^3 R_2 d^3 r d^3 r' v^8$. The dependence of this term on $R_0$ is given by $R_0^{-15}$ and the dependence on $(T_c - T)$ is given by $(T_c - T)^{5/2}$. The second derivative $d^2 \left\langle F_{EI}^2 \right\rangle / dT^2$ varies as $(T_c - T)^{1/2}$ and consequently the term $\left\langle F_{EC}^2 \right\rangle$ does not contribute to singular behavior of specific heat at $T_c$.

We have now established that through second order terms in a perturbation expansion, only the cross term given by Eqs. (6.35) is anticipated to contribute a singularity in specific heat at $T_c$. The other terms will be neglected in the analysis that follows, which aims to elucidate critical behavior. Later we will explain the physical meaning of the cross term, but next we will further develop the statistical mechanical treatment of the model that includes vortex ring interactions.

### *6.5 The superfluid wind*

The partition function $Z_N$ in Eq. (6.23) will be treated here in the model where only the cross term $Y_{IN,B}$ in the quadratic interaction is retained. It should be noted that the quartic interaction $Y_{IN,C}$ can be readily accommodated in this formalism also. It is useful to introduce a vector $\boldsymbol{U}$ as follows in order to simplify the formulas and lay some groundwork for physical interpretation of the theory,

$$\boldsymbol{U} = U\hat{\boldsymbol{M}} = U_B\hat{\boldsymbol{M}} \tag{6.96}$$

$$\begin{aligned} U_B &= -\beta \frac{M^3}{4\pi} \frac{(N-1)}{V} a_B 2n_0 k B \frac{1}{R_0^3} \\ &= -\beta \frac{M^3}{4\pi} \frac{(N-1)}{V} a_B n_0 m \alpha a_0^{-3} (T_c - T)^{\frac{3}{2}} . \end{aligned} \tag{6.97}$$

Using definitions and results given earlier, one can readily verify that $U_B$ has the dimensions of velocity and that $Y_{IN}$ can be written in the following form,

$$Y_{IN} = -N\beta P_{MI} U \ . \tag{6.98}$$

Near $T_c$, where $U$ is very small, one can express $Z_N$ as follows, where we use the condition $N$ >> 1,

$$Z_N = e^{-\beta \bar{F}_{N0}} \left( e^{-\beta P_{MI} U} \right)^N \ . \tag{6.99}$$

Using Eqs. (5.3), (6.24), and (6.99), one can write $Z_N$ as

$$Z_N = \frac{1}{N!}\left[Vh^{-3}\int d^3P\, e^{-\beta\left(F_1(P)+P_{MI}U\right)}\right]^N \ . \tag{6.100}$$

From Eq. (6.100) one can see that the partition function $Z_N$ is like that for a system of vortex rings that do not interact with each other, but where each ring is in what we will call a superfluid wind $\boldsymbol{U}$ that acts only on each core. It turns out that $U$ is a negative number, and so $\boldsymbol{U}$ is in a direction opposite to $\hat{\boldsymbol{M}}$ according to Eq. (6.96).

The superfluid wind is created by statistical correlations for vortex pairs, those correlations being due to the external portion of the interaction free energy for any pair. In some sense the correlations are like a polarization effect that each vortex ring induces on the other rings in the system.

The treatment of vortex ring interactions we are considering is based on second order perturbation theory, and is not simply a mean field theory subject to the idealization or limited accuracy usually associated with this designation. This point is emphasized here because there is an alternative method for deriving the results in Eq. (6.100) that in some aspects resembles a mean field theory. In fact that alternative method was used originally in this model, before the second order perturbation method was worked out.

The alternative method will be described briefly here because it illuminates the origin and physical interpretation of the superfluid wind. This method involves a two-stage, self-consistent calculation. In the first stage a canonical distribution for a system of vortex rings is constructed wherein the self-energy is taken into account, but only the external interaction free energy is present. Using this distribution function, one then evaluates the expectation value of $\hat{\boldsymbol{M}}_i \bullet \left( \sum_j{}' \boldsymbol{v}_j \left(\boldsymbol{R}_i\right) \right)$, where the prime indicates that $j \neq i$ in the summation. This expectation value is the component of average superfluid velocity in direction $\hat{\boldsymbol{M}}_i$ at the location of the core of the $i$th vortex ring, and we will represent that velocity by $\boldsymbol{U}$, the superfluid wind velocity. Terms that do not contribute to singular behavior in specific heat are identified and then discarded at this juncture. The second stage of the calculation consists of evaluating the canonical partition function $Z_N$ for a system of vortex rings while taking into account self-energy and the interaction of each vortex ring core with the superfluid wind $\boldsymbol{U}$. The result for $Z_N$ found in this way coincides with that in Eq. (6.100).

### *6.6 Grand partition function and thermodynamic properties for interacting vortex rings*

The grand partition function $\Omega$ is given by

$$\Omega = \sum_{0 \le N \le \infty} e^{\beta\mu N} Z_N \tag{6.101}$$

and the grand canonical potential $W$ is given by

$$W = -kT \ln \Omega \ , \tag{6.102}$$

where $Z_N$ is given by Eq. (6.100) for interacting vortex rings. It is implied in Eq. (6.101) that the limit $\mu \to 0$ is taken finally to obtain observable properties. Because the superfluid wind $U$ in Eq. (6.100) is a function of $N$, as indicated by Eqs. (6.96) and (6.97), the sum in Eq. (6.101) cannot be evaluated by the simple direct method of summation used in obtaining the grand partition function for non-interacting vortex rings in Eq. (5.5). Therefore we must find an alternative method for evaluating $\Omega$ in Eq. (6.101).

The essence of the alternative method is in identifying expectation values of certain quantities with maximum values under conditions where $N$ is large. Our analysis begins with establishing certain correspondences in treating a model of non-interacting vortex rings. Let $a$ be defined as follows:

$$a \equiv h^{-3} V \int d^3P \, e^{-\beta F_1(P)} \ . \tag{6.103}$$

The partition function $Z_{N0}$ and grand canonical potential $\Omega_0$ are given by Eqs. (5.3) and (5.4), respectively; however, when Stirling's approximation in the form

$$N! = \left( \frac{N}{e} \right)^N \tag{6.104}$$

is used in those functions, we shall call them $\bar{Z}_{N0}$ and $\bar{\Omega}_0$. Then

$$\bar{\Omega}_0 = \sum_N e^{\beta\mu N} \bar{Z}_{N0} = \sum_N \left( \frac{e}{N} a e^{\beta\mu} \right)^N = \sum_N P(N) \ . \tag{6.105}$$

One can readily show that

$$\frac{\partial P(N)}{\partial N} = e^{N \ln\left( \frac{e}{N} a e^{\beta\mu} \right)} \left[ \ln\left( \frac{e}{N} a e^{\beta\mu} \right) - 1 \right] \tag{6.106}$$

and

$$\frac{\partial^2 P(N)}{\partial N^2} = e^{N \ln\left( \frac{e}{N} a e^{\beta\mu} \right)} \left\{ \left[ \ln\left( \frac{e}{N} a e^{\beta\mu} \right) - 1 \right]^2 - \frac{1}{N} \right\} \ . \tag{6.107}$$

Let $N_0$ be the value of $N$ where $\partial P(N)/\partial N = 0$. Using Eq. (6.106) one finds

$$N_0 = a e^{\beta\mu} \ . \tag{6.108}$$

Then one finds that

$$\frac{\partial^2 P(N)}{\partial N^2} = -\frac{1}{N_0} \ , \tag{6.109}$$

which shows that $P(N)$ is a relative maximum at $N = N_0$.

Next we shall replace $\bar{\Omega}_0$ in Eq. (6.105) by its maximum term, and represent the new function by $\tilde{\Omega}_0$. Then

$$\tilde{\Omega}_0 = P(N_0) = \left( \frac{e}{N_0} \partial e^{\beta\mu} \right)^{N_0} = e^{N_0} \tag{6.110}$$

and

$$\tilde{W}_0 = -\frac{1}{\beta} \ln \tilde{\Omega}_0 = -\frac{1}{\beta} \partial e^{\beta\mu} \ , \tag{6.111}$$

where in the last step we have used Eqs. (6.108) and (6.110).

Using Eqs. (6.103) and (6.108) one can see that $N_0$ coincides with $< N >$ in Eq. (5.7). Further, one can see that $\tilde{W}_0$ in Eq. (6.111) coincides with $W_0$ in Eq. (5.6). Therefore all thermodynamic functions determined by derivatives of $W_0$, such as entropy *S*, specific heat *C*, and mean square fluctuation in number of vortex rings $(N - N_0)^2$, will have the same values whether we use $W_0$ or $\tilde{W}_0$.

For non-interacting vortex rings, results based on the maximum term method are the same as results obtained by direct evaluation of the sum in the grand canonical partition function. This is the basis of our generalization to the case where interactions are present and the direct summation cannot be carried out but the maximum term method can be readily implemented.

Turning attention to the evaluation of $Z_N$ in Eq. (6.99) using the maximum term method, we begin by replacing $(N - 1)$ by $N$ in Eq. (6.97) and introduce a parameter *b* so that the exponent involving vortex wind in Eq. (6.99) can be expressed as follows,

$$-\beta P_{MI} U = bN \ . \tag{6.112}$$

Then following the notation introduced earlier for non-interacting vortex rings but dropping the zero subscript on the partition function, etc, because the rings are now interacting, we obtain

$$\bar{\Omega} = \sum_N e^{\beta\mu N} \bar{Z}_N = \sum_N \left( \frac{e}{N} \partial \, e^{\beta\mu} e^{bN} \right)^N = \sum_N P(N) \ . \tag{6.113}$$

Taking derivatives of the function $P(N)$ defined by Eq. (6.113), one obtains

$$\frac{\partial P(N)}{\partial N} = e^{N \ln\left(\frac{e}{N} a e^{\beta\mu} e^{bN}\right)} \left[ \ln\left( \frac{e}{N} a e^{\beta\mu} e^{bN} \right) - 1 + bN \right] . \tag{6.114}$$

At the stationary point where $\partial P(N)/\partial N = 0$, we find

$$\tilde{N}_0 = a e^{\beta\mu} e^{2b\tilde{N}_0} . \tag{6.115}$$

The stationary point is a relative maximum at $\tilde{N}_0$ provided $\left(\partial^2 P(N)/\partial N^2\right)_{\tilde{N}_0} < 0$ and this condition will be met provided $\tilde{N}_0 < 1/(2b)$. This condition can be met only if $b \geq 0$. From Eq. (6.112) one infers that $U < 0$ is necessary for this theory to be valid and Eqs. (6.96) and (6.97) indicate that $U$ meets this condition. Replacing the sum over $P_N$ in Eq. (6.113) using the maximum term method, we obtain $\tilde{\Omega}$, as follows.

$$\tilde{\Omega} = \left(e^{\left(1-b\tilde{N}_0\right)}\right)^{\tilde{N}_0} . \tag{6.116}$$

The grand canonical potential $\tilde{W}$ is then given by

$$\tilde{W} = -\frac{1}{\beta} \ln \tilde{\Omega} = -\frac{1}{\beta} \tilde{N}_0 \left(1 - b\tilde{N}_0\right) . \tag{6.117}$$

Using Eqs. (6.115) and (6.117), one can readily show that

$$-\frac{\partial \tilde{W}}{\partial \mu} = \tilde{N}_0 \tag{6.118}$$

and the mean square fluctuation of $N$ is given by

$$\left\langle \left(N - \tilde{N}_0\right)^2 \right\rangle = -\frac{1}{\beta} \frac{\partial^2 \tilde{W}}{\partial \mu^2} = \frac{\tilde{N}_0}{1 - 2b\tilde{N}_0} . \tag{6.119}$$

Provided $2b\tilde{N}_0 << 1$, the relative fluctuations in $N$, given by

$$\frac{\left[\left\langle \left(N - \tilde{N}_0\right)^2 \right\rangle\right]^{\frac{1}{2}}}{\tilde{N}_0} \simeq \tilde{N}_0^{-\frac{1}{2}} \tag{6.120}$$

are small for large values of $\tilde{N}_0$.

When $b\tilde{N}_0 << 1$ and $N_0 \simeq \tilde{N}_0$ for $N_0$ given by Eq. (6.108), one finds using Eqs. (6.115) and (6.117) that through first order terms in $b$

$$\tilde{W} = -\frac{1}{\beta} a e^{\beta\mu}\left(1 + b a e^{\beta\mu}\right) \ . \tag{6.121}$$

This shows that the interaction term involving *b* lowers the grand canonical potential when *b* > 0 and *b* is sufficiently small.

For weakly interacting vortex rings, where Eq. (6.121) is accurate, the entropy *S* is given by

$$S = -\frac{\partial \tilde{W}}{\partial T} = k\left\{ a(1 + ba) + T\left[\frac{\partial a}{\partial T} + a^2 \frac{\partial b}{\partial T} + 2ab\frac{\partial a}{\partial T}\right]\right\} \ . \tag{6.122}$$

Every term in Eq. (6.122) can be readily evaluated using results found earlier in this paper, but only the term $kTa^2\left(\partial b/\partial T\right)$ produces a singularity in the specific heat at $T_c$. We are interested here in studying the critical properties at $T_c$, and so we shall neglect the other smoothly varying terms, regarding them as part of the background, and write for *T* near $T_c$,

$$S = kTa^2 \frac{\partial b}{\partial T} \ . \tag{6.123}$$

The singular part of the heat capacity at constant volume, $C_V$, is then given by

$$C_V = T\frac{\partial S}{\partial T} = kT^2 a^2 \frac{\partial^2 b}{\partial T^2} \ . \tag{6.124}$$

In Eqs. (6.123) and (6.124) we will evaluate all quantities at $T_c$, except in the case of the parameter *b*, where the temperature variation occurs in the combination ($T_c$ – *T*). We shall treat $\tilde{W}$ in a similar manner and write it as

$$\tilde{W} = -\frac{1}{\beta} a^2 b \ . \tag{6.125}$$

A factor of $e^{2\beta\mu}$ has been omitted since it does not affect *S* or $C_V$ where we take the limit $\mu \to 0$. A little later we will numerically evaluate the coefficients in the superfluid wind $U = U_B$, but for the present we will use Eqs. (6.123) – (6.125) and just examine their behavior near $T_c$ as temperature shape factors, $\tilde{W}_{SF}$, $S_{SF}$, $C_{V,SF}$. Then for $T < T_c$

$$\tilde{W}_{SF} = -\left(T_c - T\right)^{\frac{3}{2}} \tag{6.126}$$

$$S_{SF} = -\frac{3}{2}\left(T_c - T\right)^{\frac{1}{2}} \tag{6.127}$$

$$C_{V,SF} = \frac{3}{4}\left(T_c - T\right)^{-\frac{1}{2}} . \tag{6.128}$$

These shape factors are plotted in Figs. 24a – 24c. We note here that for $\mu = 0$, $W$ is the same as the Helmholtz potential $F$ where the sum over states in the partition function also includes the sum over number of vortex rings. This accounts for the function $F$ in Fig. 24a. For $T > T_c$, $b = 0$ and only smoothly varying background terms contribute to thermodynamic functions for some temperature range in a finite container.

The plot $F_{SF}$ vs $T$ in Fig. 24a shows that the interaction contribution to $F$ is negative, implying that the Helmholtz potential for the liquid is lowered by that interaction, but the amount of lowering approaches zero at $T = T_{c-}$.

The plot $S_{SF}$ vs $T$ in Fig. 24b shows that the interactions produce a more ordered state than the randomly oriented non-interacting state. This is indicated by $S_{SF} < 0$ in that plot. But the order tends to zero as the temperature increases and approaches $T_c$. Although $S_{SF}$ approaches zero at $T_c$, the slope of $S_{SF}$ tends to infinity as $T \rightarrow T_{c-}$ as can be seen from Eq. (6.127).

The plot $C_{V,SF}$ vs $T$ in Fig. 24c shows that the interaction part of specific heat increases as $T \rightarrow T_{c-}$ and Eq. (6.128) shows that it is divergent at $T_{c-}$.

Analysis of these plots along with other results found earlier in this paper is summarized below in a description of the physical mechanism of the λ transition that is indicated by this vortex ring model for the liquid in complete thermodynamic equilibrium.

## 7. The λ transition in liquid $^4$He at the He I – He II phase boundary

Calculations in previous Sections provide a basis for discussion and analysis in this Section of the role of vortices in the behavior of the liquid both below and above the λ transition. The liquid in complete thermodynamic equilibrium is treated in Sec. 7.1. Approximations used in Sec. 6 while evaluating the external free energy $F_{EN}$ in Eq. (5.1) will be seen to limit the temperature range where calculated results are very accurate to a crossover region. This limitation is indicated in the title of Sec. 7.1.1. The principal approximation responsible for this limitation on the calculated properties is in the retention of only one and two index terms in evaluating $F_{EN}$ in Eq. (5.1). The limitation will be explained in Sec. 7.1.2.

Results given in Secs. 7.1.1 and 7.1.2 point the way to a more complete model, considered in Sec. 7.1.3. This model includes an accurate treatment of $F_{EN}$ in Eq. (5.1), and takes into account all of the $N$ vortices simultaneously. A significant feature of this complete model is that it generates a clear picture of critical properties of liquid $^4$He that are very similar to those that have been discussed for models of liquid-gas critical behavior and magnetic systems near their Curie points in other theories. The role of vortices in ordering that occurs in He II is also explained in our model in Sec. 7.1.3.

Sections 7.1.4 - 7.1.6 treat line vortices and large vortex rings near $T_c$, the superfluid transition temperature for a model that contains only Landau excitations consisting of phonons and interacting rotons. The distinction between $T_c$ and $T_\lambda$, the temperature at which the peak specific heat occurs, is explained in more detail in Sec. 7.1.3. Section 7.1.5 treats a temperature range just below $T_c$, including the range just below $T_\lambda$. Section 7.1.6 treats a temperature range just above $T_c$, including a temperature where a catastrophic increase in the number of vortex lines and large vortex rings occurs.

Section 7.2 treats the liquid in metastable equilibrium, where pinned line vortices are present in the liquid. The experimental and theoretical reasons for introducing pinned line vortices in our theory are discussed in Sec. 7.2.1. The specific heat for liquid in the metastable state is treated in Sec.7.2.2.

The physical mechanism of the λ transition is explained in Sec. 7.3 in terms that are applicable to complete models treated in Secs. 7.1.3 and 7.2.2. Section 7.4 contains comments on scaling and Landau-Ginzburg-Wilson Hamiltonian based theory viewed from the perspective of our microscopic based theory.

### *7.1 Liquid in complete thermodynamic equilibrium*

#### *7.1.1 Physical mechanism of the λ transition in the crossover model*

The vortex ring model, as we have formulated it in Sec. 6, indicates that vortex rings are thermally excited in liquid $^4$He. Small rings are statistically dominant below $T_c$ and for a small temperature range above $T_c$ in a finite container. Because of their large core energy, vortex lines and vortex rings having large radii are excited with such low number density that they can be neglected. At any finite temperature, surrounding each vortex ring core is a mantle where the liquid is unstable against small perturbations in superfluid velocity according to ordinary thermodynamic stability criteria. However, because quantization of circulation imposes a constraint on disturbances of superfluid velocity, the mantle region can still contribute to the *self-energy* of the vortex rings. But the mantle does not contribute to the interaction free energy that would provide a restoring force against perturbation of the relative configuration of pairs of rings. This chameleon-like behavior has been formulated as a "*Principle of Constrained Instability*." The velocity $v_0$ of the liquid at the outer boundary of the mantle is the velocity at which the Helmholtz potential for the flowing superfluid has an inflection point, and $v_0$ is the order parameter of this theory. The order parameter is positive below $T_c$ and decreases to zero at $T_c$, and $v_0 \sim (T_c - T)^{1/2}$. There is no inflection point for $T > T_c$ and so there is no physically meaningful value of the order parameter for $T$ $> T_c$. Near $T_c$, $v_0$ is proportional to $\rho_s(0)^{1/2}$, where $\rho_s(0)$ is the superfluid density evaluated at $v_s = 0$ when phonons and interacting rotons account for the normal fluid.

The distance $R_0$ from the center of a small vortex ring at which the superfluid velocity reaches the value $v_0$ is the correlation length in this theory, and $R_0 \sim (T_c - T)^{-1/6}$. $R_0$ is the radius to the mantle boundary, which is approximately spherical near $T_c$ for small rings.

There is a statistical correlation in the relative configuration between vortex rings that induces a net superfluid velocity at the core of each vortex ring. That velocity may be regarded as a superfluid wind. The statistical correlation depends on the interaction free energy associated with regions outside the mantles, and it can be viewed as a polarization effect. The strength of the superfluid wind varies as $(T_c - T)^{3/2}$ *near* $T_c$.

The superfluid wind is in a direction opposite to the dipole moment of the vortex ring. The dipole moment direction is the same as the direction of momentum $\boldsymbol{P}_{MI}$ associated with atoms inside the core of a vortex ring. The core boundary is determined by the Landau critical velocity, the velocity at which the flowing ground state becomes degenerate with respect to a single roton state that is also in a state of flow where the flow is treated as if it is uniform throughout the liquid. There is a negative contribution to the vortex ring free energy accounted for by the scalar product of $\boldsymbol{P}_{MI}$ with the vortex wind velocity and this interaction varies as $t^{3/2}$, where $t = (T_c - T)$. The main contribution of vortex rings to specific heat near $T_c$ varies as the second derivative, $C_V \sim d^2(t^{3/2})/dt^2 = t^{-1/2}$. This produces a singularity in specific heat at $T_c = T_\lambda$. The other terms due to vortex ring interactions vary continuously with $T$ near $T_c$, and those terms can be treated as part of the background for the critical behavior. There is no contribution from the superfluid wind interaction for $T > T_c$, and the specific heat drops to the background level at $T_{c+}$.

The stated dependence of $R_0$ on $(T_c - T)$ implies that the correlation length tends to $\infty$ as $T$ approaches $T_c$. From the foregoing discussion one can see that the interaction free energy tends to zero as $T \to T_{c-}$, and so the vortex rings become increasingly uncorrelated as $T \to T_{c-}$. This explains why the vortex rings become more uncorrelated as the correlation length increases and tends to infinity as temperature increases and approaches $T_c$.

### *7.1.2 Numerical values of coefficients in superfluid wind and specific heat*

The expressions for the coefficients given by Eqs. (6.96) and (6.97) for superfluid wind and Eqs. (6.112) and (6.124) for specific heat can be evaluated numerically using parameters found earlier in this paper. The values given in this Section for those parameters are for essentially zero pressure and constant average atomic density $n_0 = 0.0218$ Å$^{-3}$ in the liquid at all temperatures treated. Results of variational calculations and statistical mechanical calculations that take into account interactions among interacting rotons are used here in fixing these parameters. The statistically dominant value of vortex ring radius $R$ = 2 Å is used in evaluating the dipole moment $M$ of a vortex ring.

The vortex ring number density $N_0/V$ is treated as constant in the temperature range near $T_c$ considered here. The statistical mechanical treatment of non-interacting vortex rings gives $N_0/V = 0.654\times10^{24}$ m$^{-3}$ at $T_c$ = 2.172 K. Further, the value of $\alpha$ is given in Eq. (6.30), the value of $a_B$ is given by Eq. (6.93) and the value of $a_0$ is given by Eq. (6.34). The value of $P_{MI}$ calculated by numerical integration of Eq. (6.8) for a vortex ring having $R$ = 2 Å and using the density profile inside the core determined by the variational calculation is $P_{MI}$ = 0.610 x $10^{-24}$ kg ms$^{-1}$.

The results found using these parameter values are as follows,

$$U = U_B = -2.163\times10^{-3}\left(T_c - T\right)^{\frac{3}{2}} \text{ ms}^{-1} \ . \tag{7.1}$$

The formula for specific heat (per unit mass) found using Eq. (6.124) is

$$c_V = \frac{C_V}{n_0 mV} = \frac{1}{k}P_{MI}\left(\frac{N_0}{V}\right)^2\left(\frac{m^3}{4\pi}\right)a_B\,\alpha\,a_0^{-3}\left(\frac{3}{4}\right)\left(T_c - T\right)^{-\frac{1}{2}} \ . \tag{7.2}$$

Using the parameter values that we have specified, one finds the following result for $c_V$:

$$c_V = 9.70\times10^{-9}\left(T_c - T\right)^{-\frac{1}{2}} \quad \text{J g}^{-1}\,\text{K}^{-1} \ . \tag{7.3}$$

The values of the parameters used in obtaining the coefficients are estimated to be applicable in the temperature range 2.072 K < $T \leq$ 2.172 K. The superfluid wind speed $|U|$ is predicted to be at most about 7 x $10^{-5}$ ms$^{-1}$ in that temperature range. The specific heat $c_V$ contribution due to vortex ring interaction is also very small even at ($T_c - T$) = $10^{-6}$ K, where $c_V \simeq 10^{-5}$ J g$^{-1}$ K$^{-1}$. Such a small contribution would be undetectable with present technology. Taken at face value, theoretical evaluation based on the stated values of parameters predicts that for the liquid in complete thermodynamic equilibrium, liquid He II will have a singularity in specific heat that is not measurable in practice and that the experimentally measured specific heat will appear to be continuous through $T_c$.

It should be noted that the values of the parameters used in evaluating the coefficients rely on output of variational calculations for vortex core energy, and that this energy in turn strongly affects the vortex ring radius that is most important in statistical mechanical treatment of specific heat and other thermodynamic properties of the rings. It is informative to consider the following conditions to study the sensitivity of the specific heat coefficient to core energy and dominant ring radius. Suppose the vortex core energy $\varepsilon_C/k$ is 12 K instead of the 18.7 K found in the existing variational calculation. Also suppose that the dominant ring radius $R$ is 4 Å instead of the 2 Å based on existing calculations. Then using the existing calculations as a guide, I estimate that the coefficient in $c_V$ would be increased by about a factor of $10^5$, and $c_V$ would then be in an experimentally measurable range. Although I believe that such large changes as those in the postulated parameter set are unlikely, the nature of variational calculations carried out so far is such that one cannot rule out changes that are that large, or larger, with confidence. Clearly, further variational calculations and possibly establishing both upper and lower bounds on the core energy would be useful in reducing uncertainty in predictions of $c_V$.

A conservative prediction of the theory is that the specific heat will diverge at $T_c$ as $(T_c - T)^{-1/2}$ and then drop to the level of the background for some finite temperature range above $T_c$ for a liquid that is in complete thermodynamic equilibrium and in a finite container. The coefficient in the divergent term may be so small that in practice the specific heat appears to be continuous at $T_c$. In Sec. 8.2, an experiment is proposed to study the prediction.

The dependence of the correlation length $R_0$ on $T_c - T$ is shown in Table 7.1 for a 2 Å radius vortex ring. The average spacing $L$ between rings near $T_c$ was found to be about 145 Å in Sec. 5.33 (see also Fig. 21.) Comparing $R_0$ with $L/2$, one finds that using average vortex ring spacing $L$, the regions of instability of two-vortex rings start to overlap near $T_c - T = 10^{-7}$ K. These results indicate that our approximate evaluation of $F_{EN}$ in Eq. (5.1) where we kept only two index terms in the external interaction part $F_{EIN}$ (see Eqs. (6.6) and (6.11) - (6.15)) becomes increasingly inaccurate as $T_c - T$ decreases in value. That inaccuracy is in large part due to neglect of regions of instability around other rings when we took account of the excluded regions indicated by the prime notation in Eqs. (6.14) and (6.15). For temperatures at least moderately far from $T_c$, where $R_0 << L/2$, our evaluation of $F_{EIN}$ is expected to be fairly accurate. This is the crossover region to critical behavior. Data in Table 7.1 suggest that the crossover region is in the approximate temperature range $10^{-2} \le T_c - T \le 10^{-4} K$. However, above the indicated range an accurate theory of the λ transition must take into account all of the $N$ vortex rings simultaneously when the contribution $F_{EIN}$ to $F_{IN}$ in Eqs. (6.19) and (6.21) is evaluated. That problem is treated in a more complete theory in Sec. 7.1.3.

**Table 7.1** Temperature dependence of the correlation length $R_0$ for a 2 Å radius vortex ring based on Eqs. (6.33) and (6.34). The average spacing $L$ between vortex rings is about 145 Å.

| $T_c - T$ (K) | $10^{-2}$ | $10^{-3}$ | $10^{-4}$ | $10^{-5}$ | $10^{-6}$ | $10^{-7}$ | $10^{-8}$ |
|---|---|---|---|---|---|---|---|
| $R_0$ (Å) | 9.0 | 13.1 | 19.3 | 28.3 | 41.6 | 61.0 | 89.53 |

*7.1.3 Physics of the λ transition in the complete model*

A complete model for critical properties of liquid $^4$He near $T_c$ must take into account all quantized vortices simultaneously in the evaluation of $Z_N$ in Eq. (6.21). Equations (6.19) - (6.21) are written in a form suitable for dealing with this situation. There are two main differences from what we did in treating the crossover model. The first is that instead of using approximations to the free energy density $f(\boldsymbol{r})$ in Eq. (6.2) and in turn to $F_{IN}$ in Eqs. (6.11), (6.19) and (6.21), we now use the exact expression for $f_N(\boldsymbol{r})$ for all of the vortices simultaneously. The second is that the integration over $d^3r$ for the external contribution $F_{EIN}$ to $F_{IN}$ is now limited to the region of stability as determined by the principle of constrained instability for the circumstance where this stable region can be represented to good approximation as the regions outside the individual mantles for all of the vortices. The stable regions are indicated as unshaded areas in Fig. 25. The unstable regions are indicated as shaded areas in Fig. 25. The contribution of the core interaction free energy $F_{CIN}$ to $F_{IN}$ is the same as in the crossover model. The term $\left\langle \frac{1}{2}\left(-\beta F_{IN}\right)^2 \right\rangle$ in Eq. (6.23) and the cross-term $2F_{CIN}F_{EIN}$ are the terms that we must consider in the complete theory. This judgment is based on the observation that the complete model must merge continuously with the crossover model in a temperature range moderately far below $T_\lambda$.

Now we must be careful to distinguish $T_c$ from $T_\lambda$. $T_c$ is the temperature where $\rho_s(0) = 0$ and where the order parameter $v_0$ vanishes. Here $v_0$ is the inflection point in an isotherm for $F/Nk$ versus $v_s$, as in Fig. 15, where only phonons and interacting rotons are taken into account. We will see that $T_c$, which marks the upper boundary temperature for ordinary superfluidity, is above $T_\lambda$ where the peak in specific heat occurs. This is contrary to what was found in the crossover model where superfluidity ended at the temperature $T_c = T_\lambda$ where specific heat is peaked. We will retain the value $T_c = 2.172$ in what follows, but in our complete model the theoretical value of $T_\lambda$ will be slightly less than the experimental value 2.172 K. Fig. 25 is helpful in elucidating this situation. Regions of instability in the integral $F_{EIN}$ are represented schematically for a fixed configuration of $N$ vortex rings. In the unshaded regions the superfluid is stable and contributes to the restoring force for fluctuations in the orientations or positions of the vortex rings from their thermal equilibrium configuration as determined by local thermodynamic equilibrium throughout the liquid. It is relevant to note that second sound can propagate as an oscillation in the stable regions. We shall call the liquid in the stable regions He II for present purposes and we shall call the liquid in unstable regions He I. This avoids introduction of new nomenclature into our discussion, but is not a quite accurate use of the term He II commonly used to refer to superfluid liquid that contains phonons and rotons as well as any quantized vortices that may be present and in referring to the He II - He I phase boundary. Fig. 25a represents the situation far enough below $T_c$ that the unstable liquid region in the mantle about each quantized vortex ring is well separated from all of the other mantles. Fig. 25b represents the situation close to $T_c$ where $v_0$ is small, but far enough below $T_c$ that some stable liquid regions exist between the mantles about the vortex rings. Fig. 25c is a magnified view of one of the stable regions in Fig. 25b that indicates more clearly how it is formed.

When the temperature of the liquid increases from that assumed in Fig. 25a, the mantles expand and mantles of some individual rings overlap. Those larger regions of instability, each having $v_0$ at its outer boundary, correspond to clusters in percolation models [105,106]. Eventually there is no region of stability that stretches continuously across the liquid, from wall to wall, when the vortex rings are spaced apart about evenly. This corresponds to the connectivity threshold in percolation models [106]. At a temperature slightly above the connectivity threshold, the regions of stability, shown as unshaded, are entirely surrounded by unstable regions, shown as shaded. As temperature increases further, but still somewhat below $T_c$, the situation appears qualitatively as in Fig. 25b.

We propose the following physical interpretation of the conditions just described. The temperature where the two qualitatively different situations meet is $T_\lambda$, the λ temperature where the specific heat peak occurs. At least roughly there is symmetry in the set of configurations where He I "bubbles", referred to as clusters earlier, appear in a matrix of He II below $T_\lambda$ and the set of configurations where He II "drops" appear in a matrix of He I above $T_\lambda$. Symmetry of these configurations is reflected in the entropy contributions to the free energy contributions by the vortex ring interactions. The rapid change of entropy with temperature associated with the different configurations of "bubbles" below $T_\lambda$ and "drops" above $T_\lambda$ account for the λ - type specific heat anomaly at the He I - He II phase boundary. The symmetry between configurations of "bubbles " and "drops" account for the symmetry in the specific heat above and below $T_\lambda$.

The situations represented in Fig. 25 are similar to those that appear in idealized models for liquid-gas critical points, and ferromagnets near their Curie temperatures. Those models commonly rely on short-range interactions between entities such as atoms or spins that have been treated in lattice cell models. Fig. 1.5 in Stanley's book [72] is based on a lattice cell model. The resemblance of that figure to Fig. 25 and its evolution as $T$ approaches $T_\lambda$, as we have described it above, is evident.

There seems to be a widely held view that short-range interactions involving neighboring objects (e.g. spins) propagated over long distances are responsible for all λ - type transitions (see for example Refs. [106a,106b]). Therefore, it is important to note that in our vortex model for the λ - type transition in liquid $^4$He the basic interactions between vortices are long-ranged. For example, the interaction $F_{CI}(i,j)$ in Eq. (6.10) varies as $\left|\boldsymbol{R}_i - \boldsymbol{R}_j\right|^{-3}$ and vortex rings separated by large distances compared to the average spacing between rings give important contributions to the cross term that occurs in $F_{IN}^2$ in Eq. (6.23).

The Helmholtz potential that determines $T_\lambda$ in our theory is a function of $v_0$. And $v_0$ is a function of the density of the liquid and density is a function of pressure. Elementary excitation (roton and phonon) energies and momenta are known to be functions of density and pressure from neutron scattering measurements [99]. The Helmholtz potential in our theory depends on elementary excitation energies and momenta. This explains why there is a line of critical points in liquid $^4$He near the superfluid transition instead of an isolated critical point as in other λ - type transitions for one-component fluids at liquid-gas criticality.

The statistical correlation among vortex rings that produces a superfluid wind at each vortex ring still exists in the complete model. This correlation represents the order that exists in the liquid for $T < T_c$.

Second sound in some range of wavelengths can propagate as oscillations in the liquid for any $T$ such that $T < T_c$. That wavelength range extends to ∞ for $T < T_\lambda$. However, for $T_\lambda < T < T_c$ the wavelength range for oscillations is restricted by the condition that $\lambda < \xi$, where $\xi$ is a linear dimension of the largest superfluid "drop."

Our model suggests that there will be two branches of the second sound dispersion curve, $\omega$ vs $k$, at temperatures in the critical region. The second sound in a high frequency - short wavelength branch propagates in the He II between neighboring "bubbles" of He I below $T_\lambda$ and in the He II embedded in the He I matrix above $T_\lambda$. In this branch the oscillations involve only normal fluid density, due to phonons and rotons, and superfluid density. These densities are similar to those for the liquid that is outside the critical range of temperatures. On the other hand, second sound in the low frequency - long wavelength branch propagates at lower velocity than in the short wavelength branch. The "bubbles" of He I reduce the average superfluid density in the stable regions of the liquid when the average is determined for spatial volume elements large enough to contain many vortex rings. The linear dimensions of such volume elements can be much less than the long wavelength of the second sound. Also, the oscillating superfluid velocity induces motion of the vortex cores and the mantles that surround them. In effect, this increases the normal fluid density involved in the second sound, and decreases the superfluid density. The average superfluid density described here may be called "renormalized" because of the effects of the "bubbles" and the vortex cores.

Second sound measurements using a low frequency heater and a detector placed on opposite walls of the container will probe the low frequency branch and the signal will vanish at the connectivity threshold, which occurs at $T_\lambda$ in our model. Electromagnetic wave scattering and neutron scattering could be used to study both branches of second sound dispersion provided that second sound at the relevant $\omega$ and $k$ couples strongly enough to the density fluctuations that are generated directly in the scattering processes.

### *7.1.4 Line vortices and large vortex rings*

We will focus on line vortices and neglect effects that small closed vortex rings may have on free energy density in the presence of superfluid flow in this Section. Consider a rectilinear vortex near $T_c$ and located on the axis of a cylindrical container that has radius $R$ and length $L$. In the theory as formulated here, the vortex core energy per unit length is independent of temperature and the variational calculation for $T$ = 0 K described earlier in this paper gives the following result, where $E_C$ is the core energy measured above the ground state energy $E_0$ for a liquid with constant atomic density everywhere:

$$\frac{E_C}{Lk} = 1.68 \quad \text{K Å}^{-1} \ . \tag{7.4}$$

For $T > 0$, the energy of the vortex core, measured above the free energy of the liquid at rest everywhere, contains a term $E_D$ that is added to $E_C$. $E_D$ is due to the $\mathcal{N}_C$ atoms that *remain* in the core, and

$$\frac{E_D}{Lk} = -\frac{\mathcal{N}_C}{L} A \ . \tag{7.5}$$

This term is present because we must enforce the condition that the same number of atoms are present in the liquid with and without the vortex. We are treating the atoms that are missing from the core region as if they are spread uniformly over the region outside the core where the free energy per atom measured with respect to $E_0$ and expressed in Kelvin is $A$ (see also Eq. (6.1) and Fig. 15). For the core density profile $n(\rho)/n_0$ resulting from the variational calculation, where $N$ = 0.50 in Eq. (3.2),

$$\frac{n(\rho)}{n_0} = \sin\frac{\pi\rho}{2r_0} \ , \tag{7.6}$$

one finds that $\mathcal{N}_C$ is given by the following formula,

$$\frac{\mathcal{N}_C}{L} = n_0 2\pi \int_0^{r_0} d\rho\, \rho \sin\frac{\pi\rho}{2r_0} = \frac{8}{\pi} n_0 r_0^2 \ . \tag{7.7}$$

For $r_0$ = 2.80 Å and $n_0$ = 0.0218 atomÅ$^{-3}$, we find $\mathcal{N}_C/L$ = 0.435 atom Å$^{-1}$. Computer calculations we made based on Eq. (4.3) give the following result for $A$ evaluated at $T = T_c$ = 2.172 K:

$$A = \frac{F(v_s = 0)}{\mathcal{N}\, k} = -0.2335 \quad \text{K atom}^{-1} \ . \tag{7.8}$$

Then using Eq. (7.7) we find

$$\frac{E_D}{Lk} = 0.10 \quad \text{K Å}^{-1} \tag{7.9}$$

and

$$\frac{E_{core}}{Lk} = \frac{1}{Lk}\left(E_C + E_D\right) = 1.78 \quad \text{K Å}^{-1} \ . \tag{7.10}$$

The free energy contribution $F_E$ by liquid external to the core can be evaluated using the superfluid velocity $\boldsymbol{v}(r) = (\kappa/2\pi\rho)\hat{\theta}$, $n_0$ = 0.0218 atomsÅ$^{-3}$, and a free energy density $f(\boldsymbol{r})$ per atom measured above the background $A$. Here $f(\boldsymbol{r})$ is the velocity dependent free energy at $\boldsymbol{r}$ constructed in the pattern of Eq. (6.2) but using the full expression for $F(v_s)$ based on Eq. (4.3) instead of the truncated expression in Eq. (6.1). Then $F_E$ is given by

$$F_E = n_0 \int_{V_E} d^3r \ f(\boldsymbol{r}) \tag{7.11}$$

where $V_E$ is the volume external to the core.

The following decomposition of $F_E$ is useful:

$$F_E = F_{E1} + F_{E2} \tag{7.12}$$

where

$$F_{E1} = n_0 \int_{V_E} d^3r \left[ f(\boldsymbol{r}) - \frac{1}{2}\frac{\rho_s(0)}{n_0} v_s^2(\boldsymbol{r}) \right] \tag{7.13}$$

$$F_{E2} = \int_{V_E} d^3r \left[ \frac{1}{2}\rho_s(0) v_s^2(\boldsymbol{r}) \right] . \tag{7.14}$$

The advantage of the decomposition in Eq. (7.12) is that the integrand in Eq. (7.13) is a short range function of cylindrical radius $r$ (like the $Cv_s^4(\boldsymbol{r})$ term in Eq. (6.1)), and the value of $F_{E1}$ does not depend on the radius of the container. The superfluid density for the liquid at rest is $\rho_s(0)$. $F_{E2}$ is a function of the radius of the container.

The integrand in Eq. (7.13) is a slowly varying function of $T$ for a small temperature range near $T_c$, and we will assign $F_{E1}$ its value at $T_c$ in the analysis that follows. Numerical evaluation of $F_{E1}$ at $T_c$ gives the result

$$\frac{F_{E1}}{Lk} = -0.78 \quad \text{K Å}^{-1} \ . \tag{7.15}$$

The integral in Eq. (7.14) can be evaluated as follows:

$$
\begin{aligned}
\frac{F_{E2}}{Lk} &= 2\pi\left(\frac{1}{2}\rho_s\left(0\right)\right)\int_{r_0}^{R} d\rho\,\rho\left(\frac{\kappa}{2\pi\rho}\right)^2 \\
&= \frac{1}{4\pi}\rho_s\left(0\right)\kappa^2\ln\frac{R}{r_0} \qquad (7.16)\\
&= \frac{\kappa^2 n_0 m\alpha\left(T_c - T\right)\ln\frac{R}{r_0}}{4\pi k}\,.
\end{aligned}
$$

In the last step we have used Eq. (6.29). The result $\alpha = 3.56\ \mathrm{K}^{-1}$ is given in Eq. (6.30).

The total free energy $F_{RV}$ of a rectilinear vortex on the axis of a cylindrical container is the sum of contributions from a short-range part $F_{SR}$ and from an external part $F_{E2}$:

$$F_{RV} = F_{SR} + F_{E2} \qquad (7.17)$$

where

$$\frac{F_{SR}}{Lk} = \frac{1}{Lk}\left(E_{core} + F_{E1}\right) = 1.00\ \ \mathrm{K\,Å^{-1}}\ . \qquad (7.18)$$

Because $F_{SR}$ depends only on short range functions of $r$, which have support only in and near the vortex core, this term will be applicable to any vortex line that is not close to the container walls. Simple estimates suggest that 100 Å separation from the walls would be sufficient to render image effects negligible for this term for a typical laboratory container with $R$ = 1 cm.

The term $F_{E2}$ is positive for $T < T_c$, zero at $T = T_c$, and negative for $T > T_c$. From Eq. (7.12) one can deduce that for $T < T_c$ where $\rho_s(0) > 0$, $F_{E2}$ is not only positive when the vortex line is located on the cylindrical axis and its image is at infinity, but also is positive for any position or orientation of the line where $\boldsymbol{v}_s(\boldsymbol{r})$ includes the velocity contribution from images.

### *7.1.5 Vortices in the liquid below $T_c$*

From these observations, one can see that for $T \le T_c$ when the liquid is in complete thermodynamic equilibrium, $F_{SR}/Lk$ in Eq. (7.18) is a lower bound on free energy per unit length of any vortex line in the bulk liquid away from container walls. The result in Eq. (7.18) then implies that in a cylindrical container 1 cm high, the excitation energy for a single vortex line parallel to the axis would be at least as great as $10^4$ eV (equivalent to about $10^8$ K), which we will call $F_M$.

One can make a crude estimate for an upper bound on the canonical probability for a vortex line parallel to the cylindrical axis but located anywhere in the liquid. It seems unlikely that the density of states factor will be greater than $\mathcal{D} = \mathcal{R}^2/a^2 \cong 10^{16}$, and we shall use this value for estimating an upper bound $\mathcal{P}$. Then $\mathcal{P} = \mathcal{D}\, e^{-\beta F_M}$ for $T = T_c$, and $\mathcal{P} = \left(10^{16}\right)\exp\left[-10^8/2.172\right]$, a number that is almost incomprehensibly small, and therefore justifying neglect of vortex lines in calculating properties of bulk liquid that is in complete thermodynamic equilibrium when $T \le T_c$.

*7.1.6 The catastrophe temperature*

Something interesting happens in a finite container as the temperature increases above $T_c$ when the liquid starts from a condition of complete thermodynamic equilibrium below $T_c$. Recall that $F_{E2}$ is negative for $T > T_c$ according to Eq. (7.16). There is a threshold temperature $T_t$ that is attained when the following condition is met,

$$F_{RV} = F_{SR} + F_{E2} = 0 \quad , \tag{7.19}$$

and one obtains the following result for $T_t$, which is independent of the length of the vortex line:

$$\left(T_t - T_c\right) = \frac{4\pi k\left(\dfrac{F_{SR}}{Lk}\right)}{\kappa^2 n_0 m\,\alpha \ln\dfrac{\mathcal{R}}{r_0}} = 0.3385\left(\ln\frac{\mathcal{R}}{r_0}\right)^{-1} \text{K} \quad . \tag{7.20}$$

For $\mathcal{R}$ = 1 cm, we find $(T_t - T_c) \simeq 19 \times 10^{-3}$ K.
For $\mathcal{R}$ = 1 μ, we find $(T_t - T_c) \simeq 41 \times 10^{-3}$ K.

There will be equal probability of the liquid being everywhere at rest and in a state with one rectilinear vortex line on the axis at the temperature $T_t$ where the two states have the same free energy. We shall call this $T_t$ the *catastrophe temperature* for reasons that will be explained shortly. We note that as the radius $\mathcal{R}$ tends to infinity, the threshold temperature $T_t$ approaches $T_c$.

A similar analysis can be carried out for large vortex rings for $T > T_c$. The total free energy $F_{CR}$ of a circular vortex ring of radius $R_V$ can be expressed as

$$F_{CV} = F_{CSR} + F_{CE2} \quad , \tag{7.21}$$

analogous to $F_{RV}$ in Eq. (7.17). Here $F_{CSR}$ is the free energy contribution from the short-range function having support in and near the vortex core. To good approximation $F_{CSR}$ is given by

$$\frac{F_{CSR}}{k} = 2\pi R_V \frac{F_{SR}}{Lk} \tag{7.22}$$

where $F_{SR}/Lk$ is given by Eq. (7.18). $F_{CE2}$ is a contribution from the long-range part of the external free energy, and is given by simple adaptation of a well-known formula [107] in semi-classical hydrodynamics, as follows:

$$\begin{aligned}\frac{F_{CE2}}{k} &= \frac{1}{2k}\rho_s(0)\kappa^2 R_V\left[\ln\frac{8R_V}{r_0}-2\right] \\ &= \frac{1}{2k}n_0 m\kappa^2 R_V\,\alpha\left(T_c-T\right)\ln\left[\frac{8R_V}{r_0}-2\right].\end{aligned} \tag{7.23}$$

The threshold value ($T_{Ct}$ – $T_c$) where $F_{CV}=0$ is given by

$$\begin{aligned}\left(T_{Ct}-T_c\right) &= \frac{4\pi k\left(\dfrac{F_{SR}}{Lk}\right)}{\kappa^2 n_0 m\alpha\ln\left[\dfrac{8R_V}{r_0}-2\right]} \\ &= 0.3385\ln\left[\frac{R_V}{r_0}+0.079\right]^{-1}\ \mathrm{K}.\end{aligned} \tag{7.24}$$

In arriving at Eq. (7.24) we have neglected effects of images on $F_{CE2}$, and this neglect can be justified only if the vortex radius $R_V$ is significantly smaller than $R$, the container radius, for a ring coaxial with the container. We will represent this rather indefinite condition as $R_V$ less than about 0.5 $R$. Subject to this assumption, one finds from Eqs. (7.24) and (7.20), that the temperature $T_{Ct}$ is such that $T_{Ct} > T_t$. This implies that a vortex line should appear on the axis at a lower temperature than a coaxial vortex ring in the bulk liquid, at least according to the criterion of line or ring having free energy equal to that of the liquid at rest everywhere. However, $T_{Ct} \simeq T_t$, and in some earlier Sections of this paper we have not made a distinction between them.

It seems likely that vortex lines and large vortex rings will proliferate rapidly once the threshold temperature is reached, and that the profusion will persist at all higher temperatures. At and above $T_t$, interactions will cause the vortices to distort in shape and move with irregular motion. This characterization is reminiscent of Feynman's [56] picture of the normal state; however, Feynman's picture was based on different arguments.

It is interesting to speculate what happens if the temperature of the liquid starts above $T_t$ and is then lowered. There may be a long relaxation time for the vorticity to subside as $T$ crosses $T_t$, and the profusion of vorticity may exist in metastable states that dominate thermodynamics of the liquid for the time scales relevant to experimental measurements that have been made so far.

### 7.2 Liquid $^4$He in metastable equilibrium

Liquid $^4$He in complete thermodynamic equilibrium was studied in Sec. 7.1. An estimate was made in Sec.7.1.2 that the strongly critical region would begin around $\left(T_\lambda - T\right) = 10^{-4}$ K. That estimate is based on the crossover model, where $T_c = T_\lambda$. On the other hand, experimental data on specific heat in liquid $^4$He at *SVP* indicate that the onset of strong criticality is around $\left(T_\lambda - T\right) = 2\times10^{-2}$ K [69]. Our theoretical estimate for onset ultimately depends heavily on our variational calculations of vortex ring energies and the momenta based on the variational core structure. If we assume that our variational results are at least moderately accurate and that our estimate for the onset of strong criticality is about correct, then there must be a mechanism at work in the liquid samples that have been studied experimentally that accounts for the discrepancy in onset temperature for strong criticality. In Sec. 7.2.1 and 7.2.2 we will study the possibility that metastable liquid containing pinned-line vortices is the origin of the mechanism that accounts for the discrepancy.

#### 7.2.1 Vortices in metastable states below $T_c$

Experiments by Awschalom and Schwartz [109] suggest that pinned vortex lines persist in the liquid, in metastable states, for long time periods as the temperature is lowered far below $T_\lambda$. Other experiments [110-112] have been performed to study relaxation of assumed vortex line density as temperature is lowered through $T_\lambda$. One part of an experiment proposed later in this paper also addresses this issue.

The density of pinned vortex lines observed by Awschalom and Schwartz seems much too small to reasonably be expected to account directly for a significant contribution to the specific heat at any temperature. However, in the analysis that follows it will be argued that the effect of pinned-line vortices on specific heat is multiplied greatly near $T_c$ (and near $T_\lambda$) through catalytic action that generates large vortex rings. Interaction between these large vortex rings is then studied in Sec.7.2.2 as the mechanism that possibly accounts for the specific heat in almost all existing measurements.

The core of a pinned vortex line at temperatures well below $T_c$ is fixed, with a definite length, and the core energy is constant. The contributions of energy, $F_{E1} + F_{E2}$, external to the core can be estimated using Eqs. (7.15) and (7.16). The specific heat for a system of $N$ non-interacting straight vortex lines can then be calculated, and the results show that there is no anomaly in the specific heat. If the approximate temperature dependence of specific heat $C_V \sim \ln\left|T_\lambda - T\right|$ observed in existing experiments is associated with the metastable pinned line vortices, then interaction between vortices must be taken into account.

Suppose vortex ring interactions contribute to the anomaly in specific heat. Pinned line vortices will affect the temperature dependence of the vortex ring correlation length that depends on the boundaries of the parts of the liquid that would be unstable against small perturbations in $v_s$ according to ordinary thermodynamic stability criteria. Furthermore, the pinned vortices can act as a catalyst for production of vortex rings and the location, character, and behavior of the resulting rings may be significantly influenced by the pinned vortex lines. The basis for these statements will be explained next.

For a rectilinear vortex on the axis of the container, just below $T_c$ the cylindrical radius $R_L$ at which $v_s(R_L) = v_0$ is given by

$$R_L = \frac{\kappa}{2\pi}\left(-\frac{B}{6C}\right)^{-\frac{1}{2}} = \frac{\kappa}{2\pi}\left(-\frac{m\alpha}{12kC}\right)^{-\frac{1}{2}}\left(T_c - T\right)^{-\frac{1}{2}} . \tag{7.25}$$

Then

$$R_L = a_L\left(T_c - T\right)^{-\frac{1}{2}} \tag{7.26}$$

where

$$a_L = 5.71 \text{ Å}\,\text{K}^{\frac{1}{2}} . \tag{7.27}$$

Comparing $R_L$ as given by Eqs. (7.26) and (7.27) with the correlation length $R_0$ for a small vortex ring in Eqs. (6.33) and (6.34), one can see that, per unit length of vorticity, the unstable region for a single vortex line would fill the volume of a finite container at a much lower temperature than would the unstable region for a small vortex ring having $R$ = 2 Å. Both $R_0$ and $R_L$ influence the strength and temperature dependence of the superfluid wind interaction for small vortex rings.

Arguments given shortly suggest that in the *metastable* state the distribution of vortex ring radii may not be sharply peaked and may include large radii. This must be taken into account in treating statistical correlations in vortex ring behavior induced by interactions among them. All of these considerations may affect the temperature dependence of the effective superfluid wind experienced by the rings, and therefore affect the dependence of specific heat on ($T_c$ – $T$), changing it from the specific heat for complete thermodynamic equilibrium.

Consider now the mechanisms by which pinned vortex lines may act as a catalyst for production and maintenance of a possibly large population of vortex rings when $T$ is near $T_c$. First we recall some analysis and observations by Awschalom and Schwartz [109] in their work on pinned line vortices. Direct evidence for such vortex lines was found at temperatures below about 1.7 K, where ions could be trapped on vortex cores. To simplify our discussion, we will suppose at first that $T$ << 1.7 K. Awschalom and Schwartz considered a pair of almost parallel, but slightly bowed, pinned vortex lines. In our discussion we will assume that the vortex lines have opposite circulation. In this instance, the lines must be bowed toward each other, as shown in Fig. 26a, in order for them to be in mechanical equilibrium when they are at rest. The curved configuration is such that the self-induced velocity [109a] of a vortex line everywhere cancels the nonlocal field [109]. The curvature of the pinned lines is greatly exaggerated in Fig. 26 for clarity. One can show that this bowed configuration has lower free energy than a strictly parallel straight-line configuration by considering the following formula [113] for the kinetic energy $E_K$ of a vortex pair in a liquid having uniform density $\rho$,

$$E_K = \frac{1}{2}\rho\int d^3R\, v^2\left(\boldsymbol{R}\right) = \frac{1}{2\pi} L\rho\kappa^2 \ln\frac{d}{r_0} \quad . \tag{7.28}$$

$L$ is the length of each line, $d$ is the distance between the pairs and $r_0$ is the radius of the assumed hollow cores. Using Eq. (7.28), one can see that the bowed lines have a smaller average value of $d$ than would straight lines, and therefore lower kinetic energy. The change in length $L$ of a vortex line due to a slight amount of bowing gives a negligible correction to $E_K$ when $d << L$.

The formula in Eq. (7.28) can be applied to vortex lines at finite temperatures provided that $\rho$ is replaced by $\rho_s(0)$ and $E_K$ is interpreted as a portion of the free energy external to vortex cores, similar to $F_{E2}$ in Eq. (7.14) where a single vortex line was treated. Then at, say 1.4 K where $\rho_s(0) \simeq \rho$ (see Fig. 16), numerical evaluation of Eq. (7.28) yields the following result

$$\frac{F_{E2}}{Lk} = 1.66 \ln\frac{d}{r_0} \;\; \mathrm{K\,Å^{-1}} \quad . \tag{7.29}$$

$F_{E2}$ in Eq. (7.29) refers to a vortex pair. For $d/r_0 \simeq 10^7$, which is about the ratio that Awschalom and Schwartz estimated for their experiment where flat plates were separated by 1 cm, one finds using Eq. (7.29) that $F_{E2}/Lk \simeq 27\ \mathrm{K\,Å^{-1}}$. The short-range contribution to the free energy for the pair under the stated conditions is roughly $F_{SR}/Lk \simeq 3\ \mathrm{K\,Å^{-1}}$. So $F_{E2}/Lk$ gives the dominant contribution to the free energy of the vortex pair with opposite circulation at low temperatures.

Now suppose that pinned vortex lines are reduced in length, the subtracted line length being transferred to vortex rings as shown schematically in Fig. 26b. The increase in the average value of $d$ would tend to increase the external free energy for the pinned lines. The vortex rings would change the kinetic energy locally and depending on vortex size may tend to either increase or decrease the external free energy. In general, the two configurations shown in Figs. 26a and 26b would *not* be approximately degenerate in the sense of having the same free energy, at least when $T << T_c$. Also, the free energy of the liquid would be high, and the probability of occurrence of such configurations would be negligible.

However, for $T$ near $T_c$, $\rho_s(0)$ approaches zero, and $F_{E2}$ for the pair tends to zero. Under these conditions, only short-range contributions to $F_{E2}$ for lines and rings are important in statistical mechanics. The short-range *energy per unit length* for an individual rectilinear vortex is then the same as for a large ring, and the configuration of vorticity in Figs. 26a and 26b have about the same free energy when the length of the vortex lines plus vortex ring circumferences is about the same in both cases. In fact, other configurations containing a multiplicity of large rings and less bowed vortex lines exist that also have about the same free energy as the highly bowed lines when no vortex rings are present. Here it is assumed that the vortex rings can configure themselves somewhat like that in Fig. 26b so that the velocity field that they induce at the positions of the less bowed lines still permits the condition of mechanical equilibrium to be satisfied at the locations of the vortex line elements.

All of the vortex configurations described here have free energy so high that their probability of occurrence even near $T_c$ is negligible when the liquid is in complete thermodynamic equilibrium. But near $T_c$ the vortices in these configurations are nearly degenerate with each other and that is statistically important for the metastable state with pinned lines. The multiplicity of states for the vortex rings together with interaction of rings with other rings and the pinned vortex lines may then dominate the thermodynamic properties of the liquid in the metastable state near $T_c$ and possibly account for the specific heat anomaly near $T_c$.

### *7.2.2 Theory of specific heat for liquid $^4$He in metastable states*

The complete model that was used in Sec. 7.1.3 in treating the $\lambda$ transition for complete thermodynamic equilibrium can be extended to treat the metastable liquid with pinned-line vortices. We will explain how that extension may be accomplished next.

We argued in Sec.7.2.1 that a number of different configurations of pinned vortex lines will be present in metastable equilibrium. Now we will focus on that particular configuration that causes the system of pinned lines and associated large vortex rings that have been created in the manner described in Sec. 7.2.1 to produce the minimum contribution to the interaction free energy term in Eq. (6.19). For brevity we will call this the *minimum* configuration. The interaction term is treated essentially the same as in the complete model in Sec. 7.1.3 where all of the vortex rings are taken into account simultaneously.

Positions of the pinned lines in the minimum configuration are assumed to be slowly varying with temperature, and we will consider them as fixed in our analysis for temperatures near $T_c$ and $T_\lambda$. To keep our discussions simple and clear, we will not represent those positions explicitly in the formulas but just state that they introduce implicit parameters in the interaction free energy of the vortex rings. The boundaries of mantles of unstable liquid about the pinned lines depend on $v_0$ and expand as $T$ increases toward $T_c$. Those mantles contribute to the regions of unstable liquid that are indicated as excluded from the vortex ring interactions in the integrals by the prime rotation in Sec. 6, and explained just below Eq. (6.41). Now the excluded regions referred to below Eq. (6.41) must be applied to the entire set of $N$ vortex rings and all of the pinned vortex lines. The unstable mantle regions make no contribution to the restoring forces that counteract small fluctuations in the orientations or positions of the vortex rings away from conditions of local thermodynamic equilibrium throughout the liquid. Those conditions also hold for the minimum configuration we are considering now. Very close to $T_c$ the mantle boundaries will be almost spherical for the large rings we are considering because a dipole approximation can be used for the superfluid velocity field. The dipole approximation was also used to establish the mantle boundaries in the crossover model and complete model treated in Sec.7.1. In particular, Eq. (6.26) is still applicable for calculating $R_0$ very near $T_c$, provided that the dipole moment *M* is evaluated using the vortex radius *R* for each large ring that is excited.

A superfluid wind at the core of each vortex ring will be generated by statistical correlations among the vortex rings, just as in the complete model for complete thermodynamic equilibrium. The vortex rings will be ordered below $T_c$ in the sense that the rings produce the statistical correlations referred to here. An interaction free energy among vortices due to the vortex wind will be generated in the same manner as was found for the complete model with liquid in complete thermodynamic equilibrium. But now that interaction free energy may be much larger than that treated in Sec. 7.1 because of the larger vortex rings that produce it. Enhancement of the vortex ring interaction strength due to involvement of *large* vortex rings is the main consideration in our proposal that metastable states containing pinned line vortices account for critical properties in essentially all existing experiments on criticality in liquid $^4$He at the He I - He II phase boundary. The strength of that interaction free energy decreases as temperature increases toward $T_c$. However, the interaction free energy does not vanish at $T_\lambda$. Instead it vanishes at $T_c$, which is somewhat above $T_\lambda$. The entropy associated with this interaction does not vanish at $T_\lambda$, but the rate of change of entropy associated with the interaction term near $T_\lambda$ accounts for the specific heat anomaly. This is the same behavior we described in Sec. 7.1.3 for the liquid in complete thermodynamic equilibrium. The physical mechanism responsible for the λ transition in the metastable state is basically the same as that discussed in Sec. 7.1 for the crossover model. To make clear the physical mechanism of the λ transition taking into account these considerations, we state it again for the complete models treated in Sec. 7.1 and 7.2.

### *7.3 Physical mechanism of the λ transition for complete models*

The following description is applicable to conditions of complete thermodynamic equilibrium and metastable equilibrium and summarizes results found in Secs. 7.1 and 7.2.

The vortex ring model, as we have formulated it, indicates that vortex rings are thermally excited in liquid $^4$He. Small closed rings are statistically dominant below $T_c$ for the liquid in complete thermodynamic equilibrium. Larger, open rings that are much smaller than the dimensions of the container are statistically dominant below $T_c$ for the liquid in metastable equilibrium, where there are pinned line vortices. Because of their large core energy, other vortex rings are assumed to be negligible.

At any finite temperature, each vortex ring is surrounded by a mantle where the liquid is unstable against small perturbations according to ordinary thermodynamic stability criteria. However, because quantization of circulation imposes a constraint on superfluid velocity, the mantle region can still contribute to the *self-energy* of the vortex rings. But the mantle does not contribute to the interaction free energy that would provide a restoring force against perturbation of the relative configuration of the vortex rings with respect to each other for the liquid *at least* in local thermodynamic equilibrium throughout the liquid. This chameleon - like behavior has been formulated as a "*Principle of Constrained Instability*." The velocity $v_0$ of the liquid at the outer boundary of the unstable mantle region is the velocity at which the Helmholtz potential for the flowing superfluid has an inflection point, and $v_0$ is the order parameter of this theory. The order parameter is positive below $T_c$ and decreases to zero at $T_c$, and $v_0 \sim \left(T_c - T\right)^{1/2}$. There is no inflection point for $T > T_c$ and so there is no physically meaningful value of the order parameter for $T > T_c$. Near $T_c$, $v_0$ is proportional to $\rho_s\left(0\right)^{1/2}$, where $\rho_s\left(0\right)$ is the superfluid density evaluated at $v_s = 0$ when phonons and interacting rotons account for the normal fluid. We will call $\rho_s\left(0\right)$ the unrenormalized superfluid density evaluated at $v_s = 0$. $R_0$ is the distance from the center of a vortex ring at which the unrenormalized superfluid velocity reaches $v_0$, and $R_0 \sim \left(T_c - T\right)^{-\frac{1}{6}}$. The mantle region determined by $R_0$ is approximately spherical.

There is a statistical correlation in the relative configuration of the vortex rings that induces a net superfluid velocity at the core of each vortex ring. That velocity may be regarded as a superfluid wind. The statistical correlation depends on the interaction free energy associated with regions external to all of the mantles, and that correlation can be viewed as a polarization effect. The mantles of any pinned line vortices that may be present must be taken into account in determining the external region.

The superfluid wind is in a direction opposite to the dipole moment of a vortex ring. The direction of the dipole moment is the same as the direction of momentum $\boldsymbol{P}_{MI}$ associated with atoms inside the core of a vortex ring. The core boundary is determined by the Landau critical velocity, the velocity at which the flowing ground state becomes degenerate with respect to a single roton state that is also in a state of flow where the flow is treated as if it is uniform throughout the liquid. A negative contribution to the vortex ring free energy is accounted for by the scalar product of $\boldsymbol{P}_{MI}$ with the vortex wind velocity.

For $T$ far enough below $T_c$, the mantles of individual rings rarely overlap and the configuration of the stable and unstable regions may be described metaphorically as He I "bubbles" embedded in a matrix of He II. See Fig. 25. For higher temperatures that are just below $T_c$, the configuration of the stable and unstable regions may be described metaphorically as He II "drops" embedded in a matrix of He I. If the temperature starts well below $T_c$ and then increases, the mantle of each vortex ring expands. At some temperature below $T_c$, the two qualitatively different types of configurations meet, and that is the condition that determines the $\lambda$ temperature $T_\lambda$. The interaction free energy among vortex rings and the associated entropy are continuous through $T_\lambda$. However, the specific heat is peaked there due to rapid changes with temperature in configuration of the stable regions that contribute to the interaction free energy near $T_\lambda$.

The peak in the specific heat at $T_\lambda$ is just a conjecture at this stage of development of the theory. However, scaling and other methods that are frequently applied to idealized models in the study of critical phenomena are available to study the validity of this conjecture. The next section includes some further comments on this problem. Also, the interaction free energy may be susceptible to direct evaluation, at least in a good approximation.

Line vortices will be pinned at different sites and in different configurations and their volume density will be somewhat different for different passages of the liquid through the catastrophe temperature $T_t$, which is somewhat above $T_c$. Therefore the absolute value of $T_\lambda$, which depends on these details, may be slightly different from one experimental run to another. However, other characteristics of the $\lambda$ transition, and in particular the critical exponents which depend on $|T - T_\lambda|$, may to be almost independent of these details of pinned line vortex characteristics.

### *7.4 Comments on other theories of the λ transition*

A number of existing measurements [70,70a,114,115] of specific heat in liquid $^4$He have been fitted to functional forms with several adjustable parameters consistent with predictions of scaling [72] and of field theoretic calculations based on a phenomenological Landau-Ginzburg-Wilson (LGW) Hamiltonian and renormalization group methods [116-118]. The fits are found to be very good in some instances, but particularly at high pressures, some experimental results deviate appreciably from theoretical predictions [119-121]. These deviations provide an incentive for seeking a more fundamental theory, as we have done in this paper.

From the viewpoint of the microscopic theory that has been elaborated in this paper, there are some puzzling questions that confront us in the scaling and LGW Hamiltonian based theory. For example, those latter approaches treat the order parameter $\psi$ as a complex scalar field where $\psi(x,y,z) = \eta e^{i\varphi}$ [122] is defined in such a way that $\rho_s = m|\psi|^2$ and $\boldsymbol{v}_s = (\hbar/m)\nabla\varphi$. The magnitude $\eta$ and phase $\varphi$ of the order parameter are treated as independent variables in those approaches, and that in turn implies that the superfluid density and the superfluid velocity are independent variables in the critical region near $T_\lambda$. On the other hand, microscopic theory that accounts well for the two-fluid model, the superfluid density, the entropy, and the specific heat at least up to around the beginning of the critical region indicates that superfluid density is a function of superfluid velocity. It seems quite improbable that this functional dependence does not persist into the critical region.

Furthermore, the approach based on the LGW Hamiltonian ignores the quantization of circulation in $^4$He in its treatment of the velocity potential that occurs as the phase of the order parameter. It seems unlikely that quantization of circulation is not operative in the critical region but is operative in the range below that region.

The interaction free energy $F_{IN}$ in the microscopic based theory that is treated in Secs. 6, 7.1.3, 7.2, and 7.3 may be a key to resolving this seeming conflict between the different approaches. An explanation was given in Secs. 7.1.3, 7.2, and 7.3 for how $F_{IN}$ may account for a $\lambda$ type phase transition. $F_{IN}$ depends on $v_0 \sim \rho_s(0)^{1/2}$, where $\rho_s(0)$ is the superfluid density evaluated at $v_s = 0$ when only phonons and rotons are taken into account in determining the normal fluid density. Clearly $v_0$ and $\rho_s(0)$ have no velocity dependence. These observations suggest that when $F_{IN}$ is treated quantitatively that the scaling relations among critical point exponents will be satisfied and that the values of the exponents themselves may be computed using $F_{IN}$.

The critical exponents are for behavior of functions near $T_\lambda$, not the $T_c$ where $\rho_s(0)$ referred to above vanishes. This points to the importance of taking into account that the superfluid density used in the scaling relations is determined by methods that probe the low frequency branch of second sound. The velocity of second sound vanishes at $T_\lambda$ for that branch, as discussed in Sec. 7.1.3. The superfluid density for that branch takes into account vortex rings and the unstable liquid in the mantles surrounding them along with the normal fluid density due to phonons and rotons, and the renormalized superfluid density will vanish at $T_\lambda$, not $T_c$. We further conjecture that $F_{IN}$ for the metastable state can be represented approximately by the LGW Hamiltonian or some slightly modified version of it.

It is noted here that Williams' vortex theory [75p,75q] of the λ transition is fundamentally different from the vortex theory that we have developed in this paper. A critical discussion of his approach is contained in Appendix D.

## 8. Proposed specific heat experiment

Our vortex model for the λ transition departs far from conventional ideas, and further experimental support of our theory would be useful. A proposed experiment with this motivation is described next. The experiment consists of measuring specific heat of the liquid under conditions of complete thermodynamic equilibrium and also under extreme conditions of non-equilibrium, in metastable states. The background, procedure, and possible consequences of the experiment are discussed in what follows.

*8.1 Summary of background for the experiment*

The theory of liquid $^4He$ developed in this paper makes the following predictions for conditions of *complete thermodynamic equilibrium*. The reader is reminded here that $T_c$ = $T_\lambda$ in the crossover model that is the basis of theoretical estimates in item (i) below.

(i) The specific heat varies as $(T_\lambda - T)^{-1/2}$ in the crossover region where there are initial signs of deviation from a smooth background as temperature increases. The crossover region merges smoothly with the critical region and the onset of critical behavior where the divergent behavior changes is much closer to $T_\lambda$ than the onset of critical behavior observed in existing experiments. The onset of strongly critical behavior is estimated to be near $|T_\lambda - T| \simeq 10^{-4}$ K for the vortex model but near $|T_\lambda - T| \simeq 10^{-2}$ K for existing experimental data. Calculations based on the vortex model applied in the crossover region indicate that the coefficient of the term that accounts for the diverging behavior may be so small that this term cannot be detected with present technology, and if that is the case, specific heat will appear to follow a smooth curve, the "background," at the transition from He II to He I.

(ii) Vortex lines and *large* vortex rings are so sparse as to be negligible below $T_c$, and only *small* vortex rings are thermally excited with significant number density there.

(iii) Above $T_c$, for liquid in a finite container, the specific heat follows a smooth background for a small but finite temperature interval until a temperature is reached where vortex lines and large vortex rings are thermally excited in great numbers in a catastrophic event, and there is a profusion of vortices at temperatures higher than the catastrophe temperature.

The explanation proposed for the discrepancy between these predictions and what has been so far observed experimentally is that in previous experimental measurements of specific heat, the liquid was in some metastable states associated with pinned line vortices.

*8.2 Procedure for the proposed experiment*

The experiment consists of the following two parts.

*8.2.1 PART I*

Starting from a temperature above $T_\lambda$ and a pressure above about 25 atmospheres (see Fig. 27) the liquid is cooled below $T_\lambda$ (Fig. 27, path 1) until it completely solidifies. Solidification will remove all vortices from the $^4He$. The $^4He$ is carefully annealed just above the solidification pressure to remove defects from the solid. The temperature is then increased slowly (Fig. 27, path 2) so that the $^4He$ melts, and the temperature is increased further until the liquid passes through the λ line that has been located (temperature and pressure) in previous measurements where solidification did not occur. If we assume for example that the temperature increase occurs at almost constant pressure, the temperature should be raised perhaps 100 millikelvin above $T_\lambda$, to observe a predicted catastrophic event, if the container radius is about 1 cm. The specific heat of the liquid is measured throughout the time where the temperature is increasing.

The annealing is intended to eliminate possible sudden bursts of energy release that might accompany melting of regions in the solid where defects may cause slippage and stress relief or other extraordinary behavior. In the stage where the temperature of the liquid is increasing, heat should be added slowly to minimize internal flow rates that might themselves result in creation of vortex lines or large vortex rings in the liquid through intrinsic nucleation. The walls of the container should be smooth in order to minimize pinning sites for any vortices that might have been produced by such flow rates or other causes. Of course, the rate of temperature increase should be slow enough to permit a close approximation to thermodynamic equilibrium of the liquid. Passage of charged particles through the liquid, which might nucleate vortex lines, should be eliminated. This may require conducting the experiment in a deep underground laboratory to eliminate cosmic ray flux.

#### *8.2.2 PART II*

Starting from a point well above $T_\lambda$ (above the supposed catastrophe temperature), the liquid is rapidly cooled below $T_\lambda$. The cooling rate should be fast enough for the profusion of vortices not to relax substantially. At some temperature below $T_\lambda$, say about 200 microkelvin below $T_\lambda$, a stage of heating the liquid is initiated and the specific heat of the liquid is measured as temperature rises up to and beyond $T_\lambda$.

The surplus of unrelaxed vorticity should change the specific heat relative to that measured under conditions of complete thermodynamic equilibrium. For fast increase of temperature, the specific heat is expected to be increased by the unrelaxed vorticity. However, there will be competition between relaxation and generation of new vorticity as heat is added to the liquid, so that the change of specific heat will depend on the rate of heating. Part II does not require high pressures, but carrying out the experiment under different pressures may be useful since the way pressure affects relaxation rates of vorticity is not known.

It would be useful to measure the specific heat as the liquid is cooled through the superfluid transition. If such a procedure could be devised it would probably be preferable to the procedure based on rising temperature through the transition point.

### *8.3 Discussion of possible outcomes of the experiment*

A positive result for Part I, i.e., either $c_V \sim \left(T_\lambda - T\right)^{-1/2}$ in the crossover region and onset of critical behavior near $10^{-4}$ K, or smooth behavior through the previously identified transition temperature, would provide significant support to the theory developed here. If the predicted behavior is not observed, further measures should be considered for eliminating non-equilibrium vorticity. A negative outcome could also instigate a search for where the theory possibly went wrong and perhaps motivate others to develop a theory closer to the truth. A good candidate for improving the theory in any case would be a first principles derivation of interactions among rotons at high roton densities.

A positive result for Part II, i.e., observation of changed behavior of specific heat during reheating following rapid cooling of the liquid below $T_\lambda$, would give useful information about the possible role of vortices on critical behavior near $T_\lambda$ and on relaxation rates of vorticity in liquid $^4$He. A positive result in Part II would also support ideas about cosmological strings and their possible role in galaxy formation in the universe. This subject is considered in the next part of this paper.

## 9. Superfluidity, neutron stars, cosmology, and strings

Theories that link superfluidity to neutron stars, cosmology, and strings have been proposed and investigated by many researchers during about the last 35 years. We will briefly describe two of those lines of investigation and suggest how our approach to superfluidity may be useful in advancing them. Then we will consider a phenomenon that would possibly be observable by astronomers if the universe were superfluid and the distant nebulae were rotating.

### *9.1 Neutron stars*

A review of theoretical studies of superfluidity in neutron stars and references to original authors have been given by Donnelly [123]. An abbreviated account of those studies is given below as background for possible further work related to CBF theory of superfluids.

According to theory, a neutron star may appear near the end of the natural evolution of a star as it depletes its thermonuclear fuel and progresses successively from white dwarf, to neutron star, to black hole. It has been postulated that pulsars are rotating neutron stars. In one theoretical model, a neutron star has a shell structure that includes superfluid neutrons and superfluid protons. Measured periods of pulsars are usually steady, and they range from 1.6 ms to a few seconds. However, glitches, or jumps to higher rotation rates, have been observed in at least two cases where $\Delta\Omega/\Omega \sim 10^{-8}$ and $10^{-6}$, with subsequent decay times of order approximately 7 days and 1 year. Depinning of quantized vortices in the superfluids has been proposed and studied as a possible mechanism that produces the glitches.

It seems plausible that new insight may be obtained into behavior of neutron and proton superfluids and quantized vortices in pulsars through extension of the CBF theory of superfluidity and quantized vortices that has been developed for liquid $^4$He. The microscopic theory of a charged boson system developed by Lee and Feenberg [124,125] for the ground state and excited states using CBF formalism could be one cornerstone for this extension. A CBF theoretical treatment of pairing gaps in nucleonic superfluids by Chen, Clark, Davé, and Khodel [126] could provide another cornerstone. A two-fluid model based on those results can be developed using methods applied by Jackson [20-23] in the theory of liquid $^4$He. That method included application of statistical mechanics followed by application of a variational principle due to Eckart [19] in establishing a bridge between microscopic theory and the two-fluid equations. Vortices can be treated with methods similar to those in earlier Sections of the present paper. Meservey [127] has shown how electromagnetic fields can be included when applying Eckart's variational principle to a charged superfluid. However, care must be taken to identify a correct set of completely independent variables when applying that variational principle. Otherwise the theory may raise objections [82,83] of the kind that for many years plagued Zilsel's [17,18] phenomenological treatment of liquid $^4$He. Jackson [20,21] showed how the objectionable features could be avoided.

### *9.2 Cosmology and strings*

Many years ago Kibble [128,129] proposed and analyzed the possibility that topological structures in the form of cosmic strings provide the inhomogeneities responsible for formation of galaxies in a universe that evolved from a state that was nearly homogeneous shortly after the hot big bang. If Kibble's ideas are correct, then it may be that cosmic strings are also responsible for inhomogeneities that have been observed in the cosmic microwave background.

Recognizing correspondences between the Landau-Ginzburg theory applied to superfluid $^4$He and a relativistic equation that occurs in field theory, Zurek [130] suggested that vortices in liquid helium may provide a useful model for cosmic strings in field theory like the theory that Kibble had discussed. One of the main observations underlying Zurek's suggestion is that vortices and strings correspond to static solutions of their respective Lagrange equations of motion. For the rectilinear case, the solutions take the form $\psi = |\psi(\rho)| e^{i\theta}$. In both types of models a symmetry breaking second order phase transition can occur if the coefficients in the potential have certain temperature dependence as postulated in the theories, and this transition is associated with launching the cosmic strings into the superfluid phase.

The ideas and work of Kibble and Zurek inspire new thoughts on how our approach to superfluidity in liquid $^4$He may be brought to bear on a much broader range of issues and problems than we had anticipated. We now turn to consideration of some of them.

Quantum mechanics accounts for the behavior of nuclei and atoms on a microscopic scale, and also accounts for the behavior of superfluid $^4$He on a macroscopic scale, as we have discussed earlier in this paper. Zurek suggested that superfluidity may be important on a cosmological scale, where general relativity reigns. These observations then suggest that superfluidity may provide a useful path to a study of quantum gravity. We will try to identify some basic questions and ideas that may aid in this study as well as in other areas.

The most direct translation of our theory of superfluidity in liquid $^4$He into field theory in high energy physics and cosmology would be to postulate a ground state for Planck atoms, which we will call *p*-atoms, that corresponds to the vacuum for all excitations in field theory. For illustrative purposes, the dimensions of *p*-atoms and the spaces between them may be assumed to be comparable to the Planck length, about 1.615 x $10^{-35}$ m. This field theory vacuum is analogous to the ground state of $^4$He atoms, which is also the vacuum for excitations consisting of phonons, rotons, and vortices. For field theory and high-energy physics, we will strip the model down to a bare form, where internal variables such as spin, isospin, and color are disregarded.

Phonons and rotons are density fluctuation excitations in liquid $^4$He and we shall suppose that these correspond to leptons in field theory. In regard to topological excitations, small closed vortex rings in liquid $^4$He may correspond to quarks in field theory; and large open vortex rings in liquid $^4$He may correspond to cosmological strings in field theory.

In field theory we picture two levels of nothingness. In one level, there is the ground state consisting of *p*-atoms, and the entropy of that ground state is zero. In this instance, nothingness refers to the absence of excitation. Nevertheless there is energy in that ground state; that energy corresponds to the energy in the ground state of liquid $^4$He at zero pressure. In field theory, the other level of nothingness corresponds to "really nothing", i.e., the density of *p*-atoms is zero and in liquid $^4$He this corresponds to null density of atoms. This second level of nothingness is attained along the singular line of topological excitations in both types of theory. We shall suppose that spacetime is flat in the field theory ground state, so that the metric tensor of Minkowski spacetime [131] holds everywhere.

An advantage of this approach is that it sets up direct correspondences between superfluids in liquid $^4$He and in field theory. However, a disadvantage of this approach is that it postulates entities that are not presently known to exist, viz *p*-atoms. An alternative model that avoids this disadvantage can be considered where the ground state in field theory does not have an underlying discrete p-atomic structure, but rather is truly a continuum. In this model, all excitations would be topological. For example, then quarks and leptons may be different versions of small, closed strings.

Cosmological strings, corresponding to large open vortex rings, would be present in both theories. Many basic questions are common to both the p-atom and continuum models. One interesting question is: "What is it in field theory and general relativity that determines the string core radius, like degeneracy of the ground state wave function associated with Landau critical velocity determines the vortex core radius in liquid $^4$He?" Bending of the ground state in the core of a string represents curvature in the metric, i.e., deviation from the almost flat spacetime outside the core, and is expected to involve considerations in general relativity. A second interesting question is: "What is it in field theory, string theory, and general relativity that corresponds to superfluid velocity and quantization of circulation in the theory of liquid $^4$He?"

### *9.3 The inflationary universe and inhomogeneities*

Even in the absence of answers to such questions as we have just posed, we may still draw from analogies in the theory of liquid $^4$He for guidance in considering issues in field theory and cosmology. Both the p-atom and continuum models for field theory may be compatible with scenarios for evolution of the early universe considered in what follows, but the p-atom model will be adopted in the account considered next. As noted earlier, this model is more directly supported by calculations for liquid $^4$He. In this scenario the cosmological phase transition in an inflationary model marking a boundary of the grand unification theory (GUT) era [132] at approximately $10^{-35}$ s after ignition of the big bang is the event that we propose to associate with the catastrophe phenomenon in liquid $^4$He in the discussion that follows.

Reasoning from analogies with the theory of liquid $^4$He treated in this paper, we might speculate that soon after the hot big bang there was a large population of strings on all length scales that fit into the "fire ball" that was the very early universe. However, as the universe expanded and cooled, it eventually approached the condition like the one we associated with the catastrophe temperature in liquid $^4$He. In our theory of liquid $^4$He in complete thermodynamic equilibrium, the catastrophe temperature was associated with a narrow temperature range where the population of large open vortex rings increased dramatically when the liquid's temperature increased and the positive core energy of each large ring was counterbalanced and then exceeded in magnitude by the negative free energy external to the core. The growth of the $TS$ term, involving entropy, in free energy $F = U - TS$ for the region external to the core was indirectly responsible for the catastrophe. That external $TS$ term is associated with low energy excitations, mainly with phonons and rotons. Small closed vortex rings would also make a contribution to this $TS$ term, but in liquid $^4$He the closed rings would make a small contribution, and we neglected that in our theory. (In the continuum model of field theory, the small closed strings would be the only low energy excitations present and therefore could not be neglected)

In the cosmological model, for a while shortly after the big bang, the universe cools as it expands and is near thermal equilibrium for each instantaneous size. Just below the catastrophe temperature only the leptons and the very small strings (quarks) and the very large open strings are statistically favored due to free energy considerations. Thermal equilibrium conditions are not expected to be maintained exactly because of relaxation times needed for many of the larger strings to adjust to a reduced population below the catastrophe temperature. If the expansion rate is fast compared to the relaxation rate, then some of the large strings will survive when the universe cools even further, through the superfluid transition temperature, and a reduced population of large open strings will be launched into a metastable state where essentially none of the large strings would exist under conditions of complete thermodynamic equilibrium. Some of the leptons and quarks will survive due to thermal excitation even in equilibrium below the superfluid transition. We note here that Part II of the experiment proposed in Sec. 8 of this paper is aimed at studying the analogous situation in liquid $^4$He. There are fewer mechanisms for string population readjustment, e.g., by reconnection processes, as the population decreases when the universe cools further. Therefore metastable large, cosmological strings may survive long enough to serve as a source of galaxy generation as the strings relax further. The scenario just described is consistent with ideas for an inflationary universe and galaxy formation that were proposed and studied by Kibble [128,129].

### *9.4 Superfluid universe with rotation*

We conclude with some speculations on a superfluid universe having boundary conditions fixed by rotating distant nebulae. (The term “distant nebulae” is used loosely here, but is suggested by the language that occurs in Feynman’s [133] discussion of Mach’s Principle.) An analogue is superfluid $^4$He in a container having rotating walls, where quantized vortex lines parallel to the axis of rotation form when the rotation rate exceeds a certain critical value. This suggests that cosmic strings parallel to the axis of rotation of the distant nebulae may occur in the universe. The ends of the strings may be on the rotating nebulae that are considered to be the rotating boundary of the universe.

Much is known about vortex lines in rotating liquid $^4$He from both theory and experiment [62,75,76]. It might be interesting to work out the theory in the cosmological case within the framework of general relativity. If the theory is formally successful, then one might devise an astronomical procedure to observe the predicted cosmic strings, e.g., by observing stars accumulated near them. If such strings were detected, one may be able to locate the center of rotation and the rate of rotation of the nebulae. Interpretation of the result in connection with Mach’s Principle would also be of fundamental interest.

## Acknowledgements

The author has benefited from extensive discussions with Dr. Martin Barmatz in the course of this work. His insightful and probing questions and observations, and sound advice have greatly improved the paper, but I accept fully the blame for any mistakes in it. Also, I wish to thank the Jet Propulsion Laboratory for its support in preparation of the manuscript for publication.

## Appendix A: Variational calculations for vortex rings

This appendix contains details of variational calculations for circular vortex rings based on Eqs. (3.6) – (3.17). The Hamiltonian is assumed to be in the form shown in Eq. (2.1) and the trial functions are given by Eqs. (3.6), and (3.7), where $\psi_0$ is real-valued.

The energy in the variational calculation is measured relative to the ground state energy $E_0$ of bulk liquid $^4$He at rest (see Eq. (2.6)). We specify that the same number of atoms are in the liquid with and without the vortex. Atoms displaced from the partly empty core are assumed to be evenly distributed in the region exterior to the core. However, because the liquid is nearly at rest in almost all of that exterior region, in the variational calculations the displaced atoms are treated as if they are all located in regions where the liquid is at rest. The density of atoms in the external region is changed by a negligible amount by the displaced atoms.

The variational procedure that we shall use is aimed at locating the minimum value of the excitation energy $\varepsilon$ in Eq. (3.17) when the adjustable parameters in our trial functions are varied. Toward that end we shall next establish a result that allows $\varepsilon$ in Eq. (3.17) to be expressed in a simpler form.

Elementary vector calculus gives the following two equations:

$$\nabla_1 \bullet \left(F^* \psi_0^2 \nabla_1 F\right) = \psi_0^2 \left(\nabla_1 F^*\right) \bullet \left(\nabla_1 F\right) + 2F^* \psi_0 \left(\nabla_1 \psi_0\right) \bullet \left(\nabla_1 F\right) + F^* \psi_0^2 \Delta_1 F \qquad \text{(A.1)}$$

and

$$\left(F^* \psi_0\right)\left[\Delta_1 \left(F \psi_0\right)\right] = \left(F^* \psi_0\right)\left[F \Delta_1 \psi_0 + \psi_0 \Delta_1 F + 2\left(\nabla_1 \psi_0\right) \bullet \left(\nabla_1 F\right)\right] \ . \qquad \text{(A.2)}$$

Subtracting Eq. (A.1) from (A.2), one finds the result

$$F^* \psi_0 \Delta_1 \left(F \psi_0\right) \rightarrow |F|^2 \psi_0 \Delta_1 \psi_0 - \psi_0^2 \left(\nabla_1 F^*\right) \bullet \left(\nabla_1 F\right) \ . \qquad \text{(A.3)}$$

The arrow means equivalent under the integral sign when one uses Gauss' divergence theorem and specifies boundary conditions where the surface integral vanishes at the boundary of the normalization volume.

We have carefully studied possible problems at interior points and surfaces of the liquid, e.g., at vortex core boundaries for wave functions we have used to describe open as well as closed rings, and found that Eq. (A.3) is applicable there.

Taking into account that

$$\hat{H} \psi_0 = \left[-\frac{\hbar^2}{2m} \sum_j \Delta_j + \frac{1}{2} \sum_{i \neq j} V\left(r_{ij}\right)\right] \psi_0 = E_0 \psi_0 \ , \qquad \text{(A.4)}$$

one can then readily show with the aid of Eq. (A.3) that

$$\int \left(F\psi_0\right)^* \left[\hat{H} - E_0\right] F\psi_0 \, d^3r_1 \, d^3r_2 \cdots d^3r_N = \frac{\hbar^2}{2m} \sum_j \int \left|\nabla_j F\right|^2 \psi_0^2 \, d^3r_1 \, d^3r_2 \cdots d^3r_N \quad . \qquad \text{(A.5)}$$

The number of atoms per unit volume $n(\boldsymbol{r}_1)$ evaluated at $\boldsymbol{r}_1$ is given by

$$n\left(\boldsymbol{r}_1\right) = \frac{N \int |\psi|^2 \, d^3r_2 \, d^3r_3 \cdots d^3r_N}{\int |\psi|^2 \, d^3r_1 \, d^3r_2 \cdots d^3r_N} \quad . \qquad \text{(A.6)}$$

From these results one finds that $\varepsilon$ can be expressed as

$$\varepsilon = \frac{\hbar^2}{2m} \int d^3r_1 \, \frac{n\left(\boldsymbol{r}_1\right)}{\left|\varphi\left(\boldsymbol{r}_1\right)\right|^2} \left|\nabla_1 \varphi\left(\boldsymbol{r}_1\right)\right|^2 \quad . \qquad \text{(A.7)}$$

Using Eqs. (3.8) and (3.9) one can readily show that

$$\frac{\nabla_1 \varphi\left(\boldsymbol{r}_1\right)}{\varphi\left(\boldsymbol{r}_1\right)} = \left[\frac{\nabla_1 P\left(\boldsymbol{r}_1\right)}{P\left(\boldsymbol{r}_1\right)} + \frac{im\boldsymbol{v}\left(\boldsymbol{r}_1\right)}{\hbar}\right] . \qquad \text{(A.8)}$$

Now $\varepsilon$ in Eq. (A.7) can be expressed as

$$\varepsilon = \frac{\hbar^2 n_0}{2m} \int d^3r_1 \, \frac{n\left(\boldsymbol{r}_1\right)}{n_0} \left[\left(\frac{\nabla_1 P\left(\boldsymbol{r}_1\right)}{P\left(\boldsymbol{r}_1\right)}\right)^2 + \frac{m^2}{\hbar^2} v^2\left(\boldsymbol{r}_1\right)\right] . \qquad \text{(A.9)}$$

In Eq. (A.9) and throughout this Appendix we will make use of cylindrical symmetry about the $z$-axis and sometimes use $P(\boldsymbol{r}_1)$ to represent $P(\rho_1,\chi_1)$, and similar notation for other variables.

Next we turn to the problem of evaluating the quantities that occur in the integral in Eq.(A.9). Equations (3.6) – (3.8) and (A.6) imply that $n(\boldsymbol{r}_1)$ can be expressed as follows:

$$n\left(r_1\right) = \frac{N \int \prod_i P^2\left(\boldsymbol{r}_i\right) \psi_0^2 \, d^3r_2 \, d^3r_3 \cdots d^3r_N}{\int \prod_j P^2\left(\boldsymbol{r}_j\right) \psi_0^2 \, d^3r_1 \, d^3r_2 \cdots d^3r_N} \quad . \qquad \text{(A.10)}$$

Utilizing a method employed in the classical theory of a non-uniform liquid [134], we will define a pair distribution function $g(\boldsymbol{r}_1, \boldsymbol{r}_2)$ by

$$n\left(\boldsymbol{r}_1\right) n\left(\boldsymbol{r}_2\right) g\left(\boldsymbol{r}_1, \boldsymbol{r}_2\right) = \frac{N\left(N-1\right) \int |\psi|^2 \, d^3r_3 \, d^3r_4 \cdots d^3r_N}{\int |\psi|^2 \, d^3r_1 \, d^3r_2 \cdots d^3r_N} \quad . \qquad \text{(A.11)}$$

In our further work in this Appendix we shall approximate $\psi_0$ by a BDJ function as follows

$$\psi_0 = \prod_{1 \le i < j \le N} \exp\left[\frac{1}{2}U\left(r_{ij}\right)\right] . \tag{A.12}$$

Next, operate on $n(\boldsymbol{r}_1)$ in Eq. (A.10) with $\nabla_1$, and use Eqs. (A.11) and (A.12). After some simple algebraic manipulation the following result can be obtained:

$$\frac{\nabla_1 P\left(\boldsymbol{r}_1\right)}{P\left(\boldsymbol{r}_1\right)} = \frac{1}{2}\left[\frac{\nabla_1 n\left(\boldsymbol{r}_1\right)}{n\left(\boldsymbol{r}_1\right)} - \int d^3 r_2 \; n\left(\boldsymbol{r}_2\right) g\left(\boldsymbol{r}_1, \boldsymbol{r}_2\right) \nabla_1 U\left(r_{12}\right)\right] . \tag{A.13}$$

As an approximation we shall take

$$g\left(\boldsymbol{r}_1, \boldsymbol{r}_2\right) = g\left(r_{12}\right) \tag{A.14}$$

where $g(r_{12})$ is the pair distribution function for the uniform liquid.

For the *exact* ground state of a uniform liquid, one can replace $\psi$ by $\psi_0$ in Eqs. (A.6) and (A.11) and show that

$$\frac{n_0}{N-1}\int d^3 r_1 \; \nabla_1 g\left(\boldsymbol{r}_{12}\right) = 0 \; . \tag{A.15}$$

For an approximate form of $\psi_0$ given by Eq. (A.12) one can use Eqs. (A.6) and (A.11) and show that

$$\frac{n_0}{N-1}\int d^3 r_1 \; \nabla_1 g\left(r_{12}\right) = n_0 \int d^3 r_2 \; g\left(r_{12}\right) \nabla_1 U\left(r_{12}\right) \; . \tag{A.16}$$

We have verified by numerical integration that the pair distribution function $g(r_{12})$ calculated by Miller and Woo [20,135] using a BBGKY procedure and a BDJ function with $U(r_{ij})$ specified later in these notes is such that Eqs. (A.15) and (A.16) are satisfied to very high accuracy. We have used Miller and Woo's values of $g(r_{12})$ in evaluating formulas in the variational method discussed here, and in our further derivations in this appendix we shall use the conditions expressed by Eqs. (A.15) and (A.16). Incorporating these results in Eq. (A.13), one obtains

$$\frac{\nabla_1 P\left(\boldsymbol{r}_1\right)}{P\left(\boldsymbol{r}_1\right)} = \frac{1}{2}\left[\frac{\nabla_1 n\left(\boldsymbol{r}_1\right)}{n\left(\boldsymbol{r}_1\right)} - \int d^3 r_2 \left[n\left(\boldsymbol{r}_2\right) - n_0\right] g\left(r_{12}\right) \nabla_1 U\left(r_{12}\right)\right] . \tag{A.17}$$

For the trial wave functions that we consider, the factor $[n(\boldsymbol{r}_2) - n_0]$ is non-zero only inside the vortex core, and this condition immensely increases the efficiency of numerically evaluating the integral in Eq. (A.17).

Equations (A.9) and (A.17) hold for both open and closed vortex rings, but in the further preparation of these two equations for numerical studies, we will treat open and closed rings separately.

*Open vortex rings*

Evaluation of $\nabla P/P$ using Eq. (A.17) will enable us to calculate $\varepsilon$ using Eq. (A.9), and therefore we will consider Eq. (A.17) next. The variables $P$ and $n$ depend on $\rho$ and $\chi$, but not on $\theta$. It is convenient to derive formulas for gradients of $P$ and $n$ in the $x$–$z$ plane and then use cylindrical symmetry about the $z$-axis in evaluating $\varepsilon$. The geometry and notation in the $x$–$z$ plane for an open ring is shown in Fig. 5a.

Taking into account that $n_0$ is constant in space, we can express Eq. (A.17) as

$$\frac{\nabla_1 P(\boldsymbol{r}_1)}{P(\boldsymbol{r}_1)} = \frac{1}{2}\left[\nabla_1 \ln\frac{n(\boldsymbol{r}_1)}{n_0} - \boldsymbol{Q}\right] \tag{A.18}$$

where $\boldsymbol{Q}$ is given by

$$\boldsymbol{Q} = \int d^3 r_2 \left[n(\boldsymbol{r}_2) - n_0\right] g(r_{12}) \nabla_1 U(r_{12}) \ . \tag{A.19}$$

Using Eq. (3.14) one can show that for $\rho_1 < r_0\ (\chi_1)$:

$$\nabla_1 \ln\frac{n(\boldsymbol{r}_1)}{n_0} = 2N \frac{\cos\left(\dfrac{\pi\rho_1}{2r_0(\chi_1)}\right)}{\sin\left(\dfrac{\pi\rho_1}{2r_0(\chi_1)}\right)} \nabla_1\left(\frac{\pi\rho_1}{2r_0(\chi_1)}\right) . \tag{A.20}$$

Observe that for any function $f(\rho_1,\chi_1)$

$$\nabla_1 f(\rho_1,\chi_1) = \hat{\boldsymbol{\rho}}_1 \frac{\partial f(\rho_1,\chi_1)}{\partial \rho_1} + \hat{\boldsymbol{\chi}}_1 \frac{1}{\rho_1}\frac{\partial f(\rho_1,\chi_1)}{\partial \chi_1} \tag{A.21}$$

where $\hat{\boldsymbol{\rho}}_1$ and $\hat{\boldsymbol{\chi}}_1$ are unit vectors. Then Eq. (A.20) can be expressed as

$$\nabla_1 \ln\frac{n(\boldsymbol{r}_1)}{n_0} = N\frac{\pi}{r_0(\chi_1)} \frac{\cos\left(\dfrac{\pi\rho_1}{2r_0(\chi_1)}\right)}{\sin\left(\dfrac{\pi\rho_1}{2r_0(\chi_1)}\right)} \left[\hat{\boldsymbol{\rho}}_1 - \hat{\boldsymbol{\chi}}_1 \frac{r_0'(\chi_1)}{r_0(\chi_1)}\right] \tag{A.22}$$

where $r_0'(\chi_1) \equiv dr_0(\chi_1)/d\chi_1$ . Next express $\hat{\boldsymbol{\rho}}_1$ and $\hat{\boldsymbol{\chi}}_1$ in Cartesian coordinates as follows:

$$\begin{aligned} \hat{\boldsymbol{\rho}}_1 &= \hat{\boldsymbol{x}}\cos\chi_1 + \hat{\boldsymbol{z}}\sin\chi_1 \\ \hat{\boldsymbol{\chi}}_1 &= -\hat{\boldsymbol{x}}\sin\chi_1 + \hat{\boldsymbol{z}}\cos\chi_1 \ . \end{aligned} \tag{A.23}$$

Then Eq. (A.22) can be expressed as

$$\nabla_1 \ln \frac{n(\boldsymbol{r}_1)}{n_0} = \begin{cases} \dfrac{N\pi}{r_0(\chi_1)} \operatorname{ctn}\left(\dfrac{\pi\rho_1}{2r_0(\chi_1)}\right) \left[ \begin{array}{l} \hat{\boldsymbol{x}}\left(\cos\chi_1 + \dfrac{r_0'(\chi_1)}{r_0(\chi_1)}\sin\chi_1\right) \\ +\hat{\boldsymbol{z}}\left(\sin\chi_1 - \dfrac{r_0'(\chi_1)}{r_0(\chi_1)}\cos\chi_1\right) \end{array} \right] \\ \text{for } 0 \le \rho_1 \le r_0(\chi_1) \\ = 0 \quad \text{for } \rho_1 > r_0(\chi_1) \end{cases} . \tag{A.24}$$

Next we turn to evaluation of $\boldsymbol{Q}$ in Eqs. (A.18) and (A.19). The function $U(r)$ that we shall use is given by

$$U(r) = \left(\frac{a}{r}\right)^5 \tag{A.25}$$

where $a$ = 2.965 Å. Miller and Woo (MW) [135, 136] used that $U(r)$ to compute $g(r)$, the pair distribution function we used in numerical calculations. Operating on $U(r_{12})$ with $\nabla_1$ gives

$$\nabla_1 U(r_{12}) = -\frac{5U(r_{12})}{r_{12}} \nabla_1 r_{12} \ . \tag{A.26}$$

Now we shall choose $\theta_1 = 0$ and express $\boldsymbol{r}_1$ and $\boldsymbol{r}_2$ as

$$\begin{aligned} \boldsymbol{r}_1 &= \hat{\boldsymbol{x}}(R + \rho_1 \cos\chi_1) + \hat{\boldsymbol{z}}\,\rho_1 \sin\chi_1 \\ \boldsymbol{r}_2 &= \hat{\boldsymbol{x}}(R + \rho_2 \cos\chi_2)\cos\theta_2 + \hat{\boldsymbol{y}}(R + \rho_2 \cos\chi_2)\sin\theta_2 + \hat{\boldsymbol{z}}\,\rho_2 \sin\chi_2 \ . \end{aligned} \tag{A.27}$$

Next, write $\boldsymbol{r}_{12}$ as

$$\boldsymbol{r}_{12} = \boldsymbol{r}_1 - \boldsymbol{r}_2 = \hat{\boldsymbol{x}} X_{12} + \hat{\boldsymbol{y}} Y_{12} + \hat{\boldsymbol{z}} Z_{12} \ , \tag{A.28}$$

where expressions for the components $X_{12}$, $Y_{12}$, $Z_{12}$ can be found using Eq. (A.27). Then one can readily evaluate $r_{12}$ and $\nabla r_{12}$, which occur in Eq. (A.26), and then one can easily show that $\boldsymbol{Q}$ is given by

$$\boldsymbol{Q} = \hat{\boldsymbol{x}} Q_x + \hat{\boldsymbol{z}} Q_z \tag{A.29}$$

where

$$Q_x\left(\rho_1,\chi_1\right)=2n_0\int_{\chi_2=0}^{2\pi}d\chi_2\int_{\rho_2=0}^{r_0(\chi_2)}d\rho_2\left[\frac{n\left(\rho_2,\chi_2\right)}{n_0}-1\right]$$
$$\times\left[R+\rho_2\cos\chi_2\right]\rho_2\int_0^{\pi}d\theta_2\;g\left(r_{12}\right)\left[-\frac{5U\left(r_{12}\right)}{r_{12}^{\;2}}\right]X_{12}\;, \tag{A.30}$$

$$X_{12}=R+\rho_1\cos\chi_1-\left[R+\rho_2\cos\chi_2\right]\cos\theta_2\;, \tag{A.31}$$

and

$$Q_z\left(\rho_1,\chi_1\right)=2n_0\int_{\chi_2=0}^{2\pi}d\chi_2\int_{\rho_2=0}^{r_0(\chi_2)}d\rho_2\left[\frac{n\left(\rho_2,\chi_2\right)}{n_0}-1\right]$$
$$\times\left[R+\rho_2\cos\chi_2\right]\rho_2\int_0^{\pi}d\theta_2\;g\left(r_{12}\right)\left[-\frac{5U\left(r_{12}\right)}{r_{12}^{\;2}}\right]Z_{12}\;, \tag{A.32}$$

$$Z_{12}=\rho_1\sin\chi_1-\rho_2\sin\chi_2\;. \tag{A.33}$$

The $y$-component, $Q_y$, is zero, a result that was established analytically, but which is also obvious from the fact that $n(\rho,\chi)$ and $P(\rho,\chi)$ have no $\theta$ dependence.

Equations (3.14), (A.18), (A.24), and (A.29) – (A.33) provide expressions suitable for numerical evaluation of $\nabla_1 P(\boldsymbol{r}_1)/P(\boldsymbol{r}_1)$.

The Biot-Savart formula [137] for $\boldsymbol{v}(\boldsymbol{R})$ is

$$\boldsymbol{v}\left(\boldsymbol{R}\right)=\frac{K}{4\pi}\int\frac{d\boldsymbol{s}'\times\left(\boldsymbol{R}-\boldsymbol{R}'\right)}{\left|\boldsymbol{R}-\boldsymbol{R}'\right|^3}\;, \tag{A.34}$$

where the line integral follows the vortex line in the positive sense. This formula was used in our numerical work for evaluating the quantity $v^2(\boldsymbol{r})$ in Eq. (A.9) and for locating the radius $r_0(\chi)$ of the core boundary. This method of evaluation is in principle exact and is more easily applied numerically than evaluation of $\boldsymbol{v}(\boldsymbol{R})$ based on the velocity potential. The expressions described here and cylindrical symmetry about the $z$-axis were used in variational calculations for the optimum wave function and energy for open vortex rings. The procedure consisted of numerically evaluating $\varepsilon$ in Eq. (A.9) for different values of $N$ in the set of trial densities given in Eqs. (3.14) and (3.13) and locating the minimum value of $\varepsilon$ at each ring radius $R$. A discrete set of $N$ values was used, and this limited the accuracy of locating the exact minimum of $\varepsilon$.

Each of the terms in the square bracket in Eq. (A.9) diverges near the singularity, i.e., when $\rho_1 \to 0$. Near the singularity, series expansions and subsequent analytic and numerical integration over a region near the singularity yielded finite results. Beyond that region the integral can be evaluated using straightforward numerical integration methods. Treatment of the region near the singularity involved lengthy derivations and is not very interesting; therefore we will spare the reader from the details. However, we have checked our method and results by verifying that the sum of the integrals for the two regions is insensitive to the precise separation radius, which we will call $B$, provided that the separation radius is small and the meshes used in numerical integrations are sufficiently fine.

Using the procedure described above, we found that the contribution $\varepsilon_B$ to $\varepsilon$ in Eq. (A.9) by the region where $0 < \rho_1 \leq B$ and $B << R$ is as follows:

$$\varepsilon_B = \frac{\hbar^2 n_0}{2m}(2\pi R)\frac{B^{2N}}{2N}\int_0^{2\pi} d\chi_1 \left(\frac{\pi}{r_0(\chi_1)}\right)^{2N}\left[\frac{1}{4}H^2(\chi_1)+1\right] \tag{A.35}$$

where

$$H^2(\chi_1) = 4N^2\left[1+\left(\frac{r_0'(\chi_1)}{r_0(\chi_1)}\right)^2\right] . \tag{A.36}$$

<u>*Closed vortex rings*</u>

Closed vortex rings are those where $v_s > v_c$ for some portion of the vortex ring axis. Closed rings occur in the size range where $R$ less than about 8 Å when $v_c$ = 58 m/s. Figures 6 - 8 exhibit examples of vortex core external boundaries for open and closed rings. Figures 5b and 5c indicate the meaning of the variables $r_0(\chi)$, $Z(\chi)$, $z_a$, and $A$ that appear in the trial functions for $n(\rho,\chi)/n_0$ that occur in Eqs. (3.15) and (3.16) for closed rings.

We want to evaluate the excitation energy $\varepsilon$ in Eq. (A.9) for closed rings. A first step is to find a simple expression for $\nabla_1 P(\boldsymbol{r}_1)/P(\boldsymbol{r}_1)$, and we can use Eqs. (A.18) and (A.19) when Eq. (3.15) is substituted into both of these equations. Now in evaluating $\nabla_1 n(\boldsymbol{r}_1)/n_0$ in Eq. (A.18) we must deal with regions 1 and 2 individually. For region 1, the result coincides with that in Eq. (A.24). For region 2, the following results are obtained:

$$\nabla_1 \ln\frac{n(\boldsymbol{r}_1)}{n_0} = \nabla_1 \ln S(\chi_1) + \nabla_1 \ln\frac{n_a(\boldsymbol{r}_1)}{n_0} , \tag{A.37}$$

where

$$\nabla_1 \ln S(\chi_1) = \frac{A\dfrac{\pi}{z_a}\sin\dfrac{\pi Z_0(\chi_1)}{z_a}}{1-\dfrac{A}{2}\left(1+\cos\dfrac{\pi Z_0(\chi_1)}{z_a}\right)}\left(\hat{\boldsymbol{x}}\frac{R}{\rho_1}\frac{\sin\chi_1}{\cos^2\chi_1} - \hat{\boldsymbol{z}}\frac{R}{\rho_1}\frac{1}{\cos\chi_1}\right) \tag{A.38}$$

and

$$Z_0(\chi_1) = -R\tan\chi_1 \ . \qquad \text{(A.39)}$$

The result for $\nabla_1 n_a(\boldsymbol{r}_1)/n_0$ is the same as in Eq. (A.24).

Formulas for $\boldsymbol{Q}$ given by Eqs. (A.29) – (A.33) still hold for closed vortex rings provided that the expression for $n(\boldsymbol{r})/n_0$ given by Eq. (3.15) are used at $\boldsymbol{r} = \boldsymbol{r}_2$ in these equations. The Biot-Savart formula can be used to evaluate $v^2(\boldsymbol{r}_1)$ in Eq. (A.9) for closed rings, just as was the case for open rings.

For closed rings the contribution to $\varepsilon$ by the region near the singularity can be evaluated by a method similar to that used for open rings, but now regions 1 and 2 must be treated individually. The contribution $\varepsilon_B$ to $\varepsilon$ by the region $0 \le \rho \le B$, for $B << R$, is given by the following formulas where $\varepsilon_{B1}$ refers to region 1 and $\varepsilon_{B2}$ to region 2:

$$\varepsilon_B = \varepsilon_{B1} + \varepsilon_{B2} \ . \qquad \text{(A.40)}$$

For region 1, where $r_0(\chi)$ is determined by $v_s = v_c$:

$$\varepsilon_{B1} = \frac{\hbar^2 n_0}{2m}(2\pi R)\frac{B^{2N}}{N}\int_0^{\chi_0} d\chi_1 \left(\frac{\pi}{2r_0(\chi_1)}\right)^{2N}\left[\frac{1}{4}H^2(\chi_1)+1\right] \qquad \text{(A.41)}$$

and $H^2(\chi_1)$ is the same as in Eq. (A.36). The angle $\chi_0$ in Eq. (A.41) is given by

$$\tan\chi_0 = \frac{z_a}{R} \quad \text{with} \quad \frac{\pi}{2} \le \chi_0 \le \pi \ . \qquad \text{(A.42)}$$

For region 2, where $r_0(\chi)$ is given by $r_0(\chi) = R/\cos\chi$:

$$\varepsilon_{B2} = \frac{\hbar^2 n_0}{2m}(2\pi R)\frac{2B^{2N}}{N}\int_{\chi_0}^{\pi} d\chi_1 \left(\frac{\pi}{2r_0(\chi_1)}\right)^{2N}\left[\frac{1}{4}K^2(\chi_1)+1\right] . \qquad \text{(A.43)}$$

Here $K^2(\chi_1)$ can be evaluated using the following formulas.

$$\boldsymbol{K}(\chi_1) = \boldsymbol{J}(\chi_1) + \boldsymbol{H}(\chi_1) \qquad \text{(A.44)}$$

$$\boldsymbol{J}(\chi_1) = \frac{A\dfrac{\pi}{z_a}\sin\dfrac{\pi Z_0(\chi_1)}{z_a}}{1-\dfrac{A}{2}\left(1+\cos\dfrac{\pi Z_0(\chi_1)}{z_a}\right)}\frac{R}{\cos\chi_1}\left(\hat{\boldsymbol{x}}\frac{\sin\chi_1}{\cos\chi_1}-\hat{\boldsymbol{z}}\right) \qquad \text{(A.45)}$$

$$\boldsymbol{H}(\chi_1) = 2N\left\{\hat{\boldsymbol{x}}\left[\cos\chi_1 + \frac{r_0'(\chi_1)}{r_0(\chi_1)}\sin\chi_1\right] + \hat{\boldsymbol{z}}\left[\sin\chi_1 - \frac{r_0'(\chi_1)}{r_0(\chi_1)}\cos\chi_1\right]\right\} . \qquad \text{(A.46)}$$

These results enabled complete evaluation of $\varepsilon$ in Eq. (A.9) in the case of closed vortex rings. We found that results for $\varepsilon$ are insensitive to the value assigned to $B$ provided that $B \ll R$ and sufficiently fine meshes are used in numerical integrations.

## Appendix B: Understanding and visualizing rotons

Landau [14] originally introduced rotons as excitations that involve vorticity in the liquid based on a quantum hydrodynamics theory that he developed. In a second paper, about six years later, Landau [16] abandoned the claim of vorticity associated with rotons and instead treated them phenomenologically as density fluctuation excitations with short wavelength of the order of spacing between atoms. In that new picture, the roton energy spectrum merged continuously with the spectrum for long wavelength phonons.

Several years later Feynman [32] treated low excited states in liquid $^4$He from a microscopic point of view based directly on quantum mechanics and the many-particle Schrödinger equation. He showed how excitations of the sound field, phonons, could be viewed as collective excitations embracing many atoms. Feynman also analyzed several models that involved short distances and few atoms. One of those models involved an excited atom in a cage formed by its nearest neighbors.

A variational procedure was applied to select the "best" model from the class of wave functions that he had proposed, and the cage model is well suited to interpreting the result of that variational approach. The Bijl-Feynman energy spectrum $\varepsilon(k) = \hbar^2 k^2 / (2mS(k))$ (see Fig. 1) emerged as an approximation to Landau's spectrum from that analysis and calculation by Feynman. The shorter wavelength part of the Bijl-Feynman spectrum, now associated with rotons, is linked by a continuous curve to the long wavelength density fluctuations associated with phonons. The link is a result of the formalism. But a central question remains unanswered at this point, viz, "What is it in the cage model of a roton that involves density fluctuations in the liquid?" The following analysis is aimed toward answering this question and aiding in visualization of a roton state. Much of the emphasis in this analysis is on excitations near the relative minimum and relative maximum of the elementary excitation spectrum. The other rotons may be understood roughly through interpolation of results obtained here.

Keesom and Taconis [138] measured the liquid structure function $S(k)$ for liquid $^4$He by x-ray diffraction in 1937 and tried to interpret their result in terms of an atomic arrangement shown in Fig. 28 [139]. This arrangement is for a crystal lattice. Of course they realized that the diffuseness of the x-ray pattern excludes a crystal structure, but the arrangement does show how the inferred nearest neighbor distance (6 nearest neighbors) of about 3.16 Å could be compatible with the mean atomic spacing $d = (V/N)^{\frac{1}{3}} = 3.58$ Å. If one supposes that the fluid contains mainly large open spaces and small open spaces of about the same dimensions shown in Fig. 28, then one can propose a physical picture of rotons that involves density fluctuations, as follows.

Consider a $^4$He atom in a momentum state $k = 2\pi/\lambda$ that exists in the same volume as that where the liquid is located. If the liquid were not there, then the selected atom would be in a plane wave state $\varphi = e^{i\mathbf{k}\cdot\mathbf{r}}$ that has equal probability density over that volume. (If a narrow wave packet were formed, one could follow the motion of this atom, but we are not adopting that viewpoint here. Instead we are postulating that the atom is in a broad wave packet that approximates an eigenstate of energy and momentum.) Now suppose the atoms of the liquid are present with many regions where the atomic arrangements and spacings are about the same as in Fig. 28. The momentum eigenstate of the selected atom, which for this discussion we consider as distinguishable, will still be characterized with a wave number $k = 2\pi/\lambda$, but in general the spatial distribution of the selected atom will be affected by the presence of the other atoms. Consider the following cases.

1. *Roton at the relative minimum* [79] *of the Landau dispersion curve.*

In this case $k_1 = 1.91$ Å$^{-1}$ and $\lambda_1 = 2\pi/k_1 = 3.28$ Å. When the selected atom having this wavelength encounters an arrangement in the liquid with spacing 3.16 Å (see Fig. 28), the liquid structure will act somewhat like a leaky cage. The selected atom will bounce back and forth in this structure, resonating in that cage, and thereby cause the probability density of the selected atom to be greater in that region than elsewhere. (Classically we would say that the bouncing back and forth causes the atom to spend more time in that region than elsewhere.) Of course, according to our assumption, there will be many such arrangements having about these dimensions throughout the liquid, so the probability density of the excited atom will be enhanced in each of these arrangements. This is the physical picture that we propose for a roton at the minimum in the dispersion curve, and it clearly relies on variability of the density, i.e., density fluctuations, in the liquid.

2. *Roton at the relative maximum* [79] *of the Landau dispersion curve.*

In this case $k_2 = 1.1$ Å$^{-1}$ and $\lambda_2 = 2\pi/k_2 = 5.71$ Å, and $\lambda_2$ will produce resonance in the leaky cage where the spacing is about 6.3 Å (see Fig. 28) and corresponding higher probability density for the excited atom where that spatial arrangement occurs in the liquid.

The view that we are taking here is that the spatial arrangements that are predominant in the ground state wave function for the liquid provide the environment in which excited atoms having specified wavelengths can resonate. The resonance conditions that we have considered are not perfect, but are close enough to be plausible. We have not taken into account the hard sphere radius of the atoms and the leakiness of the cages, which should affect the approximate conditions for resonance. Also, the liquid structure function $S(k)$ and the inferred related pair distribution function $g(r)$ have been determined by a number of experimentalists, and the results are in fair, but not perfect, agreement with each other. For these and other reasons it should be recognized that there is an element of uncertainty in the pictures of rotons proposed here.

It should be mentioned again that the model of a roton proposed here is compatible with Feynman's [32] model of a roton consisting of an excited atom in a cage formed by its nearest neighbors. Our model is simultaneously compatible with Feynman and Cohen's [33] model of a roton consisting of an excited atom in something like a plane wave state having a dipolar backflow component. Feenberg [140] has shown that a Fourier transform of a dipolar backflow term produces terms in the energy similar to those that occur in the second order perturbation calculation of Jackson and Feenberg [47]. Furthermore, our model of a roton is also compatible with further perturbation corrections of a plane wave such as occurs in the third order perturbation calculation of Lee and Lee [48]. Also, our model is not limited by these correction terms but is compatible with other "local structure" of a roton that at this point is unknown in detail, but that is clearly needed [49] to bring existing theoretical models into agreement with neutron scattering measurements of the Landau dispersion curve (see Fig. 1).

## Appendix C: Mantle surrounding vortex core for $T < T_c$ and second sound absorption

The boundary radius $R(\theta)$ in Eq. (6.25) and the correlation length $R_0$ in Eq. (6.26) define a mantle around each vortex ring where the number density of rotons is high and the liquid is unstable with respect to small perturbations in $v_s$ according to ordinary thermodynamic stability criteria [102,103]. The parameters $R(\theta)$ and $R_0$ hold for a vortex ring of any size for temperatures very close to $T_c$. However, for large open vortex rings at temperatures well below $T_c$, the boundary of that unstable region is approximately a toroid that surrounds the vortex core. For a vortex line, the mantle is approximately a cylinder that surrounds the core. The core radius is determined in our theory by the condition that the Landau critical velocity is reached, and that radius does not depend on temperature. On the other hand, the size of the mantle region grows with temperature so that it would eventually fill the container as temperature increases and approaches $T_c$. Second sound will not propagate as an oscillating wave inside the mantle region because there is no restoring force for the superfluid disturbance there. Experiments [108-110] that attempt to study the amount of vortex line per unit volume near $T_\lambda$ or $T_c$ may yield misleading results if effects of the mantle are not properly accounted for. Also, we venture to suggest that any statistically significant part of temperature dependence of the "vortex core parameter" measured by Rayfield [98] and by Steingart and Glaberson [88,89] in ion propagation experiments is determined by the mantle boundary and not by the vortex core boundary as we have defined them. On the other hand, measurements of the pressure dependence of the "vortex core parameter" were made by Rayfield [98] at 0.601 K and by Steingart and Glaberson [88] at 0.368 K. At those low temperatures the mantle radius will be almost equal to the core radius, and this we propose as the explanation of their observations. In the words of Steingart and Glaberson: "The vortex core radius increases with pressure following the decrease in the roton energy and corresponding Landau critical velocity rather than the increase in the speed of sound." This is consistent with the explanation we have given for the pressure dependence. We further suggest that the growth of the mantle region about vortex lines with temperature may be a significant contributing factor to the temperature dependence of the extra attenuation of second sound in rotating $^4$He experiments by Hall and Vinen [59], although it is clear from their results that other factors must make even larger contributions for temperatures well below $T_\lambda$.

## Appendix D: Discussion of Williams' theory of the $\lambda$ transition

Williams claims that his vortex theory successfully accounts for the experimentally observed superfluid density exponent and the specific heat exponent at $T_\lambda$. Observations discussed in this Appendix support the view that this success that he claims for his theory is of questionable validity and that his theory is based on many seriously flawed assumptions and approximations that are linked together. Also, his theory contains an obvious error that is central to his calculations, and even that error alone vitiates his results for both superfluid density and specific heat.

(1) Williams [141] modified results for properties of a vortex ring obtained by Roberts and Grant [95a] who used a highly idealized model of a Bose liquid. Williams then assumed that the modified results were applicable for a realistic description of liquid $^4$He. I believe that his assumption is flawed and will lead to serious deficiencies in his theory of the $\lambda$ transition. Some of my reasons for this belief will be explained in what follows.

Roberts and Grant used a Hartree approximation in calculating properties of a large circular vortex ring in a Bose condensate at $T = 0$ K. All of the atoms were assumed to be in the condensate and a delta function interatomic interaction was assumed. A core radius in this condensate model was identified as a healing length. The healing length varies inversely as the square root of the condensate density $\rho_\infty$, where $\rho_\infty$ is the mass density far from the vortex core.

Williams' modification, at least in his initial work [141,142], consisted of replacing the condensate density $\rho_\infty$ by the superfluid density $\rho_{s0}$ in the formulas that Roberts and Grant calculated for the vortex ring energy and for the core radius. The bare superfluid density $\rho_{s0}$ takes into account [142] just the phonons and rotons that are present at finite temperatures, as in the Landau model. The superfluid density $\rho_{s0}$ is evaluated for the limit condition where the superfluid velocity magnitude $v_s$ approaches zero. In the limit of zero temperature $\rho_{s0}$ is equal to the mass density $\rho$ of liquid $^4$He.

Experiment shows that at a fixed temperature, for increasing pressure the mass density and superfluid density increase. Then according to Williams' modification of the theory of Roberts and Grant, the healing length decreases. This implies that the core radius decreases with increasing pressure. This is opposite to the pressure dependence of the core radius that Steingart and Glaberson [88] found in their experiments. This result indicates that the above stated modification of vortex ring formulas assumed by Williams and which are important input to his theory of the $\lambda$ transition are not suitable for a realistic model for treating liquid $^4$He.

(2) Williams' theory does not take into account the velocity dependence that occurs in the superfluid density and other thermohydrodynamic functions at any finite temperature when the Landau model is assumed to be applicable. One important consequence of this velocity dependence is that rotons cannot exist in the region where the Landau critical velocity is exceeded. Williams' theory does not take into account this exclusion of rotons from that region near the vortex singularity. The modified version of the vortex ring energy that Williams used implicitly assumes that the energy density outside the healing-length-determined core is approximately $1/2\,\rho_{s0}v_s^2$ in the flow field due to the vortex. One can infer this by studying the theory due to Roberts and Grant and taking into account Williams' modification of their theory. On the other hand, for any finite temperature the Helmholtz free energy density $f(v_s)$ should be used instead of $1/2\,\rho_{s0}v_s^2$ in regions where the particle density and velocity magnitude vary slowly enough for the Landau model to be applicable. A formula for $f(v_s)$ can be obtained using Eq. (4.3) in our paper to find a formula for $F(v_s) - F(v_s = 0)$ and then dividing by the volume. Proper treatment of these conditions will greatly affect the energy of a vortex ring as $T_\lambda$ is approached, and in turn change the thermal distribution of vortices from what Williams calculated.

(3) Williams used the value of the vortex core energy $C$ computed by Roberts and Grant. The Hartree theory and delta function interatomic potential used by Roberts and Grant does not take into account interatomic correlations in the liquid. However, interatomic correlations contribute a significant amount to the core energy in more realistic models of the liquid. This result was indicated in Sec. 3.2 of our paper.

The model of a vortex ring and the value of core energy calculated in our paper gain support from the good agreement with experimental results found by Rayfield and Reif and by Steingart and Glaberson. It is noted in Secs. 3.3.5 and 5.3.1 that the liquid would be flooded with vortex rings of all sizes well below the superfluid transition if the free energy density outside the core were treated properly and the core energy found by Roberts and Grant were used as an approximation in place of the much larger core energy calculated in our paper. The large core energy calculated in our paper is responsible for avoiding the excessively large proliferation of vortices below the superfluid transition.

On the other hand, Williams [75p] used a scheme that involved vortex rings (which he referred to as dipoles) of large size being screened by vortex rings of smaller size, and predicted a proliferation of vortex rings that drive the superfluid density to zero at the transition temperature. If Williams had taken proper account of velocity-dependent roton contributions to free energy while still inappropriately using the core energy of Roberts and Grant, a great proliferation of vortex rings of all sizes would have been found even below the superfluid transition without using his scheme for screening large rings by smaller rings.

(4) Williams [142] used an incorrect sign in the term that takes into account the interaction of a vortex ring with an applied superfluid velocity field. It is clear that this is not just a typographical error because that same sign is used in his expression [142] for the polarizability of a large ring, as one will find by working through the derivation of his formula for polarizability. That same sign is also used in his later papers [75p,143]. A possible source of this error in Williams' work can be traced by starting with a reference to Fetter [76a] that was cited by Williams in Ref. [143]. In Eq. (3.2.93) of Ref. [76a], Fetter wrote a formula for energy of a vortex ring in an applied superfluid velocity field that is directed opposite to the self-induced velocity of the ring. The velocity dependent term in Eq. (3.2.93) is preceded by a minus sign and that term does not contain a dot indicating a scalar product. Only the magnitudes of $\boldsymbol{p}$ and $\boldsymbol{v}_s$ should appear in that equation. The print used in the article [76a] makes it hard to distinguish bold letters (vectors) from unbold (scalars). Williams has treated that interaction term as if there were a scalar product there.

Fetter [76a] cited work of Iordanskii [144] as one example where the energy of a vortex ring in an oppositely directed superfluid velocity field occurs. In Iordanski's [144] paper the energy of a vortex ring in the presence of a superfluid velocity field having arbitrary direction with respect to the ring momentum occurs with a plus sign instead of the minus sign that Williams used. Iordanskii did not insert a dot to indicate a scalar product (a frequent practice in the Russian literature) in the velocity dependent term in his Eq. (1.1), and that may have been the cause of some confusion. However, in Iordanskii's Eq. (1.3), where the minimum value of the energy of a vortex ring in a flow field $\boldsymbol{v}_s$ is calculated, a minus sign appears in the velocity dependent term. This shows that Iordanskii's Eq. (1.1) is intended to involve a scalar product.

Furthermore, below Eq. (3.2.93) in Fetter's [76a] paper, he says that $E$ is the energy of an excitation and $\boldsymbol{p}$ is its corresponding momentum. The vortex ring is just treated as an excitation there. The energy of an excitation in a velocity field $\boldsymbol{v}_s$ of arbitrary orientation with respect to the momentum $\boldsymbol{p}$ of the excitation is written by Khalatnikov [15] with a plus sign before the scalar product $\boldsymbol{p} \bullet \boldsymbol{v}_s$. Khalatnikov's formula in Eq. (1-11) of Ref. [15] is a well-known result that is easily obtained by both classical and quantum mechanical [32] calculations. This sign error in Williams' theory invalidates the screening scheme on which his theory relies. *Williams' entire theory is developed using an incorrect sign for the interaction term that is a central feature of his work*. This error adds to my belief that Williams' theory of the $\lambda$ transition is not credible.

(5) Williams used a model that included the features described in items 1 through 4 above as input to a real-space renormalization calculation and found a superfluid density exponent of approximately 0.5. This differs from the experimentally determined exponent of approximately 0.67. In an effort to produce the experimental result, Williams [75p] adjusted his model for the vortex core but retained many other features of his previously modified version of the results found by Roberts and Grant. This new, adjusted model postulated a "crinkled" vortex ring, schematically represented in Fig. 2 of Ref. [75p]. I believe that this "crinkled" ring model is not credible because using the localized induction approximation (see Sec. 1.6 of Ref. [76]) one can show that parts of the singular line having greatly different curvatures would move with greatly different velocities. The "crinkled" structure would be evanescent in time and would not be of sufficient permanence for representing a vortex ring in statistical mechanics calculations of thermodynamic properties.

(6) Williams [75p] also made other adjustments to his original modified version of the results of Roberts and Grant. These adjustments included treating the bare superfluid density taking into account only the phonons. The rotons were then treated as the vortex rings of smallest size, having ring radius around 2 Angstroms. On the other hand, the theory of Roberts and Grant treated only large vortex rings having, according to them, core radius of typically a few Angstroms. An important assumption in the calculations of Roberts and Grant was that the core radius is much less than the ring radius. Therefore Williams' assumption that rotons have energy and momentum properties that resemble those in the formulas of Roberts and Grant has no theoretical foundation. Furthermore, Williams’ theory still implicitly used the polarizability derived for conditions where the velocity is small in treating rotons even in the high velocity region near the vortex singular line. Nevertheless, Williams claims that his adjusted theory yields a superfluid density exponent of about 0.67, in good agreement with experimental results. In my opinion, his theoretical calculation of this exponent is not credible.

(7) Williams' [75q] calculation of the specific heat exponent relies on the same model and assumptions as those described above for the superfluid density. Therefore his theory for the specific heat exponent is not credible either, in my opinion.

(8) There are additional features of Williams' theory that I believe are seriously flawed. However, discussing them seems to be superfluous.

**Figure Captions**

Fig. 1. Theoretical and experimental determination of the phonon-roton spectrum of liquid $^4$He at $n_0$ = 0.0218 atom Å$^{-3}$. A, the Bijl-Feynman [32] spectrum; B, the Jackson-Feenberg [47] spectrum; C, the Lee-Lee [48] spectrum; and N, the experimental neutron scattering spectrum [79]. The theoretical spectra were evaluated by Campbell and Pinski [49] using an optimized Bijl-Dingle-Jastrow function [50] (after Campbell and Pinski [49]) .

Fig. 2. Atomic density $n/n_0$ versus radial distance $\rho$ at the core of a rectilinear vortex, based on Eq. (3.2) and a set of values of the exponent $N$ in the trial density function.

Fig. 3. Energy versus variational parameter $N$, calculated from Eqs. (3.2) and (3.4). The dressed energy $\varepsilon_C$ was calculated with the function $P(\rho)$ that is a solution of Eq. (3.5). The bare energy $\varepsilon_0$ was calculated with $P_0(\rho)$ from Eq. (3.3) in place of $P(\rho)$ and neglects effects of atomic pair correlations associated with $g(r)$.

Fig 4. Radial functions versus radial distance $\rho$ for the core of a rectilinear vortex. The variational optimum value $N$ = 0.50 that minimized the energy $\varepsilon_C$ in Fig. 3 was used in evaluating all three functions. $P_0(\rho)$ from Eq. (3.3) is the radial wave function factor for a bare vortex line. $P(\rho)$ is the radial wave function factor for a dressed vortex, and is the solution of Eq. (3.5). $n(\rho)/n_0$ is the scaled atomic density for both bare and dressed vortex lines.

Fig. 5. Coordinate systems and parameters for variational calculations. (a) Open vortex rings. (b) Closed vortex rings. (c) $S$ versus $z$ (Eq. (3.16)) where $z = Z_0(\chi)$.

Fig. 6. Vortex core boundaries, determined by the condition $v_s = v_c$, for three sizes of circular vortex rings, plotted to scale showing true relative sizes and shapes. The black dots indicate positions of vorticity for each ring.

Fig. 7. Vortex core boundary for a closed circular vortex ring having radius $R$ = 8.0 Å, just below the threshold radius for closure. The black dots indicate positions of vorticity.

Fig. 8. Vortex core boundary for an open circular vortex ring having radius $R$ = 9.0 Å, just above the threshold radius for closure. The black dots indicate positions of vorticity.

Fig. 9. Energy versus radius $R$ for circular vortex rings, calculated with the variational method based on minimizing $\varepsilon$ in Eq. (3.17) using trial functions for open rings in Eq. (3.14)

and for closed rings in Eqs. (3.15) and (3.16). The fluid external to the core contributes the energy $\varepsilon_E$, and is the same as for a hollow core vortex ring; $\varepsilon_C$ is the core energy (see text for details), and $\varepsilon_V$ is the total energy of the vortex ring at $T = 0$ K.

Fig. 10. Wave number $k$ versus radius $R$ for circular vortex rings having momentum $P = \hbar k$ and radius $R$, calculated using the mass defect method and optimal trial functions found with the variational procedure described in the text.

Fig. 11. Energy spectrum for circular vortex rings based on variational calculations described in the text and compared with the Landau spectrum for phonons and rotons determined from neutron scattering measurements [79].

Fig. 12. Energy versus radius $R$ for circular vortex rings. Total vortex ring energy $\varepsilon_V$ is evaluated using Eq. (3.30). The energy external to the core, $\varepsilon_E$, is given by Eq. (3.30) with $\alpha_c = 0$. The core energy per unit length, $\varepsilon_C/2\pi R$ = 1.68 K Å$^{-1}$, is the same as the variational result for a rectilinear vortex (see Fig. 3).

Fig. 13. Velocity versus radius $R$ for circular vortex rings evaluated using Eq. (3.35) (see text for justification in use of Eq. (3.35)).

Fig. 14. Velocity versus energy for circular vortex rings determined by theory through combining results in Figs. 12 and 13, compared with experimental measurements made by Rayfield and Reif [60]. Theoretical energy is evaluated with Eq. (3.30), and velocity with Eq. (3.35). See text for justification in use of Eq. (3.35).

Fig. 15. Hehmholtz free energy per particle as a function of superfluid velocity $v_s$ when $v_n$ = 0. Each curve is an isotherm calculated from Eq. (4.3), but $E_0$ has been subtracted out of that formula. The occupation numbers $n_i$ are given by Eqs. (4.4) and (4.5), where $\gamma = 0.840$ is the roton-roton coupling strength. Also, $\alpha$ is a roton chemical potential determined by the requirement that the set of roton occupation numbers satisfy Eq. (4.6) with $\mathcal{N}$ = 0.2838 $N$. The calculations are for essentially zero pressure. Excitation energies $\varepsilon_i$ are from neutron scattering measurements [79], Landau critical velocity $v_c \simeq 58$ m/s, and atomic number density $n_0$ = 0.0218 atom Å$^{-3}$. The calculated curves take into account only phonons and interacting rotons.

Fig. 16. Superfluid density/mass density, $\rho_s/\rho$, at $v_s$ = 0 evaluated as a function of temperature using Eqs. (4.8) and (4.9). The occupation numbers $n_0(p)$ are evaluated using Eqs. (4.4), (4.5), and (4.6) under conditions of essentially zero pressure, where values for $\varepsilon(p)$, $\gamma$, $\alpha$, and $n_0$ coincide with those given at Fig. 15. Experimental values of $\rho_s/\rho$ are from ref. [101a]. The theoretical curve takes into account only phonons and interacting rotons.

Fig. 17. Specific entropy and specific heat at constant volume versus temperature evaluated using Eqs. (2.9) and (4.14) at $v_s$ = 0 for non-interacting, unconstrained Landau elementary excitations. Occupation numbers $n_i$ are evaluated using $\gamma = 0$, and $\alpha = 0$ in Eqs. (4.4) and (4.5). The $\varepsilon_i$ are from neutron scattering measurements of the Landau spectrum [79]. Experimental values for entropy and specific heat are from ref. [101b].

Fig. 18. Specific entropy and specific heat at constant volume evaluated as functions of temperature using Eqs. (2.9) and (4.14) at $v_s$ = 0 for phonons and interacting rotons. Occupation numbers $n_i$ are evaluated using the parameter set specified at Fig. 15. The constraint on roton number is met with the roton chemical potential $\alpha = 0$ for the temperature range for theoretical curves shown here. Experimental values for entropy and specific heat are from Ref. [101b]. Values for specific heat in the temperature range $2.070 < T \le 2.170$ are from Ref. [69].

Fig. 19. Free energy versus Radius for a single circular vortex ring. $F_1$ is the total free energy evaluated with Eq. (5.2). $F_{C1}$ is the core free energy and $F_{E1}$ is the external free energy evaluated using results in Fig. 15. Results for $T < T_\lambda = 2.172$ are displayed in a and b. Results for $T > T_\lambda$ are displayed in c. The calculated curves take into account phonons and interacting rotons.

Fig. 20. Logarithm of vortex ring distribution $D(R)$ as a function of vortex ring radius $R$ evaluated using Eqs. (5.11) and (5.12) for non-interacting vortex rings. The calculated curve takes into account phonons and interacting rotons.

Fig. 21. Average length spacing between vortex rings as a function of temperature, evaluated for non-interacting vortex rings using Eqs. (5.8) and (5.10). The calculated curve takes into account phonons and interacting rotons.

Fig. 22. Specific entropy $S$ for non-interacting vortex rings as a function of temperature evaluated using Eq. (5.14) with $W_0$ from Eq. (5.6) when $\mu$ = 0. The calculated curve takes into account phonons and interacting rotons.

Fig. 23. Geometry for evaluating the vortex ring quadratic interaction integral $J$ in Eq. (6.44) that results from application of Gauss' theorem in Eq. (6.41). The limits on variables for integration over surfaces $\mathcal{S}_1$ and $\mathcal{S}_2$ can be deduced for the two cases where the correlation

length $R_0$ is less than or greater than the spacing $u$ between vortex rings at positions 1 and 2. There is cylindrical symmetry about the $z$-axis.

Fig. 24. Contributions of vortex ring interactions to thermodynamic functions near $T_c$ showing shape factors only for a simplified model (crossover model). These factors are for: (a) Helmholtz potential, evaluated using Eq. (6.126) (see text for explanation); (b) entropy based on Eq. (6.127); and (c) specific heat at constant volume based on Eq. (6.128). The calculated curves take into account phonons and interacting rotons.

Fig. 25. Schematic representation of He I (shaded) and He II (unshaded) configurations near the critical range of temperatures. (a) $T < T_\lambda$: He I "bubbles" in a matrix of He II. (b) $T_\lambda < T \leq T_c$: He II "drops" in a matrix of He I. (c) $T_\lambda < T \leq T_c$: Magnified view of a small region of the space outlined by the dashed square in (b). This view shows how the He II "drop" is formed in the "opening" in the He I matrix. The He I matrix is formed by overlapping mantles around the individual vortex rings.

Fig. 26. (a) Pinned line vortices at temperatures far below $T_c$. The lines shown have opposite circulation and are bowed inward to establish mechanical equilibrium. (b) Pinned line vortices at temperatures slightly below $T_c$. Near $T_c$ the configuration with less bowed lines and two vortex rings has about the same free energy as two more bowed lines having greater length but no rings as in configuration (a). Both configurations are in mechanical equilibrium. The total length of vortex line is about the same for the two cases in the real world (but not in the schematic representations in (a) and (b)).

Fig. 27. Phase diagram of $^4$He showing the initial path 1 for cooling and then solidifying the helium to eliminate pinned line vortices, and the subsequent path 2 for observing specific heat in the λ transition for liquid $^4$He in complete thermodynamic equilibrium.

Fig. 28. Cross section through the $T_d^2$ structure proposed by Keesom and Taconis. Three subsequent layers of this model are shown. The helium atoms are pictured by circles having diameter 2.30 Å. Each atom has six closest neighbors at a distance of 3.16 Å, drawn according to scale. The upper plane of atoms are spaced apart by about 6.3 Å. The mean atomic spacing 3.58 Å for this model coincides with that of liquid $^4$He at SVP and $T$ near 0 K (after London, Ref. [139]).

**Figures**

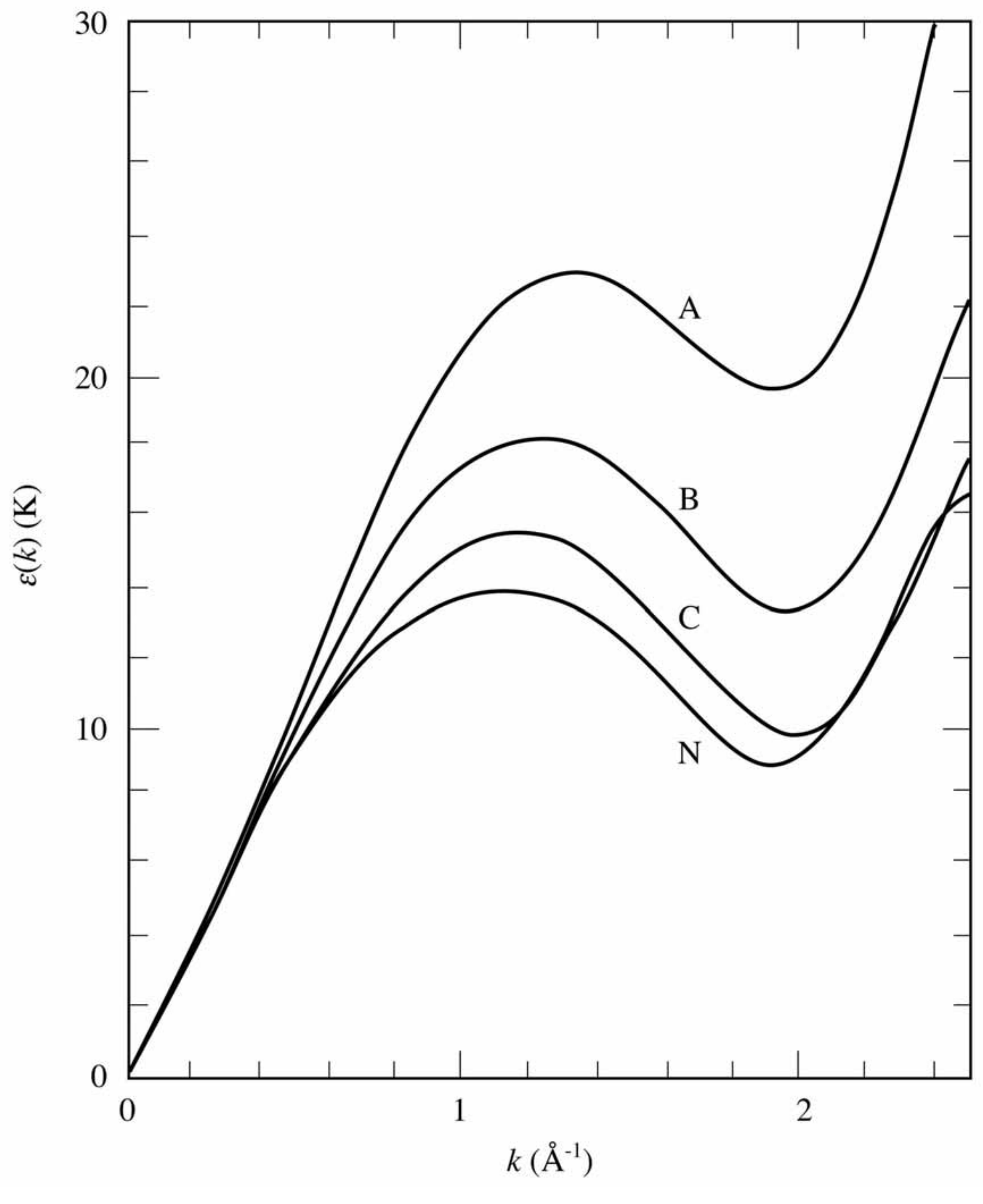

Fig. 1

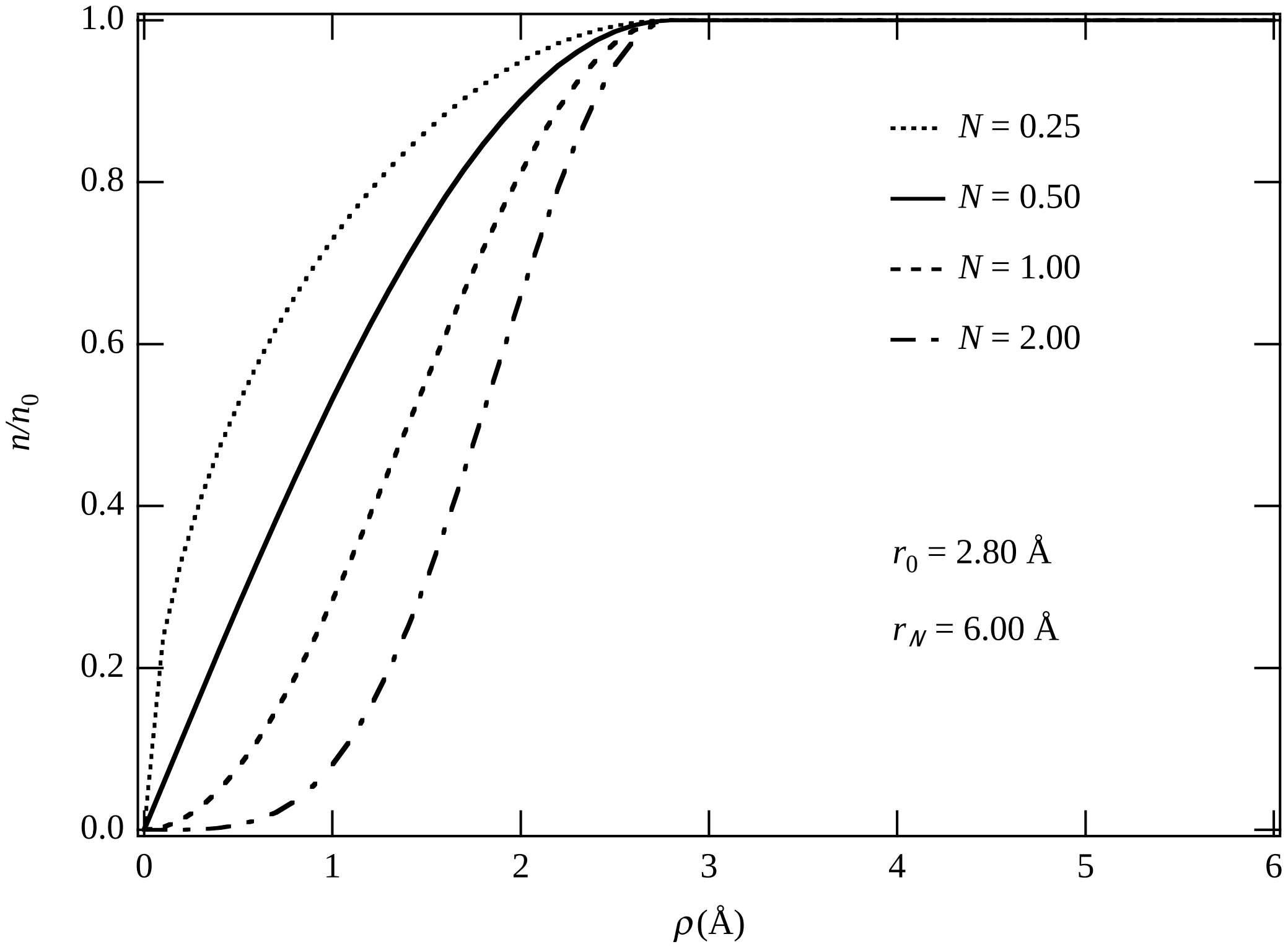

Fig. 2

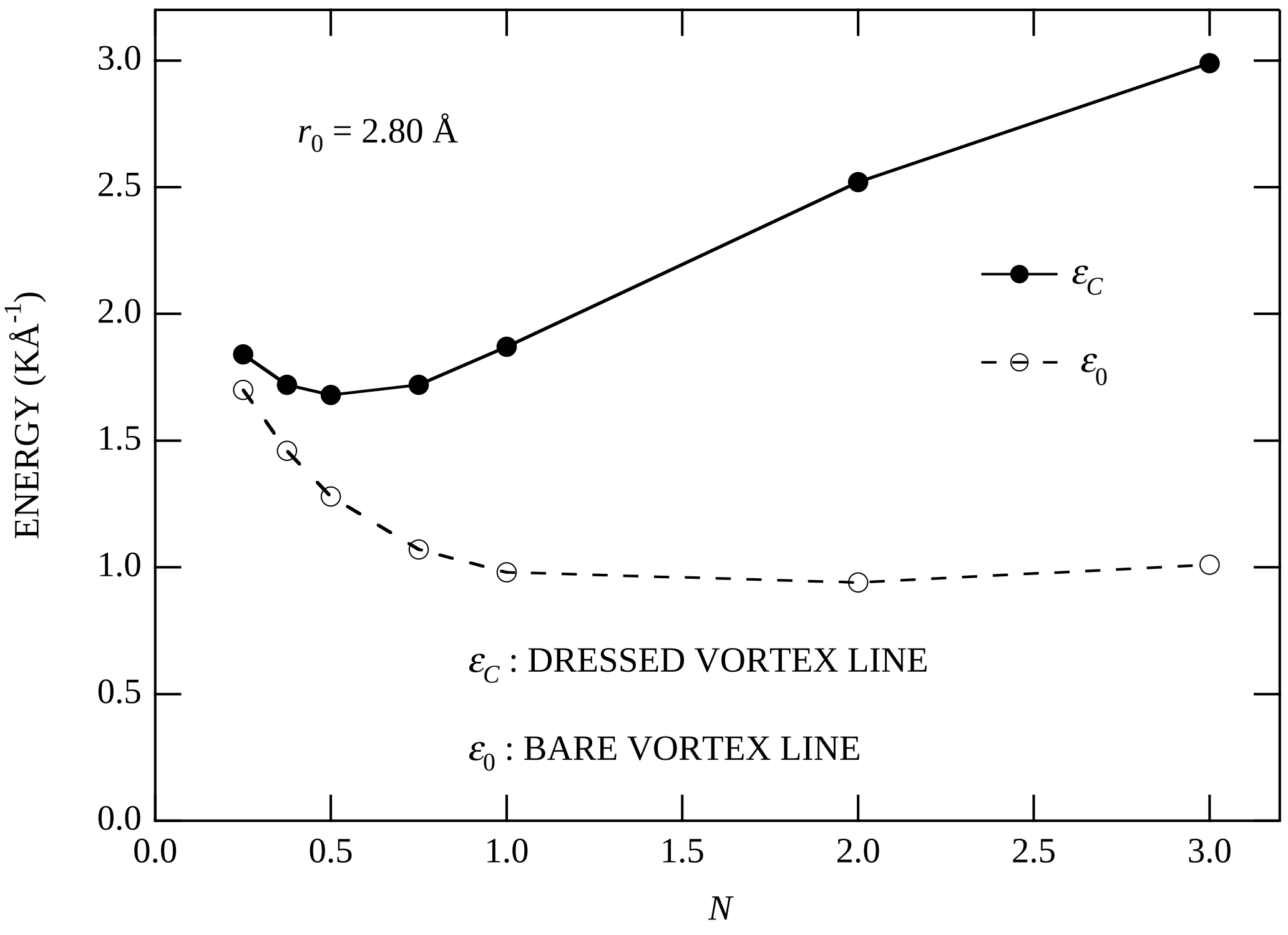

Fig. 3

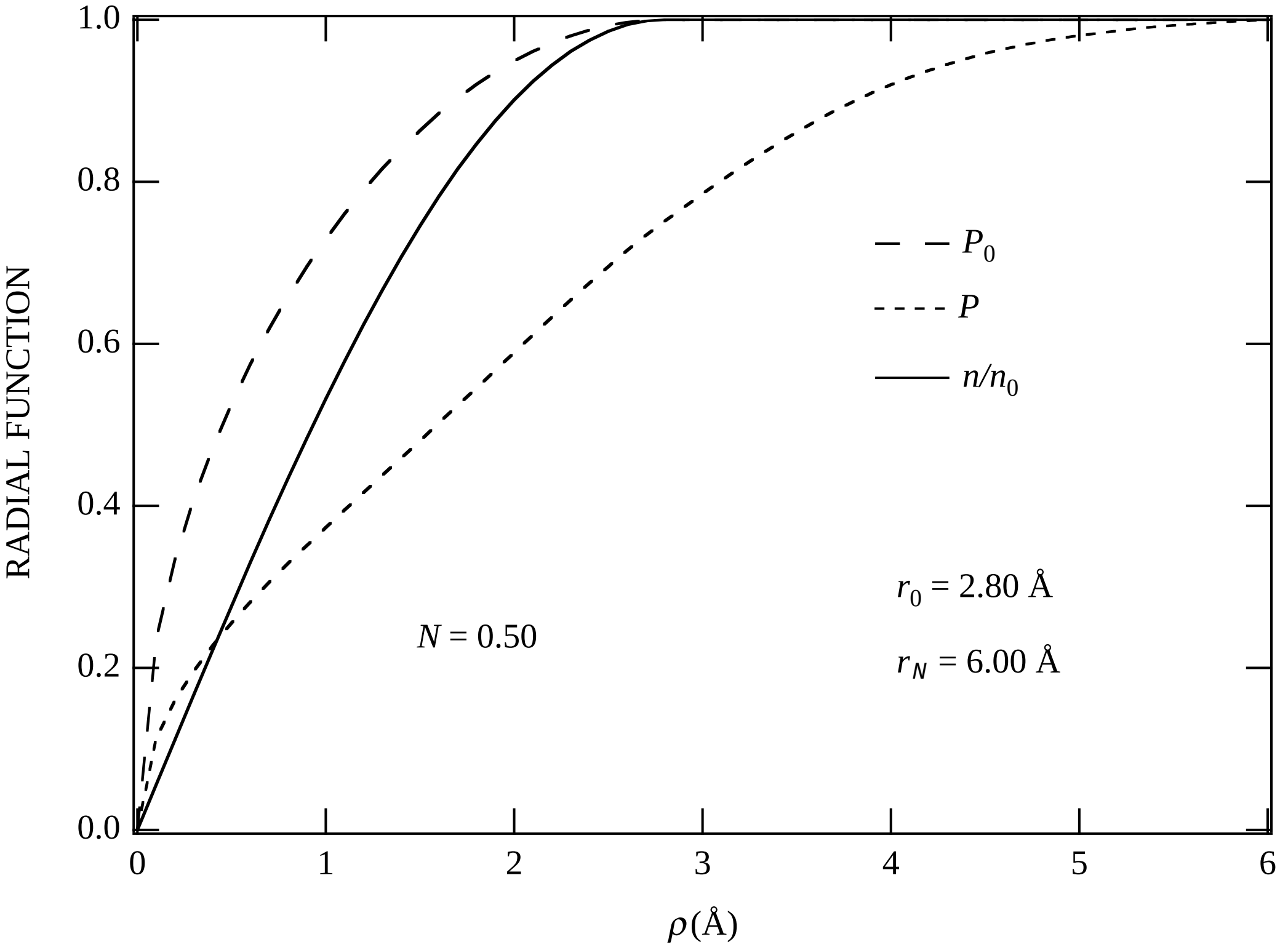

Fig. 4

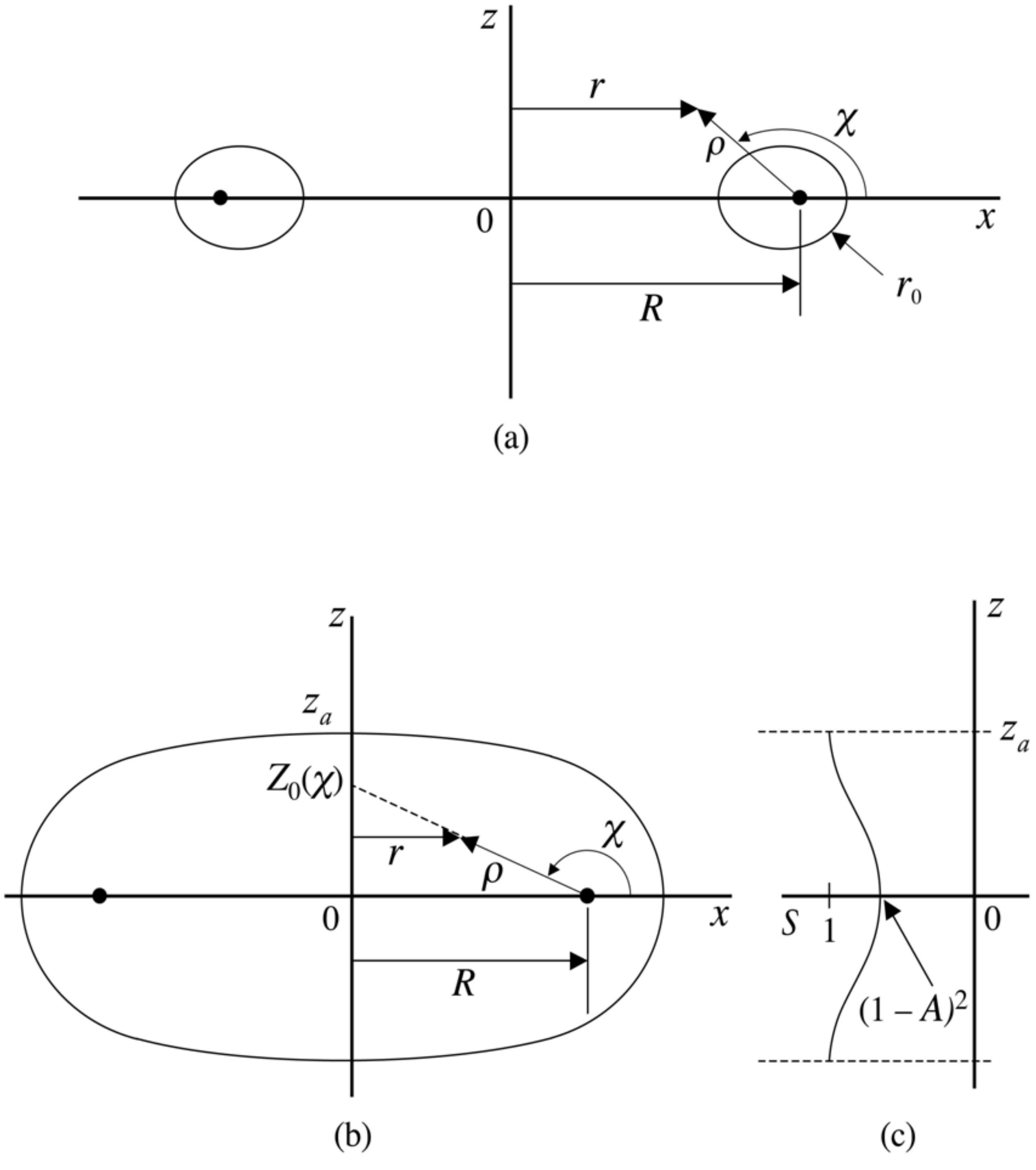

Fig. 5

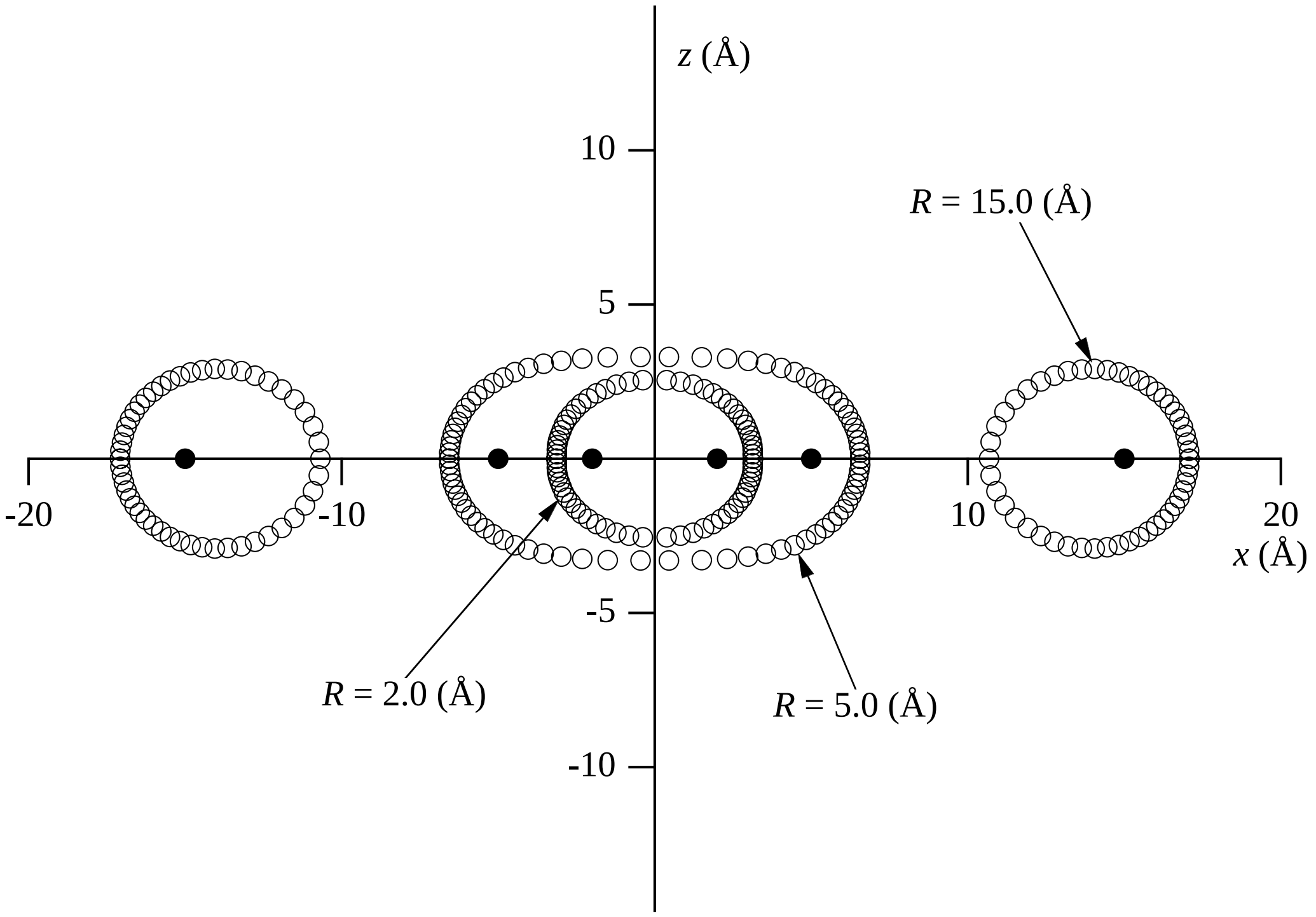

Fig. 6

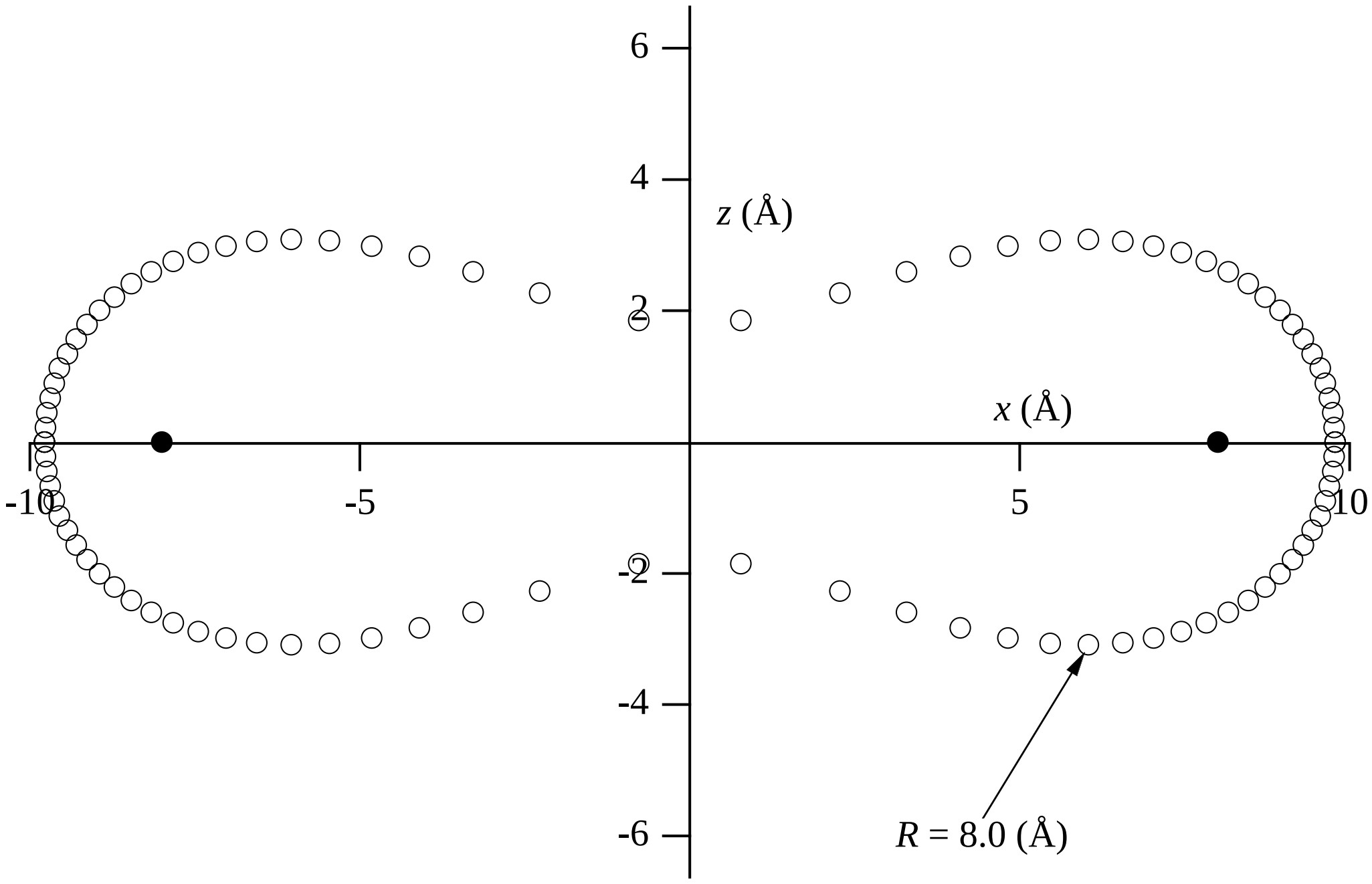

Fig. 7

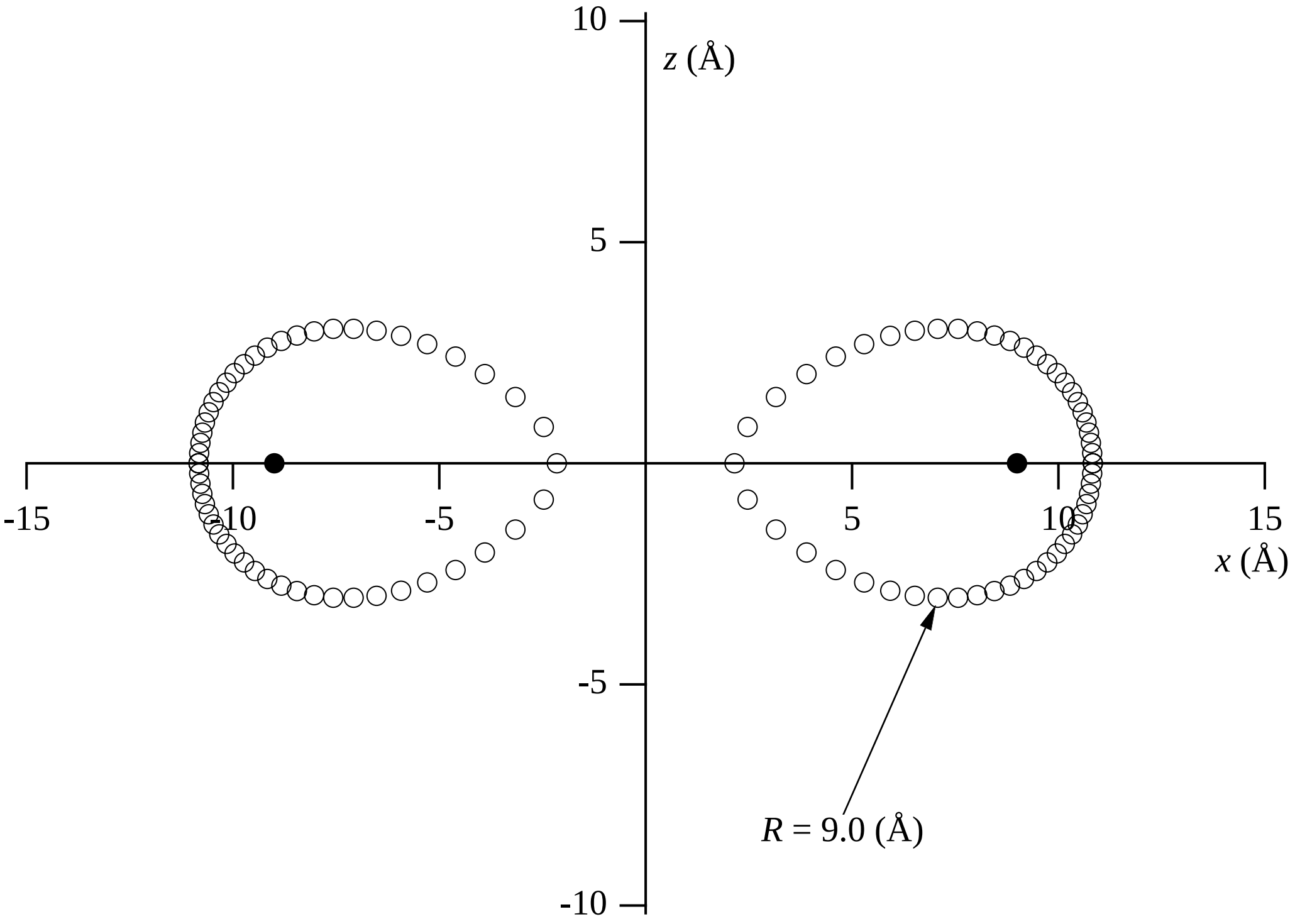

Fig. 8

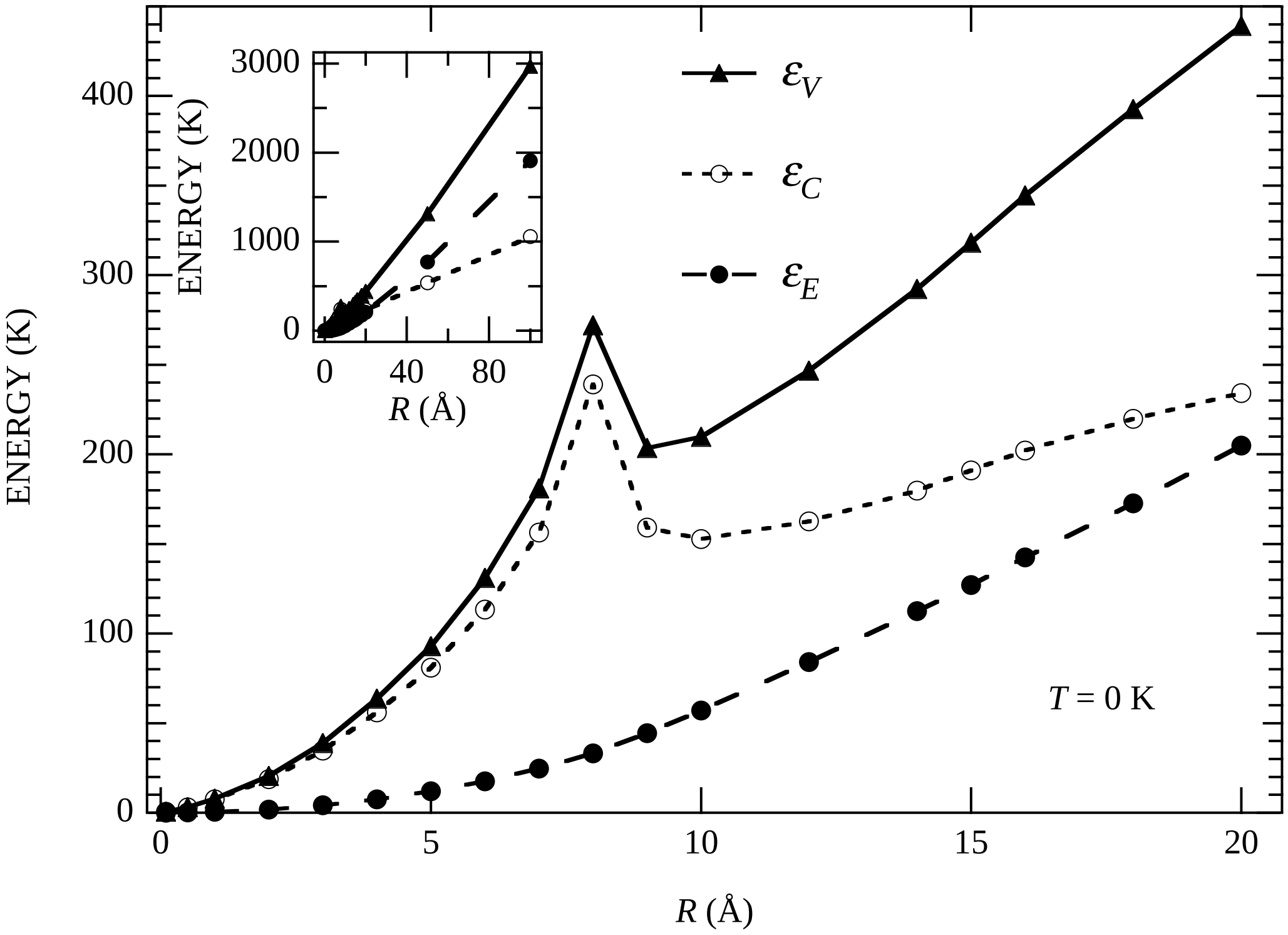

Fig. 9

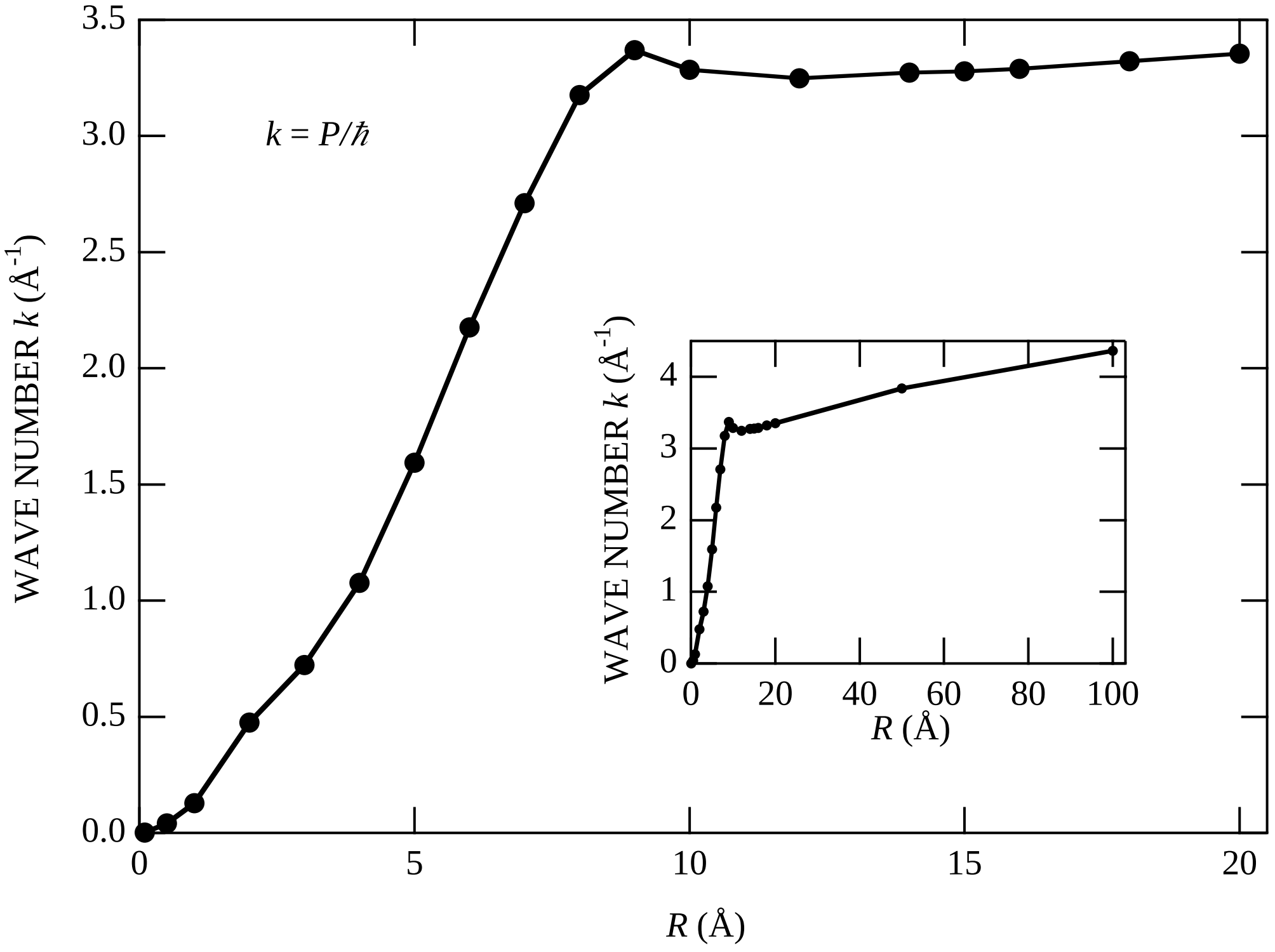

Fig. 10

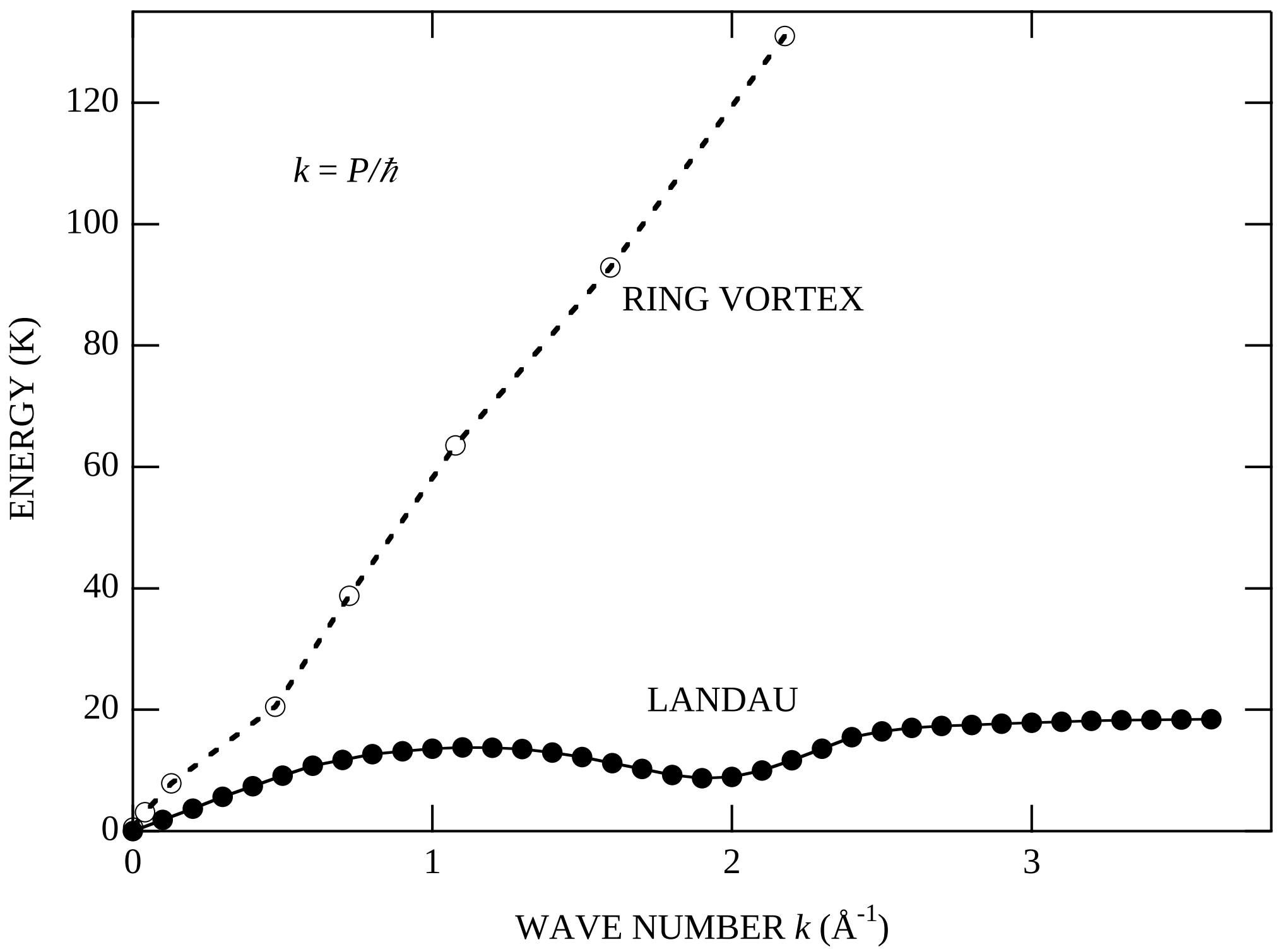

Fig. 11

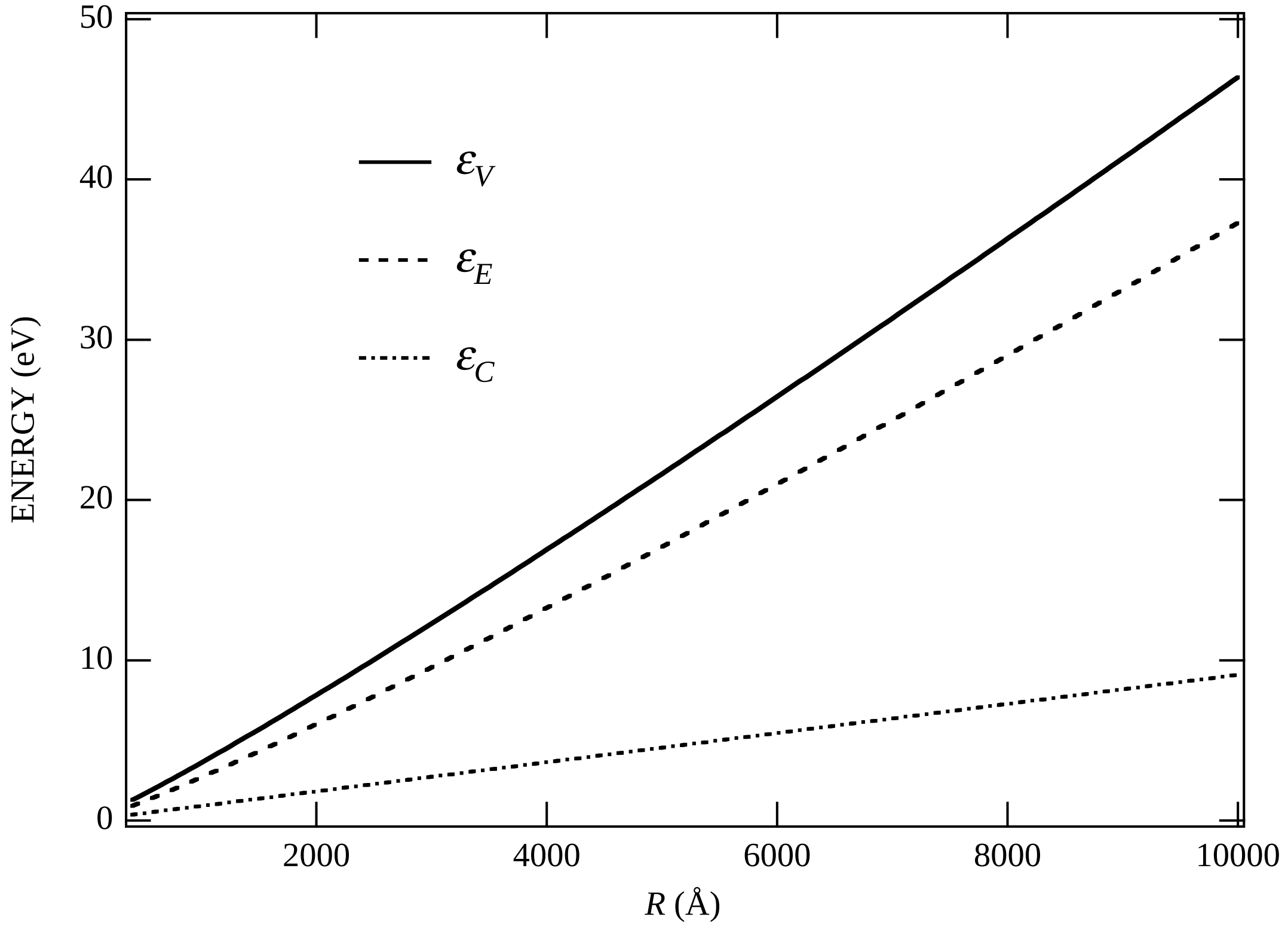

Fig. 12

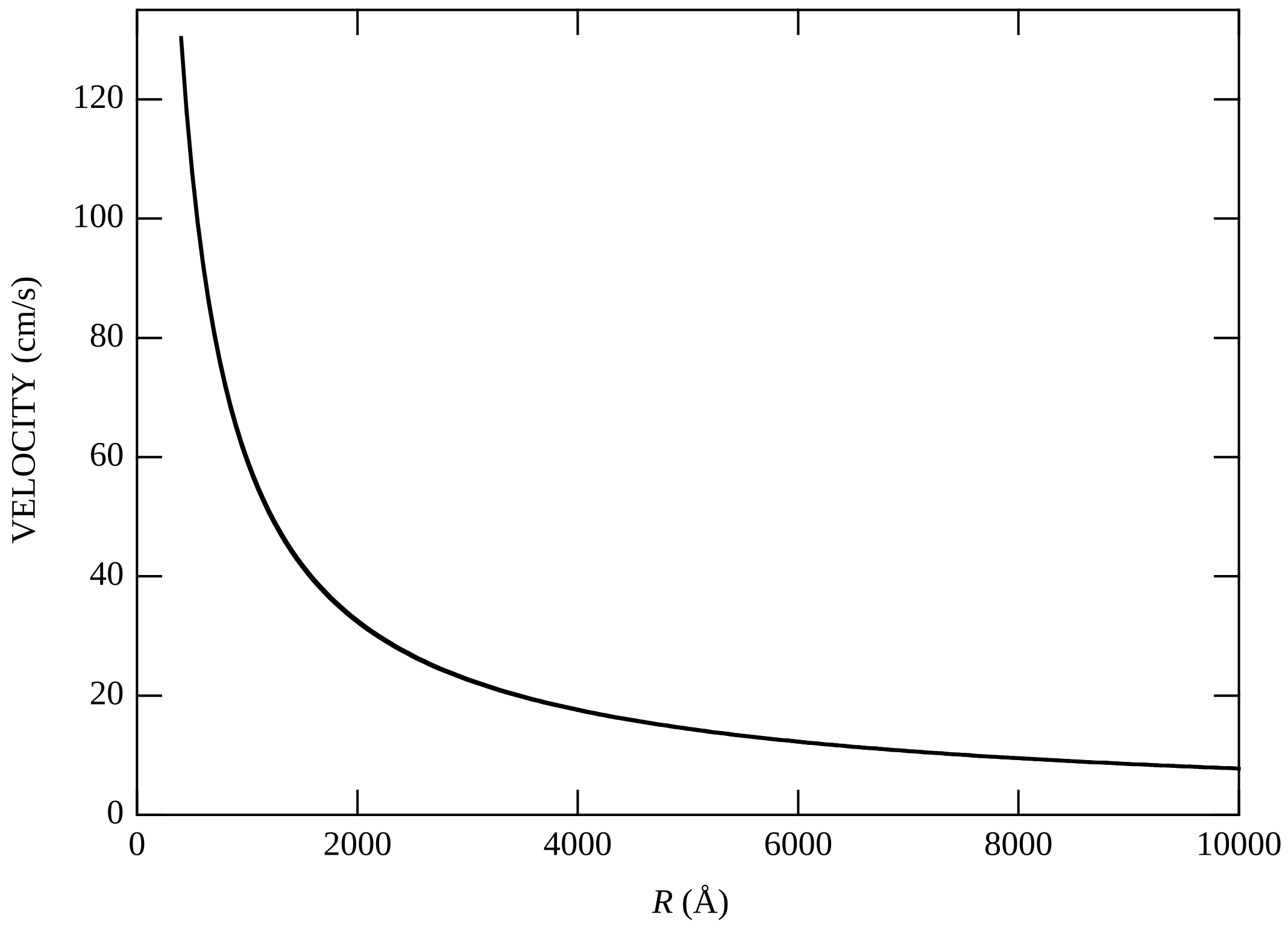

Fig. 13

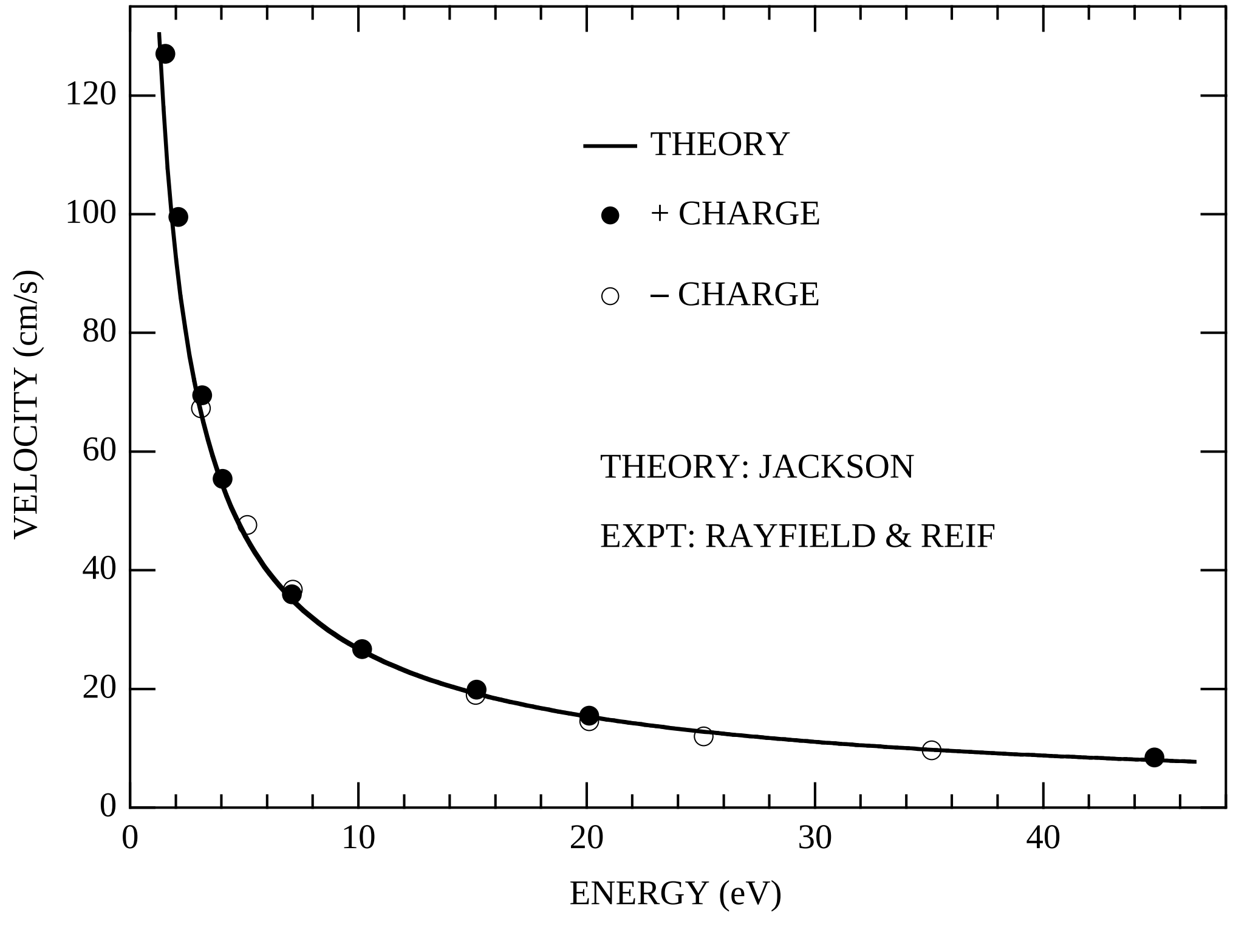

Fig. 14

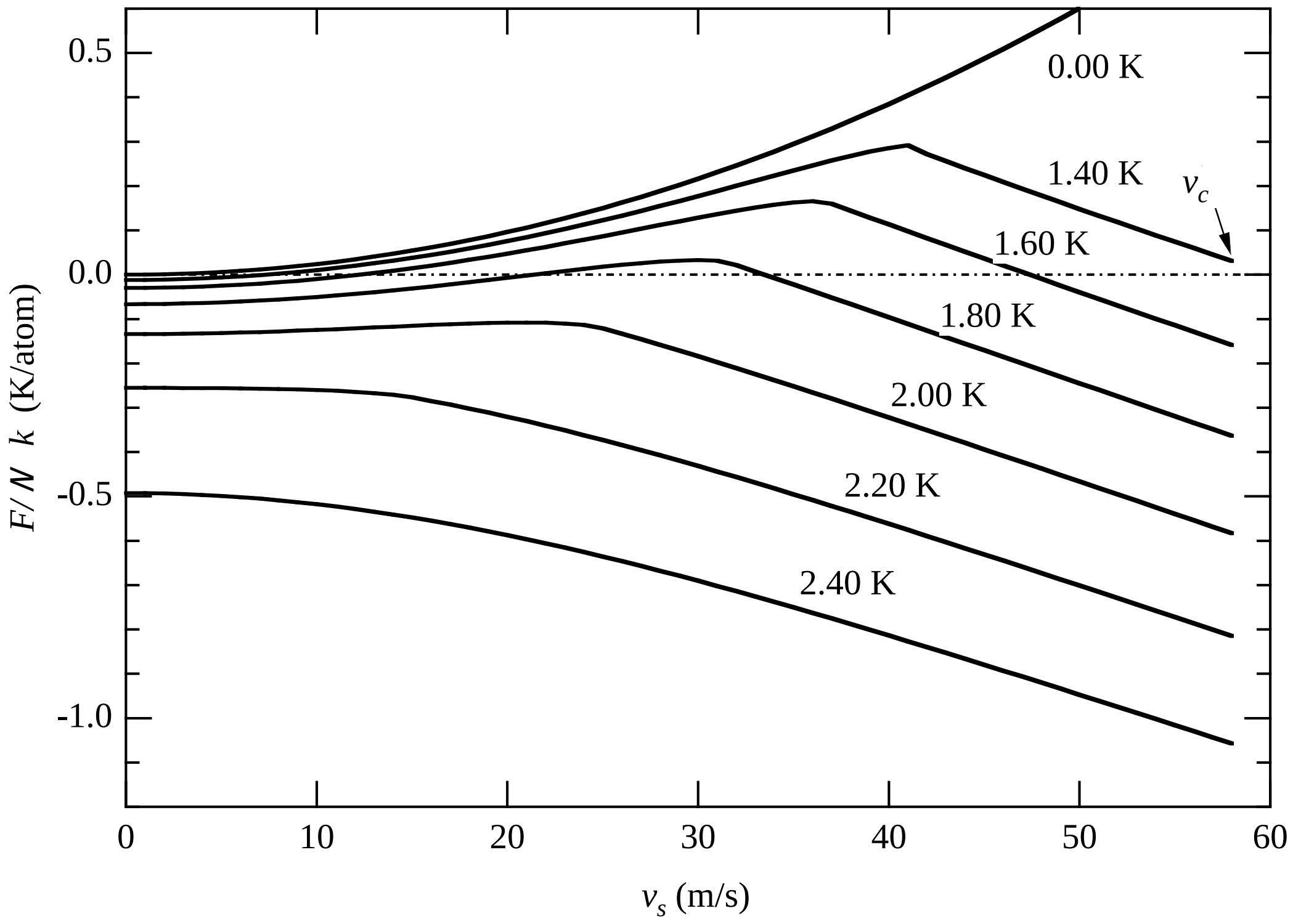

Fig. 15

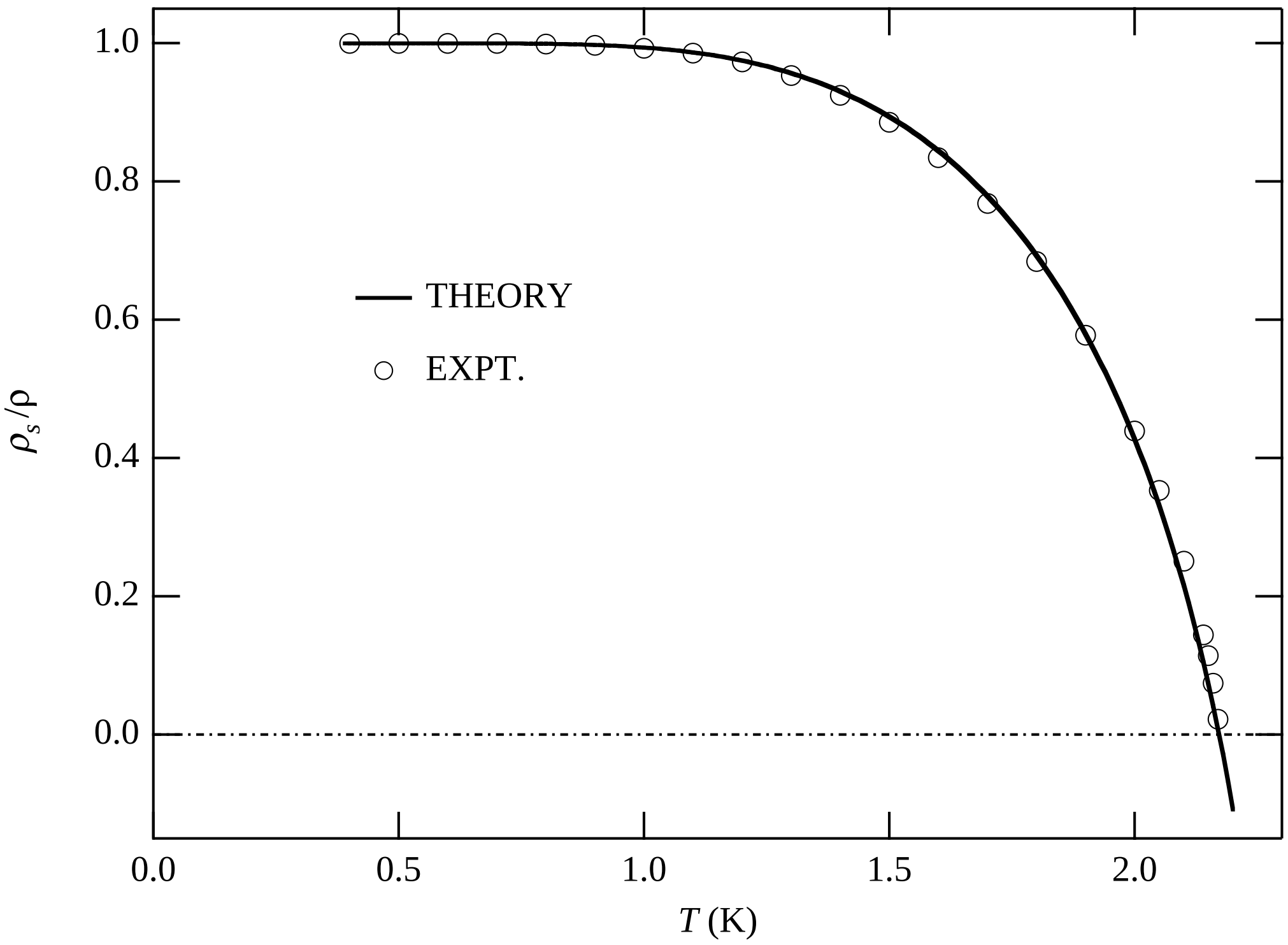

Fig. 16

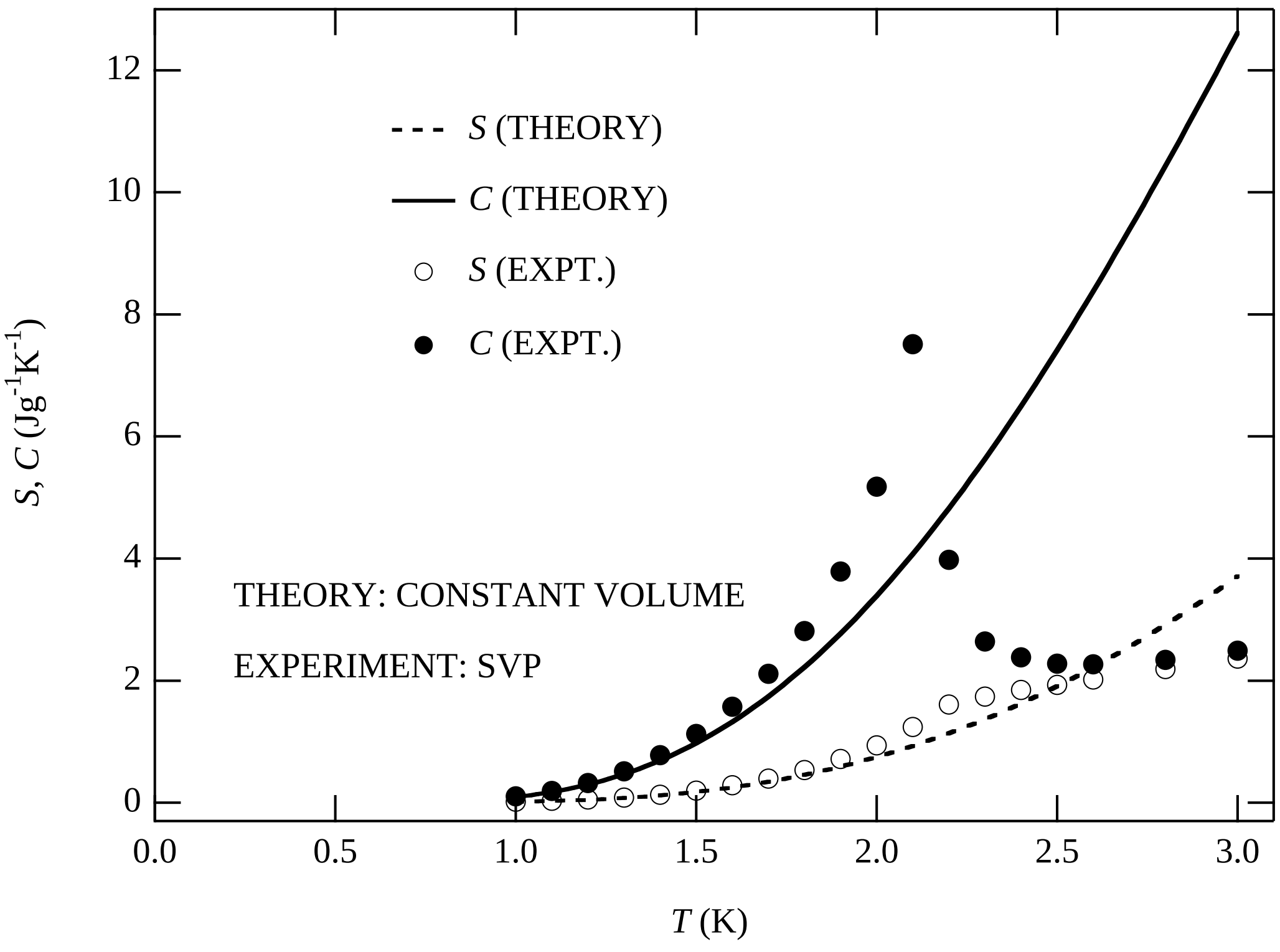

Fig. 17

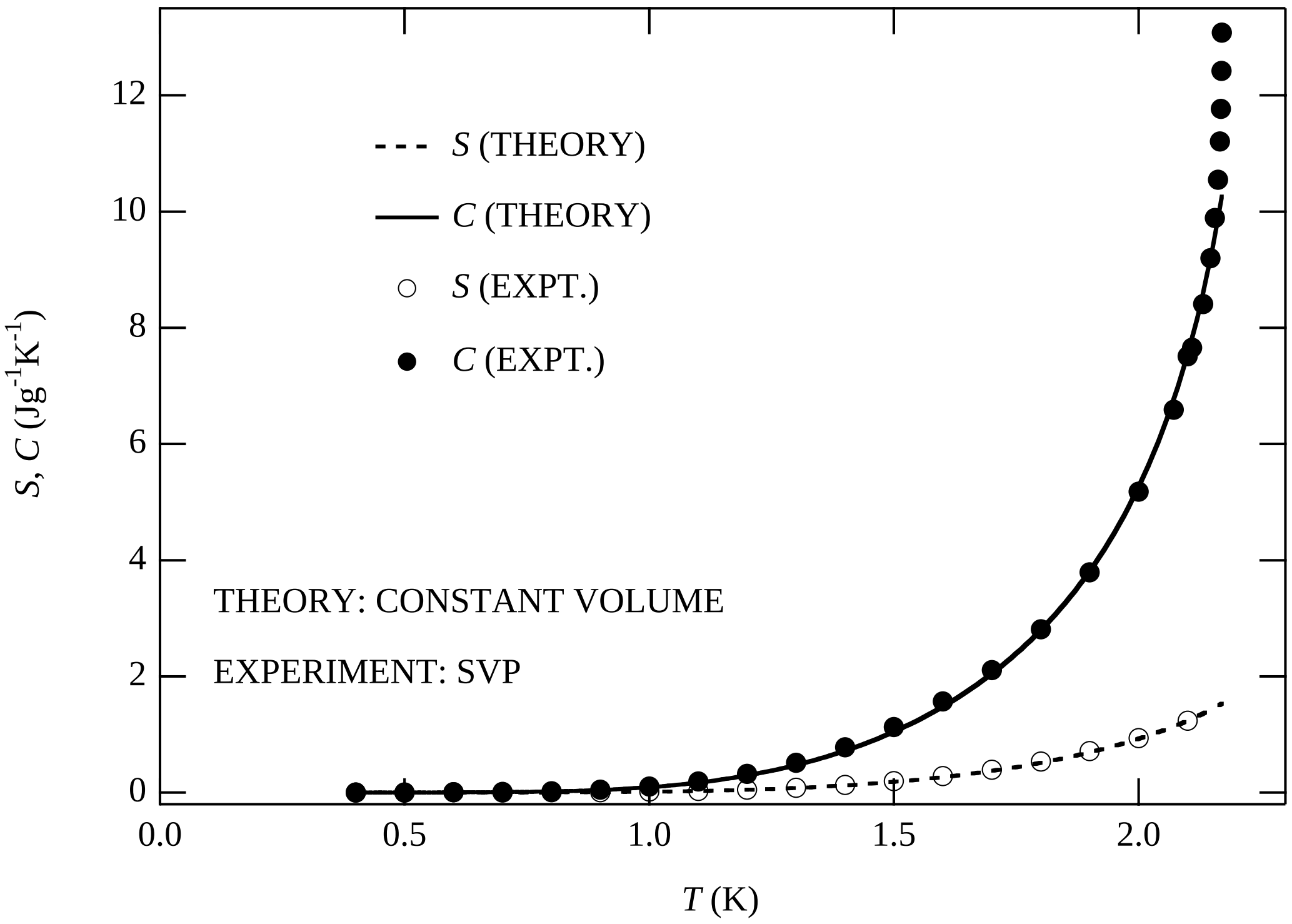

Fig. 18

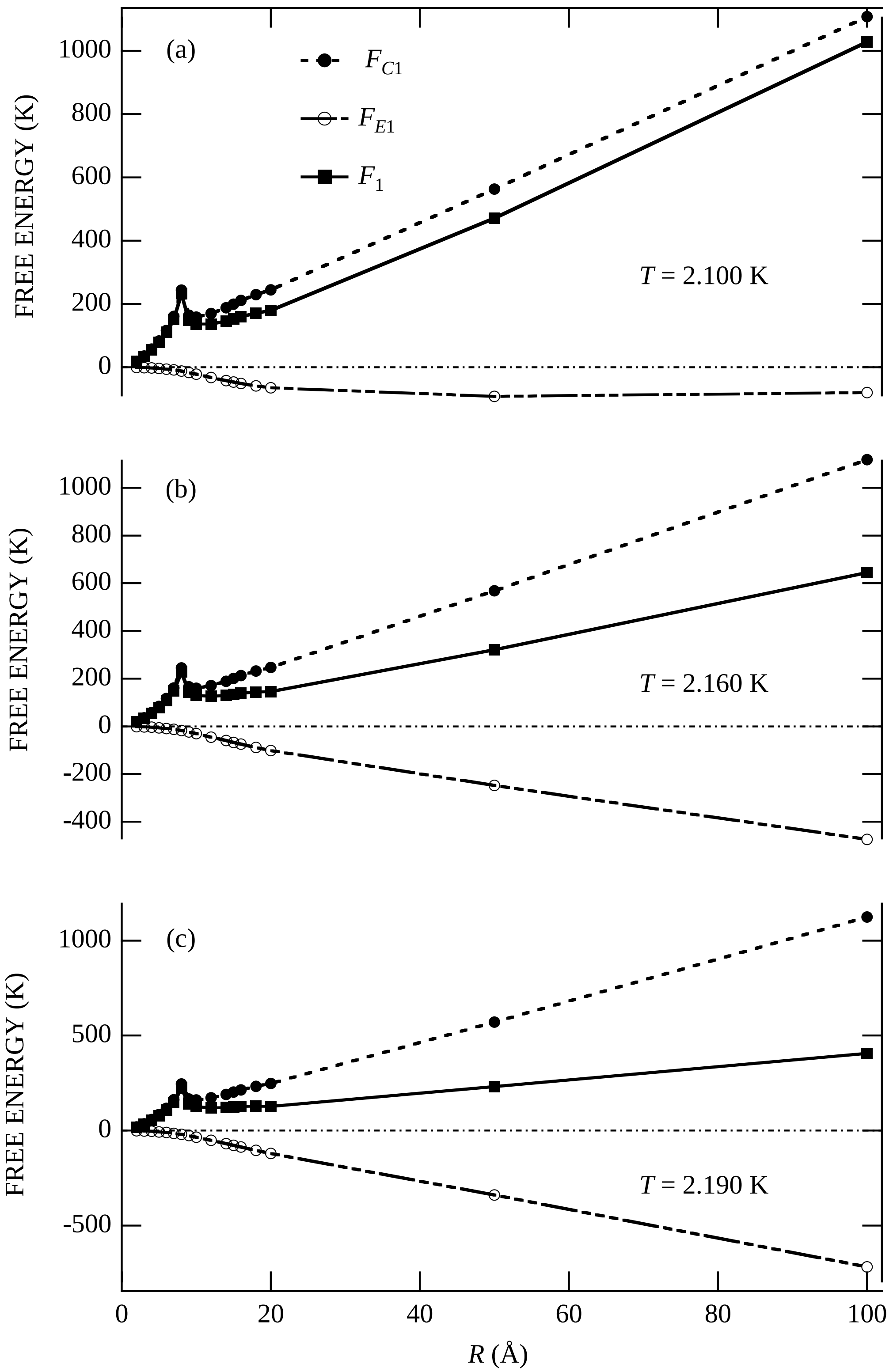

Fig. 19

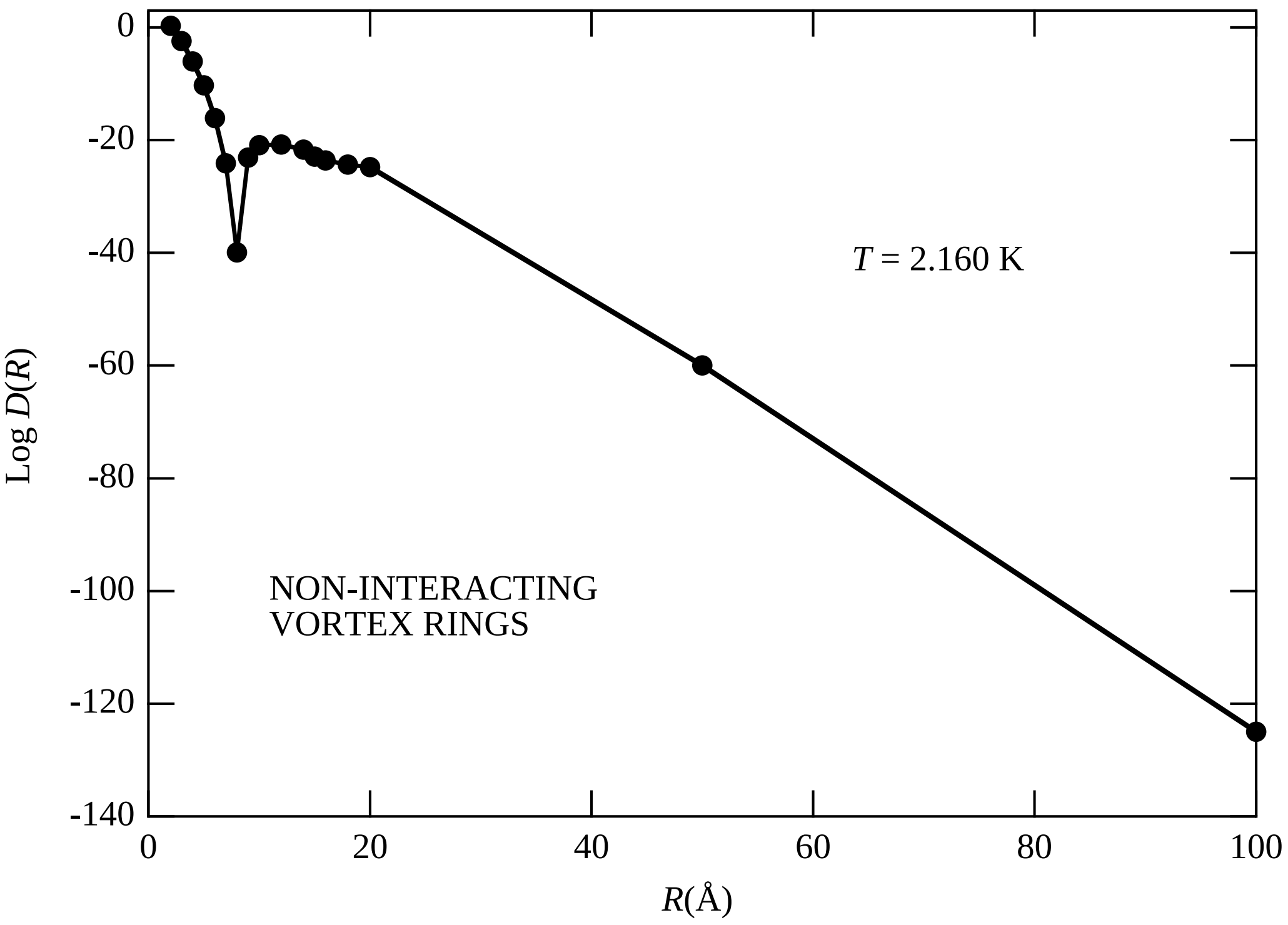

Fig. 20

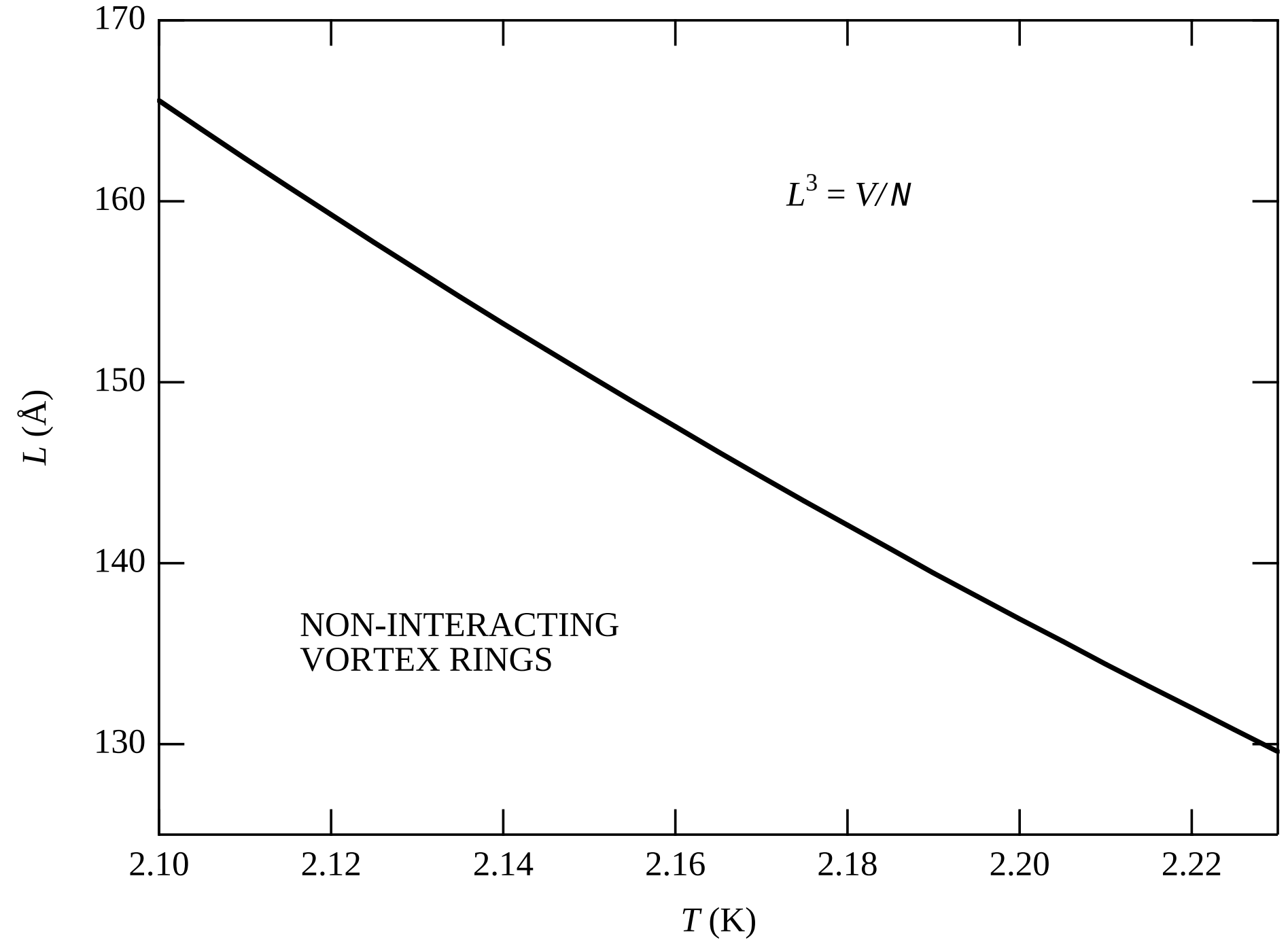

Fig. 21

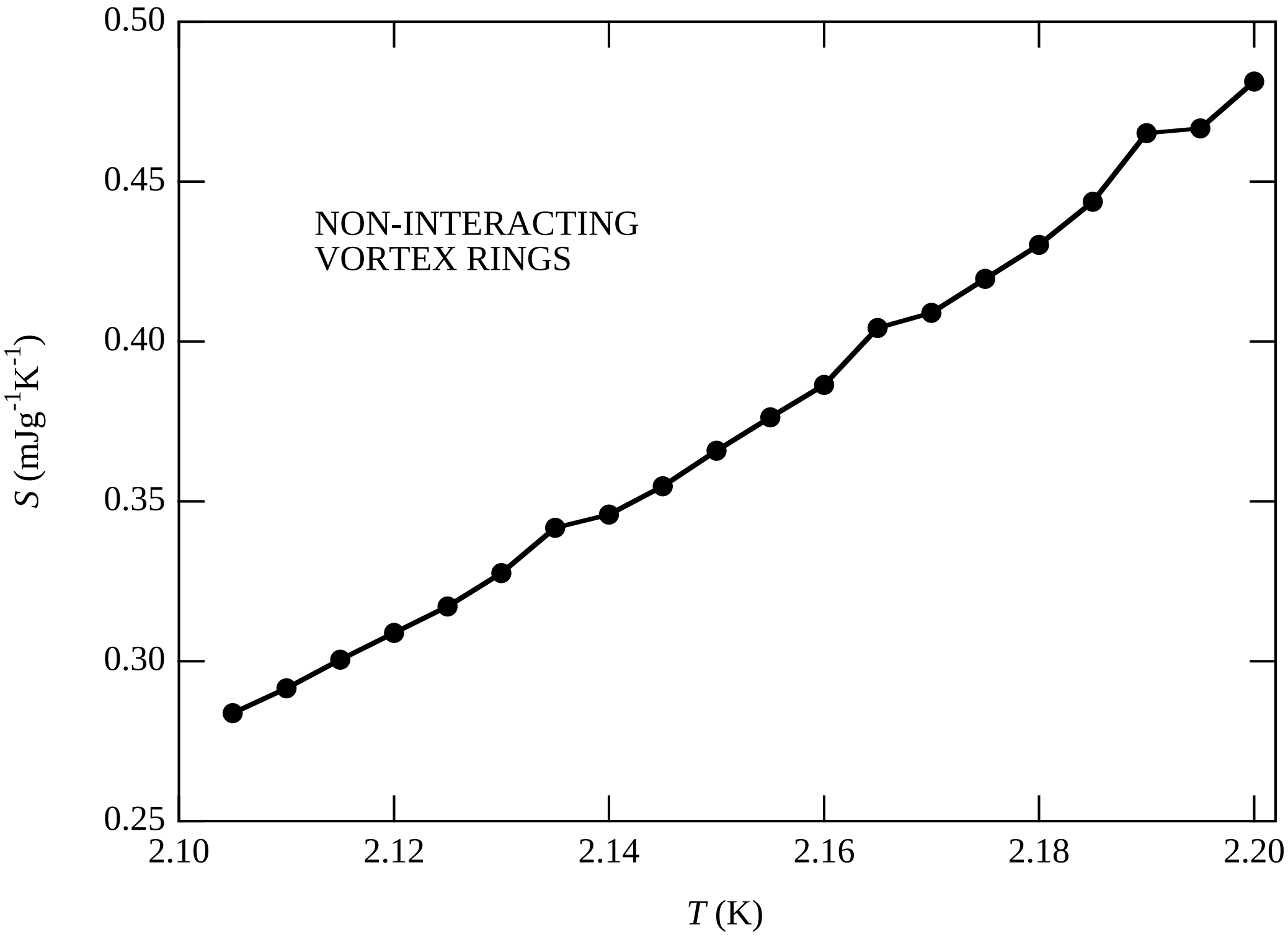

Fig. 22

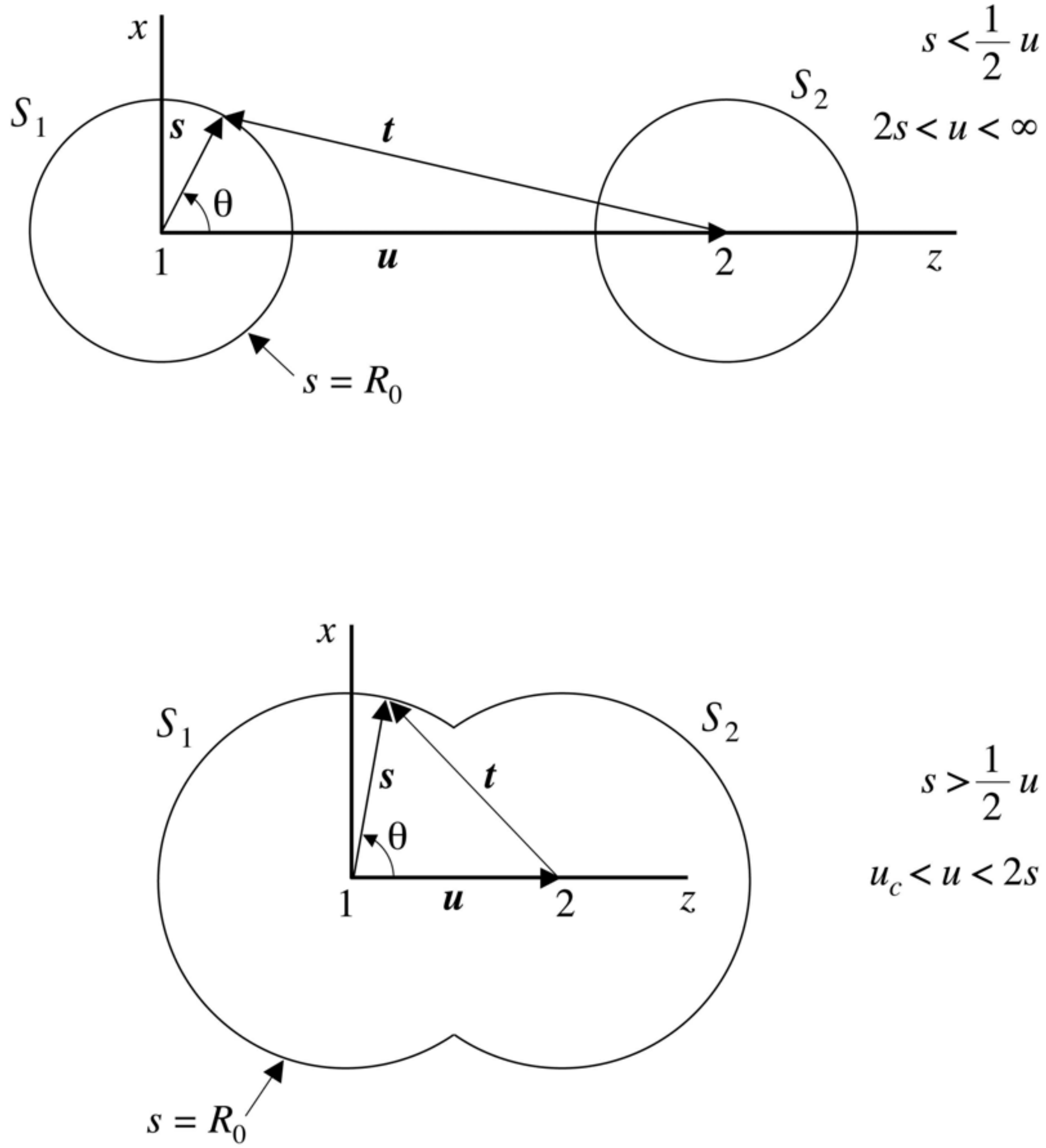

Fig. 23

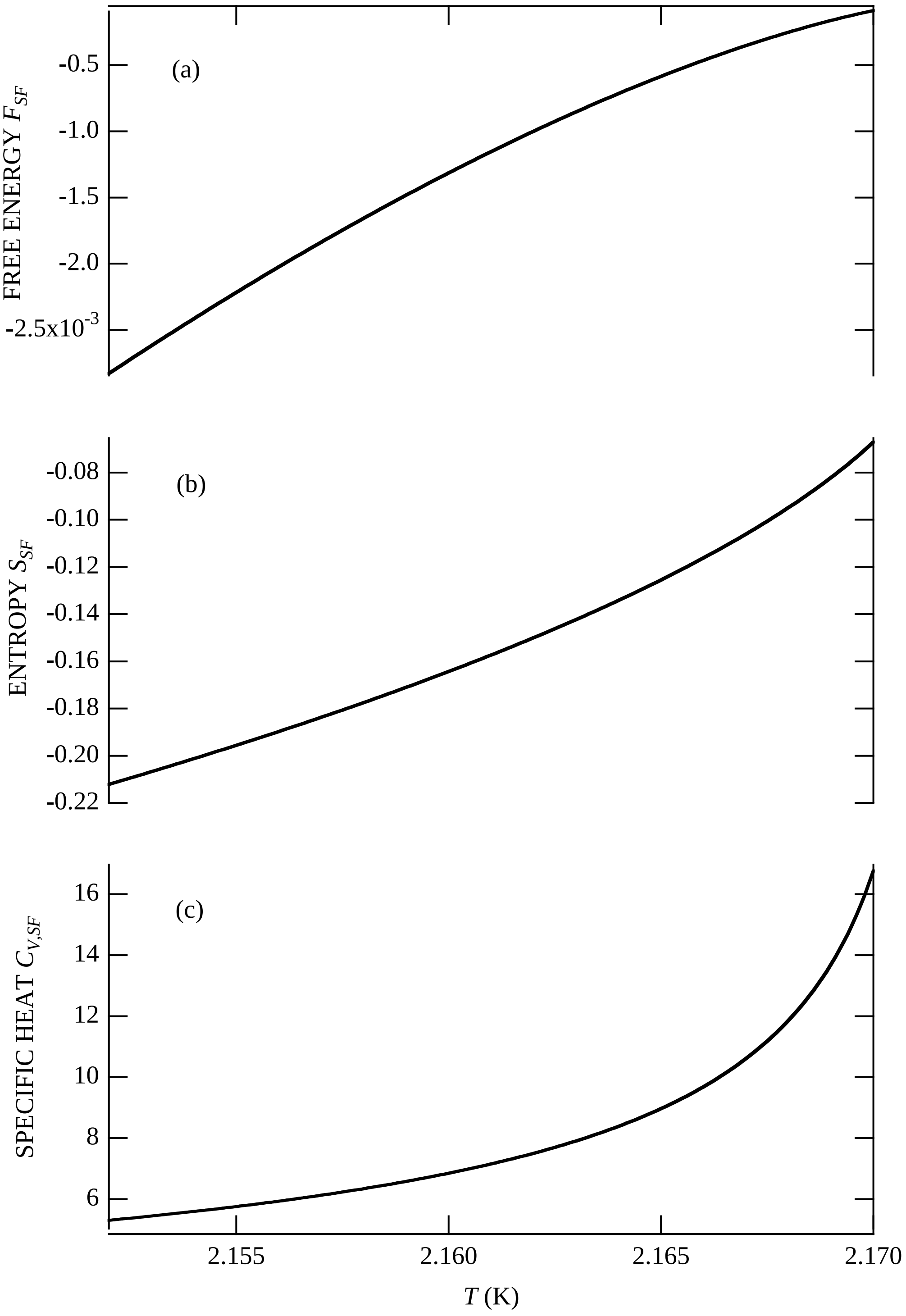

Fig. 24

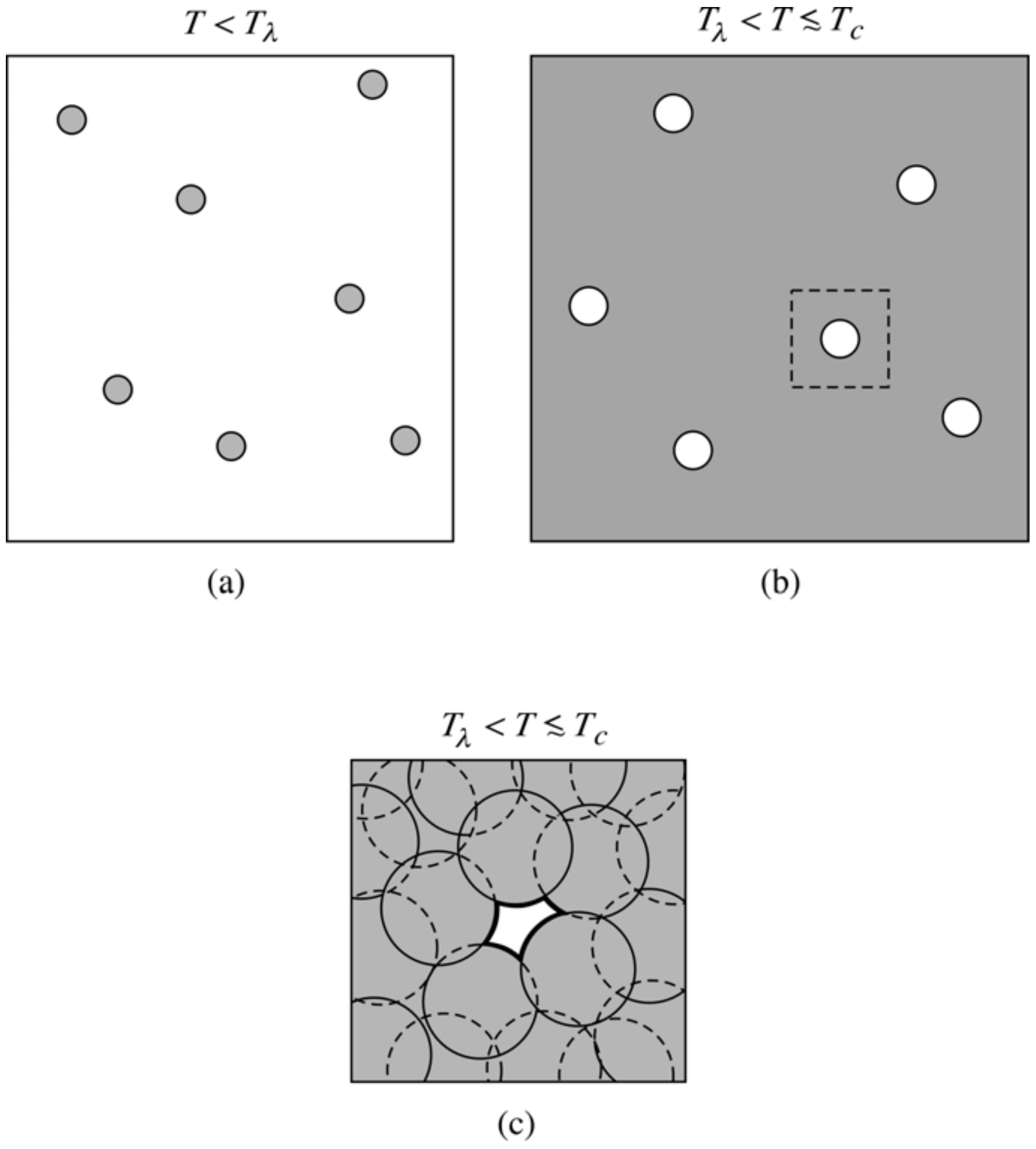

Fig. 25

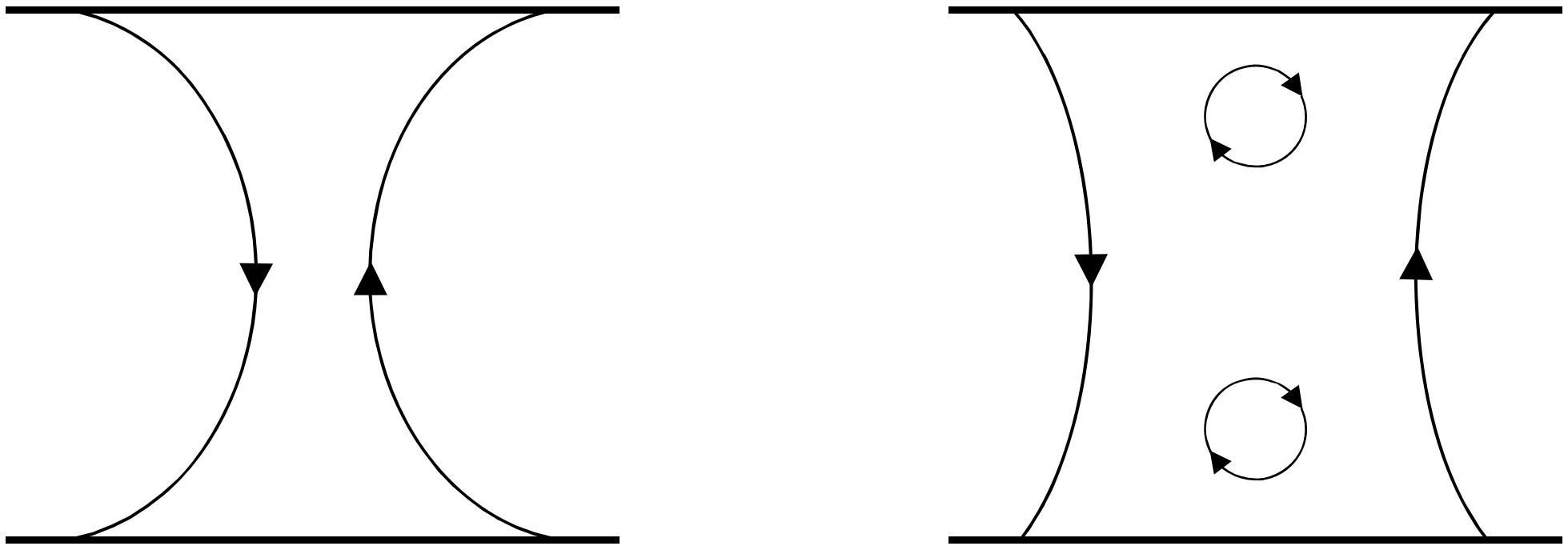

(a) (b)

Fig. 26

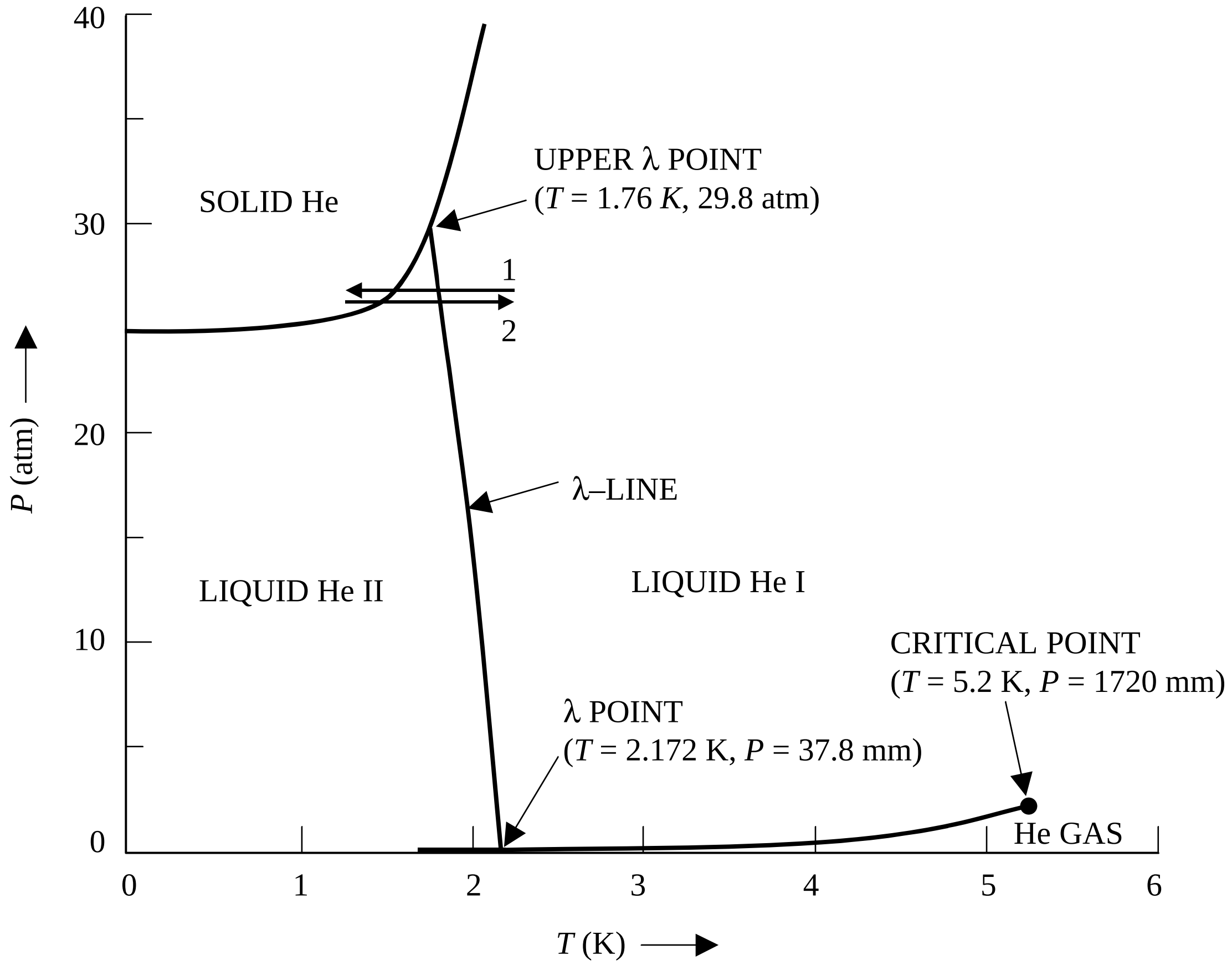
40
30
20
10
0
P (atm)
SOLID He
UPPER λ POINT
(T = 1.76 K, 29.8 atm)
1
2
λ–LINE
LIQUID He II
LIQUID He I
CRITICAL POINT
(T = 5.2 K, P = 1720 mm)
λ POINT
(T = 2.172 K, P = 37.8 mm)
He GAS
0
1
2
3
4
5
6
T (K)

Fig. 27

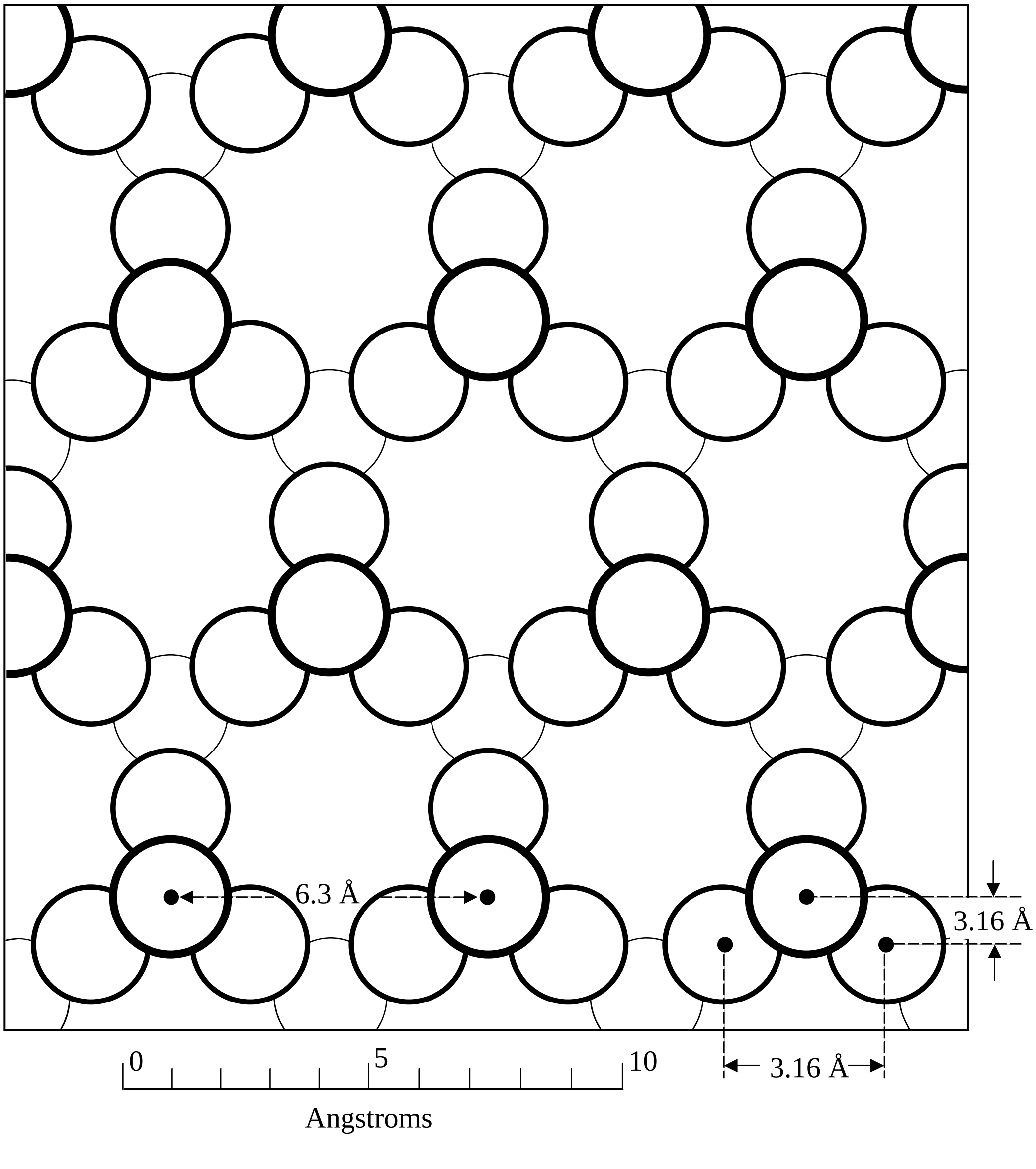

Fig. 28